\magnification=\magstep0       
%
%
\vsize=19.5truecm
\hsize=12.0truecm
\voffset=2.0truecm
\hoffset=2.0truecm
\parskip=0.0truecm
\parindent=1.0truecm
\baselineskip=12pt       
%
%
\font\huge=cmr10 scaled \magstep4
\def\ref#1{{\par\noindent \hangindent=3em \hangafter=1 #1\par}}
\def\arcm  	{$^{\prime}$} 
 
\def\arcs       {$^{\prime\prime}$} 
 
\def\bily       {${ {10^9}}$~yr}
\def\bv         {{\bf v}}
\def\br         {{\bf r}}
\def\cen 	{$\omega$\thinspace Cen}
\def\cent 	{$\omega$\thinspace Centauri}
\def\conc       {$c$ = log ($r_t/r_c$)}

\def\deg        {$^{\circ}$}
\def\elli       {$\varepsilon$}
\def\ellim      {$\langle \varepsilon \rangle$}
\def\etal       {et\thinspace al.}
\def\ergs       {erg\thinspace s$^{-1}$}
\def\feh        {[Fe/H]}
\def\gsim	{$\vcenter{\hbox{$>$}\offinterlineskip\hbox{$\sim$}}$}

\def\Ho         {$H_\circ$}
\def\kms        {km\thinspace s$^{-1}$}
\def\lsim	{$\vcenter{\hbox{$<$}\offinterlineskip\hbox{$\sim$}}$}
\def\lsun       {$L_{\odot}$}
\def\lx         {$L_X$}
\def\mbar       {\langle m\rangle}

\def\mdot       {$\dot M$}
\def\milm       {${ {10^6 M_{\odot}}}$}
\def\mily       {${ {10^6}}$~yr}
\def\msup       {$m_{sup}$}
\def\minf       {$m_{inf}$}
\def\ml         {$M/L$}
\def\mlv        {$M/L_V$}
\def\mlvso      {($M/L_V$)$_\odot$}
\def\ms         {m\thinspace s$^{-1}$}
\def\MJ         {$M_J$}
\def\msun       {$M_{\odot}$}
\def\mtot       {$M_{tot}$}
\def\mv         {$m_V$}
\def\Mv         {$M_V$}

\def\Mvint      {$M_V^{int}$}

\def\omegac     {$\Omega_c$}
\def\pdot	{$\dot P$}
\def\pe         {$p_e$}
\def\pmm        {$\pm$}
\def\ra         {$r_a$}
\def\rc         {$r_c$}
\def\rh         {$r_h$}
\def\rmax       {$r_{max}$}
\def\rt         {$r_t$}

\def\roo	{$\rho_{\circ}$}
\def\rhoo       {$\rho_{\circ}$}
\def\rhoh       {$\rho_{h}$}
\def\rhot       {$\rho_{t}$}
\def\sig        {$\sigma$}

\def\sigccf     {$\sigma_{\rm CCF}$}

\def\siglos     {$\sigma_{los}$}

\def\sigp       {$\sigma_p$}
\def\sigpo      {$\sigma_p$(0)}
\def\sigref     {$\sigma_{\rm ref}$}

\def\Sn         {$S_N$}
\def\Tc         {$T_{c}$}
\def\tcr        {$t_{cr}$}
\def\tev        {$t_{ev}$}
\def\trc        {$t_{rc}$}
\def\trh        {$t_{rh}$}
\def\Vr         {$V_{r}$}

\def\Voso       {$V_{\circ}/\sigma_{\circ}$}
\def\Vrot       {$V_{rot}$}
\def\Vrotmax    {$V_{rot}^{max}$}
\def\today      {\number\day\enspace
                \ifcase\month\or January\or February\or March\or 
		April\or May\or June\or July\or August\or September\or
		October\or November\or December\fi , \enspace\number\year}
\def\tuc 	{47\thinspace Tuc}
\def\tuca 	{47\thinspace Tucanae}

\def\x          {$\times$}
%
%
\def\endtable{\endgroup}
\def\tableheight{\vrule width 0pt height 8.5pt depth 3.5pt}
{\catcode`|=\active \catcode`&=\active 
    \gdef\tabledelim{\catcode`|=\active \let|=\vbar
                     \catcode`&=\active \let&=\nobar} }
\def\table{\begingroup
    \def\twidth{\hsize}
    \def\tablewidth##1{\def\twidth{##1}}
    \def\defaultheight{\vrule width 0pt height 8.5pt depth 3.5pt}
    \def\heightdepth##1{\dimen0=##1
        \ifdim\dimen0>5pt 
            \divide\dimen0 by 2 \advance\dimen0 by 2.5pt
            \dimen1=\dimen0 \advance\dimen1 by -5pt
            \vrule width 0pt height \the\dimen0  depth \the\dimen1
        \else  \divide\dimen0 by 2
            \vrule width 0pt height \the\dimen0  depth \the\dimen0 \fi}
    \def\spacing##1{\def\defaultheight{\heightdepth{##1}}}
    \def\nextheight##1{\noalign{\gdef\tableheight{\heightdepth{##1}}}}
    \def\end{\cr\noalign{\gdef\tableheight{\defaultheight}}}
    \def\zerowidth##1{\omit\hidewidth ##1 \hidewidth}    
    \def\hline{\noalign{\hrule}}
    \def\hdoubleline{\hline \skip{2pt} \hline}
    \def\skip##1{\noalign{\vskip##1}}
    \def\bskip##1{\noalign{\hbox to \twidth{\vrule height##1 depth 0pt \hfil
        \vrule height##1 depth 0pt}}}
    \def\header##1{\noalign{\hbox to \twidth{\hfil ##1 \unskip\hfil}}}
    \def\bheader##1{\noalign{\hbox to \twidth{\vrule\hfil ##1 
        \unskip\hfil\vrule}}}
    \def\spanloop{\span\omit \advance\mscount by -1}
    \def\extend##1##2{\omit
        \mscount=##1 \multiply\mscount by 2 \advance\mscount by -1
        \loop\ifnum\mscount>1 \spanloop\repeat \ \hfil ##2 \unskip\hfil}
    \def\vbar{&\vrule&}
    \def\nobar{&&}
    \def\hdash##1{ \noalign{ \relax \gdef\tableheight{\heightdepth{0pt}}
        \toks0={} \count0=1 \count1=0 \putout##1\end 
        \toks0=\expandafter{\the\toks0 &\end} \xdef\piggy{\the\toks0} }
        \piggy}
    \let\e=\expandafter
    \def\putspace{\ifnum\count0>1 \advance\count0 by -1
        \toks0=\e\e\e{\the\e\toks0\e&\e\multispan\e{\the\count0}\hfill} 
        \fi \count0=0 }
    \def\putrule{\ifnum\count1>0 \advance\count1 by 1
        \toks0=\e\e\e{\the\e\toks0\e&\e\multispan\e{\the\count1}\leaders\hrule\hfill}
        \fi \count1=0 }
    \def\putout##1{\ifx##1\end \putspace \putrule \let\next=\relax 
        \else \let\next=\putout
            \ifx##1- \advance\count1 by 2 \putspace
            \else    \advance\count0 by 2 \putrule \fi \fi \next}   }
\def\tablespec#1{
    \def\vdimens{\noexpand\tableheight}
    \def\tabby{\tabskip=0pt plus100pt minus100pt}
    \def\r{&################\tabby&\hfil################\unskip}
    \def\c{&################\tabby&\hfil################\unskip\hfil}
    \def\l{&################\tabby&################\unskip\hfil}
    \edef\templ{\noexpand\vdimens ########\unskip  #1 
         \unskip&########\tabskip=0pt&########\cr}
    \tabledelim
    \edef\body##1{ \vbox{
        \tabskip=0pt \offinterlineskip
        \halign to \twidth {\templ ##1}}} }
%
%
\hyphenation{
	af-fect ASP astro-phy-sics an-iso-tro-py
	clus-ter  con-di-tion  cor-re-la-tion
	clus-ters con-di-tions cor-re-la-tions
	dif-fe-rent Djor-govski dy-na-mi-cal
	Eq exact 
	form func-tion func-tions
	ga-la-xy ga-la-xies glo-bu-lar
	km
	LMXBs 
	Mar-chant Mer-ritt mem-bers mo-dels 
	ob-ser-va-tion 
	ob-ser-va-tions
	obtains 
	pa-pers pre-sent	
	Reijns re-fe-ren-ce re-fe-ren-ces
	se-cond
	si-mu-la-tions sta-tionary stars strag-glers stu-dies smooth 
	the-re theo-re-ti-cal
	va-ria-tion va-ria-tions
	}
%
\phantom{start}
\vskip 3truecm
\parindent=0.0truecm
{\bf \huge {Internal dynamics of globular clusters}}
	\footnote{}{\baselineskip=10pt
	{\sevenrm {\sl Correspondence to}}: G. Meylan \hfill Version:~ \today
        \smallskip}

\vskip 30pt
\parindent=1.0truecm
\noindent
{\bf G. Meylan$^1$ and D.C. Heggie$^2$}
\vskip 10pt

\noindent
{\sevenrm 
$^1$ European Southern Observatory, 
Karl-Schwarzschild-Strasse 2, 
D-85748 Garching bei M\"unchen, Germany

\noindent
$^2$ Department of Mathematics and Statistics, King's
Buildings, University of Edinburgh, Edinburgh EH9 3JZ, United Kingdom
}

\vskip 20pt
\noindent
Received 3 August, 1996

\vskip 30pt
\noindent
{\bf Summary.} Galactic globular clusters, which are ancient building
blocks of our Galaxy, represent a very interesting family of stellar
systems in which some fundamental dynamical processes have taken place
on time scales shorter than the age of the universe.  In contrast with
galaxies, these clusters represent unique laboratories for learning
about two-body relaxation, mass segregation from equipartition of
energy, stellar collisions, stellar mergers, and core collapse. In the
present review, we summarize the tremendous developments, as much
theoretical as observational, that have taken place during the last
two decades, and which have led to a quantum jump in our understanding
of these beautiful dynamical systems.


\vfill\eject 

\vskip 30pt
\noindent
{\bf Contents} 

\vskip 20pt
\noindent
1. Introduction \hfill p.~3  

\vskip 5pt
\noindent
2. Definition of globular clusters  \hfill p.~6 

\vskip 5pt
\noindent
3. Internal dynamics: a brief historical summary \hfill p.~8

\vskip 5pt
\noindent
4. Characteristics of the globular clusters in our Galaxy \hfill p.~11

\item{} {\sl 4.1 The radial distribution} \hfill p.~12

\item{} {\sl 4.2 The clusters away from, and near to, the galactic
	bulge} \hfill p.~13

\item{} {\sl 4.3 Comparison with stellar populations} \hfill p.~14

\item{} {\sl 4.4 Age spread among galactic globular clusters} \hfill p.~14

\item{} {\sl 4.5 Implications for the formation and evolution of our Galaxy 
	\hfill \break \hglue 6mm
	and its globular clusters} \hfill p.~16

\item{} {\sl 4.6 Correlations between various properties of galactic globular 
	\hfill \break \hglue 6mm
	clusters} \hfill p.~17

\vskip 5pt
\noindent
5. Formation of globular clusters  \hfill p.~23

\item{} {\sl 5.1 Luminosity function of a globular cluster 
	system} \hfill p.~24

\item{} {\sl 5.2 Specific frequency of a globular cluster 
	system} \hfill p.~25

\item{} {\sl 5.3 Globular cluster formation models} \hfill p.~26

\item{} {\sl 5.4 Collapse, fragmentation, and 
	initial mass function} \hfill p.~29

\item{} {\sl 5.5 Early stellar evolution and violent relaxation phase}
	\hfill p.~31

\item{} {\sl 5.6 Formation of globular clusters in the Magellanic Clouds} 
	\hfill p.~32

\item{} {\sl 5.7 Formation of globular clusters in other nearby galaxies} 
	\hfill p.~35

\vskip 5pt
\noindent
6. Observations providing dynamical constraints \hfill p.~36

\item{} {\sl 6.1 Star counts and surface brightness profiles from
	photo-
	\hfill \break \hglue 6mm
	graphic plates, photomultipliers, and CCD images} \hfill p.~37

\item{} {\sl 6.2 Proper motions, stellar radial velocity dispersion, 
	and 
	\hfill \break \hglue 6mm
	velocity dispersion from integrated-light spectra} \hfill p.~41

\item{} {\sl 6.3 Initial and present-day mass functions} \hfill p.~46

\item{} {\sl 6.4 The possibility of dark matter 
	in globular clusters} \hfill p.~52

\vskip 5pt
\noindent
7. Quasi-static equilibrium: slow pre-collapse evolution \hfill p.~54

\item{} {\sl 7.1 The relaxation time} \hfill p.~54

\item{} {\sl 7.2 Energy equipartition and mass segregation} \hfill p.~57

\item{} {\sl 7.3 Evaporation through escaping stars} \hfill p.~60

\item{} {\sl 7.4 Tidal truncation} \hfill p.~64

\item{} {\sl 7.5 Theoretical models} \hfill p.~67

\item{} {\sl 7.6 Observational evidence of cluster rotation} \hfill p.~73

\item{} {\sl 7.7 Model fitting: parametric and non-parametric methods}
	\hfill p.~77

\vskip 5pt
\noindent
8. Evolutionary models \hfill p.~82

\item{} {\sl 8.1 N-body integrations} \hfill p.~83

\item{} {\sl 8.2 Fokker-Planck methods} \hfill p.~87

\item{} {\sl 8.3 Conducting gas models} \hfill p.~91

\vfill\eject 

\vskip 5pt
\noindent
9. Towards catastrophic phases of evolution~? \hfill p.~92

\item{} {\sl 9.1 The gravothermal instability and core collapse} \hfill p.~92

\item{} {\sl 9.2 Observational evidence of core collapse through the
	\hfill \break \hglue 6mm
	density profile and concentration} \hfill p.~96

\item{} {\sl 9.3 Observational evidence of core collapse through the 
	\hfill \break \hglue 6mm
	velocity dispersion profile} \hfill p.~101

\item{} {\sl 9.4 Physical interactions} \hfill p.~105

\item{} {\sl 9.5 Dynamics and formation of binaries} \hfill p.~107

\item{} {\sl 9.6 Observational evidence of binaries in globular clusters 
	\hfill \break \hglue 6mm
	(compared to the field)} \hfill p.~112

\item{} {\sl 9.7 Influence of dynamical evolution on stellar populations}
	\hfill p.~117

\item{} {\sl 9.8 Observational evidence of possible 
	products of stellar 
	\hfill \break \hglue 6mm
	encounters (blue stragglers, high-velocity stars, X-ray  
	\hfill \break \hglue 6mm
	sources, and pulsars.)} \hfill p.~121

\vskip 5pt
\noindent
10. Late phases of evolution and disruption \hfill p.~131

\item{} {\sl 10.1 Gravothermal oscillations; post-collapse evolution}
	\hfill p.~131

\item{} {\sl 10.2 Disruption} \hfill p.~135

\vskip 5pt
\noindent
11. Future directions \hfill p.~137

\vskip 5pt
\noindent
12. References \hfill p.~139

\vskip 30pt
\noindent
{\bf 1. Introduction} 

\vskip 20pt
\noindent
Till the late ninety seventies, globular clusters were thought to be
relatively static stellar systems.  This was partly due to the fact
that most observed surface-brightness profiles of globular clusters
(obtained from aperture photometry in the central and intermediate
parts, and star counts in the outer parts) were successfully fitted by
equilibrium models.  Some of these models are based on lowered
maxwellians and commonly known as King models (King 1966); they are
the simplest dynamical models which incorporate the three most
important elements governing globular cluster structure: dynamical
equilibrium, two-body relaxation, and tidal truncation.

It had been already known, since the early sixties, that globular
clusters had to evolve dynamically, even when considering only
relaxation, which causes stars to escape, consequently cluster cores
to contract and envelopes to expand.  But dynamical evolution of
globular clusters was not yet a field of research by itself, since the
very few theoretical investigations had led to a most puzzling
paradox: core collapse (H\'enon 1961, Lynden-Bell \& Wood 1968, Larson
1970a,b, Lynden-Bell \& Eggleton 1980).  It was only in the early
eighties that the field grew dramatically.  On the theoretical side,
the development of high-speed computers allowed numerical simulations
of dynamical evolution. Nowadays, Fokker-Planck and
conducting-gas-sphere evolutionary models have been computed well into
core collapse and beyond, leading to the discovery of possible
post-collapse oscillations.  In a similar way, hardware and software
improvements of N-body codes provide very interesting first results
for 10$^4$-body simulations (Makino 1996a,b, Spurzem \& Aarseth 1996),
and give the first genuine hope, in a few years, for 10$^5$-body
simulations.  On the observational side, the manufacture of
low-readout-noise Charge Coupled Devices (CCDs), combined since 1990
with the high spatial resolution of the Hubble Space Telescope (HST),
allow long integrations on faint astronomical targets in crowded
fields, and provide improved data analyzed with sophisticated software
packages.

It is not an exaggeration to say that our vision of globular cluster
dynamics has significantly been altered during the last decade.
Globular clusters are not dormant stellar systems.  Their apparent
smoothness, regularity, and symmetry are hiding everything but
simplicity.  Because of typical individual masses of a few $10^5
M_{\odot}$, intermediate between open clusters and dwarf galaxies,
globular clusters are of crucial importance in stellar dynamics:
fundamental dynamical processes (such as relaxation, mass segregation,
core collapse) take place in these systems on time scales shorter than
the Hubble time.  Recent theoretical and observational studies of
high-concentration globular clusters, with \conc\ \gsim\ 2, where \rt\
and \rc\ are the tidal and core radii, have confirmed what was
strongly suspected: stellar and dynamical evolutions are intimately
connected.  Observational studies concerning individual stars as well
as those devoted to integrated properties of stellar distributions
(e.g., color and population gradients) show that stellar encounters,
collisions, and mergers complicate and enrich the dynamical study of
globular clusters.

In this review we describe the present status of our knowledge of the
internal dynamics of globular clusters, from both theoretical and
observational points of view.  It is structured as follows: 

Section 2 gives a tentative definition of globular clusters;

Section 3 gives a brief historical summary of the study of globular
cluster dynamics;

Section 4 describes the general characteristics of the globular
clusters in our Galaxy and discusses a few astrophysical properties of
this cluster system;

Section 5 summarizes what is known (and above all unknown) about the
formation of globular clusters;

Section 6 describes the different kinds of observations providing
dynamical constraints;

Section 7 describes clusters in terms of quasi-static equilibrium,
i.e., especially in the pre-collapse regime;

Section 8 describes the different kinds of evolutionary models;

Section 9 describes the evolution towards catastrophic phases,
provides the existing observational evidence for core collapse, and
discusses the influence on stellar populations of the high stellar
density resulting from dynamical evolution;

Section 10 describes the late phases of the evolution and disruption;

finally, 

Section 11 discusses possible future directions, from both theoretical
and observational points of view.

\vskip 20pt
\noindent
Hereafter follows, for the interested reader, a nonexhaustive list of
some of the most important and already published reviews, monographs,
and proceedings related to the dynamics of globular clusters.
	 
\vskip 20pt
\noindent
First, two extensive reviews about dynamical evolution and binaries in
globular clusters, respectively:
\vskip 10pt
\item{$\bullet$} Lightman, A.P., Shapiro, S.L., 1978, 
	{\sl Dynamical Evolution of Globular Clusters}, 
	Rev. Mod. Phys., 50, 437;
\item{$\bullet$} Hut, P., McMillan, S.L.W., Goodman, J., Mateo, M., 
	Phinney, E.S., 
	Pryor, C., Richer, H.B., Verbunt, F., Weinberg, M., 1992, 
	{\sl Binaries in Globular Clusters}, PASP, 104, 981.

\vskip 20pt
\noindent
Second, seven reviews published in Annual Review Astronomy \&
Astrophysics and related to globular cluster dynamics:
\vskip 10pt
\item{$\bullet$} Michie, R.W., 1964, {\sl The Dynamics of Star Clusters,} 
	ARA\&A, 2, 49;
\item{$\bullet$} Harris, W.E., Racine, R., 1979, 
	{\sl Globular Clusters in Galaxies,} ARA\&A, 17, 241;
\item{$\bullet$} Freeman, K.C., Norris, J., 1981, {\sl The Chemical 
	Composition, Structure, and Dynamics of Globular Clusters,} 
	ARA\&A, 19, 319;
\item{$\bullet$} Elson, R.A.W., Hut, P., Inagaki, S., 1987, 
	{\sl Dynamical Evolution of Globular Clusters,} ARA\&A, 25, 565;
\item{$\bullet$} Valtonen, M., Mikkola, S., 1991, 
	{\sl The Few-Body Problem in Astrophysics,} ARA\&A, 29, 9;
\item{$\bullet$} Harris, W.E., 1991, {\sl Globular Cluster Systems in Galaxies 
	Beyond the Local Group,} ARA\&A, 29, 543;
\item{$\bullet$} Bailyn, C.D., 1995, {\sl Binary Stars, Blue Stragglers, and 
	the Dynamical Evolution of Globular Clusters,} ARA\&A, 33, 133.

\vskip 20pt
\noindent
Third, three fundamental books:
\vskip 10pt
\item{$\bullet$} Saslaw, W.C., 1987, 
	{\sl Gravitational Physics of Stellar and Galactic Systems,} 
	(Cambridge: Cambridge University Press);
\item{$\bullet$} Spitzer, L., 1987, 
	{\sl Dynamical Evolution of Globular Clusters,}
	(Princeton: Princeton University Press);
\item{$\bullet$} Binney, J., Tremaine. S., 1987, {\sl Galactic Dynamics,}
	(Princeton: Princeton University Press).

\vskip 20pt
\noindent
Fourth, the proceedings of thirteen workshops and conferences related to
globular cluster dynamics, all of them containing excellent reviews:
\vskip 10pt
\item{$\bullet$} Hayli, A., ed., 1975, 
	{\sl Dynamics of Stellar Systems}, 
	IAU Symp. 69, (Dordrecht: Reidel);
\item{$\bullet$} Hesser, J.E., ed., 1980, 
	{\sl Star Clusters}, 
	IAU Symp. 85, (Dordrecht: Reidel);
\item{$\bullet$} Goodman, J., Hut, P., eds., 1985, 
	{\sl Dynamics of Star Clusters}, 
	IAU Symp. 113, (Dordrecht: Reidel);
\item{$\bullet$} de Zeeuw, T., ed., 1987, 
	{\sl Structure and Dynamics of Elliptical Galaxies}, 
	IAU Symp. 127, (Dordrecht: Reidel);
\item{$\bullet$} Grindlay, J.E., Philip, A.G.D., eds., 1988, 
	{\sl The Harlow-Shapley Symposium on 
	Globular Cluster Systems in Galaxies}, IAU Symp. 126,
	(Dordrecht: Kluwer);
\item{$\bullet$} Merritt, D., ed., 1989, 
	{\sl Dynamics of Dense Stellar Systems},
	(Cambridge: Cambridge University Press);
\item{$\bullet$} Valtonen M.J., ed., 1988, 
	{\sl The Few Body Problem}, IAU Coll. 96., 
	(Dordrecht: Kluwer);
\item{$\bullet$} Janes, K., ed., 1991, 
	{\sl The Formation and Evolution of Star Clusters},
	ASP Conference Series, Vol. 13, (San Francisco: ASP);
\item{$\bullet$} Smith, G.H., Brodie, J.P., eds., 1993, 
	{\sl The Globular Cluster - Galaxy Connection}, 
	ASP Conference Series, Vol. 48, (San Francisco: ASP);
\item{$\bullet$} Djorgovski, S.G., Meylan, G., eds., 1993, 
	{\sl Structure and Dynamics of Globular Clusters}, 
	ASP Conference Series, Vol. 50, (San Francisco: ASP);
\item{$\bullet$} Saffer, R.A., ed., 1993, 
	{\sl Blue Stragglers}, 
	ASP Conference Series, Vol. 53, (San Francisco: ASP);
\item{$\bullet$} Milone E.F., Mermilliod J.-C., eds., 1996, 
	{\sl The Origins, Evolution, and Destinies of 
	Binary Stars in Clusters}, 
	ASP Conference Series, Vol. 90, (San Francisco: ASP);
\item{$\bullet$} Hut P., Makino J., eds., 1996, 
	{\sl Dynamical Evolution of Star Clusters: Confrontation
	of Theory and Observation}, 
	IAU Symp. 174, (Dordrecht: Kluwer).

\vskip 20pt
\noindent
In addition, some articles and extensive lists of references are found
in the triennial Transactions of the International Astronomical Union,
{\sl Reports on Astronomy}.  

\vskip 30pt
\noindent
{\bf 2. Definition of globular clusters} 

\vskip 20pt
\noindent
The usual definition of a globular cluster describes it as an old star
cluster (with an age $\tau$ larger than about 10~Gyr) found in the
bulge and halo regions of the Galaxy.  A precise determination of the
absolute age of the oldest galactic globular clusters is still an
elusive cosmological problem. From both observational and theoretical
arguments, Walker (1992) and Chaboyer (1995) reach a similar
conclusion: the absolute ages of the oldest globular clusters are found
to lie in the range 11-21~Gyr.  A mean age of about $\tau$ $\sim$
15~Gyr is generally accepted, but calibrations through stellar
evolution models are uncertain: e.g., Shi \etal\ (1995) and Shi (1995)
shows that adopting an initial helium abundance of $Y$ = 0.28 or a
mass loss rate \mdot\ $\sim$ 10$^{-11}$ \msun~yr$^{-1}$ near the main
sequence turn-off region lowers the current age estimate from 15~Gyr
to about 10-12~Gyr.  See also Mazzitelli \etal\ (1995) for an
investigation of globular cluster ages with updated input physics and
van den Bergh (1995a,b,c), Sarajedini \etal\ (1995), and Chaboyer
\etal\ (1996a,b) for interesting discussions.

Contrary to absolute ages, the relative ages of some galactic globular
clusters are more precisely known.  They are obtained by comparison of
their color-magnitude diagrams, which display clear differences in age
of about 3~Gyr (Bolte 1989).  Chaboyer \etal\ (1996c), on new age
estimates for 43 globular clusters, argue that their sample has a
statistically significant age spread of at least 5~Gyr.

The above age definition ($\tau$ \gsim\ 10~Gyr) suits the kind of
globular clusters which are the main subject of the present review.
Nevertheless, other galaxies contain younger stars clusters among
which some may be the progenitors of stellar clusters similar to the
galactic globular clusters.

It is also worth mentioning that, already in our Galaxy, globular
clusters differ strongly from one to the other, e.g., in integrated
absolute magnitude and total mass, which range from \Mvint\ = --10.1
and \mtot\ = 5 \x\ \milm\ (Meylan \etal\ 1994, 1995) for the giant
galactic globular cluster \cent\ down to \Mvint\ = --1.7 and \mtot\
$\simeq$ $10^3 M_{\odot}$ for the Lilliputian galactic globular
cluster AM-4 (Inman \& Carney 1987).  AM-4 is located at $\simeq$ 26
kpc from the galactic centre, and at $\simeq$ 17 kpc above the
galactic plane and cannot be considered to be an old open cluster.
The uncertainties on the above total mass estimates, perhaps as large
as 100\%, do not alter the fact that, in our Galaxy, the individual
masses of globular clusters range over three orders of magnitude.  It
is not known to what extent these mass differences are ``congenital''
or due to subsequent pruning by dynamical evolution.  

Although most galactic globular clusters are located within 20 kpc
from the galactic centre, it is worth mentioning the existence of a
few very remote galactic clusters.  The distance record is held by
AM-1 (Aaronson \etal\ 1984, Madore \& Freedman 1989) which is located
at about 120 kpc from the galactic centre, i.e., more than twice the
distance to the Large Magellanic Cloud.

The most recent update of observational and structural parameters of
all known galactic globular clusters may be found in the appendices
and tables of the proceedings of the 1992 Berkeley workshop edited by
Djorgovski \& Meylan (1993).

Globular clusters are observed in other galaxies of the Local Group
and beyond (cf. Harris 1991 and references therein for globular
cluster systems in galaxies).  The major difference with the galactic
globular clusters resides in the fact that the above definition based
on the age only ($\tau$ \gsim\ 10~Gyr) is no longer sufficient.  Rich
stellar systems with ages smaller than 10~Gyr are observed.  E.g., in
the Magellanic Clouds, the two dwarf irregular companion galaxies of
ours, there are star clusters with ages 10$^6$ \lsim\ $\tau$ \lsim\
10$^9$~yr.  There is still debate about the status of the richest of
these star clusters: are they the progenitors of genuine old globular
clusters?  Should the previous definition, related to age only, be
relaxed in order to include, e.g., clusters of different ages (a car
is called a car, independently of the fact that it is a new or used
one)?  Following van den Bergh (1993d), the most powerful discriminant
between open and globular clusters is their luminosity function: the
globular clusters have a gaussian luminosity function whereas the open
clusters have a luminosity function increasing monotonically towards
faint luminosities.  Is this discriminant totally independent of the
definitions adopted for sorting between open and globular clusters?
See recent interesting discussions by van den Bergh (1995a,b,c).

Not every globular cluster has a mass of about \milm.  Not every
globular cluster has an age larger than 10~Gyr.  Consequently, there
is no simple (one-parameter) definition of globular clusters which
would apply to every globular cluster around any galaxy.  In a few
cases, the classification between globular and open clusters remains
unclear (e.g., see Ortolani \etal\ 1995).  The discussion about a
clear definition of globular/open clusters may look semantic after
all, but it becomes essential when, e.g., luminosity functions of
systems of globular and open clusters are used for constraining the
importance of galaxy mergers in cluster formation (see, e.g., van den
Bergh 1995b, 1996 and references therein).  Fortunately, in the
framework of the present review, a perfect definition of globular
clusters is not essential since an overwhelming fraction of the
high-quality dynamical observations of globular clusters concerns only
the nearby rich (\mtot\ equal a few $10^5 M_{\odot}$) galactic
globular clusters.  All these observed clusters have ages $\tau$
larger than about 10~Gyr, and their large numbers of member stars make
them interesting from a dynamical point of view.  It is to these
observations that theoretical models are fitted, and it is against
these stellar systems that our theoretical understanding of the
internal dynamics of globular clusters is tested.

\vskip 30pt
\noindent
{\bf 3. Internal dynamics: a brief historical summary} 

\vskip 20pt
\noindent
Apart from the catalog of Charles Messier (1784), which mentions 28
galactic globular clusters visible from Europe, the first scientific
description of globular clusters --- which clearly identified them as
huge swarms of stars of regular symmetrical appearance --- was
published by William Herschel (1814).  A few decades later, this work
was extended to the southern hemisphere by his son John Herschel
(1847).

Mere descriptions of visual observations were superseded gradually,
during the second half of the 19$^{\rm th}$ century, by more useful
and efficient observations thanks to the development, and numerous
sophisticated improvements till not long ago, of photographic
techniques applied to astronomy.  It is from photographic observations
of two globular clusters --- \cent\ and \tuca\ --- that Bailey (1893)
made what were probably the first extensive star counts, which
represent the oldest observational constraint for the study of
globular clusters.  Bailey's counts added to some new material
concerning other clusters were used by Pickering (1897) in the first
important comparisons between observed and theoretical profiles in
order to study the radial distribution of stars in clusters.  A few
years later, W.E. Plummer (1905) and von Zeipel (1908) showed, in
studies of M3, M13, \tuca, and \cent, how the radial space
distribution of stars may be deduced numerically from the observed
projected density profile.  Von Zeipel compared these profiles with
those to be expected for a spherical mass of gas in isothermal
equilibrium.  At most a little physics was present in such studies.

In parallel with the improvement of observational techniques, the
second half of the 19$^{\rm th}$ century experienced also dramatic
progress in theoretical physics, with the invention of the new fields
of thermodynamics and statistical mechanics, in order to describe
gases with molecules of infinitesimal size.  Maxwell (1860) wrote down
the now famous maxwellian law of the distribution of velocities in a
work which gave birth to the kinetic theory of gases, and it was
further developed by Boltzmann (1896), among others.

In the early years of the 20$^{\rm th}$ century, some parallels were
drawn between a molecular gas and star clusters: the stars were
considered as mass points representing the molecules in a
collisionless gas.  The analogy between a gas of molecules and a gas
of stars is subject to criticisms, since the mean free path of a
molecule is generally quite small compared with the size or scale
height of the system, whereas the mean free path of a star is much
larger then the diameter of the cluster; in addition molecules travel
along straight lines, while stars move along orbits in the
gravitational potential of all the other stars of the stellar system.
Stellar collisions in clusters were studied by Jeans (1913), who
remarked that they might be important in such stellar systems.  The
problem was then to seek the possible spherical distribution of such a
gas in a steady state.  H.C. Plummer (1911, 1915) pursued the search
for a physical basis on which the distribution of stars in globular
clusters could be established, a search followed by a flurry of
essential theoretical contributions by Eddington (1913, 1915a, 1915b,
1916) and by Jeans (1913, 1915, 1916a, 1916b).

This amazing burst of fundamental papers was followed by a relatively
dormant period which ended with another major era in cluster theory,
containing the essential theoretical contributions by Ambartsumian
(1938), Spitzer (1940), and Chandrasekhar (1942, his \S5.8), who
investigated the consequences of stellar encounters.  The next burst
of fundamental papers took place in the late fifties and early
sixties, with the contributions by King (1958b, 1962, 1965, 1966) and
by Michie (1961, 1963a,b,c,d), among others.  At that time, two
clusters, namely, M3 and \cent, were the subjects of studies by Oort
\& van Herk (1959) and Dickens \& Woolley (1967), respectively.  These
two papers initiated the modern interplay of observation and
model-building that still continues today.  The paper by Gunn \&
Griffin (1979) was another notable landmark in these developments.

Already before, and also after, the pioneering work of von Hoerner
(1960), who made the first $N$-body calculations with $N$ = 16, it was
realized that computation of individual stellar motions could be
replaced by statistical methods.  The structure of a globular cluster
is defined at each moment by a distribution function in a phase space
with 7 dimensions (positions, velocities, and time).  Unfortunately,
the numerical study of such a general form is intractable.  It is
necessary to make some simplifying hypotheses, e.g., spherical
symmetry of the cluster, or quasi-static equilibrium.  The major
simplification consists in considering separately the problems of
structure and evolution.


A series of works had studied the structure of globular clusters
without taking into account their evolution (e.g., Plummer 1911,
Eddington 1915a,b, 1916, Jeans 1915, 1916a,b, Chandrasekhar 1942, Camm
1952, Woolley \& Ro- bertson 1956, H\'enon 1959, Michie 1963a, King
1966).  Depending on further simplifications varying from one
author to another, the distribution function may have any form, as
long as it satisfies Jeans' theorem.  King's studies (1966) have shown
``lowered maxwellian'' energy dependence to be a good approximation to
the solution of the Fokker-Planck equation describing the phase-space
diffusion and evaporation of stellar systems like globular clusters.
These models fit the density profiles of globular clusters rather
well.  Nevertheless, this agreement need have no deep physical
meaning, given the ad~hoc hypotheses simplifying the fundamental
equations, even if there is some dynamical justification for King's
choice.

On the contrary, other studies had considered the evolution of
globular clusters with a fixed structure (e.g., Spitzer 1940,
Chandrasekhar 1943a,b,c, King 1958a,b,c, Spitzer \& H\"arm 1958, von
Hoerner 1958, Agekian 1958, H\'enon 1960a, King 1960, Michie 1961).  In
most cases, the cluster was supposed to be homogeneous with a uniform
gravitational potential.  The results obtained --- essentially the
escape rate of stars from the clusters --- were rather different, once
again because of simplifying hypotheses.

In reality, structure and evolution cannot be dissociated: they are
intimately linked, determined the one by the other.  H\'enon (1961,
1965) made the first attempt to solve the structure and evolution
equations simultaneously, in the simplified case of a self-similar
evolution with a distribution function depending only on the total
energy (isotropy of the velocity dispersion) and with all stars having
the same mass.  Even to other theorists his model looked pathological:
it had infinite central density and a flux of energy emerged from the
central singularity.  But H\'enon showed that a cluster {\sl without}
such a singularity would evolve into one that did, and he realised
that, in a real system, the flux of energy might well be supplied by
the formation and evolution of binary stars.  H\'enon was right, and
his results had given him a first glance at what was to become the
Holy Grail of globular cluster dynamics: core collapse.

The whole concept of core collapse, linked to the gravothermal
instability which may develop in a gravitational system because of its
negative specific heat, was first investigated by
Antonov (1962), and Lynden-Bell \& Wood (1968).  What was then called
the gravothermal catastrophe was eventually recognized as being not so
catastrophic after all, since the cluster core does not collapse for
ever but bounces back towards lower stellar density phases.  Again it
was H\'enon, this time in his 1975 paper, who showed theorists the way
past the apparent impasse of core collapse into the post-collapse
phase of evolution.

This brief history is far from being exhaustive but brings us to the
seventies.  It is the further modern theoretical and observational
developments which are the subject of the present review.

\vskip 30pt
\noindent
{\bf 4. Characteristics of the globular clusters in our Galaxy} 

\vskip 20pt
\noindent
After having provided observational indications of the extended
structure of our Galaxy and the eccentric position of the sun with
respect to the galactic centre (Shapley 1930), the globular cluster
system of our Galaxy has been long recognized as an interesting tool
to study the early dynamical and chemical evolution of the galactic
halo (Trumpler 1930).  Over the last few decades, analyses have tried
to show evidence of a metal-rich disk subsystem, complementary to the
metal-poor halo subsystem (e.g., Baade 1958, Kinman 1959, Morgan 1959,
Woltjer 1975, Harris 1976, Hartwick \& Sargent 1978, van den Bergh
1979, Zinn 1980, Frenk \& White 1980, 1982, Zinn 1985, Hesser \etal\
1986, Armandroff \& Zinn 1988, Armandroff 1988, Armandroff 1989,
Thomas 1989, Minniti 1995).  There is now clear evidence indicating
that the globular cluster system of the Galaxy consists of two
separate subsystems, a slowly rotating halo subsystem and a rapidly
rotating disk subsystem.  A detailed knowledge of these subsystems is
essential for understanding the fact that some internal properties of
clusters, e.g., the concentration parameter, correlate well with
global variables such as the galactocentric distance.  This suggests
that some external effects strongly influence the internal dynamical
evolution of a globular cluster (Chernoff \& Djorgovski 1989).

\vskip 6.3truecm
\item\item{\sevenrm{{\bf Fig.~4.1.} \hskip 2mm Projected distribution 
of the 143 known globulars in galactic coordinates (from Djorgovski \&
Meylan 1994, Fig.~1).  The symbol size scales with the logarithm of
the luminosity.  The strong central concentration is obvious.}}
\vskip 0.5truecm

The galactic globular system consists of 143 confirmed globular
clusters (Djorgovski \& Meylan 1993b).  Fig.~4.1 shows the distribution
of clusters on the galactic sky (Djorgovski \& Meylan 1994).  The
well-known strong central concentration is the most obvious feature of
the distribution.  The absence of an obvious zone of avoidance near
the galactic plane immediately suggests no large numbers of clusters
are missing due to obscuration.  Nevertheless, it is very likely that
some clusters are still missing, lost in the obscured areas near the
galactic plane or in the outer parts of the halo.  The latest addition
to the list consists of the new globular cluster C J0907--372 (Pyxis)
recently discovered by Weinberger (1995) and confirmed by Da Costa
(1995) and Irwin \etal\ (1995).  The kinematics and dynamics of the
galactic globular cluster system have been studied by, e.g., Frenk \&
White (1980, 1982), Innanen \etal\ (1983), and Thomas (1989).

\vskip 30pt
\noindent
{\sl 4.1 The radial distribution} 

\vskip 20pt    
\noindent
It is possible to parameterize the radial distribution of the galactic
globular clusters by using a simple power law with a core:
$$
\rho(r) ~= ~\rho_0 ~\left( 1 ~+ ~{{r}\over{r_c}} \right) ^{-\alpha}
\eqno(4.1)
$$
This approach is purely empirical, and it is not meant to imply any
physical meaning of the distribution given by Eq.~4.1 (Djorgovski \&
Meylan 1994).  Such a simple fit neglects the disk-halo dichotomy, and
many other fine details.  The purpose is simply to estimate the
number of clusters which may be missing in the central parts of the
Galaxy.

Taking into account the probable incompleteness of the data and the
distance errors, which could be rather substantial for the heavily
obscured clusters at small galactocentric radii, Djorgovski \& Meylan
(1994) perform fits to the data with model curves of various values of
the core radius $r_c$ and the power-law exponent $\alpha$.  Good
matches are found for the values of $\alpha$ $\sim$ 3.5-4.  A
generally quoted value in the literature is 3.5 (see, e.g., Harris
1976, Zinn 1985).  The core radii $r_c$ $\sim$ 0.5-2 kpc.  The
flattening of the distribution near the centre -- into a core -- is
probably due to a combination of three effects: smearing due to the
distance errors, genuine clusters which are missing due to
obscuration, and the real flattening of the distribution.  The latter
may reflect the initial conditions, but also possible dynamical
effects, viz., a more effective tidal destruction of clusters near the
galactic centre.  For the faint clusters towards the galactic centre,
there is also the mere problem of classification: e.g., NGC~6540,
previously considered as an open cluster, has been recently recognized
as a globular cluster (Bica \etal\ 1994).

The steep observed slope of this distribution differs significantly
from the density law of the dark halo, $\rho(r) \sim r^{-2}$, which
results in a flat rotation curve.  It is hard to imagine an
evolutionary process which could convert, over an Hubble time, a
$r^{-2}$ distribution into a $r^{-3.5}$ one, for the globular clusters
but not for the dark halo material.  This implies a different origin
for the globular cluster system (and presumably the visible stellar
halo), and the dark halo, whatever its constituents are.

The apparent core radii found by Djorgovski \& Meylan (1994) ($\sim
0.5 - 2$ kpc) are considerably larger than the characteristic radii
for the stellar distribution in the bulge: Blanco \& Terndrup (1989)
give $r_c = 0.11 \pm 0.04$ kpc for the bulge light.  This is
reminiscent of the situation seen in M87, where the core radius of the
globular cluster system is some 13 times larger than that of the
underlying galaxy's light (Lauer \& Kormendy 1986).

The approach by Djorgovski \& Meylan (1994) and other alternatives
show that the number of missing clusters at low latitudes and/or near
the galactic centre is perhaps of the order of 10.  Similar
conclusions have been reached by Racine \& Harris (1989), who
performed a more detailed analysis, and also by Woltjer (1975) and
Oort (1977).  In conclusion, there is probably a slight selection
effect, leading to an incompleteness of $\sim$ 5\% of the total number
of clusters.

\vskip 30pt
\noindent
{\sl 4.2 The clusters away from, and near to, the galactic bulge}

\vskip 20pt    
\noindent
[Fe/H] values, compiled by Zinn \& West (1984), Zinn (1985), and
Armandroff \& Zinn (1988), exist for 119 of the 143 galactic globular
clusters.  Since there is no clear spatial division between halo and
disk populations, it is generally admitted, from the distribution of
IRAS sources, that the galactic bulge extends to an angular distance
$\omega$ of approximately 15\deg\ from the galactic centre, with
$\omega$ = 15\deg\ being the dividing line between bulge and non-bulge
regions (Zinn 1990, 1996).  There is also no clear division between
disk globular clusters and open clusters: from the color-magnitude
diagram of Lyng\aa~7, Ortolani \etal\ (1993) and Tavarez \& Friel
(1995) observe that this cluster, previously classified as an open
cluster, might be a metal rich globular cluster or, alternatively, the
oldest open cluster so far detected.

\vskip 20pt
\noindent
{\sl The clusters away from the galactic bulge ($\omega$ $>$ {\rm
15}\deg).}~~ The observational evidence (see, e.g., Fig.~1 in Zinn
1990) indicating that distinct halo and disk subsystems exist among
the clusters with $\omega$ $>$ 15\deg\ comes from:

\vskip 5pt
\noindent
(i) the distribution of cluster [Fe/H] values, which is bimodal with
peaks at [Fe/H] = --1.6 and --0.6;

\vskip 5pt
\noindent
(ii) the distribution of distances from the galactic plane
$\vert$Z$\vert$, which shows that while the metal-poor clusters are
scattered over a large range in distance, the clusters more metal-rich
than [Fe/H] = --1 are all at less than 4 kpc from the galactic plane;

\vskip 5pt
\noindent
(iii) the metal-poor and metal-rich clusters, which have very
different values for the \Vrot/\siglos\ ratio;

\vskip 5pt
\noindent
(iv) the most metal-rich bin, which has both the largest value of
rotational velocity, \Vrot\ = 172 \pmm\ 26 \kms, and the smallest
value of the line-of-sight velocity dispersion, \siglos\ = 60 \pmm\ 14
\kms.

\vskip 20pt
\noindent
{\sl The clusters near the galactic bulge ($\omega$ $<$ {\rm
15}\deg).}~~ The distribution of cluster [Fe/H] values is also
bimodal, with two peaks at approximately the same values as the peaks
for clusters with $\omega$ $>$ 15\deg.  However, the relative
amplitudes of the peaks differ in the sense that the percentage of
clusters that are metal rich ([Fe/H] $>$ --0.8) is about 43\% in the
$\omega$ $<$ 15\deg\ sample, whereas it is only 16\% in the $\omega$
$>$ 15\deg\ sample.  Liller~1 seems to be the most metal rich globular
cluster known, with \feh\ = +0.25 \pmm\ 0.3 (Frogel \etal\ 1995).  For
the clusters near the galactic bulge, there is no clear separation
between the metal-poor and metal-rich clusters.  The metal-rich and
metal-poor clusters have velocity dispersions \siglos\ = 77 \pmm\ 14
and 126 \pmm\ 20 \kms, respectively.  There is no evidence, however,
for a more rapid rotation of the metal-rich clusters.

\vskip 30pt    
\noindent
{\sl 4.3 Comparison with stellar populations}

\vskip 20pt
\noindent
In both giant and dwarf elliptical galaxies, there is considerable
evidence that their globular clusters do not have the same
metallicities, spatial distributions, and kinematics as their stellar
populations.  In our Galaxy, on the contrary, the four following
comparisons suggest that the clusters and the stellar populations away
from the galactic bulge region are very similar.  First, Armandroff
(1989) has shown that the metal-rich disk subsystem has approximately
the same [Fe/H], \Vrot, \siglos, and scale height as the so called
thick disk stellar population that has been identified in a large
number of studies (cf. Gilmore 1989 for a review).  Second, studies of
the number densities of globular clusters and RR Lyrae variables as a
function of the distance to the galactic centre $R$ have shown that
they both approximate $R^{-n}$ fall-offs, with $n$ in the range of 3
to 3.5 (Saha 1985, Zinn 1985). There is additional evidence that the
clusters and halo stars have similar distributions.  Third, the [Fe/H]
distributions of the subdwarfs and globular clusters have very nearly
the same mean values (Laird \etal\ 1988).  Fourth, the
horizontal-branch morphology of the field is not grossly different
from that of globular clusters lying in the same zone of $R$ (Kraft
1989).

While the study of the galactic bulge is still in its infancy, there
is little question that at least its composition is different from
that of the globular clusters within the same area of the sky.  But
away from the galactic bulge, there is no strong evidence to suggest
that the globular clusters and the halo and thick disk stellar
populations have different properties.  See Zinn (1996) for a recent
review.

\vskip 30pt
\noindent
{\sl 4.4 Age spread among galactic globular clusters}

\vskip 20pt
\noindent
A precise knowledge of absolute ages of galactic globular clusters,
which would provide a lower limit to the age of the universe and hence
an upper limit to the Hubble constant, is still out of reach, due to
uncertainties in theoretical models of stellar evolution and in basic
calibrations (e.g., absolute luminosities of subdwarfs).  According to
standard pictures for the formation of the Galaxy (Eggen, Lynden-Bell,
\& Sandage 1962), the system of globular clusters formed
during the rapid dynamical collapse of the protogalactic cloud, a
process which should have lasted no more than 1~Gyr (cf. Sandage 1990
for the exact meaning of ``rapid'').  Fortunately, the relative ages
of the galactic globular clusters are more precisely known than their
absolute ages, providing evidence for a spread in age among them.

The two galactic globular clusters NGC~288 and NGC~362 are central to
recent claims (Bolte 1989, Green \& Norris 1990, VandenBerg \etal\
1990, Sarajedini \& Demarque 1990) of large age differences ($\sim$
3~Gyr) between galactic globular clusters.  But see Stetson \etal\
(1996). The claimed age differences are derived from stellar evolution
models using assumed CNO abundances, whose uncertainties of about a
factor of three could account for an apparent 2-Gyr age difference.
Dickens \etal\ (1991) have accurately measured abundances in red
giants in NGC~288 and NGC~362 and find that the Fe abundance and the
sum of the C, N, and O abundances are essentially the same in every
star studied, thus eliminating composition differences and confirming
the reality of the age spread.

There are a few clusters --- Pal~12, Ruprecht~106, Arp~2, Terzan~7,
and IC~4499 --- which seem to be unambiguously younger than most
globular clusters.  With regard to Pal~12 (\feh\ $\simeq$ --1.1) and
Ruprecht~106 (\feh\ $\simeq$ --1.6), how {\sl young} they are depends
on what the metal abundances really are, but an age {\sl difference}
of $\sim$~3~Gyr compared to clusters of similar [M/H] seems required
(Bolte 1993).  The complexity is further increased by the report by
Buonanno \etal\ (1994) of the existence of ``young'' metal-poor and
metal-rich galactic globular clusters.  Arp~2, with \feh\ $\simeq$
--1.8, is $\simeq$ 3~Gyr younger than the group of the metal-poor
clusters (Buonanno \etal\ 1995a), while Terzan~7, with \feh\ $\simeq$
--0.49, is $\simeq$ 4~Gyr younger than \tuca, another galactic
globular of similar metallicity (Buonanno \etal\ 1995b).  IC~4499 is
the most recently studied such young globular (Ferraro \etal\ 1995b).
See also van den Bergh (1993c) and Stetson \& West (1994) about
NGC~6287, which could be the oldest galactic globular cluster.

Another interesting point comes from the 5 best studied clusters with
\feh\ $\sim$ --2, viz. M68, M92, NGC~6397, M3, and M13, which show a
remarkable similarity in age with all values within 0.3~Gyr
(VandenBerg \etal\ 1990; Bolte 1993).

Richer \etal\ (1996) use the 36 globular clusters with the most
reliable age data. These clusters span galactocentric distances from 4
through 100 kpc and cover a metallicity range from \feh\ $\simeq$
--0.6 to --2.3.  They find that the majority of the globular clusters
form an age distribution with a dispersion $\sigma(t)$ $\simeq$ 1 Gyr,
and a total age spread smaller than 4 Gyr.  Clusters in the lowest
metallicity group (\feh\ $<$ --1.8) have the same age to well within 1
Gyr, at all locations in the Galaxy halo, suggesting that star
formation began throughout the halo nearly simultaneously in its
earliest stages. Richer \etal\ (1996) find no statistically
significant correlation between mean cluster age and galactocentric
distance (no age gradient) from 4 to 100 kpc.

The above facts would favor the scenario of Searle \& Zinn (1978), in
which galaxies are built from the hierarchical merging of smaller
subunits in a formation process characterized by the chaotic nature of
the collapse, and occurring over a period of a few billion years,
several times longer than in the original Eggen \etal\ (1962) model
(but see Sandage 1990 and \S5 below).  Depending on the adopted
value of the Hubble constant \Ho, there is a potential conflict
between the age of the Universe and the age of the globular clusters
(see, e.g., Bolte \& Hogan 1995).

\vskip 30pt	
\noindent
{\sl 4.5 Implications for the formation and evolution of our Galaxy
	and its globular clusters} 

\vskip 20pt    
\noindent
The properties of the halo cluster system that have the largest impact
on the theories of the formation and evolution of the Galaxy and its
globular clusters are: 

\vskip 5pt
\noindent
(i) the low mean \Vrot\ of the halo;

\vskip 5pt
\noindent
(ii) the weakness of the [Fe/H] gradient with $R$;

\vskip 5pt
\noindent
(iii) the wide range in [Fe/H] at every $R$;

\vskip 5pt
\noindent
(iv) the lack of correlation between \Vrot\ and [Fe/H];

\vskip 5pt
\noindent
(v) the range in age of several billion years (Gyr) between clusters of
the same [Fe/H];

\vskip 5pt
\noindent
(vi) the systematic variation in horizontal-branch morphology with $R$.

\vskip 5pt
\noindent	
Recent investigations (e.g., Deliyannis \etal\ 1990, Lee \etal\ 1990,
1994) have cast doubt on the viability of the second-parameter
candidates other than age (although challenged by Stetson \etal\
1996).  See Fusi Pecci \etal\ (1996) for a review. A case has been
made for the importance of stellar density (Buonanno 1993) for the
morphology of the horizontal branch: see Fig.~9.8 below from Buonanno
\etal\ (1985a).  If age is the second parameter, then point (vi) above
indicates that the Galaxy evolved from the inside out (Searle \& Zinn
1978).  Approximately the inner 8-kpc volume of the halo is then older
on average by a few Gyr and much more homogeneous in age than the
outer halo.  Did the inner halo undergo the kind of rapid collapse
envisioned by Eggen \etal\ (1962, see Sandage 1990), while the outer
halo was built over several Gyr by the merger of several dwarf
galaxies, as argued by Searle \& Zinn (1978)?

The properties of the disk globular clusters that are most important
from the point of view of galactic evolution are their ages and
metallicity gradients with $R$ and $\vert$Z$\vert$.  These are nearly
open questions, however, for very few disk globular clusters have been
precisely dated.

\vskip 30pt
\noindent
{\sl 4.6 Correlations between various properties of galactic 
	globular clusters}

\vskip 20pt
\noindent
A fundamental problem in globular cluster study lies in the
determination of the extent to which their properties are either
universal or dependent on characteristics of the parent galaxy.  The
identification of correlations and trends between various properties
of galactic globular clusters provides clues which can be used to test
and constrain theoretical models of cluster formation and evolution.
Earlier work includes the pioneering study by Brosche (1973), along
with the studies by Peterson \& King (1976), Brosche \& Lentes (1984),
Chernoff \& Djorgovski (1989), Djorgovski (1991), and Covino \&
Pasinetti Fracassini (1993), among others.  Djorgovski \& Meylan
(1994) gives the most extensive and up-do-date study of this kind.
They use a set of 13 cluster parameters, viz.:
\hfill \break
-- the absolute visual magnitude, $M_V$; \hfill \break
-- the concentration parameter, \conc; \hfill \break
-- the log of the core radius in parsec, $r_c$; \hfill \break
-- the log of the half-light radius in parsec, $r_h$; \hfill \break
-- the central surface brightness in the $V$ band, 
	$\mu_V (0)$; \hfill \break
-- the average surface brightness in the $V$ band within 
	$r_h$, $\langle \mu_V \rangle _h$; \hfill \break
-- the log of the central luminosity density in 
	$L_{\odot V}$/pc$^3$, $\rho _0$; \hfill \break
-- the log of the central relaxation time in years, 
	$t_{rc}$; \hfill \break
-- the log of the half-mass relaxation time in years, 
	$t_{rh}$; \hfill \break
-- the metallicity, [Fe/H]; \hfill \break
-- the log of the central velocity dispersion in \kms, 
	$\sigma$; \hfill \break
-- the log of the distance from the galactic centre in kpc, 
	$R_{gc}$; and \hfill \break 
-- the log of the distance from the galactic plane 
	in kpc, $Z_{gp}$. \hfill \break
The definitions of these quantities, error estimates, and other
details can be found in the following data bases: Djorgovski \& Meylan
(1993b), Peterson (1993a), Pryor \& Meylan (1993), Trager \etal\
(1993), and Djorgovski (1993b).  Among the 13 quantities mentioned
here, only 9 are measured independently: $\langle \mu_V
\rangle _h$ is derived from the $M_V$ and $r_h$; $\rho _0$ is derived
from the $\mu_V (0)$, $c$, and $r_c$; $t_{rc}$ is derived from the
$M_V$, $c$, and $r_c$; and $t_{rh}$ is derived from the $M_V$, and
$r_h$.  This may cause spurious correlations.

The first striking thing about the data on globular clusters is the
vast range they span in many of their properties, e.g.  luminosity and
density, more so than either elliptical or dwarf galaxies
(cf. Djorgovski 1993a for comparisons).  Most core or central
parameters span a larger range than the corresponding half-light
($\sim$ half-mass) quantities. Qualitatively, this may be understood
as a consequence of dynamical evolution which operates faster at core
scales, where the reference (``relaxation'') time scales are shorter,
by up to a factor of a hundred.  It is worth noting that clusters tend
to increase the range of their properties as time proceeds.  This
``stretching of properties'' of clusters is inevitable: even if the
distributions of cluster properties started as $\delta$-functions,
some spread would occur over a Hubble time, already for no other
reason than the different tidal effects from one cluster to the other.

The observed fact that the half-mass relaxation times span over two
orders of magnitude, and the central relaxation times over some five
orders of magnitude, practically guarantees that the globular cluster
population will contain a range of objects at all stages of dynamical
evolution.  Using cluster ellipticities and orientations from White \&
Shawl (1987), Djorgovski \& Meylan (1994) find that these two
quantities do not correlate with any other cluster parameters.  They
also used the ratios $Z_{gp}/R_{gc}$, which are a statistical measure
of the orbit inclinations, without obtaining any new insights.

{\sl Luminosity correlations.}~~ Luminosity is perhaps the most
fundamental observed quantity characterizing a stellar system.  For a
set of old stellar systems it is a good relative measure of the
baryonic mass.  Many other properties correlate with luminosity for
elliptical and dwarf galaxies; not so for globular clusters
(cf. Djorgovski 1993a for comparisons).  The only good correlation
with luminosity is that with the velocity dispersion.  The only other
discernible trends are with the concentration and central surface
brightness (or equivalently, central luminosity density).  More
luminous clusters tend to have higher concentrations and denser cores,
but there is a large scatter at every luminosity (see also van den
Bergh 1994).  Interestingly, neither $r_c$ nor $r_h$ correlates with
luminosity; this is in a marked contrast with both elliptical and
dwarf galaxies, for which the corresponding correlations are excellent
(see, e.g., Kormendy 1985).

It has recently been found (Bellazzini \etal\ 1996) that the
correlation between luminosity and core parameters is stronger for
clusters lying outside the solar circle than for those inside, which
is consistent with the hypothesis that the correlation is primordial,
but has been erased by subsequent evolution where evolution time
scales are short enough.

{\sl Trends with the position in the Galaxy.}~~ Globular clusters live
in the tidal field of the Galaxy, and are subject to tidal shocks due
to both bulge and disk passages (Ostriker \etal\ 1972; Chernoff \&
Shapiro 1987; Aguilar \etal\ 1988; Chernoff \& Weinberg 1990, and
references therein).  Moreover, properties of newly formed clusters
may well depend on their position in the proto-Galaxy (e.g., Fall \&
Rees 1977, 1985; Murray \& Lin 1992).  It is thus reasonable to expect
that some correlations of cluster properties with $R_{gc}$ and/or
$Z_{gp}$ will be found.  A generic expectation is that clusters closer
to the galactic centre will be more dynamically evolved, as tidal
shocks accelerate their internal evolution towards the core collapse
or dissolution.  Chernoff \& Djorgovski (1989) analysed the frequency
of occurrence of collapsed clusters as a function of position in the
Galaxy, and found them to be highly concentrated towards the galactic
centre and plane.  This trend continues for non-collapsed clusters, in
order of decreasing concentration.  Djorgovski \& Meylan (1994)
confirm and extend their findings by looking at the correlations of
core parameters with $R_{gc}$ and $Z_{gp}$.  Clusters at smaller
$R_{gc}$ tend to have smaller and denser cores and higher
concentrations, and thus also shorter central relaxation times.
Similar trends are seen when $Z_{gp}$ is used instead of $R_{gc}$.
Indeed, the theory by Fall \& Rees (1985) predicts a radial trend of
the mean cluster densities, bound by the scaling laws given by the
thermal instability ($\rho_h \sim R ^{-1}$) and by tidal truncation
($\rho_h \sim R ^{-2}$, following the density law of the dark halo),
although other theoretical explanations are certainly possible
(cf. Surdin 1995).

\vskip 12truecm
\item\item{\sevenrm{{\bf Fig.~4.2.} \hskip 2mm Correlations between 
the core parameters (from Djorgovski \& Meylan 1994, Fig.~10).
Clusters with smaller cores have higher concentrations, higher central
surface brightness and luminosity densities, and therefore also
shorter central relaxation times.  This is as expected from a family
of objects with a roughly constant initial core mass, evolving towards
core collapse.  All three principal parameters, $c$, $r_c$, and
$\mu_V(0)$, are measured independently, and correlations are thus
real.  The correlation involving $t_{rc}$, as in the lower left panel,
is entirely artificial, by mere definition of $t_{rc}$.}}
\vskip 0.5truecm

{\sl Correlations of core properties.}~~ Some of the best correlations
of globular cluster properties are those between the various core
parameters and concentrations.  They are displayed in Fig.~4.2.  All
three principal quantities $r_c$, $\mu_V (0)$, and $c$ are measured
independently; correlations among them are real.  The correlation
between the core radius, $r_c$, and the central surface brightness,
$\mu_V (0)$, has been noted by Kormendy (1985).  On the other hand,
the spectacular correlation between $t_{rc}$ and $r_c$ is entirely
artificial: it reflects the derivation of $t_{rc}$, which depends on
$r_c ^{3/2}$ (close to the apparent slope of the correlation), and
other quantities which also correlate with $r_c$.

\vskip 12.5truecm
\item\item{\sevenrm{{\bf Fig.~4.3.} \hskip 2mm Velocity dispersion 
correlations (from Djorgovski \& Meylan 1994, Fig.~12).  These are the
best non-trivial correlations known for globular clusters.  The
corresponding scaling laws are indicated in the upper left of each
panel, and the Pearson ($r$) and Spearman rank ($s$) correlation
coefficients are listed in the lower right of each panel.  The lower
right panel shows a bivariate correlation, where core radius is used
as a ``second parameter'' to improve the corresponding correlation
shown in the upper right panel. }}
\vskip 0.5truecm

{\sl Metallicity non-correlations.}~~ Unlike elliptical and dwarf
galaxies, globular clusters show no correlations between metallicity
and luminosity or velocity dispersion.  The standard explanation for
these correlations in galaxies is self-enrichment in the presence of
galactic winds.  It is thus natural to conclude that globular clusters
are not self-enriched systems.  This conclusion is also supported by
the extreme internal chemical homogeneity of most globular clusters
(e.g., Suntzeff 1993; but see Norris \& Da Costa 1995 in the
exceptional case of \cent\ where stars have --1.8 $<$ \feh\ $<$
--0.8).  Even a single supernova exploding in a still gaseous
proto-globular cluster would deposit $\sim 10^{51}$ erg of kinetic
energy, which is comparable to the binding energies of globular
clusters today, $E_{bind} \sim 10^{50} - 10^{51}$ erg.  Thus, with a
possible exception of the most massive systems such as \cent, where
some chemical inhomogeneities are seen (Dickens \& Bell 1976, 
Fran\c cois \etal\ 1988, Mukherjee \etal\ 1992), a still gaseous
proto-cluster could be immediately disrupted.  Obviously, the exact
outcome would depend on many of the as-yet poorly known details of the
physics of globular cluster formation.  For self-enrichment of
globular clusters see, e.g., Smith (1986, 1987) and Morgan \& Lake
(1989).

{\sl Velocity dispersion correlations.}~~ Aside from the correlations
of core properties, the best non-trivial correlations of globular
cluster properties are between the velocity dispersion and luminosity
or surface brightness.  They are displayed in Fig.~4.3.  The
corresponding scaling laws are indicated in the upper left of each
panel.  The central surface brightness expressed in linear units is
$I_0$, the average surface brightness within $r_h$ is $I_h$, and the
total luminosity is $L$ (all in the $V$ band).  Since $M_V$ and $r_h$
are not correlated, the correlation between $\sigma$ and $I_h$ is not
simply a consequence of the $L - \sigma$ relation, although they are
obviously related.  These correlations probably reflect the formation
processes of globular clusters more than their subsequent dynamical
evolution, and therein lies their significance.  In the case of the
galactic globular clusters, the relation between velocity dispersion
and luminosity ($L - \sigma$) has been already discussed by Meylan \&
Mayor (1986), Paturel \& Garnier (1992), and Djorgovski (1991, 1993a),
and the relation between velocity dispersion and surface brightness
($\sigma - \mu$) by Djorgovski (1993a).

The slope of the $L - \sigma$ relation for globular clusters, viz., $L
\sim \sigma ^{5/3}$, is significantly different from the Faber-Jackson
(1976) relation for ellipticals, or its equivalent for dwarf galaxies,
viz., $L \sim \sigma ^4$.  The slope of the $\sigma - \mu$ relation
for globular clusters has the {\sl opposite sign} from the
corresponding relation for ellipticals and is significantly tighter.
The origin of these correlations is not well understood, but they may
well reflect initial conditions of cluster formation, and perhaps even
be used to probe the initial density perturbation spectrum on a $\sim
10^6~M_\odot$ scale.  Core radii and concentrations play a role of a
``second parameter'' in these correlations.

\vskip 20pt
{\sl A Multivariate Data Analysis approach: The manifold of globular
clusters.}~~ Any globular cluster system suffers numerous evolutionary
processes, of which some may be connected in very complex ways (like,
e.g., the apparent dependence of internal dynamical evolution towards
core collapse on the position of the cluster within the Galaxy).
While the simple approach of examining individual monovariate
correlations of globular cluster parameters provides a useful first
look at the system properties, the complexity of the situation calls
for a more sophisticated approach.  Dealing with a multidimensional
data set, subsets of several observables may be connected in
multivariate correlations.  Simple, monovariate correlations are only
a very special and rare case.  A multivariate statistical analysis may
be used to reveal correlations of a more complex nature (e.g., see
Fig.~4.3 lower-right panel; see also Djorgovski \& Meylan 1994 and
Djorgovski 1959).

The data points occupy a volume in an $N$-dimensional parameter space,
where $N$ is the number of input quantities. If any of the input
quantities are derived from the others, the data will occupy a volume
of dimension $M$, where $M$ is the number of $independent$ input
quantities. $N = 13$ and $M = 9$ in Djorgovski \& Meylan (1994).  If,
in addition, any correlations are present in the data, the
dimensionality of the volume occupied by the data points, also called
the data manifold, will be reduced further.  The effective statistical
dimensionality of the data manifold, $D \leq M$, gives the number of
independent factors which fully describe the data (see the monograph
by Murtagh \& Heck 1987).

The global manifold of cluster properties has a large statistical
dimensionality ($D > 4$), and can be interpreted as a product of many
distinct evolutionary processes shaping the observed properties of
globular clusters at the present day.  A less daunting and more
practical approach is to restrict the analysis to some heuristic
subsets of variables, where a significant reduction of dimensionality
may be found.  Consider only the photometric, structural, and
dynamical parameters of clusters, $M_V$, $c$, $r_c$, $r_h$, $\mu_V
(0)$, and $\sigma$, available for 56 clusters (Djorgovski \& Meylan
1994).  The statistical dimensionality of this manifold is clearly $D
= 3$.  This is exactly what can be expected from a family of objects
described by King (1966) models.  They require 3 input parameters: a
scaling of the core radius, a scaling of the surface brightness, and a
shape parameter.  The fact that the velocity dispersion participates
in the manifold suggests that globular clusters have uniform $(M/L)$
ratios (Djorgovski \& Meylan 1994; see also, e.g., Brosche \& Lentes
1984, \'Eigenson \& Yatsuk 1986, 1989, Fusi Pecci \etal\ 1993a, and
Djorgovski \etal\ 1993).

The statistical dimensionality of globular clusters is greater than
that of elliptical galaxies, for which most global properties form a
statistically two-dimensional manifold.  However, field elliptical
galaxies could be more heterogeneous with a higher dimensionality --
three or four (de Carvalho \& Djorgovski 1992).  Santiago \&
Djorgovski (1993) have used multivariate statistical analysis to study
the relation between the globular cluster content of early-type
galaxies and a number of their observed properties.

\vskip 20pt
{\sl The fundamental plane correlations for globular clusters.}~~ In
the parameter space defined by a radius (core or half-light), a
surface brightness (central or averaged within the half-light radius),
and the central projected velocity dispersion, globular clusters lie
on a two-dimensional surface, a plane if logarithmic quantities are
considered (Djorgovski 1995).  This is analogous to the fundamental
plane of elliptical galaxies (Djorgovski \& Davis 1987, Faber \etal\
1987, Bender \etal\ 1992; see also Djorgovski \& Santiago 1993,
Schaeffer \etal\ 1993).  For the core parameters \rc, $\sigma$, and
$\mu_V(0)$, Djorgovski (1995) obtains a bivariate least-square
solution \rc\ = f($\sigma$,$\mu_V(0)$) which corresponds to the
following scaling law:
$$
r_c\ \sim\ \sigma^{1.8\pm0.15}\ I_0^{-1.1\pm0.1}.
\eqno(4.2)
$$
Alternatively, a bivariate least-square solution $\mu_V(0)$ =
f($\sigma$,\rc), provides a more stable and better fit through surface
brightness (Djorgovski 1995) which corresponds to the following
scaling law:
$$
r_c\ \sim\ \sigma^{2.2\pm0.15}\ I_0^{-1.1\pm0.1}.
\eqno(4.3)
$$
The average of these two solutions is remarkably close to the scaling
law expected from the virial theorem:
$$
r_c\ \sim\ \sigma^{2}\ I_0^{-1}\ (M/L)^{-1}.
\eqno(4.4)
$$
Thus, Eqs.~4.2 and 4.3 are consistent with globular cluster cores
being virialized homologous systems with a constant \ml\ ratio. The
corresponding scaling laws on the half-light scale are different, but
are nearly identical to those derived from the fundamental plane of
elliptical galaxies.

Consequently, the characteristic radii, surface brightness, and
central velocity dispersion for globular clusters form statistically
two-dimensional manifolds, both on the core and half-light scales.
This fundamental plane of globular cluster properties produces the
best correlations known for these stellar systems.

{\sl Correlations for globular clusters in M31.}~~ Similar
correlations, involving $\sigma$, $M_V$, $\mu_V (0)$, and $\langle
\mu_V \rangle _h$, have been obtained recently for a sample of 21
globular clusters in our neighboring galaxy M31, the Andromeda galaxy
(Djorgovski \etal\ 1996).  These globular clusters follow the same
correlations between velocity dispersion and luminosity, central, and
average surface brightnesses, as do their galactic counterparts.  This
suggests a common physical origin for these correlations.  They may be
produced by the same astrophysical conditions and processes operating
at the epoch of globular cluster formation in both galaxies.  The very
existence of these excellent correlations, and their quantitative form
as scaling laws, represent challenges and constraints for theories of
globular cluster formation (Djorgovski \etal\ 1996).

\vskip 30pt
\noindent
{\bf 5. Formation of globular clusters}

\vskip 20pt
\noindent
The origin of globular clusters requires a physical explanation in any
cosmological picture.  Globular cluster formation is intricately
linked to galaxy formation and evolution, in a way that is difficult
to disentangle given the potential multiplicity of simultaneously
operating formation scenarios.  Because of their great ages, spatial
distributions, kinematics, and metallicities, globular clusters stand
out as observable clues of the process of galaxy formation and
evolution. The standard picture for galaxies --- they formed from the
gravitational collapse of primordial density fluctuations --- may not
be applicable in the case of globular clusters, since in some
low-redshift galaxies they still appear to be at the stage of
formation.  At variance with our Galaxy, globular clusters in other
galaxies are not always roughly coeval.  Any model should be able to
explain the formation not only of single clusters, but also of systems
of globular clusters, consisting possibly of a few successive
generations.  See the extensive review about ``Galaxy Formation and
the Hubble Sequence'' by Silk \& Wyse (1993).

\vskip 30pt
\noindent
{\sl 5.1 Luminosity function of a globular cluster system}

\vskip 20pt
\noindent
Analysis of the globular cluster luminosity function is important from
the perspective of (i) distance estimates (i.e., the globular cluster
luminosity function itself is taken to be a ``standard candle'') and
(ii) galaxy formation models and the dynamical evolution of globular
cluster systems (Harris 1991, 1996).

In a given galaxy, the number of globular clusters per unit of
magnitude interval, $\phi(m)$, is the luminosity function of the
globular cluster system, which can also be described in term of
absolute magnitude, $\phi(M)$. The globular cluster system luminosity
functions now available for several galaxies show that $\phi(m)$ can
be simply and accurately described by a gaussian distribution,
$$ 
\phi(m)~dm = A~\exp[-(m-m_0)^2/2\sigma^2]~dm, \eqno(5.1) 
$$
where $A$ is the simple normalization factor representing the total
population $N_t$, $m_0$ is the mean or peak (turnover) magnitude of
the distribution, and $\sigma$ is the dispersion.

To first order, globular cluster system luminosity functions in
different galaxies can then be compared through the two parameters
$M_0$, the absolute magnitude at the turnover, and $\sigma$, the
dispersion.  Over a broad range of systems (Hubble type), the turnover
absolute magnitude $M_0$ is nearly independent of parent galaxy size
and environment.  For 138 galactic globular clusters, Abraham \& van
den Bergh (1995) obtain $\langle M_0\rangle$ = --7.41 \pmm\ 0.11 mag
and $\sigma$ = 1.24 mag.  An unweighted mean for the $M_0$ of the
galaxies in Table~2 of Harris (1991) yields $\langle M_0\rangle$ =
--7.13 \pmm\ 0.43 mag.  The intrinsic dispersion $\sigma$ may be
systematically a bit larger for the giant ellipticals (for which the
best functional fits are reached consistently at $\sigma$ $\simeq$
1.4) than for the other systems (for which $\sigma$ $\simeq$ 1.2 seems
preferable).  The uniformity in $\langle M_0\rangle$ is all the more
remarkable when we consider that the galaxies studied represent at
least three distinguishable different processes of galaxy formation
(dwarf ellipticals, giant ellipticals, and the spheroids of disk
galaxies) as well as a large range of dynamic erosion mechanisms.

The first order similarity of the globular cluster luminosity function
from galaxy to galaxy has become increasingly well justified from a
purely observational point of view.  The physical processes which
might have dictated the formation and evolution of a ``universal''
globular cluster luminosity function are not yet understood, but
several possibilities on theoretical grounds now exist (e.g., Fall \&
Rees 1988, Harris 1991, Jacoby \etal\ 1992).

\vskip 30pt
\noindent
{\sl 5.2 Specific frequency of a globular cluster system}

\vskip 20pt
\noindent
A useful quantity allowing the intercomparison of the globular cluster
populations around different galaxies is defined by Harris \& van den
Bergh (1981). It is called the ``specific frequency \Sn'' of globular
clusters and represents the total number of globular clusters per unit
\Mv\ = -- 15 of host galaxy luminosity, 
$$ 
S_N \equiv\ N_t~10^{0.4(M_v+15)}, \eqno(5.2) 
$$
where $N_t$ is the total number of clusters integrated over the entire
globular cluster luminosity function and \Mv\ is the absolute visual
magnitude of the galaxy.

Although specific frequencies show very large galaxy-to-galaxy
variations, there is a clear tendency for \Sn\ to increase along the
sequence from late-type to early-type galaxies.  Characteristic values
of \Sn\ range from \Sn\ \lsim\ 1 for spiral and irregular, to \Sn\
$\simeq$ 2-3 for normal elliptical galaxies in low-density
environments, to \Sn\ $\simeq$ 5-6 for ellipticals located in rich
clusters; cD galaxies located at the centres of rich clusters have the
largest known specific globular cluster frequencies, typically \Sn\
$\simeq$ 10-20.  The prototype of these high-\Sn\ galaxies is NGC~4486
$\equiv$ M87, the central cD in the Virgo cluster, which has a
population of at least 15,000 globular clusters, with \Sn\ = 14.  Our
Galaxy has \Sn\ = 0.5.  The fact that the \Sn\ values of elliptical
galaxies in rich clusters are systematically higher than those of
their counterparts in low-density regions, suggests that the local
galactic environment plays a key role, along with the galaxy type, in
determining globular cluster frequencies.  See Harris (1991).

Several scenarios have been proposed to explain the origin of the
observed systematic variations in globular cluster specific frequency
(see van den Bergh 1993d, 1995b, and Hesser 1993 for interesting
discussions):

(1) Harris (1981) and van den Bergh (1982) suggest that elliptical
galaxies might have been, for some unspecified reason, more efficient
than spiral galaxies at forming globular clusters, and that the higher
\Sn\ values for ellipticals in rich clusters compared to those in the
field may be accounted for by assuming that the latter had experienced
a greater number of past mergers with low-\Sn\ spiral galaxies.  

(2) Fabian \etal\ (1984) and Fall \& Rees (1985) suggest that globular
clusters might form from gas which condenses out of cooling flows in
the dense cores of rich galaxy clusters.  This accounts for the high
\Sn\ values of some cD galaxies in galaxy cluster cores, but fails to
explain why some cD galaxies which appear to sit in the middle of
large cooling flows have ``normal'' \Sn\ values while other high-\Sn\
galaxies reside in rich clusters which have no cooling flows at
present.

(3) Considering that central cluster galaxies, cD galaxies in
particular, are thought to have grown by cannibalism and/or mergers,
and since normal cluster galaxies have much lower \Sn\ values than
central galaxies, van den Bergh (1984, and references therein) argues
that such mergers will reduce the \Sn\ values of the central galaxies.
The specific frequencies of central cluster galaxies must therefore
have been even greater originally than they are at present.  Van den
Bergh suggests that ``central galaxies in rich clusters were special
{\sl ab initio}''.

(4) Muzzio (1987, and references therein) explored the possibility
that globular clusters might be stripped from the outer halos of
galaxies in the dense environments of rich clusters, and captured
later on by massive galaxies residing at the bottom of the galaxy
cluster potential well.  However, their $N$-body simulations indicate
that the magnitude of this effect is rather small.  See also West
\etal\ (1995).

(5) There is increasing observational evidence (e.g., Schweizer 1987,
Holtzman \etal\ 1992, 1996) which supports the hypothesis that
globular clusters form during the interactions or mergers of galaxies
(Ashman \& Zepf 1992, Zepf \& Ashman 1993, and references therein).
Elliptical galaxies presumably underwent more frequent merging than
spiral galaxies, and accordingly would be expected to have more
abundant globular cluster systems.

(6) Zinnecker \etal\ (1988) and Freeman (1990, 1993) suggest that many
of the globular clusters seen around high-\Sn\ galaxies may actually be
the surviving cores of nucleated dwarf elliptical galaxies.  Tentative
support for this view comes from observational similarities --- such
as luminosities, integrated colors, and velocity dispersions ---
between nucleated cores and globular clusters.  See the numerical
simulations by Bassino \etal\ (1994). 

Interestingly enough, there is plethora of scenarios for forming
globular cluster systems; nevertheless, none of them is able to
explain without major ad hoc tuning why the faintest galaxy known to
have globular clusters has, by far, the highest known specific
frequency value, viz. \Sn\ = 73.  It is the Fornax dwarf spheroidal
galaxy (dSph), which has five globular clusters for an absolute
magnitude \Mv\ = --12.3 (Harris 1991).  Is the Fornax dwarf spheroidal
galaxy, located at about 130 kpc from the centre of our galaxy, a
genuine and unique exception?  Is such a unique faint low-density
galaxy, which would be hardly detectable around M31, located by chance
in our galactic neighborhood?  See Minniti \etal\ (1996) for globular
clusters around dwarf elliptical galaxies.

\vskip 30pt     
\noindent
{\sl 5.3 Globular cluster formation models}

\vskip 20pt
\noindent
A fully consistent model for globular cluster formation has proven to
be a formidable theoretical challenge and is still missing.
Nevertheless, some scenarios/models have been developed during the
last few decades, and represent steps towards a general understanding
of the way globular clusters (and galaxies) form.  The studies can be
sorted into two broad families: (i) globular clusters were the first
condensed systems to form in the early universe (Peebles \& Dicke
1968) or during conditions which existed only in protogalactic epochs
(Fall \& Rees 1985), (ii) globular clusters originated in larger
star-forming systems that later merged to form the present galaxies
(Larson 1993, 1996).

The conventional picture (von Weizs\"acker 1955) describes the
globular clusters as formed along with the halo field-population stars
during the initial collapse of the protogalaxy, a collapse described
as rapid, smooth, and homogeneous in Eggen, Lynden-Bell, \& Sandage
(1962; but see Sandage 1990). 

This view was challenged by Peebles \& Dicke (1968) who consider the
expected properties of the first bound systems to have formed out of
the expanding universe.  They point out that the coincidence between,
(i) the properties of globular clusters, and (ii) the computed mass
and estimated radius of objects at the time of first fragmentation
into stars, argues strongly for the general validity of the view that
these first systems are protoglobular clusters.  The predicted masses
and radii, of order 10$^6$ \msun\ and 10 pc, are typical of the
observed values for globular clusters.  Consequently, globular
clusters would reflect the Jeans mass at recombination, when the
temperature dropped below 10$^4$ K and the mean density of the
universe was about 10$^9$ times higher than at present.  In other
words, the smallest gravitationally unstable clouds produced from
isothermal perturbations just after recombination could be identified
as the progenitors of globular clusters.

An interesting refinement of the above scenario is the possibility
that globular cluster formation might have been ``biased'' in the
sense that only those $\sim$ \milm\ peaks in the fluctuating density
field that exceeded some critical global threshold were able to form
globular clusters.  Some aspects of biased globular cluster formation
are presented in Peebles (1984), Couchman \& Rees (1986), and
Rosenblatt \etal\ (1988), although all these studies focus on the
specific case of a universe dominated by cold dark matter.  West
(1993) presents a simple model of biased globular cluster formation
which relates the efficiency of globular cluster formation to both
galaxy type and local environment.  While the magnitude of this effect
is clearly sensitive to assumptions about biasing parameters which are
poorly constrained, this study shows that, for quite reasonable
assumptions about the biasing process, it is possible to reproduce the
observed variations in globular cluster populations remarkably well.
Biased formation may also explain why other globular cluster
properties, such as the luminosity function, appear to be universal.

Fall \& Rees (1985) argue that globular clusters would form in the
collapsing gas of a protogalaxy.  The Jeans mass \MJ\ of a spherical
cloud with a temperature \Tc\ confined by an external pressure \pe\ is
roughly
$$ 
M_J \approx (k T_c / m_H)^2 G^{-3/2} p_e^{-1/2} . \eqno(5.3)
$$
A natural value for \Tc\ is 10$^4$ K, where the radiative cooling rate
drops precipitously, and a natural value for \pe\ is $\rho_g v_g^2$,
where $\rho_g$ is the mean density within a protogalaxy, and $v_g$ is
a typical collapse or virial velocity: the result is then \MJ\ $\sim$
\milm.  Their starting point is the generally accepted view that 
fragmentation and star formation can only occur when the gas is able
to cool in a free-fall time (Rees \& Ostriker 1977, Silk 1977).  Under
these conditions, any gas at the virial temperature, of order 10$^6$
K, will be thermally unstable and will develop a two-phase structure.
Fall \& Rees suggest that the condensation of cold clouds
progressively depletes the hot gas in such a way that its cooling and
free-fall times remain comparable.  The clouds, which have
temperatures near 10$^4$ K and densities several hundred times that of
the surrounding hot gas, are gravitationally unstable if their masses
are of order \milm.  They identify these objects as the progenitors of
globular clusters and speculate on their later evolution.  Such events
must occur at redshifts less than 10 because a thermal instability is
not effective when Compton scattering by cosmic background radiation
is the dominant cooling process.  Some aspects of the work of Fall \&
Rees (1985) complement the suggestion by Gunn (1980) and McCrea (1982)
that globular clusters formed in the compressed gas behind strong
shocks.  In contrast with these previous discussions, Fall \& Rees
emphasize that the clouds must cool slowly at temperatures just below
10$^4$ K to imprint a characteristic mass of order \milm.  They also
show that the heating of much smaller clouds by X-rays from the hot
gas would inhibit the formation of field stars and small clusters
during the initial collapse.  In a follow-up study, Kang \etal\ (1990)
examine in more detail the thermal history of metal-free gas overtaken
by radiative shocks with velocities characteristic of gravitationally
induced motions inside a typical protogalaxy.  See also Ashman (1990)
and Murray \& Lin (1991, 1992).

As an alternative view to the Eggen \etal\ (1962) galaxy formation
scenario, Searle \& Zinn (1978) consider protogalaxies as very lumpy
systems.  Galaxies are built from the hierarchical merging of smaller
subunits.  As a result of the more chaotic nature of the collapse in
the Searle \& Zinn (1978) scenario, the formation process occurs over
a period of a few billion years, several times longer than in the
original Eggen \etal\ (1962) model.  Within the framework of the
Searle \& Zinn (1978) model, it is plausible that the collision and
coalescence of the subunits could lead to conditions appropriate for
the Fall \& Rees (1985) model, as discussed by Kang \etal\ (1990).
However, an emphasis on the lumpy nature of protogalaxies promotes the
consideration of other ideas.  Larson (1986) notes that in observed
star-forming regions, the fraction of the parent cloud that ends up in
stars is small.  In complexes like Orion only about 10$^{-3}$ of the
original cloud will end up in a bound star cluster.  The inference is
that the progenitors of globular clusters must have had masses in
excess of 10$^8$ \msun.  Such objects may be identified with the
protogalaxy lumps of Searle \& Zinn (1978).

Globular clusters may form during the interaction or merger of
galaxies, complicating further the previous scenarios.  Ashman \& Zepf
(1992) and Zepf \& Ashman (1993) suggest that galaxies in which
globular cluster formation is currently occurring are systems which
are interacting with larger galaxies (e.g., the Large Magellanic Cloud
interacting with the Galaxy, see \S5.6 below).  They also describe
indirect evidence that some of the globular clusters in massive
galaxies were formed as the result of interaction or merger of
pre-existing disk galaxies, but such scenarios have difficulties in
explaining the sheer numbers of clusters in elliptical and dwarf
elliptical galaxies.

Some of the above studies are based, among other simplifications, on
the hypothesis that the masses of globular clusters are confined to a
remarkably narrow range, roughly 10$^5$-10$^6$ \msun.  This may be
plausible for the rich globular clusters in the inner part of the
Galaxy, but does not account easily for the properties of the globular
clusters at large galactocentric distances.  In our Galaxy, the
present (possibly dynamically evolved) masses of globular clusters
span a range of more than three orders of magnitude (see \S2).  In
recent years, the mass function of globular cluster systems, which has
an extremely similar shape in all galaxies (Harris \& Pudritz 1994,
McLaughlin \& Pudritz 1996), has emerged as a major clue to the
formation processes.  The number of clusters per unit mass is nearly
constant for masses less than $\simeq$ $10^5 M_{\odot}$, a limit
corresponding to the turnover point in the luminosity function (Harris
1991).  For clusters with higher masses, a simple power-law form $N(M)
\propto\ M^{-\gamma}$ applies extremely well, with exponent $\gamma$
$\simeq$ 1.7 - 1.9 for the spirals and dwarf ellipticals, and $\gamma$
$\simeq$ 1.6 for the giant ellipticals (Surdin 1979, Harris
\& Pudritz 1994, McLaughlin \& Pudritz 1996). Cluster dynamical
evolution (stellar mass loss and tidal shocking) must influence the
low-mass end of the distribution, but the ubiquity of the breakpoint
at $\simeq$ $10^5 M_{\odot}$ and the similar slope at higher masses,
both across a large diversity of environments, suggest that the major
part of the mass function of globular cluster systems must be a
characteristic of the formation process.  These mass-function slopes
$\gamma$ $\simeq$ 1.5 - 2.0 are also similar to the values found for
the mass functions of both the open clusters and the dense molecular
clumps in which they are born.  Harris (1996) suggests that these
observational constraints rule out the above theories in which
globular clusters, assumed to form by thermal instability, have
pregalactic origin or arose in environmental conditions present only
in protogalactic epochs (Peebles \& Dicke 1968, Fall \& Rees 1985,
Rosenblatt \etal\ 1988, Murray \& Lin 1990b, 1992, Vietri \& Pesce
1995).

The alternative is that there is nothing special about globular
cluster formation: it represents only the high-mass tail of the
general process of star cluster formation which is happening nowadays
in any galaxy which contains a decent supply of cool gas (Larson 1990,
1993, 1996, Harris \& Pudritz 1994, Patel \& Pudritz 1994).

\vskip 30pt
\noindent
{\sl 5.4 Collapse, fragmentation, and initial mass function} 

\vskip 20pt
\noindent
A major goal of studies of globular cluster formation is to
understand, through fragmentation, the spectrum of masses with which
stars form, since the initial mass spectrum plays a fundamental role
in determining the observed properties of stellar systems and their
subsequent dynamical evolution. We briefly mention hereafter a few
recent studies related to fragmentation of clouds into stars.

The fragmentation process of molecular clouds has been investigated by
Chi\`eze (1987), taking into account the observed relations $M \propto
R^2$ and $\sigma \propto R^{1/2}$, between the mass $M$, the radius
$R$ and the internal velocity dispersion \sig\ of molecular clouds,
relations first noticed by Larson (1981).  Chi\`eze (1987) shows that
interstellar molecular clouds which are close to gravitational
instability exhibit precisely the same scaling laws, provided they
interact with a constant pressure environment.  He suggests that these
conditions may trigger the fragmentation of clouds.  See also Chi\`eze
\& Pineau des For\^ets (1987) for fragmentation of low-mass molecular
clouds, de Boisanger \& Chi\`eze (1991) for formation of molecular
clumps in an inhomogeneous radiation field, and Renard \& Chi\`eze
(1993) for the behavior of critical Jeans mass close to thermal
instability.  Murray \& Lin (1989a,b) have studied proto-globular
cluster fragmentation in the case of thermal and gravitational
instabilities, respectively.  See also Di Fazio (1986) in the case of
gravitational instabilities.  Myers \& Fuller (1993) find clear
relations, in the form of simple power laws, between the line width of
a dense core observed in the 1.3 cm lines of NH$_3$ and the luminosity
and mass of the most massive stars associated with this core.  From
their study of gravitational formation times and IMF, they predict
infall times equal to 1-2, 4-8, and 1-12 \x\ 10$^5$~yr 
for stars of mass 0.3, 3, and 30 \msun, respectively.

Dynamical mixing in molecular clouds in relation to the origin of
metal homogeneities in globular clusters have been investigated by
Chi\`eze \& Pineau des For\^ets (1989) and Murray \& Lin (1990a).

The difficulty of predicting the initial mass function (IMF) comes
from the fact that a large number of different physical processes are
likely to take place during star formation, including cloud
fragmentation, fragment coalescence, mass accretion in a disk, stellar
wind mass loss, among others. How these processes combine to determine
a final stellar mass at a particular time in a cloud, or to determine
an average mass spectrum in a composite of clouds, is difficult to
simulate in any detail.  See Shu \etal\ (1987) and Adams \& Fatuzzo
(1996).

In one of the first attempts to determine the IMF, Elmegreen (1985)
uses a statistical approach which specifies from physics the mass
distribution of stars in a cloud, but not the mass of an individual
star.  The mass of each individual star is, in such a theory, the
result of a large number of independent events, all of which involve
combinations of randomly chosen parameters.  A combination of
fragmentation and accretion processes in hierarchical groupings of
forming stars may play an important role in the formation of massive
stars (Larson 1982, 1992). As reviewed by Scalo (1986), there is
considerable evidence that molecular clouds have complex hierarchical
structures and are typically filamentary in shape.  A fractal
description of star-forming clouds was first propose by Henriksen
(1986) and then further explored by Dickman \etal\ (1990) and
Falgarone \etal\ (1991).

Prescribed IMF of stars reaching the main sequence have been used by
Fletcher \& Stahler (1994a,b) in order to compute the history of the
luminosity function of young clusters still forming within a molecular
cloud.  In these models, the number of protostars rises quickly but
levels off to a nearly constant value which lasts until the dispersal
of the cloud.

From a theoretical point of view, understanding the origin of the IMF
remains a difficult task, with the result that model predictions are
still in their infancy.  The same is true from an observational point
of view: e.g., although there are some theoretical arguments
predicting that low-mass star formation may be suppressed in regions
of high-mass star formation, the observational constraints remain
inconclusive. Zinnecker (1996) shows that at least in the case of
NGC~3603 there is evidence, from adaptive optics data in $H$ and $K$
bands, that subsolar-mass stars are present. In the case of R136, the
core of the 30~Doradus nebula, the IMF could not be probed below about
2 \msun, but no cutoff has been observed down to this detection limit
(Zinnecker 1996).

\vskip 30pt
\noindent
{\sl 5.5 Early stellar evolution and violent relaxation phase}  

\vskip 20pt
\noindent
From the point of view of dynamics, the most important consequence of
the early evolution of the stars is the accompanying loss of mass,
which tends to unbind the cluster.  Usually, this is modelled as a
sudden loss of mass by each star at the end of its main sequence
evolution, and usually it is assumed that the mass is ejected
instantaneously out of the cluster.  Usually the process is handled in
terms of a lookup table which provides, for a main sequence star of a
given mass, the time and amount of mass loss.  A commonly used
prescription is that adopted by Chernoff \& Weinberg (1990), which was
based on work of Iben \& Renzini (1983) and Miller \& Scalo (1979).

After some early, but still interesting and relevant, investigations
by Angeletti \& Giannone (1977c, 1980), Applegate (1986) was among the
first to revive interest in the dynamical effects of mass loss at the
end of main sequence evolution.  He used a simple model of relaxation
to show how sufficient loss of mass (and the resulting expansion of
the cluster) either delayed the onset of relaxation processes or
exposed the cluster to the danger of disruption by tidal shocking.
More quantitative detail was added to this picture by Chernoff \&
Weinberg (1990), who did a more careful job of modelling relaxation
(by using a Fokker-Planck code), but included only a steady tide.
Their results were qualitatively similar, and showed that the mass
loss by stellar evolution would always disrupt a cluster with a
relatively flat mass function (i.e. $dN\propto m^{-\alpha} dm$ with
$\alpha = 1.5$ over the range 0.4 $<$ $m$ $<$ 15 \msun).  Clusters
with steeper mass functions would survive without disruption provided
that the initial concentration was high enough; they used King models
as initial models and found that a model with initial scaled central
potential $W_0 = 7$ would survive for $\alpha \ge 2.5$.  These results
are clearly dependent on the assumed range of the mass function, and
somewhat more generalised results will be found in Weinberg (1993a)
and Chernoff (1993), where special consideration is given to clusters
that disrupt so quickly that relaxation effects can be ignored.

More careful ($N$-body) modelling of this same problem has been
carried out by Fukushige \& Heggie (1995).  They confirm the
qualitative results of Chernoff \& Weinberg (1990), but find that the
destruction times were underestimated by factors as large as 10 in
some cases.  The problem appears to arise from the fact that the time
scale on which mass is lost by the cluster is not long enough compared
to the crossing time, and this leads to failure of the assumption (on
which earlier workers depended) that the cluster evolves through a
sequence of quasi-equilibrium models.

These models are based on instantaneous mass loss by each star (at the
end of the main sequence lifetime appropriate to its initial mass),
and are of importance for a theoretical understanding of the evolution
of globular clusters.  They may, however, be oversimplified,
especially in the first intensive phase of mass loss.  During this
early phase (roughly the first $10^7$yr), massive clusters may well
have possessed substantial quantities of unejected gas, while those of
lower mass may have generated an outflow in the form of a cluster wind
(Smith 1996).  Such a wind can have the effect of expelling the
residual gas left from the star formation process itself.  The
dynamical consequences of this expulsion have been considered mostly
in the context of open star clusters (e.g. Tutukov 1978, Hills 1980,
Mathieu 1986 and references therein, Lada \& Lada 1991 and references
therein). In the context of globulars, some $N$-body modelling of
these problems has been carried out recently by Goodwin (1996).

Large amounts of irregular mass loss may induce violent changes of the
gravitational field of a newly formed globular cluster.  This phase of
dynamical mixing changes the statistics of stellar orbits on a time
scale of the order of the crossing time ($\sim$ \mily), consequently,
this is an encounterless relaxation phenomenon.  It has been named
violent relaxation and was first addressed by Lynden-Bell (1962,
1967), H\'enon (1964), and King (1966).  The violent relaxation leads
quickly to the smooth light distribution typical of King-Michie
clusters, which are characterized by a steady dynamical evolution with
relaxation due to stellar encounters, leading slowly, after a few Gyr,
to core collapse and/or evaporation.

More recent theoretical discussions about the fundamentals of violent
relaxation are found in Shu (1978, 1987), Tremaine \etal\ (1986),
Kandrup (1987), Tanekusa (1987), Aarseth \etal\ (1988), Funato \etal\
(1992), and Spergel \& Hernquist (1992).

It is known that, in large systems like globular clusters, primordial
binaries are left largely intact by early phases of violent relaxation
(Vesperini \& Chernoff 1996).  Their importance for subsequent
evolutionary stages is one of the main themes of \S9.5.

\vskip 30pt
\noindent
{\sl 5.6 Formation of globular clusters in the Magellanic Clouds} 

\vskip 20pt  
\noindent
The spread in age among the Magellanic Clouds clusters is much larger
than for those in the Galaxy, but a recent study of Hodge~11, a
globular cluster in the Large Magellanic Cloud, points towards an age
identical to that of the galactic globular M92 (Mighell \etal\ 1996).
Consequently, the oldest star clusters in the Large Magellanic Cloud
and in the Galaxy appear to have the same age.  From the histogram of
the ages of the star clusters in the Large Magellanic Cloud (LMC), it
is as conspicuous as it is surprising to see that about one half of
all clusters are younger than $\sim$ 10$^8$~yr (van den Bergh 1981,
Elson \& Fall 1988). It looks as if the LMC managed to produce in the
last $\sim$ 10$^8$~yr as many clusters as during the last $\sim$
10$^{10}$ yr.  Are we lucky enough to witness a burst of formation of
star clusters?  Probably not.  Most of the currently forming star
clusters are not massive enough to remain in a magnitude limited
catalog for more than $\sim$ 10$^8$~yr.  Dynamical disruption as well
as fading away bring their integrated luminosity below the threshold
of present catalogs.

But there is little doubt that rich star clusters, which have no
equivalent in our Galaxy, are currently forming in the Magellanic
Clouds.  However, there has been some debate about two points:

\vskip 20pt
\noindent
{\sl First, it is not sure that the very rich and young LMC clusters
are truly the progenitors of their older globular counterparts.}  The
masses of galactic globular clusters span quite a broad range, from
less than 10$^4$ \msun\ to over 10$^6$ \msun.  Many of the young LMC
clusters fall comfortably in this range: e.g.,
\mtot(NGC~1850) = 6 $\times$ 10$^4$ \msun\ (Fischer \etal\ 1993a), 
\mtot(NGC~1866) = 6 $\times$ 10$^5$ \msun\ (Lupton \etal\ 1989), 
\mtot(NGC~1866) = 1 $\times$ 10$^5$ \msun\ (Fischer \etal\ 1992a), 
\mtot(NGC~1978) = 9 $\times$ 10$^5$ \msun\ (Meylan \etal\ 1991c), 
\mtot(NGC~1978) = 2 $\times$ 10$^5$ \msun\ (Fischer \etal\ 1992b), 
\mtot(NGC~2164) = 2 $\times$ 10$^5$ \msun\ (Lupton \etal\ 1989), and
\mtot(NGC~2214) = 4 $\times$ 10$^5$ \msun\ (Lupton \etal\ 1989). 
The differences between the estimates concerning one given cluster,
such as NGC~1978, are model dependent and come also from the fact that
radial velocities of individual stellar members of Magellanic clusters
suffer from crowding problems, leading to underestimation of the true
velocity dispersion.  A greater concern resides in the fact that the
young objects in the LMC are more closely analogous to the open
clusters of the Milky Way, as suggested by similarities in the cluster
luminosity functions (Elson \etal\ 1987a, van den Bergh 1993d,
1995a,b).  However, the conspicuous difference between young
Magellanic and open galactic cluster systems is in the presence of
young massive clusters in the LMC, such as NGC~1866 (young $\sim$
10$^8$~yr, rich $\sim$ 10$^5$ \msun, and luminous $\sim$ 10$^6$
\lsun), which has no known counterpart in the disk of the Milky Way.
This cluster looks like a genuine globular cluster which may be
similar to NGC~1835 in 10~Gyr. The LMC seems able to make one genuine
globular cluster in $\sim$ 10$^8$~yr.  Considering in the LMC the
number of young poor clusters to be about 100, there is a cluster
formation efficiency of about 1 poor cluster per 1~Myr, and 1 rich
cluster per 100~Myr.  The time scale to produce an even more massive
cluster (a few times 10$^5$ \msun) is longer. It may be very well that
there is no special epoch in the LMC history, and the age histogram of
the star clusters (of all richness) looks similar over the Gyr.  There
is just a continuous replenishment of new small clusters as those
already formed fade away, and only rarely a fairly massive cluster is
formed which manages to remain brighter than the threshold for many
years (Renzini 1991).  See Fujimoto \& Noguchi (1990) for an
interesting investigation of dynamical conditions for globular cluster
formation, in the specific case of the Magellanic Clouds, by studying
hydrodynamical collisions between gas clouds and their subsequent
coalescence.

Current observations are consistent with the idea that both the
galactic disk and the LMC are currently forming star clusters, but
only the LMC contains young clusters with masses characteristic of
globulars (see also Renzini 1991).  Dynamical simulations including
the combined effects of relaxation, and tidal and binary heatings are
consistent with suggestions that the shape of cluster luminosity
functions results from evaporation and disruption of low mass clusters
(Chernoff \& Weinberg 1989, Murali \& Weinberg 1996).  Since the less
massive Magellanic clusters are more susceptible to disruption by
various evolutionary processes, the LMC cluster luminosity function
will evolve so that it will more closely resemble the luminosity
function of the galactic globulars, whose luminosity function has
largely been shaped by dynamical selection (Murali \& Weinberg 1996).

\vskip 20pt
\noindent
{\sl Second, it has been thought that the old LMC clusters are
significantly less massive than their galactic globular cluster
counterparts.}  Meylan (1988b), Dubath \etal\ (1990) show that, in the
case of NGC~1835, a projected velocity dispersion \sigp (core) = 10.1
\pmm\ 0.2 \kms\ provides a total mass \mtot\ = 1.0 \pmm\ 0.3 $\times$
10$^6$ \msun, corresponding to a global mass-to-light ratio \mlv\ =
3.4 \pmm\ 1.0 \mlvso.  This study shows that when the same kind of
dynamical models (King-Michie) constrained by the same kind of
observations (surface brightness profile and central value of the
projected velocity dispersion) are applied to an old rich Magellanic
globular cluster, viz., NGC~1835, the results are similar to those
obtained in the case of galactic globular clusters.  Consequently, the
rich old globular clusters in the Magellanic clouds could be quite
similar (in mass and \mlv) to the rich globular clusters in the
Galaxy.

\vskip 20pt
\noindent
{\sl Is the 30~Doradus Nebula a globular cluster progenitor?}~~ If a
genuine globular cluster were forming right now in the Local Group,
there would be probably only one place where this could be happening:
within the 30 Doradus Nebula.  The LMC star cluster NGC~2070 is
embedded in the 30~Doradus nebula, the largest HII region in the Local
Group (see Meylan 1993 for a review).  The physical size of NGC~2070,
with a diameter $\sim$ 40 pc, is typical of old galactic and
Magellanic globular clusters.  The size of NGC~2070 is also comparable
to the size of its nearest neighbor, the young globular cluster
NGC~2100, which lies about 53\arcm\ southeast of 30~Dor. With an age
of $\sim$ 4 $\times$ 10$^6$~yr (Meylan 1993, Brandl \etal\ 1996),
NGC~2070 appears slightly younger than NGC~2100 which has an age of
$\sim$ 12-16 $\times$ 10$^6$~yr (Sagar \& Richtler 1991). Their masses
are also quite similar. For the 30~Dor cluster, Churchwell (1975)
estimates the mass of ionized gas larger than 3 $\times$ 10$^5$ \msun\
and the total mass contained in the stellar cluster larger than 4
$\times$ 10$^5$ \msun; extrapolations of the IMF exponent obtained for
the high mass stars give total masses from 3 $\times$ 10$^4$ to 6
$\times$ 10$^5$ \msun\ within 100\arcs\ (Meylan 1993); Malumuth \&
Heap (1994) obtain, from the Hubble Space Telescope (HST) data, a
lower limit to the mass within 17.5\arcs\ equal to 2 $\times$ 10$^4$
\msun, while Brandl \etal\ (1996), from data obtained with the ESO
adaptive optics system COME\thinspace ON+, estimate, from the total
$K$ magnitude, the mass within 20\arcs\ equal to 3 $\times$ 10$^4$
\msun, with an upper limit on this value equal to 1.5 $\times$ 10$^5$
\msun.  A star cluster with a mass of this range and a typical
velocity dispersion of $\sim$ 5 \kms\ would be gravitationally bound,
a conclusion not immediately applicable to NGC~2070 because of the
important mass loss due to stellar evolution experienced by a large
number of its stars (see Kennicutt \& Chu 1988 and below). For
NGC~2100, Lee (1991) finds a total mass \mtot\ = 5 $\times$ 10$^5$
\msun.  All mass determinations for these two very young Magellanic
clusters provide results typical of masses of old galactic globular
clusters. 

Mass segregation may have been observed in NGC~2070.  Brandl \etal\
(1996) determine, for stars more massive than 12 \msun, a mean
mass-function slope $x$ = 1.6 [$x$(Salpeter) = 1.35], but this value
increases from $x$ = 1.3 in the inner 0.4 pc to $x$ = 2.2 outside 0.8
pc.  The fraction of massive stars is higher in the centre of R136,
the core of NGC~2070.  This may be due to a spatially variable initial
mass function, a delayed star formation in the core, or the result of
dynamical processes that segregated an initially uniform stellar mass
distribution.

In their study of the formation and evolution of rich star clusters,
Kennicutt \& Chu (1988) use a simple cluster evolution and
photoionization model and show that for a cluster like NGC~1866, its
initial ionizing luminosity is consistent with the actual ionization
requirement of the 30~Dor Nebula.  Furthermore, in their later study
of the kinematic structure of this object, Chu \& Kennicutt (1994)
reach the conclusion that 30~Dor and its vicinity will evolve into a
supergiant shell as seen in nearby galaxies (see also Hunter \etal\
1995).

\vskip 30pt
\noindent
{\sl 5.7 Formation of globular clusters in other nearby galaxies} 

\vskip 20pt
\noindent
Kennicutt \& Chu (1988) have reviewed the question of the formation of
young globular clusters and their possible association with giant HII
complexes in nearby galaxies.  They define a young globular as any
object with $B-V$ $<$ 0.5 and a mass exceeding 10$^4$ \msun. For the
two massive spiral galaxies in the Local Group, the number of young
globulars is negligible (zero in the Milky Way; a few marginal
candidates in M31).  The LMC has a large number of young globulars,
whereas the young objects in the SMC are close to the adopted
luminosity threshold. Other galaxies in the Local Group seem to
contain young clusters.

Outside of the local group, even more massive clusters are apparently
forming in starburst galaxies, especially in interacting and merging
systems.  High angular resolution observations of several merging
galaxies have been obtained with the HST.  Holtzman \etal\ (1992)
discovered a population of about 60 bright blue pointlike sources
concentrated within 5 kpc from the nucleus of NGC~1275, a galaxy
thought to be the result of a recent merger.  The brightest object has
an absolute magnitude \Mv\ $\sim$ --16, with typical \Mv\ from --12 to
--14.  Ages are of the order of a few 100 \x\ \mily\ or less, with
masses between 10$^5$ and 10$^8$ \msun.  Subsequent spectroscopic data
obtain by Zepf \etal\ (1995) for the brightest of these sources give
an age of about 0.5 Gyr. Whitmore \etal\ (1993) observed in NGC~7252,
another merger remnant, a concentrated population of 40 bright blue
pointlike sources with mean \Mv\ $\sim$ --13 and mean age of about 100
\x\ \mily.  O'Connell \etal\ (1994) observed three such bright
clusters in NGC~1569 and NGC~1705; they have \Mv\ between $\sim$
--13.3 and --14.1 and ages larger then 15 \x\ \mily.  In NGC~4038/4039
(the Antennae), the prototypical example of a pair of colliding
galaxies, Whitmore \& Schweitzer (1995) observed a population of 700
bright blue pointlike sources.  The brightest objects have absolute
magnitudes \Mv\ $\sim$ --15, and the mean value is \Mv\ = --11. The
brightest bluest clusters have ages less than 10 \x\ \mily.  In M82, a
starburst galaxy, probably as a consequence of tidal interactions with
its neighbor, O'Connell \etal\ (1995) found over 100 bright blue
pointlike sources, with mean \Mv\ = --11.6.  See Holtzman \etal\
(1996) for star clusters in interacting and cooling flow galaxies and
Forbes \etal\ (1996) for star clusters in the central regions of
kinematically distinct core ellipticals.

In relation to globular clusters formation theory, it is worth
mentioning that these extremely luminous young stellar aggregates
found in all these interacting/merging galaxies have sizes and
estimated masses which overlap with those of the globular
clusters. Their luminosity functions have a power-law form similar to
those of the open clusters and the more massive globular clusters (see
\S5.3 above).  Systematic spectroscopy of these objects would help 
estimating the fraction of these bright blue sources which may evolve
into genuine old globular clusters.  Present spare spectroscopy data
strongly support the notion that they are young globular clusters
formed during interactions or mergers (Schweitzer \& Seitzer 1993,
Zepf \etal\ 1995).

\vskip 30pt
\noindent
{\bf 6. Observations providing dynamical constraints}

\vskip 20pt
\noindent
Most dynamical models can be constrained by the same kind of
observations, viz. the surface brightness profile and the velocity
dispersion profile.  These profiles can be constructed from the
following observational data: (i) density profiles from star counts,
(ii) density profiles from surface brightness measurements, (iii)
velocity dispersion profiles from proper motions, (iv) velocity
dispersion profiles from stellar radial velocities, and (v) core
velocity dispersions from integrated-light spectra.  It is worth
mentioning at this stage that two models intrinsically very different
may have very similar surface density profiles (see \S7.7).  Another
important input parameter is the mass function, which can be reliably
obtained from observations only for the upper part of the main
sequence, although the HST with WFPC2 provides significant improvement
in the sampling of the luminosity function down to stars of about 0.1
\msun\ (see, e.g., Richer \etal\ 1995 and King \etal\ 1995).

\vfill\eject 

\vskip 20pt
\noindent
{\sl 6.1 Star counts and surface brightness profiles from
photographic plates, photomultipliers, and CCD images} 

\vskip 20pt
\noindent
{\sl Star counts from photographic plates.}~~ Around the end of the
19$^{\rm th}$ century, the advance of photographic techniques applied
to astronomy gave astronomers the opportunity to shift from
descriptive work to more quantitative, scientifically more objective
studies.  It is from a photographic plate of \cent, taken at Arequipa
(Peru), with an exposure time of two hours, that Bailey (1893) made
what probably was the first extensive star count study of a galactic
globular cluster.  These data were used by Pickering (1897) in the
first important comparisons between observed and empirically guessed
theoretical profiles.  During the following decades, till the
development of CCDs, star counts from photographic plates were
intensively used in order to study the distribution of stars in
clusters.  All these data represent very heterogeneous material which
is scattered in the literature and not easily accessible.

Refinements in the theoretical understanding of cluster dynamics led
to a strong need for extensive and homogeneous star count data.  King
(1966) provided, for the first time, a grid of models with different
concentrations \conc\ that approximately incorporated the three most
important elements governing globular cluster structure: dynamical
equilibrium, two-body relaxation, and tidal truncation (\rt\ and \rc\
are the tidal and core radii, respectively; see \S7.5 below).  These
models, being spherical, isotropic, and composed of stars with a
single mass, were the simplest that might acceptably represent the
star count data.  King \etal\ (1968) demonstrated the success of these
models when they published an enormous amount of observational data,
viz., star counts for 54 galactic globular clusters.

No similar effort, in bringing a coherent and large data base for star
counts in globular clusters, has enlarged and improved the earlier
work by King \etal\ (1968), until the recent publication by Grillmair
\etal\ (1995a).  They obtained deep two-color photographic photometry in
order to examine the outer structure of 12 galactic globular clusters,
using star count analyses.  They find that most of their sample
clusters show, in their surface density profiles, extra-tidal wings
whose profiles have forms consistent with recent numerical studies of
tidal stripping of globular clusters (Grillmair \etal\ 1995b).
Two-dimensional surface density maps are consistent, for several
clusters, with the expected appearance of tidal tails, with the
allowance for the effects of orbit shape, orbital phase, and
orientation of our line of sight.  The extra-tidal material is
identified with stars still in the process of being removed from the
clusters, limiting the accuracy of the determination of the tidal
radius. Grillmair \etal\ (1995a) conclude that the stars found beyond
the best-fit values of \rt\ are probably unbound as a result of
previous and ongoing stripping episodes.  They speculate that globular
clusters in general have no observable limiting radius.

\vskip 20pt	
\noindent
{\sl Surface brightness profiles from photomultipliers.}~~ The
advantage of the large fields of photographic plates, well suited for
star counts in the outer parts of globular clusters, was
counterbalanced by the poor spatial resolution which, because of
crowding, prevented the resolution of the inner parts --- within a few
core radii --- of most globular clusters.  A way out of this dilemma
has been the observation of the integrated light, providing surface
brightness profiles.  This was made possible because of the
development of photoelectric devices that measure the surface
brightness through different apertures.  Photoelectric techniques
applied to astronomy were developed, in part, by J. Stebbins and A.E.
Whitford (see, e.g., Stebbins \& Whitford 1943), and studies of the
cores of globular clusters were published, starting in the 1950's by,
e.g., Gascoigne \& Burr (1956), Kron \& Mayall (1960), and more
recently by Illingworth \& Illingworth (1976), Da Costa (1979), Kron
\etal\ (1984), and Kron \& Gordon (1986).  

Dickens \& Woolley (1967) were the first to employ extensive
photometric data with dynamical modelling in their study of \cent.  A
composite profile made by combining a surface brightness profile for
the inner part of the cluster with star counts in the outer part
allowed Da Costa \& Freeman (1976) to show that single-mass, isotropic
King models are unable to fit the entire profile of M3.  They
generalized these simple models to produce more realistic multi-mass
models with full equipartition of energy in the centre.

\vskip 20pt
\noindent
{\sl Star counts and surface brightness profiles from CCD images.}~~
It is only with the development, in the eighties, of CCDs (Charge
Coupled Devices) for astronomical applications, coupled with software
improvement for photometry in crowded fields (e.g., DAOPHOT by Stetson
1987, DOPHOT by Schechter \etal\ 1993), that the brightest stars in
the cores of all globular clusters, even with the highest
concentrations, have been at last fully resolved, even in the inner
few seconds of arc.  The near legendary core of the globular cluster
M15 = NGC~7078, which has long been the prototype of the
collapsed-core globular clusters, unveiled at least part of its inner
structure.  The first look at the inner core, with 0.55\arcs\ seeing,
was published by Auri\`ere \& Cordoni (1981a,b) who partly resolved
the three bright central stars.  Images with a FWHM resolution of
0.35\arcs, taken by Racine \& McClure (1989) with the High-Resolution
Camera of the Canada-France-Hawaii Telescope (CFHT), and, in
particular, images with a FWHM of 0.08\arcs\ obtained with the HST, by
Lauer \etal\ (1991) and Yanny \etal\ (1993, 1994a) with the Planetary
Camera, show that most of the former central cusp in luminosity was
due to a group of a few bright stars, although post-refurbishment HST
data exhibit a star-count profile which continues to climb within
2\arcs\ (Sosin \& King 1996, and \S9.2).

Globular clusters are known to contain numerous pulsars, bright X-ray
sources, and a growing number of dim X-ray sources.  Accurate
positions are needed for providing possible counterparts to X-ray
sources (Paresce \etal\ 1992; King \etal\ 1993). The positions of
these objects within the clusters can give useful information about
their formation process, as well as about mass segregation.  Pulsars
in clusters can also be used to probe the gravitational potential of
the cluster, since changes in the observed period can be attributed to
Doppler shifts induced by gravitational acceleration of the pulsar
itself (Phinney 1992).  The use of such ways to investigate the
internal structure of a globular cluster requires the position of the
optical core of the cluster to be well defined. In several cases,
particularly for the high-concentration (collapsed) clusters and for
the highly obscured clusters towards the galactic centre, the
uncertainty in the optical position of the cluster core is now the
limiting factor in the determination of the offset between the core
and the radio or X-ray position.  E.g., Calzetti \etal\ (1993) report
a difference of 6\arcs\ between the positions of the ``dynamical'' and
``light'' centres of \tuca, a difference which is most probably due to
the two methods used.  The relative accuracy of methods for
determining the position of the centres of globular clusters has been
investigated by Picard \& Johnston (1994) using a testbed of
artificial clusters.  They also develop a new and more robust method
for determining the clusters' centres, giving now positions of the
centres with an accuracy of about 1\arcs\ (Picard \& Johnston 1996).

Apart from the Galaxy, investigation of the surface brightness
profiles of globular clusters has been done so far only in nearby
galaxies, like the Large and Small Magellanic Clouds, the Fornax dwarf
spheroidal galaxy, and M31.  In the Galaxy and M31, the effects of
interactions between the clusters and the galaxian central potential,
the disk potential, and giant molecular clouds are apparent from the
cutoff seen in the radial light and density profiles.  The definition
of the observed or theoretical tidal cutoff is not a simple issue.
The interpretation of the light profile in the outer parts of clusters
can depend on assumptions about the isotropy of the velocity
distribution and on the net angular momentum of the cluster outskirts
(see Weinberg 1993a).  The analysis of the profiles consists, most of
the time, of a comparison with single- or multi-mass King models,
which provides estimates of the core radius, tidal radius, and the
concentration.  Departures from King profiles are observed, which
hamper the quality of the fit and its interpretation (Grillmair \etal\
1995a).

Photoelectric, electronographic, and, especially, CCD observations
have allowed a systematic investigation of the inner surface
brightness profiles of 127 galactic globular clusters and the
observational confirmation of the reality of the core collapse
phenomenon (Djorgovski \& King 1986, Chernoff \& Djorgovski 1989).
They sort the profiles into two families: (i) the King model clusters
and (ii) the collapsed-core clusters.  See \S9.2 for a general
discussion of the use of surface brightness profiles in the study of
collapsed-core clusters.  A new compilation of basic data and
references for 143 galactic globular clusters, along with new deduced
King-model structural parameters for 101 of them, are contained in the
appendices and tables of the proceedings of the 1992 Berkeley workshop
(Djorgovski \& Meylan 1993).  Trager \etal\ (1995) present a catalog
(available in the AAS CD-ROM series) of surface-brightness profiles of
125 galactic globular clusters, the largest such collection ever
gathered, mostly from CCD data.  All but four of these
surface-brightness profiles have photometric zero points.  Central
surface brightness, King-model concentrations, core radii, and
half-light radii are derived.

The results of two surveys for structural parameters in the surface
brightness profiles of young and old clusters in the Magellanic Clouds
are given in Meylan \& Djorgovski (1987) and Mateo (1987).  They
emphasize the bumpy surface brightness profiles of the young clusters
and mention the possible collapsed character of three old LMC globular
clusters, viz., NGC~1916, NGC~2005, and NGC~2019 (see \S9.2).  Elson
\etal\ (1987a) present the surface brightness profiles of 10 rich star
clusters in the LMC, with ages between 8 and 300 \x\ \mily.  Most of
the clusters do not appear to be tidally truncated, and a plausible
theoretical interpretation is that expansion of a newly formed cluster
either through mass loss or during violent relaxation could lead to
the formation of a halo of unbound stars.  From calculations including
the tidal field of the LMC, they find their clusters extending beyond
their tidal radii, with up to 50\% of the total masses in unbound
halos.  In a subsequent study of 35 rich star clusters in the LMC,
with ages between 1~Myr and 10~Gyr, Elson \etal\ (1989) find that the
core radii increase from $\sim$ 0 to $\sim$ 5 pc between 1~Myr and
1~Gyr, and then begin to decrease again. They suggest that the
expansion of the cores is probably driven by mass loss from evolving
stars.  See also Elson (1991, 1992).

In contradistinction, the effects on the structure of clusters in the
less disruptive milieu of the Fornax dwarf spheroidal galaxy may be
visible, providing clues to the initial conditions of the formation of
globular clusters and to the extent to which these conditions are
mirrored in the structures of those clusters as seen today.  See
Buonanno \etal\ (1985b) for the color-magnitude diagrams of all 5
Fornax dSph globular clusters.  Rodgers \& Roberts (1994) describe the
observations of the surface brightness profiles of the five brighter
clusters in the Fornax dwarf galaxy.  They appears to fall into two
groups. Clusters \#1 and \#2 have similar core radii and follow
truncated single-mass King model profiles.  Clusters \#3, \#4, and \#5
have similar smaller core radii and extended halos not well fitted by
King models.  These groupings correlate neither with the differing
chemical compositions of the clusters nor with their horizontal-branch
morphology, adding further evidence that cluster formation and
evolution in Fornax was a complex and diverse process.  It is worth
mentioning that all five clusters in the Fornax dwarf galaxy are old
globulars ($\tau$ \gsim\ 10~Gyr) contrary to the LMC clusters studied
in Elson \etal\ (1989) which have ages between 1~Myr and 10~Gyr.  From
X-ray imaging of Fornax with ROSAT, Gizis \etal\ (1993) observe no
source in the energy range 10$^{36}$-10$^{38}$ erg~s$^{-1}$.  The
low-density environment of the dwarf galaxy evidently does not produce
a population of accreting neutron stars through star-star collisions,
and no such source is observed in the Fornax dSph globular clusters.

The HST has provided the possibility of studying the surface
brightness profiles of globular clusters in the nearby spiral galaxy
M31.  Bendinelli \etal\ (1993) and Fusi Pecci \etal\ (1994) report on
the comparison of the structure parameters of M31 globular clusters
with those of the galactic globular clusters which shows strong
similarities between the two cluster populations.  

\vfill\eject 

\vskip 30pt
\noindent	
{\sl 6.2 Proper motions, stellar radial velocity dispersion, and
velocity dispersion from integrated-light spectra}  

\vskip 20pt
\noindent
The acquisition of kinematic data brings a great deal of information
on the amount and the distribution of mass in globular clusters.
These data have been acquired much more recently than surface
brightness profiles, since a few technological challenges had to be
mastered first: although it is now the case for more than two decades
for the radial velocities, it is no yet entirely so for the proper
motions (Meylan 1996).

{\sevenrm 
{{\hsize=12truecm
{
$$
\table
\tablewidth{10.0truecm}
\tablespec{\l\c\l}
\body{
\header{{\bf Table~6.1:} Internal proper motions in galactic globular clusters}
\skip{10pt}
\hdoubleline
\skip{1pt}
\skip{1pt}
& ~~~~cluster & year     &  ~~~~reference &\end
\skip{1pt}
\skip{1pt}
\hline
\skip{1pt}
\skip{1pt}
& NGC~7078 $\equiv$ M15     & 1976 &   Cudworth, AJ, 81, 519              \end
\skip{1pt}
& NGC~6341 $\equiv$ M92     & 1976 &   Cudworth, AJ, 81, 975              \end
\skip{1pt}
& NGC~6205 $\equiv$ M13     & 1979 &   Cudworth \& Monet, AJ, 84, 774     \end
\skip{1pt}
& NGC~5272 $\equiv$ M3      & 1979 &   Cudworth, AJ, 84, 1312             \end
\skip{1pt}
& NGC~5904 $\equiv$ M5      & 1979 &   Cudworth, AJ, 84, 1866             \end 
\skip{1pt}
& NGC~6838 $\equiv$ M71     & 1985 &   Cudworth, AJ, 90, 65               \end
\skip{1pt}
& NGC~6656 $\equiv$ M22	    & 1986 &   Cudworth, AJ, 92, 348              \end
\skip{1pt}
& NGC~7089 $\equiv$ M2	    & 1987 &   Cudworth \& Rauscher, AJ, 93, 856  \end
\skip{1pt}
& NGC~6712                  & 1988 &   Cudworth, AJ, 96, 105              \end
\skip{1pt}
& NGC~6121 $\equiv$ M4      & 1990 &   Cudworth \& Rees, AJ, 99, 1491     \end
\skip{1pt}
& NGC~6626 $\equiv$ M28     & 1991 &   Rees \& Cudworth, AJ, 102, 152     \end
\skip{1pt}
&NGC~6171 $\equiv$ M107     &1992  &   Cudworth \etal, AJ, 103, 1252      \end
\skip{1pt}
& NGC~6341 $\equiv$ M92     & 1992 &   Rees, AJ, 103, 1573                \end
\skip{1pt}
& NGC~5904 $\equiv$ M5      & 1993 &   Rees, AJ, 106, 1524                \end
\skip{1pt}
& NGC~6656 $\equiv$ M22     & 1994 &   Peterson \& Cudworth, ApJ, 420, 612\end
\skip{1pt}
& NGC~6121 $\equiv$ M4      & 1995 &   Peterson \etal, ApJ, 443, 124      \end
\skip{1pt}
\skip{1pt}
\hline
}\endtable
$$
}
}}
}

\vskip 20pt		
\noindent
{\sl Proper motions.}~~ In theory, proper motions provide more
dynamical information than radial velocities, since they are two-
instead of one-dimensional (see, e.g., Wybo \& Dejonghe 1995, 1996).
The space velocities of some globular clusters are known from radial
velocities and absolute proper motions (see, e.g., Cudworth \& Hanson
1993); however, the small size of the {\sl internal} proper motions of
cluster stars has made them difficult to measure with the required
precision.  For example, for a nearby cluster at a distance of 5~kpc,
a velocity dispersion of 5 \kms\ corresponds to a displacement of 20
milliarcsec per century, which is the equivalent of 1.5 micron in 80
years on Yerkes plates.  K.M. Cudworth has been the pioneer of this
field, squeezing velocity dispersions and astrometric distances out of
the data.  But even in the best studied clusters, the errors in the
proper motions have been comparable in size to the motions themselves.
This explains why only very few studies of cluster internal proper
motions have provided dynamical information.  Apart from the Cudworth
\etal\ papers in Table~6.1, the only two dynamical studies using
published proper motions are, so far, Lupton, Gunn, \& Griffin (1987)
and Leonard \etal\ (1992), both of which examined M13
using the data of Cudworth \& Monet (1979).

It is obvious that only a very small fraction of the dynamical
information contained in globular cluster proper motions has been
extracted so far.  An expansion and reanalysis of the Yerkes data by
Rees (1992, 1993) should result in a welcome increase in the accuracy
of the motions and in the understanding of their uncertainties in
several clusters.  The report on the large proper motion study
undertaken by Reijns \etal\ (1993) for about 7,000 stars in \cent\ has
whetted our appetites, and hopefully will provide essential results in
the near future.

\vskip 20pt
\noindent
{\sl Stellar radial velocity dispersion.}~~ During the first half of
this century and later, all stellar radial velocities were acquired
from techniques using photographic plates.  The typical errors of the
best measurements were $\simeq$ 10 \kms, i.e., of the same order of
magnitude, or larger than, the velocity dispersion value expected in
globular clusters.  A catalog of such radial velocities in galactic
globular clusters has been published by Webbink (1981).

But since the pioneering work of Griffin (1967, 1974),
cross-correlation techniques have proven their exceptional efficiency
in radial velocity determination. The cross-correlation between a
stellar spectrum and a template condenses the radial velocity
information contained in the stellar spectrum into the equivalent of a
single spectral ``line'', the cross-correlation function.  With the
construction around the 70's and 80's of instruments using such
cross-correlation techniques (e.g., Baranne \etal\ 1979 and Mayor 1985
for CORAVELs; Flechter \etal\ 1982, McClure \etal\ 1985; Latham 1985,
Peterson \& Latham 1986), the typical errors on the best measurements
are $\simeq$ 0.5 \kms, providing an essential tool for investigating
the internal dynamics of globular clusters.  A new generation of
instruments, taking advantage of improved technologies applied to the
same cross-correlation techniques, brings the typical errors down to
$\simeq$ 10 \ms\ (e.g., Marcy \& Butler 1992, Mayor \& Queloz 1995),
so far, only for relatively bright (\mv\ \lsim\ 9) nearby stars.

The first dynamical study of a globular cluster using high-quality
stellar radial velocities was published by Da Costa \etal\ (1977),
who fitted a projected density profile and velocities for 11 giants
in NGC~6397 with a single-mass King model.  Two years later, Gunn \&
Griffin (1979) published the first results from their extensive study
of cluster velocity dispersions, giving velocities for 111 giants in
M3.  Density and velocity dispersion profiles were simultaneously fit
to multi-mass anisotropic dynamical models based on the King-Michie
form of the phase-space distribution function $f(\varepsilon,l)$ (see
\S7.7 below).  Today, eight other clusters have published studies
using similar models and sample sizes (between 68 and 469 stars): M92
(Lupton \etal\ 1985), M2 (Pryor \etal\ 1986), M13 (Lupton \etal\
1987), \cent\ (Meylan 1987, Meylan \etal\ 1995), \tuca\ (Mayor \etal\
1984, Meylan 1988a, 1989), M15 (Peterson \etal\ 1989 for the
velocities and Grabhorn \etal\ 1992 for the analysis), NGC~6397
(Meylan \& Mayor 1991), and NGC~362 (Fischer \etal\ 1993b).

Stellar radial velocities have been acquired in a few other galactic
globular clusters (e.g., Peterson \& Latham 1986, Pryor \etal\ 1989a,
1991), but in smaller quantities, providing weaker dynamical
constraints. In the case of M4, Peterson \etal\ (1995) publish 182
radial velocities with no dynamical study.

Initially, technological developments were driven by the need for
small errors in radial velocity measurements ($\simeq$ 0.5
\kms) as required in order to get access to the internal dynamics
of globular clusters.  The present improvements are now also driven by
the size of the samples, as nearly all of the above sets of velocity
data are too small to employ, for example, non-parametric methods (see
\S7.7 below).  Acquiring even a few hundred stellar radial velocities
one at a time is a slow and tedious job, even on 4-m class telescopes.
But the number of stellar velocities in globular clusters has recently
grown explosively because of the new technology becoming available to
make these measurements.  Fiber-fed, multi-object spectrographs like
ARGUS at Cerro Tololo, HYDRA at Kitt Peak, and AUTOFIB at the Anglo
Australian Observatory can obtain velocities about 25 times faster.
Similar gains result from using Fabry-Perot interferometers to measure
radial velocities (Gebhardt \etal\ 1995).  Four clusters have
published non-parametric studies using sample sizes from a few hundred
up to a few thousand stars: \tuca\ (Gebhardt \& Fischer 1995), NGC~362
(Gebhardt \& Fischer 1995), NGC~3201 (Gebhardt \& Fischer 1995,
C\^ot\'e \etal\ 1995), and M15 (Gebhardt \etal\ 1994, Gebhardt
\& Fischer 1995).  In the framework of the major study of \cent\
(Reijns \etal\ 1993), the radial velocities of about 3,500 stars in
this cluster have been acquired (Seitzer, pers. comm.) and
will be combined with their proper motions.

Stellar radial velocities have been obtained in a few clusters in the
Magellanic Clouds (e.g., Seitzer 1991, Mateo \etal\ 1991, Fischer
\etal\ 1992a,b).  Acquiring these stellar radial velocities is
difficult because the distance to the Clouds is about ten times larger
than that to the best-studied galactic globular clusters.  Thus the
stars are faint, and crowding and contamination by field stars create
serious problems.

\vskip 20pt
\noindent
{\sl Velocity dispersion from integrated-light spectra.}~~ The value
of the velocity dispersion within 10\arcs\ of the cluster centre is
important for the understanding of cluster dynamical evolution, since
the velocity dispersion in the core may display a power-law cusp due,
e.g., to core collapse.  At the same time, this value is very
difficult to obtain for high-concentration globular clusters from
radial velocities of individual stars because of serious crowding
problems.  A way to overcome this difficulty is to measure the Doppler
broadening in integrated-light spectra obtained from an area of a few
square arcseconds at the centre of the cluster.

For globular clusters, the first such observations are those for 10
clusters described in Illingworth (1976), along with a new method
based on Fourier power spectra for accurately determining the velocity
dispersions.  Essentially, this method involves (i) artificially
broadening suitable stellar spectra with a range of velocity
dispersions (broadening the spectra has the effect of steepening their
Fourier power spectra), and (ii) comparing these spectra with the
cluster spectrum and selecting the velocity dispersion giving the best
match in the Fourier domain.

\vskip 11.5truecm
\item\item{\sevenrm{{\bf Fig.~6.1.} \hskip 2mm Cross-correlation 
functions for five standard stars with very different metallicities
(--2.0 $\le$ [Fe/H] $\le$ 0.0, from top to bottom, respectively).  All
cross-correlation functions are (i) well approximated by gaussian
functions and (ii) have widths which are independent of the
metallicity (from Dubath \etal\ 1996, Fig.~6).}}
\vskip 0.5truecm		

Since 1987, a numerical version of the analog cross-correlation
technique used with CORAVEL spectrometers has been developed at Geneva
Observatory (Meylan \etal\ 1989, Dubath \etal\ 1990).  Instead of
doing an analog cross-correlation ``online'' at the telescope, as is
done, e.g., with CORAVEL spectrometers, integrated-light echelle
spectra (covering about 1500 \AA\ between 4000 and 7500 \AA) are
obtained and cross-correlated numerically afterwards.  This approach
has noticeable advantages: the scanning required to build the CORAVEL
analog cross-correlation function at the telescope is no longer
necessary, providing an immediate gain of about 2.5 mag, and there are
further gains due to the higher quantum efficiency of CCDs as compared
to photomultipliers.  The read-out noise of the CCD is the limiting
factor.

Because the numerical technique has been designed to be similar to the
analog technique of the CORAVEL spectrometers (similar templates and
wavelength ranges), the numerical cross-correlation functions have the
same behavior as the CORAVEL's.  CORAVEL experience shows (i) that the
cross-correlation functions are well approximated by gaussian
functions, and (ii) that the widths of these cross-correlation
functions do not depend on the metallicity, as is seen in Fig.~6.1,
which displays the cross-correlation functions of 5 standard stars
with very different metallicities (--2.0 $\le$ [Fe/H] $\le$ 0.0).
Thus, the broadening of a cluster cross-correlation function is only
produced by the Doppler line broadening present in the
integrated-light spectra because of the velocity dispersion of the
stars along the line of sight (Dubath \etal\ 1996).

\vskip 8.5truecm
\item\item{\sevenrm{{\bf Fig.~6.2.} \hskip 2mm Normalized 
cross-correlation functions of the cluster NGC~1835 (triangles) and of
the comparison star HD~31871 (dots).  The continuous lines are the
corresponding fitted gaussians.  The significant broadening of the
cluster cross-correlation function is conspicuous and allows an
immediate determination of the projected velocity dispersion in the
core of NGC~1835 (from Dubath, Meylan, \& Mayor 1990, Fig.~3).}}
\vskip 0.5truecm

An example using the LMC cluster NGC~1835 illustrates the above points
(Meylan \etal\ 1989, Dubath \etal\ 1990).  After normalizing the
cross-correlation function of the cluster to have the same depth as
the cross-correlation function of the comparison star, the significant
broadening of the cluster cross-correlation function (CCF),
\sigccf(cluster), is conspicuous, as seen in Fig.~6.2. All standard
stars have been checked by direct CORAVEL measurements to have almost
zero rotation and are used to determine the standard deviation of the
instrumental stellar cross-correlation function, \sigref.  Because of
the gaussian approximation of the cross-correlation functions, the
projected velocity dispersion in the core of NGC~1835 --- \sigp\ =
10.3 \pmm\ 0.4 \kms\ --- is immediately obtained from the following
quadratic difference:
$$ 
\sigma_p^2 = \sigma_{\rm CCF}^2({\rm cluster}) - \sigma_{ref}^2. \eqno(6.1) 
$$
%
%
In order to study globular cluster masses and mass-to-light ratios as
functions of galaxy type and environment, Meylan \etal\ (1991b) and
Dubath \etal\ (1993b, 1996) have, in the framework of their survey,
obtained integrated-light echelle spectra of the core of about 60
galactic, Magellanic, and Fornax globular clusters.  Zaggia \etal\
(1991, 1992a,b, 1993) have developed a similar technique which they
applied to seven galactic globular clusters.  Dubath \etal\ (1996)
compare their results with those obtained by Illingworth (1976) and
Zaggia \etal\ (1992a,b), for the nine globular clusters which have
core velocity dispersion determined by at least two of these three
studies.  In most cases, for a given cluster, the results are not
significantly different from each other (within one sigma).  The
remaining differences can easily be explained by the differences in
sampling areas and cross-correlation techniques.  There may be some
indications of slight underestimates of errors in a few clusters.

Pryor \& Meylan (1993) provide an extensive list of all velocity
dispersion data (from individual stars and from integrated-light
spectra) available in the literature concerning galactic globular
clusters.

In the case of globular clusters, constraints on the velocity
dispersion values have been obtained, indirectly, thanks to the
presence of pulsars.  In dense globular clusters, pulsars are so
accelerated by the mean gravitational field of the cluster that their
changing Doppler shift can overwhelm the intrinsic positive period
derivative \pdot.  The negative \pdot s provide strict limits to
cluster surface mass densities and mass-to-light ratios (Phinney 1992,
1993).  Such velocity dispersion estimates come from the use of
dynamical models.  The two globular clusters M15 and \tuca\ contain
numerous pulsars with some of them having negative \pdot s (see
Phinney 1993 for M15 and Robinson \etal\ 1995 for \tuca).

\vskip 30pt
\noindent
{\sl 6.3 Initial and present-day mass functions}

\vskip 20pt
\noindent
During the decades when only photographic data were used, no
information was available concerning luminosity and mass functions of
globular clusters, apart from the narrow mass range occupied by the
giants, subgiants, and the main-sequence stars just below the
turn-off.  The advent of CCDs, combined now with the HST, have allowed
an increasingly deeper view down the main sequence.

Scalo (1986) gives a review of the early luminosity function results
based on photographic photometry of globular clusters.  Subsequent
deep photometric studies (down to \Mv\ $\simeq$ 6) in globular
clusters, made possible by CCDs, have shown that the main-sequence
luminosity functions vary significantly from cluster to cluster.  For
example, McClure \etal\ (1986) tentatively identified, from a sample
of CCD-based luminosity functions of 7 globular clusters, a
correlation between the cluster metallicity and the main-sequence mass
function exponent.  However, Richer \etal\ (1990) and Richer
\etal\ (1991) (see also Richer and Fahlman 1992), find no correlation
between the mass function slope and the metallicity.  Capaccioli
\etal\ (1991), Capaccioli \etal\ (1993) and Djorgovski \etal\ (1993)
also find no obvious correlation from an extended sample of 17
galactic globular clusters.  They show that, (i) the dispersion in the
mass function slopes is much higher than expected from the errors,
even after correction for mass segregation effects, and (ii) the
position of a globular cluster with respect to the Galaxy acts as a
dominant parameter in its mass function slope, while the metallicity
plays a weaker role.  Capaccioli \etal\ (1993) interpret this
dependence of the mass function slope on the distance from the
galactic centre and galactic plane as evidence of a selective loss of
stars induced by cluster dynamical evolution.  Stiavelli \etal\ (1991,
1992) show, by using simple semi-analytical models, that the above
dependence can be reproduced assuming that all globular clusters are
born with identical mass functions, which then evolved through
interactions with the Galaxy, pointing towards disk shocking as the
most effective phenomenon in stripping the lightest stars.  Dauphole
\etal\ (1996) provide calculations of the orbits of 26 galactic
globular clusters, which show clearly that some clusters are hit much
harder and much more often by passages through the galactic plane then
others.

Djorgovski \etal\ (1993) use appropriate multivariate statistical
methods, applied to the sample of 17 galactic globular clusters, to
disentangle this complex situation, since the mass function slopes
depend simultaneously on more than one variable and many cluster
parameters are mutually correlated.  They confirm that the mass
function slopes in the range 0.5 \msun\ $\leq$ $M$ $\leq$ 0.8 \msun\
are largely determined by three quantities: mainly the position in the
Galaxy (distances to the galactic centre and to the galactic plane,
related to the cluster pruning along its orbit), and to a lesser
extent metallicity.  Their best fitting result gives the following
relation for the slope of the global mass function (cf. Eq.~6.3 for
the definition of $x$):
$$
x = (3.1\pm0.4)(\log R_{GC} + 0.25\log Z_{GP} - 0.13[Fe/H]) -
(3.3\pm0.5), \eqno(6.2) 
$$
where $R_{GC}$ and $Z_{GP}$ are the distances in kpc from the galactic
centre and plane, respectively.  Thus steeper mass functions are
associated with clusters which are more distant and/or more metal
poor.  Other parameters have little effect.

For globular cluster modelling, main sequence stars, white dwarfs and
other heavy remnants such as stellar black holes and/or neutron stars
have usually been estimated by simple extrapolation, based generally
on the following single power-law form for the whole mass spectrum:
$$ 
dN \propto m^{-x} d\thinspace {\rm log}(m) \eqno(6.3) 
$$
where the exponent $x$ would equal 1.35 in the case of Salpeter's
(1955) galactic initial mass function (IMF).  There is an irritating
ambiguity in the meaning of the phrase ``power-law index''.  Often
this refers to Eq.~6.3, but it also often refers to $\alpha$ in the
form $dN\propto m^{-\alpha}dm$; note that $\alpha=x+1$.

\vskip 9.5truecm
\item\item{\sevenrm{{\bf Fig.~6.3.} \hskip 2mm HST color-magnitude
diagram, summed over the four fields of WFPC2, in the galactic
globular cluster M4. Apparent $U$ magnitudes are indicated along the
right-hand ordinate, absolute ones on the left-hand. The white dwarf
cooling sequence is seen as the bluest stars in the diagram stretching
from $M_U$ $\sim$ +9 down to the limit of the data near $M_U$ $\sim$
+13.  This represents the first extensive sequence of cooling white
dwarfs seen in a globular cluster (from Richer \etal\ 1995, Fig.~1).}}
\vskip 0.5truecm

The presence or importance of stellar remnants and low-mass stars was
either ignored or governed by the upper and lower mass limits
(typically, \msup\ = 100 \msun\ and \minf\ = 0.1 \msun).  The upper
limit has no dynamical or photometric influence, because it concerns
only small numbers of stars that have already evolved into heavy
remnants: e.g., for $x$ = 1.5, the fraction of the total mass in the
form of heavy remnants varies by 0.05\% of the total mass when going
from an upper limit of 150 to 50 \msun; for $x$ = 1.0, the same
fraction varies by 0.6\%, and for $x$ = 0.5, by 4.0\% of the total
mass.  The above variations are much smaller than the uncertainty in
the total mass. The upper limit is chosen arbitrarily between 50 and
150 \msun.  The lower limit is much more controversial because of the
potential dynamical importance of a large number of low-luminosity
stars.  There is no observational constraint on the mass function for
stellar masses below $\simeq$ 0.1 \msun.  As noticed by Gunn and
Griffin (1979), this lower mass cutoff, if it is low enough, does not
significantly affect the cluster structure as traced by the giant
stars.  Variations of the total mass in the low-mass components do not
influence the quality of the fit. The individual mass of the lightest
stars is taken generally equal to 0.1 \msun.

In the case of globular clusters, present-day stellar mass functions
may reflect a mixture of both the initial conditions prevailing at the
epoch of cluster formation, and the subsequent consequences of the
dynamical evolution characterized by a selective escape of stars,
i.e., depending on the stellar mass (e.g., King 1996).  Consequently,
even on the main sequence, there is no way to observe the initial
mass function, which has been altered by stellar and dynamical
evolution.

The observational constraints related to the initial mass function of
stars which were more massive than the present turn-off mass ($\sim$
0.8 \msun) are indirect and vanishingly small.  These stars are in the
form of dark remnants.  Although the bright part of the sequence of
white dwarfs is now clearly observed in a few globular clusters
(Fig.~6.3), no quantitative parameters (e.g., mass function) can be
extracted in order to constrain dynamical models.  There is no
observational data about the mass function of heavier remnants, like
neutron stars, although, in a totally different way, one single pulsar
can be used as a dynamical probe of its host cluster (Phinney 1992,
1993).

The HST allows photometry and counting several magnitudes fainter than
with ground-based data.  For the closest clusters, luminosity
functions and mass functions can be determined down to nearly the
hydrogen-burning limit.  Fortunately, the crude approximations
represented by Eq.~6.3 become more and more outdated because of the
very deep star counts made possible with HST data (King 1996),
providing at last very deep luminosity functions for main sequence
stars and white dwarfs (see, e.g., De Marchi \& Paresce 1995a, Piotto
\etal\ 1996b, Santiago et al. 1996 for \tuca; Richer \etal\ 1995 for 
M4; De Marchi \& Paresce 1994a, Paresce \etal\ 1995, King \etal\ 1995,
Cool \etal\ 1996, Piotto \etal\ 1996b for NGC~6397; De Marchi \&
Paresce 1995b, Piotto \etal\ 1996b for M15; and Piotto \etal\ 1996b
for M30).

It should be emphasized that, in most clusters, the mass function is
reliably determined observationally only in the interval of about 0.4
\msun\ below the turn-off, i.e., between 0.4 \msun\ \lsim\ $M$ \lsim\
0.8 \msun.  Below 0.4 \msun\ the mass-luminosity relation becomes
increasingly uncertain, propagating large errors in the mass function
slope for light stars, because of the paucity and faintness of nearby
low-mass stars added to the large uncertainties of stellar evolution
models, which are in turn due to poor knowledge of stellar opacities.
See Henry \& McCarthy (1993) for a mass-luminosity relation (0.08
\msun\ $\leq$ $M$ $\leq$ 1.0 \msun) established using a combination of
long-term astrometric studies and infrared speckle imaging, and above
all D'Antona \& Mazzitelli (1996 and references therein) for
Population II mass-luminosity relations (0.09 \msun\ $\leq$ $M$ $\leq$
0.8 \msun) from stellar models of very low-mass main-sequence stars,
with a study of the dependence of the mass-luminosity relation on the
metallicity.

\vskip 7truecm
\item\item{\sevenrm{{\bf Fig.~6.4.} \hskip 2mm HST colour-magnitude 
diagram (left panel) of NGC~6397, in the filters $I_{814}$ and
$V_{555}$, and the corresponding luminosity function (right panel),
compared with the HST luminosity function by Paresce \etal\ (1995) and
the ground-based luminosity functions by Fahlman \etal\ (1989), from
King \etal\ (1996a, Fig.~1).}}
\vskip 0.5truecm

In the case of NGC~6397, the two HST-based luminosity functions by
Paresce \etal\ (1995) and by King \etal\ (1996a) are in good
agreement over the common range (Fig.~6.4).  Although in agreement at
bright magnitudes, at fainter magnitudes, however, the ground-based
luminosity function by Fahlman \etal\ (1989) rises significantly above
both the presumably more reliable HST-based luminosity functions.
Similar discrepancies, at the faint end of the luminosity function,
between HST and ground-based results are noted by Elson \etal\ (1995)
in the case of \cent.

Interesting comparisons are possible between ground-based data from
Drukier \etal\ (1993) and Piotto \etal\ (1996a) and HST data from Piotto
\etal\ (1996b).  The agreement is quite good, suggesting that, while
ground-based luminosity functions should not be relied on at very
faint magnitudes, they can be relied on at brighter magnitudes.  This
means that ground- and HST-based data provides nicely complementary
information on both ends of the main sequence.

The luminosity functions of M30 and M15 (Fig.~6.5) are very similar,
over a range of more than 6 magnitudes, while NGC~6397 is markedly
deficient in faint stars.  The above data implies that the mass
functions of M30 and M15 are very similar in the range 0.12 \msun\
$\leq$ $M$ $\leq$ 0.8 \msun.  This may be the consequence of both very
similar initial conditions and very similar evolution, a scenario
which is less contrived than assuming that the present similarity has
been created by evolution from different initial conditions. But,
since all three clusters are very similar in metallicity --- \feh\
$\sim$ --2.0 --- and morphology or dynamical status (all are
collapsed), what can be the reason why there are so many fewer
low-mass stars in NGC~6397~?  The reason is that these clusters have
very different orbits around the galactic centre.  Although the
traditional theories of tidal shocks have never been well enough
quantified and are known to be unreliable (Weinberg 1994a,b,c), the
effect of tidal shocks is certainly present in the parameters
governing the dynamical evolution of globular clusters.  NGC~6397,
much closer to the galactic plane and galactic centre than M15, is
clearly hit much harder and much more often by passages through the
galactic plane (King 1996 and Dauphole \etal\ 1996; see also \S10.2).

\vskip 8truecm
\item\item{\sevenrm{{\bf Fig.~6.5.} \hskip 2mm HST luminosity functions
in the filters $I_{814}$ (left panel) and $V_{555}$ (right panel), for
three metal-poor collapsed globular clusters, from Piotto \etal\
(1996b, Fig.~3).  The luminosity function of NGC~6397 has been extended
up to the turn-off using ground-based data.  Where omitted, the error
bars are smaller than the symbol size.}}
\vskip 0.5truecm

It is worth mentioning that \cent\ is the only cluster, of the five
already studied with HST data, for which no drop-off towards fainter
luminosities has been detected, to the limit of the existing HST
observations (Elson \etal\ 1995).  This may be linked to the strong
gravitational potential of this cluster, the most massive galactic
globular, which could be about 75 times more massive than NGC~6397
(Drukier 1995, Meylan \etal\ 1995).

Differences in the radial distributions of stars of different masses
are at last definitely observed with HST, providing conclusive
observational evidence of mass segregation (see \S7.2).

\vskip 30pt
\noindent
{\sl 6.4  The possibility of dark matter in globular clusters}

\vskip 20pt
\noindent
This question arises frequently, if only because globular clusters are
the next step down in size from the smallest objects in which firm
evidence for dark matter can be found, i.e., dwarf spheroidal galaxies
(see, e.g., the reviews by Mateo 1994 and Pryor 1994; see also the
recent contributions by Armandroff \etal\ 1995 and Olszewski \etal\
1996).  A second reason is the theoretical work of Peebles (1984), who
showed that an isothermal distribution of stars in a potential
dominated by a uniform dark background would have a profile roughly
resembling that of a star cluster.  

The construction of dynamical models (\S7.7)provides one approach to
this question.  In almost all cases, multi-mass anisotropic King-Michie
models do a satisfactory job, and for those cases in which such models
are clearly unsuccessful, an interpretation in terms of post-collapse
evolution is plausible (e.g., Grabhorn \etal\ 1992, Phinney 1993).  In
this sense, then, no dark matter is {\sl required}, except for the
modest fractions of neutron stars and white dwarfs included in such
models.

One may also ask, how much dark matter {\sl could} there be?  It is
often found that an adequate fit is obtained with a range of models with
a considerable spread of total masses.  For example (e.g. Fischer \etal\
1992b) found for the LMC cluster NGC~1978 total masses varying over a
factor of 5.  Though it may be tempting to take this to mean that as
much as 80\% of this cluster could consist of dark matter, all the
models are constructed from ordinary stars and stellar remnants, and it
is not clear how much of this could be replaced by dark matter without
degrading the fit to the surface brightness.  Using simple models, 
Heggie \etal\ (1993) and Taillet \etal\ (1995, 1996) considered how much
dark matter could be added before its effects become noticeable (see
also Fig.6.6).

The above approaches are open to the criticism that the results are too
model-dependent, which has prompted the development of non-parametric
methods (see \S7.7).  In this way Gebhardt \& Fischer (1995) have
presented quite well constrained estimates of the mass density profiles
of several clusters.  How much of this is ``dark'' may be determined, in
principle, by using deep star counts (\S6.3) to count how much is
contributed by normal stars. A preliminary study (Heggie \& Hut 1996)
suggests that up to half of the inferred mass is invisible.  On the
other hand it is not implausible that all of this is made up either of
white dwarfs (only the brightest of which can be counted at present) or
low-mass stars below about $0.1M_\odot$.

New observational techniques for potentially determining the
contribution to the mass budget by low-mass stars (which would occupy a
halo around the bright stars) are discussed by Taillet \etal\ (1995).
Moore (1996) has recently argued against the existence of such halos of
dark matter, pointing out that they would inhibit tidal stripping, in
conflict with observation.

\vfill\eject 

\vphantom{ }

~~~

\vskip 15truecm
\item\item{\sevenrm{{\bf Fig.~6.6.} \hskip 2mm Effect of dark matter 
on a simple model star cluster.  The solid line shows the profile of
rms projected velocity (upper panel) and surface density (lower panel)
in a King model with \rc\ = 1 pc, \rt\ = 100 pc and total mass 10$^5$
\msun.  The other three curves on each figure show the effect on this
``bright'' component of adding an equal mass of dark matter, the
parameters of the model being adjusted to preserve the mass and scale
radii of the bright matter.  Short dashes: dark matter particles have
same mass as ``bright'' stars; long dashes: dark matter particles each
have 1/8 of the mass of a bright star; dot-dashes: dark matter is
uniformly distributed out to the tidal radius. }}
\vskip 0.5truecm

\vskip 30pt
\noindent
{\bf 7. Quasi-static equilibrium: slow pre-collapse evolution} 

\vskip 20pt
\noindent
When the phase of violent relaxation (see \S5.5 above) comes to an
end, a cluster will have settled into a structure close to dynamical
equilibrium, except for subsequent transient disturbances as it passes
through the galactic plane.  In this quasi-equilibrium phase, the
cluster will be nearly spherically symmetric if its rotation is slow
(as is true of observed clusters in the Galaxy, see \S7.6), and if we
confine attention to parts well inside the tidal boundary.  In this
chapter we turn to the dynamical processes which begin to dominate the
evolution of a globular cluster in this long phase of existence, where
we now find them.  Thus this chapter really provides the theoretical
background for the remainder of this review, just as the previous
chapter provides the observational background.

\vskip 30pt		
\noindent
{\sl 7.1 The relaxation time} 

\vskip 20pt
\noindent
Of the various evolutionary mechanisms we discuss in this section, it
is the one which is referred to as ``collisional relaxation'' or
``two-body relaxation'' which has the longest history.  Long ago Jeans
(1929) estimated the time scale on which it acts, and later his result
was refined and developed by Chandrasekhar (1942).  Genuine collisions
are not implied (see \S9.4), but rather the purely gravitational
encounters of individual pairs of stars.  Two stars exchange energy in
an encounter, and the cumulative effect of many mild encounters
eventually produces major changes in the structure of the cluster,
without significantly disturbing its dynamical equilibrium.

The time scale on which this process becomes significant is generally
called the {\sl relaxation time}, though several different precise
definitions exist.  Among theorists the most commonly used is that of
Spitzer (1987, Eq.~2-62), who defines:
$$
t_{r} = {0.065\langle
v^2\rangle^{3/2}\over\rho\langle m\rangle G^2\ln\Lambda}.\eqno(7.1) 
$$
Here, $\langle v^2\rangle$ is the mass-weighted mean square velocity
of the stars, $\rho$ is the mass density, $\langle m\rangle$ is the
mean stellar mass, and $\Lambda\simeq 0.4 N$, where $N$ is the number
of stars in the cluster.

Eq.~7.1 stems from considering the time scale on which the cumulative
mean square value of $\Delta v_\Vert$, i.e., the component of the
velocity change which is parallel to the velocity itself, becomes
comparable with the mean square value of one velocity component.
Other definitions make use, e.g., of the perpendicular component of
$\Delta \bv$, or the time scale on which the direction of motion of a
star is deflected, by two-body encounters, through a large angle.  All
have a similar form, differing only in the numerical coefficient
and/or in the value of $\Lambda$ (see below).  Adequate though these
estimates are for many purposes, the time scale of relaxation may be
considerably altered by the existence of a spectrum of stellar masses
(cf. \S7.2 below), or by clumpiness in the spatial distribution of
stars, which may occur in young star clusters (Aarseth \& Hills 1972).

Apart from $N$, the quantities appearing in Eq.~7.1 are local, and so
the relaxation time varies from low values in the dense core to
extremely large values as the tidal radius is approached.  For rough
estimates a useful global measure of the time of relaxation
substitutes mean values for the inner half of the mass, i.e.,  within
the {\sl half-mass radius} $R_h$.  With one other approximation, based
on the virial theorem, which relates the mean square velocity to $R_h$
itself, these considerations lead to the {\sl half-mass relaxation
time}:
$$
t_{rh} = 0.138{M^{1/2}R_h^{3/2}\over\mbar G^{1/2}\ln\Lambda} \eqno(7.2)
$$
(Spitzer 1987, Eq.~2-63).  Values for galactic globular clusters range
from about $3 \times 10^7$ to about $2 \times 10^{10}$ years
(Djorgovski 1993b).

It is worth relating the relaxation time scale to the other major time
scale in the dynamics of star clusters, the {\sl crossing time}.  As
with the relaxation time this can be defined in several ways, but a
common convention is to define:
$$ 
t_{cr} = {2R\over v}, \eqno(7.3) 
$$
where $R$ is a measure of the size of the system and $v$ a measure of
the mean stellar velocity.  Thus the crossing time is a measure of the
time taken for a star to traverse the diameter of the cluster.  More
specifically, $R$ is often chosen to be the {\sl virial radius}:
$$
R_{vir} = -GM^2/(2W), \eqno(7.4) 
$$
where $M$ is the total mass of the cluster and $W$ is its potential
energy, i.e., that computed from the interactions among the stars of
the cluster (each binary being treated as a single star with a mass
equal to the combined mass of the components), and excluding the
galactic tidal field.  It is often found that $R_{vir}$ is comparable
with the half-mass radius $R_h$; for example, $R_h\sim 0.77R_{vir}$ in
the Plummer model (cf. \S7.5).  A common specific choice for $v$ is
the mass-weighted root mean square velocity of the stars, i.e., $v^2 =
2T/M$, where $T$ is the kinetic energy of the stars (binaries being
treated as in the computation of $W$).  With these choices, therefore,
the result is that:
$$ 
t_{rh}/t_{cr} = 0.138\left({R_h\over
2R_v}\right)^{3/2}{N\over\ln\Lambda}.  \eqno(7.5) 
$$ 
Duncan \& Shapiro (1982) and Hut (1989) provide instructive
introductions to this and other relations between time scales of
interest in the internal dynamics of star clusters.

The foregoing estimates are based on the generally accepted theory of
relaxation which is described, for example, in Spitzer (1987).  On
theoretical grounds, however, various modifications or alternatives
have been proposed from time to time. The theory is local, as
mentioned earlier, and the effects of this assumption have been
discussed by Parisot \& Severne (1979) and by Weinberg (1993a).  It
takes no account of the orbits of the stars in the smooth potential,
the effect of which {\sl may} be substantial (Kandrup 1983, Severne \&
Luwel 1984, Tremaine \& Weinberg 1984, and Rauch \& Tremaine 1996).
The theory also considers only the cumulative effect of many weak
interactions, and the effect of the occasional strong interaction
requires more elaborate treatment (e.g., Agekian 1959, H\'enon 1960b,
V'yuga \etal\ 1976, Retterer 1979, Ipser \& Semenzato 1983).  The
appropriate functional form of $\Lambda$ in Eq.~7.1 has been
questioned on theoretical grounds by Kandrup (1980), with numerical
support from Smith (1992), though this contradicted the earlier
conclusion of Farouki \& Salpeter (1982) (cf. also McMillan \etal\
1987).  More seriously still, it has been suggested that the
combination of relaxation with the chaotic nature of stellar orbits in
``non-integrable'' potentials (e.g., most axisymmetric potentials)
causes a great enhancement in the rate of relaxation (Pfenniger 1986,
Kandrup \& Willmes 1994).  Another suggestion which, if confirmed,
would revolutionise the theory of relaxation was made by Gurzadyan \&
Savvidy (1984, 1986; see also Gurzadyan \& Kocharyan 1987, Gurzadyan
1993), and taken up by a number of other authors (e.g., Kandrup 1988,
Sakagami \& Gouda 1991, Boccaletti \etal\ 1991).  They suggest that
relaxation is much faster than in standard theory, by a factor of
order $N^{2/3}$.  Interestingly, it is claimed that there is support
for this view on observational grounds (Vesperini 1992), though
Goodman \etal\ (1993) assert that the time scale estimated by
Gurzadyan \& Savvidy is wrong and that the mechanism they discuss is
not even a relaxation process in the usual sense.

Numerical experiments can provide independent evidence on these
debates.  Those by Standish \& Aksnes (1969) and Lecar \&
Cruz-Gonzalez (1971) gave results agreeing with those of conventional
theory, but the motions of the stars were deliberately simplified.  In
a much more realistic setting, though with a ``softened'' potential,
Huang \etal\ (1992) found close agreement between numerical
measurements of the ``diffusion time'' and the relaxation time, and
Theuns (1996) has found similar agreement, on the whole, between
numerical and theoretical diffusion coefficients.  Giersz \& Heggie
(1993a,b) also found that the results of $N$-body calculations could
be adequately explained by the traditional theory of relaxation, with
an appropriate choice of the numerical factor $\gamma$ in the
expression $\Lambda = \gamma N$ for the argument of the Coulomb
logarithm (cf. also Giersz \& Spurzem 1994, Spurzem \& Takahashi 1995,
and Fig.~7.1).  Any radical revision of the relaxation time scale
would destroy their observed consistency between $N$-body data and
conventional theory.

Relaxation affects the evolution of a stellar system in several ways,
which are discussed in detail in \S\S7.2 and 9.  In addition, however,
it regulates the anisotropy of the distribution of velocities.  It is
often argued that anisotropy should be small in parts of a cluster
where the relaxation time is short, and indeed relaxation can reduce
the global anisotropy of a system (Fall \& Frenk 1985), but it must
also be realised that relaxation by itself can create anisotropy where
none was present initially.  This has been demonstrated many times,
and is the particular topic of studies by Bettwieser \etal\ (1985) and
Bettwieser \& Spurzem (1986).

Another area in which relaxation plays a role is in the rotation of a
stellar system.  The main information comes from Fokker-Planck
simulations by Goodman (1983a) and $N$-body studies by Fall \& Frenk
(1985) and Akiyama \& Sugimoto (1989).  The relation between rotation
and escape is discussed in \S7.3.

\vskip 10truecm
\item\item{\sevenrm{{\bf Fig.~7.1.} \hskip 2mm Comparison between four 
models of the evolution of an isolated stellar system (from Giersz \&
Spurzem 1994, Fig.~1).  The initial model is a Plummer model, and all
stars have equal mass.  Lagrangian radii (i.e., the radii of spheres
containing a fixed fraction of the total mass) are plotted against
time.  Units are such that $G$ = $M$ = -4$E$ = 1, where $M$ and $E$
are the total initial mass and energy, respectively. Key: AGM --
anisotropic gaseous model, IGM -- isotropic gaseous model, FOK --
isotropic Fokker-Planck model, NBO -- average of many $N$-body models
with $N$ = 1,000.}}
\vskip 0.5truecm

\vskip 30pt
\noindent
{\sl 7.2 Energy equipartition and mass segregation} 

\vskip 20pt
\noindent
In some theories of star formation, the spatial distribution of stars
of different mass will differ at birth (Podsiadlowski \& Price 1992,
Murray \& Lin 1993, Gorti \& Bhatt 1996).  Usually, however, it is
assumed that the processes of stellar formation give rise to a cluster
in which different stellar masses are undifferentiated spatially and
dynamically.  The early processes of dynamical evolution --- mass loss
from stellar evolution, and violent relaxation --- do not change the
cluster in this respect, except for the progressive loss of the more
massive stars as they evolve internally.  Relaxation is the first
dynamical process which does differentiate stars according to their
mass.  It produces a tendency towards equipartition of kinetic energy,
and so the larger mass involved in a gravitational encounter tends to
lose kinetic energy, and then fall deeper into the potential well of
the cluster.  At the same time, stars of lower mass are driven out,
and the stars are segregated by mass.

The time scale for this process may be estimated from formulae given
by Spitzer (1987, Eq.~2-60), by computing the rate of change of the
difference in the kinetic energy of stars in a two-component system.
The result is a mass segregation time scale given by:
$$
t_{ms} = {0.028(\langle v_1^2\rangle+\langle v_2^2\rangle)^{3/2}
\over m_1m_2nG^2\ln\Lambda},    \eqno(7.6)
$$ 
where subscripts refer to the two components, and $n$ is the total
number density.  $N$-body models show that the time scale for mass
segregation (more specifically, for the growth of the half-mass radius
of the lighter species) can be well matched by a similar equation
(with a suitably chosen coefficient), and it is found empirically that
the result can be extended also to continuous mass spectra (Farouki \&
Salpeter 1982).

If dynamical friction alone is important (which is a satisfactory
approximation for the evolution of the stars of greatest mass) the
development of mass segregation can be explored with a simplified
treatment (White 1976).  In general, however, the details of the
tendency to equipartition and of mass segregation are best evaluated
with the use of a detailed dynamical evolutionary model (see \S8
below), and here we summarise the main results in the earlier phases
of core collapse (Saito \& Yoshizawa 1976; Inagaki 1983, 1985; Inagaki
\& Wiyanto 1984, Inagaki \& Saslaw 1985, Chernoff \& Weinberg 1990).
Some of these results, however, refer to idealised systems in which
stars have only two or a few possible masses, and are obtained with
isotropic Fokker-Planck or gas models.

There is first a fairly rapid phase of evolution (presumably on a time
scale comparable with Eq.~7.6) in which the different mass components
tend towards equipartition in the central regions. How closely they
reach equipartition depends on the mass spectrum.  Generally speaking,
it is most closely approached when either the range of stellar masses
is small, or else the spectrum of masses is steep (and so the heaviest
stars do not contribute much of the total mass).  In other cases there
is approximate equipartition amongst the heaviest stars only.  These
conditions for the achievement of approximate equipartition resemble
those derived on the basis of simple theory by Spitzer (1969).

Associated with the (limited) tendency towards equipartition is the
process of mass segregation.  Just as equipartition tends to be set up
only amongst the heavier masses, the spatial distribution of the
heavier masses is greatly differentiated by mass segregation, whereas
the spatial distribution of a great range of low-mass stars remains
rather similar.

An extreme population for which mass segregation would be important is
the population of stellar remnants in the form of black holes of mass
$\sim 10M_\odot$ (Larson 1984, Sigurdsson \& Hernquist 1993, Kulkarni
\etal\ 1993).  Their possible effects on clusters include enhancements
of the central velocity dispersion and stripping of the envelopes of
red giants, and there is observational evidence for this (Fusi Pecci
\etal\ 1993a).  

The foregoing remarks refer to the core.  By the time the core has
come to equipartition as closely as it ever does, there is still
little tendency towards equipartition at and beyond the half-mass
radius.  This means that observational evidence for mass segregation
should be found mostly in the core, and can be obtained by comparison
between the core and other regions within the half-mass radius.  This
may be quite problematic because observational selection effects
(crowding and faintness) have similar biases: in the dense crowded
core of a star cluster, where stars of low mass should be depleted,
faint stars are more easily missed.  This makes HST the ideal
telescope to look, quantitatively, for mass segregation in globular
clusters.

For decades, the differences in the radial distributions of stars of
different masses have been seen from the ground, significantly but
weakly, in various low-concentration or nearby galactic globular
clusters (see, e.g., Sandage 1954 and Oort \& van Herk 1959 in M3,
Richer \& Fahlman 1989 in M71, Drukier \etal\ 1993 in NGC~6397, among
many others).  With the HST, however, faint stars can be seen all the
way into the core of the clusters, providing strongly significant mass
segregation observations. Mass segregation is observed with the HST in
NGC~6752 by Shara \etal\ (1995), in \tuca\ by Paresce \etal\ (1995)
and Anderson \& King (1996), and, in a more quantitative way, by
King \etal\ (1995, 1996b) in NGC~6397.

In imaging with the HST the high-concentration (core-collapsed)
globular cluster NGC~6397, King \etal\ (1995, 1996b) find the mass
segregation effects to be enormous, compared with the marginal degree
of segregation observed in this cluster, with ground-based data, by
Drukier \etal\ (1993).  Fig.~7.2 displays the mass functions, in stars
per arcmin$^2$, obtained in NGC~6397, at radii 7\arcs\ and 4.6\arcm,
by King \etal\ (1995).  The numbers in the 7\arcs\ field are higher
than those in the 4.6\arcm\ field, because of the higher density at
the cluster centre, but the mass functions are quite different.
Relative to those of high mass, the low-mass stars are depleted at the
centre by more than an order of magnitude.

King \etal\ (1995) have carried out some dynamical modelling to verify
that the observed amount of mass segregation is in agreement with
dynamical predictions.  The use of multimass King models is reasonable
here, even though NGC~6397 is a core-collapsed cluster, as long as
only very high-concentration models are used (so high that the exact
value of the concentration does not matter).  Though it is usual to
distinguish collapsed from uncollapsed clusters in terms of those
which can be fitted with King profiles and those which cannot, this
kind of dichotomy refers to {\sl single component} King models.  There
is no evidence that fits of {\sl multi-mass} King models to
post-collapse clusters are any less satisfactory than those to
uncollapsed clusters.

\vskip 9.5truecm
\item\item{\sevenrm{{\bf Fig.~7.2.} \hskip 2mm HST mass functions
in NGC~6397, at radii 7\arcs\ and 4.6\arcm, in stars per arcm$^2$,
from King \etal\ (1995, Fig.~4).  The mass functions observed in the
7\arcs\ and 4.6\arcm\ fields are conspicuously different. The solid
lines are from a dynamical model fitted to the cluster. }}
\vskip 0.5truecm

The continuous lines in Fig.~7.2 are from such a dynamical model
fitted to the King \etal\ (1995) observations of NGC~6397: the numbers
have been fitted to the observations at 4.6\arcm\ but not at 7\arcs.
The dashed line represents the global mass function of the model.  The
model is of course chosen to fit the outer points, but there is no
requirement whatever that it fit the inner points.  The fact that the
inner points are indeed reproduced (within the errors) shows that these
observations are in satisfactory agreement with theoretical (although
somewhat crude) predictions (see also Anderson \& King 1996, King
\etal\ 1996b).

\vskip 30pt
\noindent
{\sl 7.3 Evaporation through escaping stars} 

\vskip 20pt
\noindent
The theoretical study of the rate of escape of stars from clusters has
a checkered history, as one sees even from the study of idealised
isolated systems.  One class of estimates (e.g., Ambartsumian 1938,
Spitzer 1940, Chandrasekhar 1942 (his \S\S5.3 and 5.4), 1943a,b,c,
Spitzer \& H\"arm 1958, King 1965, Danilov 1973, Johnstone 1993) have
been based on relaxation phenomena (escape by the cumulative effect of
many small disturbances) and yield a fractional rate of escapes
proportional to an inverse relaxation time, i.e., $\dot N/N\propto
-1/t_r$.  Another class of theories, based on individual two-body
encounters, was developed by H\'enon (1960a) and by Woolley \& Dickens
(1962).  These yielded results of a form similar to the first type of
method, except for differences in the numerical factor, and the
absence of the Coulomb logarithm (which enters in the definition of
the relaxation time $t_r$).  H\'enon's treatment can be applied
conveniently to any system with an isotropic distribution of stellar
velocities, and yields a simple analytical result for the Plummer
model.  He also later tabulated results for a Plummer model in which
stars have different masses (H\'enon 1969), though his model
necessarily excludes mass segregation.  Third, a somewhat different
method was adopted by Kaliberda (1969), who also treated escape as due
to discrete changes in energy rather than diffusion, but considered
the same sort of simplified potential as Spitzer \& H\"arm (1958).
Finally, a somewhat hybrid theory was presented by Spitzer \& Hart
(1971a,b), the effect of encounters during one passage through the core
being estimated from relaxation theory, and it was applied to other
models by Saito (1976).

Detailed modelling is a preferable way of investigating the escape
rate, without simplifying assumptions and, in combination with mass
segregation, provides the relative escape rates of different stellar
masses.

First we summarise some results for isolated systems with equal
masses.  Though unrealistic, this is an important simplification for
understanding the role played by different factors in the escape
process.  Over a few $t_{rh}$, modest-sized $N$-body models
(summarised in Wielen 1975) show that $\dot N t_{cr}\sim 0.1-0.2$,
where $t_{cr}$ is the crossing time (Eq.~7.3).  Fokker-Planck models
(Spitzer \& Shull 1975a; cf. \S8.2 below) revealed the added
refinement that the escape rate increases as the evolution of the
system proceeds, at least while the core is still collapsing.  Recent
$N$-body models (Giersz \& Heggie 1993a) show that this arises from
two causes: one is the increasing concentration of the core, and the
other is the growth of anisotropy, which tends to enhance the escape
rate.

Now we drop the simplifying assumptions which were introduced above.
First, in systems of stars with unequal masses, results from theory
(H\'enon 1969) and $N$-body models (Wielen 1974a, 1975) show that the
overall escape rate (by number) is enhanced, by as much as a factor of
30 for a quite reasonable mass spectrum.  Furthermore the rate of
escape is heavily mass dependent, the fraction of massive stars which
escape in a given time being much smaller than the fraction of
low-mass stars. However, there is little difference between the escape
rate of stars of lowest mass and those of, say, twice the minimum
mass. The fundamental dynamical reason for the mass-dependence of the
escape rate is that it is relatively easy for a massive star to impart
a large kinetic energy to a low-mass star. This is the same mechanism
causing mass segregation, which further depresses the escape rate of
massive stars.

The next simplifying assumption to remove is the assumption that the
system is isolated, i.e., to reintroduce a steady external field.  An
important point to notice with regard to tidally-influenced systems,
however, is that the definition of what is meant by ``escape'' is
rather less clear than for isolated systems.  If the tidal field is
approximated by a spherically symmetric potential then the main effect
is that the threshold of escape is lowered, and escape is easier, but
no more complicated, than for an isolated system.  Even for a cluster
in a circular orbit about a spherical galaxy, however, the tidal field
is not spherically symmetric (Chandrasekhar 1942, his \S5.5), and
study of orbits in $N$-body models (Terlevich 1987) or smooth cluster
potentials (Jefferys 1976) shows that it is possible for stars to
remain in retrograde orbits bound for long periods to the cluster,
even though their orbits take them well beyond the conventional tidal
radius.  Furthermore, it is only in directions close to those of the
Lagrange points that one has a threshold for escape (Hayli 1967); in
the orthogonal directions the combined effect of the tidal and
centrifugal forces is to help trap stars within the cluster.  Ross
\etal\ (1996) have recently established a criterion for escape in this
problem (where simple energy considerations are insufficient.)

The relative stability of retrograde orbits has led to the conclusion
that a cluster may eventually exhibit substantial retrograde rotation
(cf. Oh \& Lin 1992).  On the other hand various authors (Agekian
1958, Shapiro \& Marchant 1976, Longaretti \& Lagoute 1996a) have
concluded that preferential escape of stars of high angular momentum,
which occurs even in the absence of a tidal field, would lead to a
{\sl decrease} of rotation and therefore of rotationally induced
flattening (if present initially).  (Actually Agekian's result was
more complicated, as he found that initially highly flattened systems
became flatter still.)  It should be mentioned that escape is probably
not the most effective process for altering the flattening of a
rotating cluster, just as it is not the most important process for
driving a system into core collapse.  In fact Fall \& Frenk (1985)
found that it is too slow to be of importance, compared with internal
processes.  Their estimate for the time scale for flattening by
internal mechanisms was comparable with that observed in Fokker-Planck
models by Goodman (1983a).  On the other hand, their study referred to
isolated systems of stars of equal mass, and their estimate of the
escape time scale was based on a simplified treatment.  Their result
may, therefore, underestimate the importance of escape.  At any rate,
it is evident that our understanding of this problem is rather patchy.

In view of these complications, care must be taken in the
interpretation of data on the escape rate.  Nevertheless, a common
approximation is to assume that a star has escaped when its radius
exceeds the conventional tidal radius (\S7.4), and $N$-body models
show that this leads to consistent results, whether or not a tidal
field is included (Giersz \& Heggie 1993b).  Results from both
Fokker-Planck (Spitzer \& Chevalier 1973) and $N$-body models (Hayli
1967, 1970a; Wielen 1968; Danilov 1985; Giersz \& Heggie 1996b)
confirm that, in systems of stars of equal mass, the presence of a
tide greatly increases the escape rate, by about an order of magnitude
in the case of the Fokker-Planck models.  (A qualitatively different
conclusion was, however, reached by Oh \& Lin (1992), using a hybrid
numerical scheme.)

In systems with a mass spectrum, the loss of stars of low mass is
relatively enhanced by mass segregation, which already places these
stars at large radii. As with mass segregation itself, however, this
does not significantly differentiate the stars of low mass from each
other.  For example, $N$-body results (Giersz \& Heggie 1996a) and
Fokker-Planck results (Chernoff \& Weinberg 1990) agree in showing
(Fig.~7.3) that, up to the time of core collapse, stars of mass 0.4
\msun\ escape only marginally faster than those of 1 \msun, in a 
system with a power law spectrum of masses in the range $0.4$ to 15
\msun. (This result would certainly be altered quantitatively in
models including stellar evolution, however.)  These simulations dealt
with systems up to the point of core collapse; the changes in the
evaporation rate in a tidally limited cluster {\sl after} core
collapse are described with the aid of Fokker-Planck simulations by
Lee \& Goodman (1995).

\vskip 11.0truecm
\item\item{\sevenrm{{\bf Fig.~7.3.} \hskip 2mm Rate of escape from 
a model cluster with a steady tidal field but no stellar evolution
(from Chernoff \& Weinberg 1990, Fig.~13).  The initial model was a
King model with scaled central potential $W_0$ = 3.  Each curve is
labelled with the mass in \msun, the initial mass function being $dN 
\propto m^{-2.5}dm$, discretised into 16 bins.  For each bin the curve
plots the remaining fraction of the original mass in that bin, against
time in units such that the initial half-mass relaxation time is about
0.019.  The collapse time is also about 0.019 unit.}}
\vskip 0.5truecm

In general terms the preferential loss of stars of low mass from
tidally bound systems is further enhanced if the effects of mass loss
(e.g., by supernova explosions) are included (Aarseth \& Woolf 1972).
The grid of Fokker-Planck models computed by Chernoff \& Weinberg
(1990) is the best starting point for grasping the combined effects of
a steady tidal field and mass-loss from stellar evolution.  Another
time-dependent process which may greatly enhance the preferential
escape of low-mass stars is tidal shocking (cf. \S6.3, and also
Weinberg 1994c).

So far this review of escape has concentrated on clusters with single
stars, but a sufficient abundance of primordial binaries (unless these
are extremely close) can greatly enhance the rate of {\sl
high-velocity} escapers (Leonard \& Duncan 1988, 1990).  Even in
clusters initially consisting of single stars, dynamical processes may
lead to the formation of binaries (see \S9.5) which then have a
substantial effect on the escape rate (Hayli 1970a, Danilov 1978), and
especially the flux of energy carried off by escapers (cf. Szebehely
1973, Giersz \& Heggie 1994).  There is a major difference here
between escape due to two-body encounters and that due to binaries;
the former actually increases the binding energy of the cluster,
whereas the latter causes a decrease.  Even in clusters with
primordial binaries, the net energy changes due to both process are
roughly comparable in magnitude (Heggie \& Aarseth 1992).  There is
also a non-negligible flux of escaping binaries.

Another class of high-energy escapers consists of neutron stars,
because of the high space velocities with which they are usually
thought to be born.  It has been estimated that at most 4\% of single
neutron stars would be retained within the modest potential well of a
typical globular cluster, though larger fractions are retained if the
neutron star is a member of a binary which is not disrupted by its
formation (Drukier 1996).

In principle the reverse of escape (capture) is possible (Peng \&
Weisheit 1992).

\vskip 30pt
\noindent
{\sl 7.4 Tidal truncation}

\vskip 20pt
\noindent
For many years the emphasis of theoretical studies was on isolated
systems, and this is one of several reasons why theoretical work has
had less influence on the interpretation of observations than should
have been the case.  On the other hand the theoretical difficulties
posed by inclusion of tidal effects, a few of which are already
mentioned in \S7.3, are non-trivial.

The motion of a star in a cluster is determined by the potential $\Phi
= \Phi_c + \Phi_g$, where the two terms refer to the cluster and the
galactic tide, respectively.  In an isolated system, $\Phi_g$ is taken
to be zero.  A first non-zero approximation for $\Phi_g$ is a
spherically symmetric concave function, in which case there is escape
of stars outside a certain limiting (tidal) radius $R_t$ at which
$d\Phi/dr = 0$.  This model leads to the idea of a {\sl cutoff
radius}, which is one of the main features of King's models for star
clusters, and their derivatives (cf. \S7.5).  On the other hand, these
models are constructed on the assumption that the potential in which
the stars move is $\Phi_c$ alone.

A next refinement is to compute more correctly the tidal field
experienced by stars in clusters.  More precisely, the motion of each
star is referred to a frame which moves, like the centre of mass of
the cluster, in a smooth orbit in an assumed galactic potential.  Then
the equations of motion of each star include inertial forces caused by
the acceleration of the reference frame, though the relatively small
size of clusters justifies the use of a linear approximation for the
relative tidal field. Even if $\Phi_g$ is steady in an inertial frame,
it may be time-dependent in the cluster frame, and the terms ``disk
shocking'' and ``bulge shocking'' refer to two situations in which
this feature is important.

Though relatively unrealistic for globular clusters, the case which
can be worked out in some detail is that of a cluster in a circular
orbit (e.g., in the equatorial plane of an axisymmetric galaxy; cf.
Chandrasekhar 1942, his \S5.5).  This complicates the construction of
models, however, because the tidal field lacks spherical symmetry: as
in problems of binary stars, the cluster is surrounded by a Roche lobe
with two Lagrangian points (in the directions of the galactic centre
and anticentre; cf. Fig.~7.4).  Their distance from the cluster is about
1.5 times that of the cutoff radius referred to above (Spitzer 1987,
Lee 1990, Heggie \& Ramamani 1995).  The asymmetry also has an effect
on the isotropy of stellar velocities (Oh \& Lin 1992), especially at
large radii (but within the tidal radius); in addition the Coriolis
force tends to deflect stars moving on nearly radial orbits.

If account is taken of the fact that the orbit of a cluster is
non-circular, this simple analysis fails.  A simple model studied by
Angeletti \etal\ (1983) and Angeletti \& Giannone (1983, 1984) showed
how the critical mean density for a bound system depended on the
eccentricity of the cluster orbit. More elaborate numerical studies
(Oh \etal\ 1992, Oh \& Lin 1992) indicate that the cluster is
truncated at a radius comparable with the theoretical tidal radius at
perigalacticon (as assumed by King 1962 and Ninkovi\'c 1985), unless
the relaxation time is sufficiently short, and then it may be
comparable to the theoretical result at apogalacticon.  The
observational position is not clear (Meziane \& Colin 1996).

The time-dependence of the tidal field reaches extreme limits in the
cases of disk shocking.  Its effects on the orbits of stars was
briefly discussed by Keenan \& Innanen (1975), using a 3-body model,
but its effect on the structure and evolution of the entire cluster
was investigated in a series of Fokker-Planck models by Spitzer \&
Chevalier (1973), and Spitzer \& Shull (1975b).  For the most part
disk shocking has been treated by computing the {\sl mean} change in
the energy of a star or cluster using an impulsive approximation, but
two recent developments have renewed interest in the process.  First,
Weinberg (1994a,b) has shown that slow (``adiabatic'') disk crossings
may be more disruptive than was previously thought.  Also, not all
effects of the shock are rapidly damped, and, in particular, it may
excite an oscillation in which the densest part rocks back and forth
within the envelope (Weinberg 1993b); this is reminiscent of the claim
by Calzetti \etal\ (1993) of the offset of the density centre from the
gravitational centre of \tuca. Second, Kundic \& Ostriker (1995) have
shown that the energies of the stars in a shocked cluster are subject
to {\sl random} changes which act rather like a relaxation mechanism,
and it may be especially important for stars beyond the half-mass
radius.  It has been aptly named ``shock relaxation'', and it is one
factor underlying a recent claim that the rate of destruction of
globular clusters in the galaxy has been underestimated (Gnedin \&
Ostriker 1996).  

\vskip 10truecm
\item\item{\sevenrm{{\bf Fig.~7.4.} \hskip 2mm An $N$-body model of 
a cluster evolving in a steady tidal field.  Though initially
consisting of $N$ = 8192 stars, the results are scaled to a cluster
with initial mass 1.5 \x\ 10$^5$ \msun, moving at speed 220 \kms\ on a
circular orbit of radius 4 kpc about a galaxy modelled as a point
mass.  The model is shown at a time which scales to about 7.7 Gyr, and
the unit of length scales to about 10 pc.  The initial model was a
King model with $W_0$ = 5 and a mass function $dN\propto m^{-3.5}dm$.
Mass loss through stellar evolution was included, and about 30\% of
the mass has been lost by a combination of all processes.  The
projection into the orbital plane is shown, the horizontal axis being
in the direction towards the galactic centre.  Stars escape in the
vicinity of the Lagrangian points, but are deflected by the Coriolis
force. From Aarseth \& Heggie (in preparation). }}
\vskip 0.5truecm

Other important time-dependent tidal processes which have been
considered include those due to interstellar clouds (Bouvier 1972;
Knobloch 1976, 1977), neighboring stellar systems (Layzer 1977) and
hypothetical massive black holes (Wielen 1987).  However the
disruptive effect of encounters with interstellar clouds is less
important than that of disk shocking (Chernoff \etal\ 1986).  
Bulge shocking has been considered by Alladin \etal\ (1976), Aguilar \&
White (1985), Spitzer (1987), Weinberg (1994b), and Charlton \& Laguna
(1995).

From a purely observational point of view, the direct measurement of
any tidal or limiting radius is extremely difficult, since it requires
the determination of a very low surface density of stars over very
large areas with potentially variable back- and foregrounds.  In
practice, and contrary to all other structural parameters, the
determinations of (idealised) tidal radius values for galactic
globular clusters, as published, e.g., by Trager \etal\ (1995) for 125
clusters, are always mere extrapolations of the
surface-brightness/star-count profiles, using King (or other
theoretical) models.

Grillmair \etal\ (1995a) provide, in the first study driven by a
purely observational approach, two-dimensional surface density maps of
the outer structure of 12 galactic clusters (see \S6.1 above).  The
extra-tidal material observed in most of their sample clusters is
identified with stars still in the process of being removed from the
clusters.  The complexity of the structures observed around these
clusters illustrates the foregoing theoretical remarks and shows the
intrinsic limiting accuracy in the process by which the tidal radius
is usually determined.

Chernoff \& Djorgovski (1989) find that, in the Galaxy, the
distribution of the collapsed-core clusters is much more concentrated
about the galactic centre than the distribution of the King model
clusters.  Within the King model cluster family, a similar trend
exists: centrally condensed clusters are found, on average, at smaller
galactocentric radii.  At fixed distance from the centre, the clusters
at smaller heights above the plane (and thus less inclined orbits) are
marginally more concentrated.  The fact that some internal properties
of clusters correlate well with global variables, such as the
galactocentric radius, suggests that some external effects are
important in cluster evolution.

Another important way in which time-dependent effects relate to tidal
truncation is in the dynamics of young star clusters.  It is possible
that as much as half of the mass of some young clusters in the LMC is
in the process of being lost by tidal overflow (Elson \etal\ 1987a).

\vskip 30pt
\noindent
{\sl 7.5 Theoretical models} 

\vskip 20pt
\noindent
In this section we assume we are dealing with the dynamics of star
clusters at a stage long after the time-dependent effects of the
initial conditions have reached dynamical equilibrium.  We have seen
that time-dependent tides have important effects on the subsequent
evolution, and intermittently disturb the assumed equilibrium.  Some
of these effects may be long-lived, but where disk shocking is mild,
which is the case for most galactic globular clusters, it may be
expected that departures from dynamic equilibrium will also be slight.
Another way of regarding the situation is to observe that the two
essential time scales for dynamical evolution are very different: the
crossing time \tcr, $\sim$ 10$^6$~yr, is much less than the relaxation
time \trh\ $\sim$ 10$^8$~yr, (and still smaller compared with the
typical evolution time \tev\ $\sim$ 10$^{10}$~yr).

The commonest way of defining a model of a star cluster is in terms of
its distribution function $f(\br,\bv,m)$, which is defined by the
statement that $fd^3\br d^3\bv dm$ is the mean number of stars with
positions in a small box $d^3\br$ in space, velocities in a small box
$d^3\bv$ and masses in an interval $dm$.  In terms of this description
a fairly general equation for the dynamical evolution is Boltzmann's
equation
$$
{\partial f\over\partial t} + \bv.\nabla_{\br}f -
\nabla_{\br}\Phi.\nabla_{\bv}f = {\partial f\over\partial t}_{enc},
$$ 
where $\Phi$ is the smoothed gravitational potential per unit mass,
and the right-hand side describes the effect of two-body encounters.
Under the above circumstances, however, the general Boltzmann equation
can be greatly simplified.  Because \tcr\ is so short, after a few
orbits the stars are mixed into a nearly stationary distribution, and
so the term $\partial f/\partial t$ is practically equal to zero. In a
similar way, because \trh\ is so long, the collision term $(\partial
f/\partial t)_{enc}$ can be ignored.  What is left, i.e.,
$$
\bv.\nabla_{\br}f - \nabla_{\br}\Phi.\nabla_{\bv}f = 0, \eqno(7.7)
$$
is an equilibrium form of what is frequently called Liouville's
equation, or the collisionless Boltzmann equation, or the Vlasov
equation.

In simple cases, the general solution of Eq.~7.7 is given by Jeans'
theorem, which states that $f$ must be a function of the constants of
the equations of motion of a star, e.g., the stellar energy per unit
mass $\varepsilon = v^2/2 + \Phi$. If not all constants of the motion
are known, such functions are still solutions, though not the most
general.  For a self-consistent solution, the distribution function
$f$ must correspond to the density $\rho$ required to provide the
cluster potential $\Phi_c$, i.e.:
$$\eqalign{
\nabla^2\Phi_c &= 4\pi G\rho \cr
               &= 4\pi G\int mfd^3\br d^3\bv dm. \cr 
}    \eqno(7.8)
$$
Many different kinds of models may be constructed with this approach.
In the first place there is considerable freedom of choice over which
integrals to include.  In the second place one is free to choose the
functional dependence of these integrals, i.e., the analytic form of
the distribution function (see, e.g., Binney 1982, and Binney \&
Tremaine 1987, Chapter~4.4).  In this section we describe those which
are of interest for a variety of reasons, while \S~7.7 concentrates on
those with important applications in the interpretation of
observational data.  See also Table~7.1.

{\sevenrm 
{{\hsize=12truecm
{
$$
\table
\tablewidth{12.0truecm}
\tablespec{\l\l\c\c\c\c\c\c\c\c}
\body{
\header{{\bf Table~7.1:} Dynamical models of globular star clusters}
\skip{10pt}
\hdoubleline
\skip{1pt}
\skip{1pt}
&&&&$\longleftarrow$&Static&$\longrightarrow$&&$\longleftarrow$&Evolutionary&$\longrightarrow$\end
&&&&                &Models &                &&                & Models & \end
\skip{1pt}
\skip{1pt}
&&          && King & Michie- & 3-Integral && Gas & Fokker- & N-Body \end
&&          &&      &  King   &            &&     & Planck  &        \end
\skip{1pt}
\skip{1pt}
\hline
\skip{1pt}
\skip{1pt}
&~~~Dynamical&Features~~~&&  &  &  &&  &  &  \end
\skip{1pt} 
\skip{1pt}
\hline
\skip{1pt}
\skip{1pt}
& Anisotropy& && ... & $\surd$ &  $\surd$ && $\surd$ & $\surd$ & $\surd$ \end
\skip{1pt}
& Rotation  & && ... &  ...    &  $\surd$ &&  ...    & $\surd$ & $\surd$ \end
\skip{1pt}
& Flattening& && ... &  ...    &  $\surd$ &&  ...    & $\surd$ & $\surd$ \end
\skip{1pt}
\skip{1pt}
\hline
\skip{1pt}
\skip{1pt}
&~~~Dynamical&Processes~~~& &  &  &  &&  &  &  \end
\skip{1pt}
\skip{1pt}
\hline
\skip{1pt}
\skip{1pt}
&Stellar  &1-body&& ...   & ...   & ...   &&$\surd$&$\surd$&$\surd$\end
&evolution&      &&       &       &       &&       &       &       \end
\skip{1pt}
\skip{1pt}
&Relaxation&2-body&&$\surd$&$\surd$&$\surd$&&$\surd$&$\surd$&$\surd$\end
\skip{1pt}
\skip{1pt}
&Tidal     &      &&       &       &       &&       &       &       \end
&Interactions,&2-body&&... & ...   & ...   && ...   &$\surd$&$\surd$\end
&Collisions&      &&       &       &       &&       &       &       \end
\skip{1pt}
\skip{1pt}
& Stellar&2-body&&$\surd$& ...   & ...   && ...   &$\surd$&$\surd$\end
&Escape  &      &&       &       &       &&       &       &       \end
\skip{1pt}
\skip{1pt}
&Primordial&3- and&& ...   & ...   & ...   &&$\surd$&$\surd$&$\surd$\end
&Binaries  &4-body&&       &       &       &&       &       &       \end
\skip{1pt}  
\skip{1pt}
&Stellar&collision-&&$\surd$&$\surd$&$\surd$&&$\surd$&$\surd$&$\surd$\end
&Motions&less      &&       &       &       &&       &       &       \end
\skip{1pt}
\skip{1pt}
&Steady&collision-&&$\surd$&$\surd$&$\surd$&& ...   &$\surd$&$\surd$\end
&Tide&less        &&       &       &       &&       &       &       \end
\skip{1pt}
\skip{1pt}
&Disk&collision-&& ...   & ...   &  ...  && ...   &$\surd$&$\surd$\end
&Shocking&less  &&       &       &       &&       &       &       \end
\skip{1pt}
\skip{1pt}
\hline
}\endtable
$$
}
}}
}
{\rm Note: under the heading ``Dynamical Process'', the second column
states what kind of physical process it is that is named in the first column}

\vskip 20pt
\noindent
{\sl 1) Systems whose distribution functions depend only on the energy
per unit mass $\varepsilon$.}  These are the most commonly used
models.  They are spherical and have an isotropic velocity dispersion
($\overline {v_r^2}$ = $\overline {v_{\theta}^2}$ = $\overline
{v_{\phi}^2}$):

$\bullet$ Isothermal sphere: this historical starting point cannot
itself serve as a realistic model because its density $\rho$ $\propto$
$r^{-2}$ at large radii, which means that the model has an infinite
mass.  Nevertheless it is of great importance for theory, and is a
useful approximation for parts of more realistic models.

$\bullet$ Plummer's model and allied models: Plummer's model is used
frequently by theorists for its analytical convenience (cf. Spitzer
1987), but several other sets of analytical models have been
investigated (e.g., Bagin 1979).  Veltmann (1983) and Dejonghe (1984)
have shown how to construct series of models which include Plummer's
model and H\'enon's isochrone model (H\'enon 1959) as special cases.

$\bullet$ King models: these are the simplest models that observers
take seriously.  They can be thought of as a modification of the
isothermal model (King 1966), with a distribution function given by
the ``lowered maxwellian'' form
$$ 
f\propto\cases{e^{\displaystyle{-2j^2\varepsilon}} -
e^{\displaystyle{-2j^2\varepsilon_t}}&if $\varepsilon<\varepsilon_t$,\cr
0 &if $\varepsilon<\varepsilon_t$,\cr}
\eqno(7.9) 
$$
where $j$ and $\varepsilon_t$ are constants.  The truncation at energy
$\varepsilon_t$ corresponds to an absence of very loosely bound stars.
The worldwide renown of the King model is probably due to King's ideal
combination of both theoretical and observational innovations,
supported by a very simple and clear presentation, and ease of
computation.  By giving results in the observational plane, King's
work provided a simple, yet essential, interface between theory and
observation.  A decade later, da Costa \& Freeman (1976) showed that
single-mass, isotropic King models are unable to fit the entire
density profile of M3.  They generalized these simple models to
produce more realistic multi-mass models with approximate
equipartition of energy in the centre.  (The construction of models
with equipartition is a non-trivial issue, actually; cf. \S7.7 and
Merritt 1981). The observational application of King models and their
variants is further described below (\S7.7).  Before passing on from
King models, however, it should be mentioned that, despite their
equilibrium nature, they have been used to investigate, in a quick but
approximate manner, the evolutionary effects of various dynamical
processes, assuming that the system evolves along the King
sequence. Examples include the studies by Prata (1971a,b), Retterer
(1980a), and Chernoff \etal\ (1986).

$\bullet$ The Wilson sphere and other variants: in this model (Wilson
1975) the distribution function differs from King's distribution in
that both the function and its gradient vanish at the boundary
$\varepsilon_t$, in contrast to King's distribution which has non-zero
gradient at this point.  Wilson spheres have more heavily truncated
distribution functions than King models, and therefore have more
extended envelopes.  Other ways of adjusting the maxwellian have been
described by Woolley (1954) and Woolley \& Dickens (1962), who
implemented a cutoff by simply truncating the maxwellian, and by
Davoust (1977), Binney (1982) and Madsen (1996).

\vskip 20pt
\noindent
{\sl 2) Systems whose distribution functions depend only on the energy
per unit mass $\varepsilon$ and the specific angular momentum $l$.}
Such models are spherical and have an anisotropic velocity dispersion
($\overline {v_r^2}$ $\not=$ $\overline {v_{\theta}^2}$ = $\overline
{v_{\phi}^2}$):

$\bullet$ Eddington models: Eddington (Shiveshwarkar 1936) took the
distribution of the isothermal sphere times $\exp(-j^2 l^2/r_a^2)$,
where \ra\ is a constant.  This anisotropy factor makes the
distribution function almost zero when $l$ is large, i.e., it
depopulates all those orbits which at large distances are almost round
(see Eq.~7.11).  The density profile of an Eddington model falls off
more rapidly than that of the equivalent isothermal sphere, but never
drops to zero.

$\bullet$ King-Michie models: they associate the ``lowered
maxwellian'' of the King model with the anisotropy factor of the
Eddington models (Gunn \& Griffin 1979, and Eq.~7.11).  They have
tidal radii that lie between the tidal radii of the corresponding King
and Wilson models.  The multi-mass King-Michie models have been the
ones most frequently used when fitting simultaneously density and
velocity dispersion profiles, and are therefore described further in
\S7.7.

$\bullet$ There are more such recent models, all of them being
variations on the theme of the two integrals of motion $\varepsilon$
and $l$.  For example, Osipkov-Merritt models have a distribution
function which depends on $\varepsilon$ and $l$ only through the
variable $Q = \varepsilon - l^2/(2r^2_a)$ (Osipkov 1979, Merritt
1985a,b).  Dejonghe (1987) has constructed a convenient series of
models which all have the same density profile as a Plummer model, but
with varying amounts of anisotropy.  Another series described by Louis
(1993) has a particularly convenient distribution function.  Other
examples are given by Batt \etal\ (1986) and Louis (1990).

\vskip 20pt
\noindent
{\sl 3) Systems whose distribution functions depend only on the energy
per unit mass $\varepsilon$ and the component of angular momentum
parallel to the rotation axis $l_z$.} Such models are elliptical and
have tangential anisotropy of the velocity dispersion ($\overline
{v_r^2}$ = $\overline {v_{\theta}^2}$ $\not=$ $\overline
{v_{\phi}^2}$).  They have been extensively studied in the context of
galactic stellar dynamics, where the subject has been efficiently
reviewed by de Zeeuw (1987).  Observations have shown that rotation is
present in globular clusters (\S7.6), even though it is weaker than in
many galaxies, and this has motivated a renewed search for suitable
models.

$\bullet$ Uniform rotation: the simplest distribution functions that
involve only $\varepsilon$ and $l_z$ have the form $f = F (\varepsilon +
\omega l_z)$, where $F$ is an arbitrary function and $\omega$ is
constant (e.g., Woolley \& Dickens 1962, Vandervoort \& Welty 1981).
In a frame rotating with angular velocity $\omega$ the distribution
function actually depends on energy alone.  Unfortunately, for any
distribution function of this form, the mean motion corresponds to
rotation at the constant angular speed $\omega$, which is quite
unrealistic.

$\bullet$ Prendergast-Tomer and Wilson models: some more realistic
models have been introduced by Prendergast \& Tomer (1970) and Wilson
(1975).  The models of Prendergast \& Tomer introduce differential
rotation not as an explicit part of the theory, but merely as a result
of the finite escape velocity interacting with a velocity distribution
that would otherwise have yielded a solid-body rotation.  In Wilson's
models, the differential rotation has been included explicitly via an
adjustable parameter.  Such models have two characteristics: (i) after
an increase in the inner part towards a maximum rotational velocity,
the rotation curve decreases towards the outer parts; (ii) the central
parts of the model are always rather spherical.  These two points make
them more suited to globular clusters than elliptical galaxies,
although these models have essentially been applied to elliptical
galaxies.

\vskip 20pt
\noindent
{\sl 4) Systems whose distribution functions depend on a third
integral of motion $I_3$, in addition to the energy per unit mass
$\varepsilon$ and the component of angular momentum parallel to the
rotation axis $l_z$.}  Although no general analytical form for a third
integral is available, the existence of an analytic third integral of
motion $I_3$ in special cases has been known for decades, since the
work by Jeans (1915).  In other cases it has been shown from numerical
orbit calculations that the motions of stars are effectively
``integrable'', which in this context means that they are governed by
an approximate third integral.  It is because the ellipticities of
globular star clusters are so modest that the motions of the stars are
unlikely to exhibit any of the usual signs of a breakdown of
integrability, such as chaotic orbits.

$\bullet$ 3-integral models: so far, only a few studies have tried to
develop 3-integral models; see, e.g., Petrou (1983a,b), Dejonghe \& de
Zeeuw (1988).  The first such study totally devoted to globular
clusters (Lupton \etal\ (1985), Lupton (1985), Lupton \& Gunn 1987,
Lupton \etal\ 1987) uses $l^2$ as a first approximation to $I_3$,
leading to a distribution function depending on $\varepsilon$, $l_z$,
and $l^2$.  Because the rotation creates a nonspherical potential,
$l^2$ is in fact only an approximate integral and Lupton \& Gunn's
distribution function does not obey the collisionless Boltzmann
equation for equilibrium (Eq.~7.7).  See Dehnen \& Gerhard (1993) for
a similar study related to oblate elliptical galaxies.

\vskip 20pt
\noindent
Models constructed in the way we have described, i.e., by use of
Jeans' theorem, are designed to satisfy Eq.~7.7 rigorously. Another
approach is to construct models satisfying moments of Eq.~7.7, i.e.,
Jeans' equations.  This is the approach taken by Bagin (1976b) in the
construction of multi-component rotating models, and by Davoust (1986)
in the single-component case, and this yielded models which he applied
to several globular clusters.  A hazard of this method is that the
resulting model {\sl may} not be realisable using positive
distribution functions $f$ (cf. Bagin 1976a).

\vskip 30pt
\noindent
{\sl Stability.}  One factor which may influence the choice of an
appropriate model, whether it is constructed from a distribution
function or from Jeans' equations, is its stability.  What is at issue
here is stability on the crossing time scale, and not on the
relaxation time scale; the latter is discussed in \S9.1.  Thus
dynamical stability is concerned with much the same issues as violent
relaxation, i.e., whether bulk motions of the matter in a stellar
system damp out or grow.  This is a large subject, and Merritt (1987a)
and Binney \& Tremaine (1987, Chapter~5) provide nice introductory
accounts.  For a full mathematical, but still very readable treatment,
Palmer (1994) is recommended.  The following remarks describe recent
work, especially that related to globular clusters (i.e., spherical or
slowly rotating systems).

For spherical non-rotating models, the most relevant instability
(Dejonghe \& Merritt 1988, Merritt 1990) is the ``radial orbit
instability'', which leads to bar formation in sufficiently
anisotropic systems.  As shown by Palmer \& Papaloizou (1987), this
instability can in principle manifest itself in systems in which the
global anisotropy is arbitrarily small, and Palmer \etal\ (1990)
extended this result to axisymmetric systems.  In physically more
reasonable models, the global anisotropy has to be sufficiently large
for the instability to occur (e.g., Weinberg 1991).  Saha (1992) has
shown how to construct anisotropic spherical models for a
single-component system and to test for their stability.  Other
instabilities affect the radial distribution of the stars without
changing the shape of the system (e.g., Stiavelli 1990), or give rise
to a displacement of the densest part of the system (e.g., Merritt \&
Hernquist 1991).  Even slowly rotating systems are subject to a
``tumbling'' instability (Allen \etal\ 1992).

\vskip 30pt
\noindent
{\sl 7.6 Observational evidence of cluster rotation} 

\vskip 20pt
\noindent
Compared with galaxies, galactic globular clusters are anomalously
spherical in shape.  The flattest elliptical galaxies observed are of
type E7, in striking contrast to the flattest galactic globular
clusters, which have type E2.  As relaxation times of galaxies greatly
exceed the Hubble time, a quasi-steady evolution cannot have altered
substantially either the initial dynamical structure or the shape of
such stellar systems.  On the contrary, central relaxation times \trc\
of globular clusters being much shorter than the age of the universe
(typically 10$^6$ \lsim\ \trc\ \lsim\ 10$^8$~yr; frequently \trh\
\lsim\ \bily), strong evolutionary changes may have transformed the
shape of the globular clusters since the time of formation.

Agekian (1958), Shapiro \& Marchant (1976), and Longaretti \& Lagoute
(1996a) all studied the way in which the angular momentum carried off
by escapers can affect the ellipticity (see \S7.3 above), and found
that the ellipticity can decrease significantly over the lifetime of a
cluster.  An age-ellipticity relation has indeed been observed by
Frenk \& Fall (1982) for clusters in the Galaxy and in the Large
Magellanic Cloud (see also Geyer \etal\ 1983 and Akiyama 1991).

An interesting study of the true shape of globular clusters is given
in Fall \& Frenk (1983), who discuss the distributions of true and
apparent ellipticities for random orientations.  The intrinsic shapes
of globular clusters in our Galaxy, M31, and in the Large and Small
Magellanic Clouds are compared by Han \& Ryden (1994; see also Ryden
1996).  They find that, for the galaxies with similar structure, mass,
and age, their globular clusters tend to have similar shapes, i.e.,
the clusters in our Galaxy and M31 are, on average, more spherical
than those in the Magellanic Clouds.  It is worth mentioning that, due
to the intrinsic differences between photographic and CCD images and
between the various ellipticity estimate techniques and definitions
(e.g., mean ellipticity or at a given radius), the measured
ellipticities for individual clusters frequently disagree.

The detection of rotation in globular clusters (from proper motions
and/or radial velocities of numerous individual stars) suffers always
from the uncertainty due to the lack of knowledge of sin~$i$, where
$i$ is the angle between the line of sight and the rotation axis of
the cluster ($i$ = 90\deg\ when equator-on, $i$ = 0\deg\ when
pole-on).

\vskip 20pt     
\noindent
{\sl Rotation from stellar proper motion measurements.}~~ So far, it
is only in the case of M22 that Peterson \& Cudworth (1994) have been
able to clearly detect rotation from proper motion data, although Rees
\& Cudworth (pers. comm.) see rotation in the \tuca\ proper motions as
well.

{\sevenrm 
{{\hsize=12truecm
{
$$
\table
\tablewidth{12.0truecm}
\tablespec{\l\c\c\c\r\r}
\body{
\header{{\bf Table~7.2:} Rotation in galactic globular clusters}
\skip{10pt}
\hdoubleline
\skip{1pt}
\skip{1pt}
&~~~ cluster & \Vrotmax & \Vrotmax / $\sigma$ & $\varepsilon$ $^{(1)}$ & number & reference~~~ &\end
\skip{1pt}
&         &  (\kms)     &                  &          & of stars &          &\end
\skip{1pt}
\skip{1pt}
\hline
\skip{1pt}
\skip{1pt}
& NGC~5272$\equiv$ M3   & 1.0 &0.12 &0.04 & 107~~~ & Gunn \& Griffin (1979)\end 
\skip{1pt}
& NGC~6341$\equiv$ M92  & 2.5 &0.30 &0.10&  49~~~ & Lupton \etal\ (1985)   \end 
\skip{1pt}
& NGC~7089$\equiv$ M2   & 5.5 &0.34 &0.11&  69~~~ & Pryor \etal\ (1986)    \end 
\skip{1pt}
& NGC~5139$\equiv$ \cen & 8.0 &0.32 &0.17& 318~~~ & Meylan \& Mayor (1986) \end 
\skip{1pt}
& NGC~104~$\equiv$ \tuc & 6.5 &0.26 &0.09& 272~~~ & Meylan \& Mayor (1986) \end 
\skip{1pt} 
& NGC~6205$\equiv$ M13  & 5.0 &0.25 &0.11& 142~~~ & Lupton \etal\ (1987)   \end 
\skip{1pt}
& NGC~6397              & 0.5 &0.11 &0.07& 127~~~ & Meylan \& Mayor (1991) \end 
\skip{1pt}
& NGC~6656$\equiv$ M22  & 3.8 &0.50 &0.14&130~~~ &Peterson\& Cudworth (1994)\end 
\skip{1pt}
& NGC~362               & 0.0 &0.01 &0.01& 208~~~ & Fischer \etal\ (1993b) \end 
\skip{1pt}
& NGC~7078$\equiv$ M15  & 1.7 &0.15 &0.05& 216~~~ & Gebhardt \etal\ (1994) \end 
\skip{1pt}
& NGC 3201              & 1.2 &0.28 &0.12& 399~~~ & C\^ot\'e \etal\ (1995) \end 
\skip{1pt}
& NGC~5139$\equiv$ \cen & 7.9 &0.41 &0.17& 469~~~ & Merritt \etal\ (1996)  \end 
\skip{3pt}
\skip{1pt}
\hline
}\endtable
$$
}
}}
}

{\sevenrm 
\vskip -7mm
~(1) all ellipticity values from White \& Shawl (1987)
}

\vskip 20pt
\noindent
{\sl Rotation from stellar radial velocity measurements.}~~
Cross-correlation techniques provide stellar radial velocities with
errors of typically 1 \kms\ or less, i.e., significantly smaller than
proper motion uncertainties, and have, consequently, allowed detection
of rotation in a few globular clusters (see Table~7.2).

The galactic globular cluster in which rotation is expected the most
is \cent, the giant southern globular cluster, which has the largest
mean ellipticity \ellim\ = 0.12 (with 0.05 $\leq$ \elli\ $\leq$ 0.17,
Geyer \etal\ 1983) and the longest central relaxation time (\trc\
$\simeq$ \bily, Meylan 1987).  The first indication of the presence of
rotation is published by Harding (1965), who uses a sample of 13 stars
(each having at least three radial velocity measurements), with the
projection of the rotation axis supposed identical to the minor axis
of the stellar distribution on the plane of the sky.  An unpublished
study by Seitzer (1983), based on 118 stars, displays the differential
rotation of \cent.  

In \cent\ Meylan \& Mayor (1986) use the radial velocities of 318
stars, scattered on the plane of the sky between 0.30\arcm\ and
22.4\arcm\ from the cluster centre.  These data reveal immediately the
presence of rotation by analysis of the radial velocities according to
the hypothesis of a projected differential rotation of the type:
$$
\langle V_r \rangle = A(r){\rm sin} (\alpha + \phi) + V_{\circ} \eqno(7.10)
$$
where $r$ and $\alpha$ are the polar coordinates of the star with
respect to the cluster centre.  This approach has been used in most of
the studies mentioned in Table~7.2.  Fig.~7.5 displays a plot of \Vr\
vs.  $\alpha$, where \Vr\ is the mean radial velocity of each star,
for all stars measured in \cent\ in a ring on the plane of the sky in
which the maximum of the rotation is reached.  The sinusoidal
distribution of the points betrays immediately the presence of
rotation, although Eq.~7.10 does not describe the real behavior of the
field of radial velocities as projected on the plane of the sky.

\vskip 10truecm
\item\item{\sevenrm{{\bf Fig.~7.5.} \hskip 2mm Radial velocities \Vr\
as a function position angles $\alpha$, for 205 stars in \cent\ with
radii $r$ between 300\arcs\ and 1200\arcs\ (from Meylan \& Mayor 1986,
Fig.~1a).  The conspicuous sinusoidal distribution of the points
reveals the presence of rotation. }}
\vskip 0.5truecm

In order to estimate the systemic rotation of the cluster as a whole
around the axis of symmetry of the ellipsoid, Meylan \& Mayor (1986)
use an ad~hoc analytic form, as general as possible, in order to mimic
any kind of rotation curve (e.g., flat or Keplerian).  This has the
advantage of reducing the dependence on any idiosyncrasies of the
model.  The analytic form depends on four free parameters.  Three of
them describe the equatorial rotation curve, namely (i) a solid
rotation in the cluster inner part, (ii) the maximum of the rotation
curve \Vrotmax\ and its distance \rmax\ from the axis of symmetry, and
(iii) a differential rotation in the outer parts; a fourth parameter
represents (iv) the decrease of the rotation in the direction of the
poles, since the cluster does not have cylindrical rotation.  For each
nonlinear least-squares fit between computed and observed velocities,
sin~$i$ (0\deg\ $\leq$ $i$ $\leq$ 90\deg) is a fixed parameter.
Rotation is definitely observed in \cent.  For $i$ = 90\deg, \Vrotmax\
= 8.0 \kms, occurring at 3-4 \rc.  This non-cylindrical differential
rotation is most important in a central torus and weak in the outer
parts (see Fig.~7.6).  The angular velocity \omegac\ inside 1 \rc\
equals 1.4 \x\ $10^{-6}$yr$^{-1}$, which corresponds to one revolution
of the core in 4.5 \x\ \mily.  The similarity between the rotation and
the ellipticity curves is impressive and suggests that the flattening
of \cent\ is due to rotation (see Fig.~2a in Meylan \& Mayor 1986).

\vskip 9truecm
\item\item{\sevenrm{{\bf Fig.~7.6.} \hskip 2mm Line-of-sight rotational
velocity field in the first quadrant of the meridional plane, from a
non-parametric estimate of the rotation in \cent\ by Merritt \etal\
(1996).  The units on both axes are arcminutes and contours are
labelled in \kms. }}
\vskip 0.5truecm

With an enlarged sample of 469 stellar members of \cent\ (Meylan
\etal\ 1995), a non-parametric (see \S7.7) estimate of the mean
line-of-sight velocity field on the plane of the sky has been
constructed by Merritt \etal\ (1996).  Feast \etal\ (1961) have shown
that some of the observed rotation is merely a perspective effect
caused by the proper motion of the entire cluster.  Merritt \etal\
(1996) have corrected for this slight effect (about 1 \kms\ in the
case of \cent).  Fig.~7.6 displays the line-of-sight rotational
velocity field in the first quadrant (Merritt \etal\ 1996).  These
results confirm, both qualitatively and quantitatively, those
displayed in Fig.~3 in Meylan \& Mayor (1986) and obtained in a
completely different way.

Mayor \etal\ (1984) provides the first detection of rotation in \tuca,
the second best studied globular cluster, for which the mean
ellipticity \ellim\ = 0.10 (with 0.08 $\leq$ \elli\ $\leq$ 0.13, Geyer
\etal\ 1983).  Meylan \& Mayor (1986) use mean radial velocities of
272 member stars scattered on the plane of the sky between 0.15\arcm\
and 14.4\arcm\ from the cluster centre.  In a way similar to the study
of \cent, rotation is definitely observed in \tuca.  For $i$ = 90\deg,
\Vrotmax\ = 6.5 \kms, occurring at 11-12 \rc.  This non-cylindrical
differential rotation is most important in a central torus and weak in
the outer parts.  The angular velocity \omegac\ inside 1 \rc\ equals
2.7 \x\ $10^{-6}$yr$^{-1}$, which corresponds to one revolution of the
core in 2.9 \x\ \mily.  Probably because of poor ellipticity data,
the similarity found in \cent\ between the rotation and the
ellipticity curves is not observed in \tuca.

All the above results obtained for \cent\ and \tuca\ depend on the
value of the angle $i$, which remains unknown.  Since these two
clusters belong to the small group of the clusters which, among the
150 galactic globular clusters, are the flattest ones, we can expect,
from a statistical point of view, that their angles $i$ should not be
very different from 60\deg\ $\leq$ $i$ $\leq$ 90\deg.  The importance
of rotation (for a given projected rotation velocity) increases as $i$
gets closer to 0\deg.  The relative importance of rotational to random
motions is given by the ratio \Voso, where $V_{\circ}^2$ is the
mass-weighted mean square rotation velocity and $\sigma_{\circ}^2$ is
the mass-weighted mean square random velocity.  For $i$ = 90\deg\ and
60\deg, in \cent\ the ratio \Voso\ = 0.35 and 0.39 and in \tuca\ the
ratio \Voso\ = 0.40 and 0.46, respectively (Meylan \& Mayor 1986).
Even with $i$ = 45\deg, the dynamical importance of rotation remains
weak compared to random motions.  The ratio of rotational to random
kinetic energies is $\simeq$ 0.1, confirming the fact that globular
clusters are, above all, hot stellar systems.  

As displayed in Table~7.2, rotation has been directly observed and
measured in ten globular clusters.  The diagram \Voso\ vs. \ellim\ has
been frequently used for elliptical galaxies and its meaning is
extensively discussed in Binney \& Tremaine (1987 Chapter~4.3).  The
low luminosity ($L$ \lsim\ 2.5 10$^{10}$ \lsun) elliptical galaxies
and spheroids have (\Voso,\ellim) values which are scattered along the
relation for oblate systems with isotropic velocity-dispersion
tensors, while the high luminosity ($L$ \gsim\ 2.5 10$^{10}$ \lsun)
elliptical galaxies have (\Voso,\ellim) values which are scattered
below the above relation, indicating the presence of anisotropic
velocity-dispersion tensors.  Given their small ellipticities (0.00
$\leq$ \ellim\ $\leq$ 0.12), globular clusters are located in the
lower-left corner of the \Voso\ vs. \ellim\ diagram, an area
characterized by isotropy or mild anisotropy of the
velocity-dispersion tensor.

\vskip 30pt
\noindent
{\sl 7.7 Model fitting: parametric and non-parametric methods}

\vskip 20pt
\noindent
During the last two decades, the purpose of building dynamical models
has been to construct various simplified mathematical descriptions of
a star cluster, each of them easily comparable with observations.  The
general principles on which such models may be built are described in
\S7.5, where many examples were summarised.  Very few of these,
however, have achieved any prominence in applications, and it is on
these that we concentrate in the present subsection.  Foremost amongst
these are King's models and their variants.

King models approximately incorporate three essential dynamical
processes (Table~7.1): (i) dynamical equilibrium, (ii) the effect of
gravitational encounters between pairs of stars, which, like
collisions in a gas, tend to set up a maxwellian distribution of
velocities, and (iii) a cutoff in energy ($\varepsilon_t$,
cf. Eq.~7.9) above which stars are deemed to have escaped; the cluster
potential $\Phi_c$ takes this value at a finite radius, which can be
interpreted loosely as the tidal radius $r_t$ (\S7.4).  (As already
mentioned in \S7.4, however, in the construction of King models, the
galactic potential $\Phi_g$ is ignored.)  The models are readily
constructed numerically, and results are conveniently tabulated in
Ichikawa (1985).

King's models depend on three dimensional parameters, which can be
taken to be the central density $\rho_c$, the central velocity
dispersion $\sigma_c$, and the tidal radius \rt.  There is one
dimensionless parameter, which can be taken to be the ratio of the
tidal to core radius, i.e., the concentration $c = r_t/r_c$, or the
dimensionless central potential (Fig.~7.3).  The definition of $r_c$
can cause confusion, but here it refers to the scaling length which
appears in the theory of the models, which satisfies $8\pi G\rho_c
r_c^2 j^2/9 = 1$.  This is often extended to other models by replacing
$3/(2j^2)$ by the central three-dimensional velocity dispersion, but
even for King models this is only an approximation.  There is no
really satisfactory theoretically-based definition for multi-component
models.

As originally described, King models are single-component models,
i.e., Eq.~7.9 makes no distinction between stellar masses.  The
construction of multi-mass variants is a matter of reasoned choice.
If the analogy is made with the kinetic theory of gases, one assumes
that the constant $j^2$ in Eq.~7.9 is proportional to mass (e.g.,
Illingworth \& King 1977, Kondrat'ev \& Ozernoi 1982).  In a
maxwellian distribution this leads to the usual equipartition of
kinetic energies, but with the lowered maxwellian used in King models,
the distribution of velocities becomes nearly independent of mass for
the lowest masses.  The specification of the mass function is the
usual compromise between convenience and realism.  It is often taken
to be a power law (see Eq.~6.3).

After the introduction of a mass spectrum, the next important variant
of King's original models deals with the isotropy of the velocity
distribution.  When Eq.~7.9 is used for the distribution function, the
distribution of velocities is isotropic. In order to include
anisotropy, the distribution function can be made to depend on the
specific angular momentum $l$.  Following Gunn \& Griffin (1979), the
most commonly chosen form is
$$ f \propto \exp(-j^2 l^2/r_a^2)
(\exp(-2j^2\varepsilon)-\exp(-2j^2\varepsilon_t))  \eqno(7.11) 
$$
for $\varepsilon < \varepsilon_t$, where $r_a$ is a constant.  This
introduces a second dimensionless parameter (the ratio of $r_a/r_c$,
for example). When anisotropy is introduced in this way along with
unequal masses, it is usual to take $r_a$ to be the same for all
masses, as it corresponds to the radius beyond which the anisotropy
becomes important.  Incidentally, the increase of anisotropy with
radius may not be all that appropriate for tidally bound models,
because of the isotropising effect of the tidal field, which is
important at large radii.

\vskip 20pt
\noindent
The construction of King-Michie models and their variants is based on
a simplified mathematical description of a star cluster.  Such models
allow the quick computation of grids of models which sample large
ranges for the values of the free parameters which are, typically, the
central potential, the mass function index, and the anisotropy radius.
The fits of these models to the observational constraints (generally,
density and velocity dispersion profiles) provides the structural
parameters, i.e. the radii \rc, \rh, and \rt, and hence the
concentration $c$, along with stellar density, total mass, and \mlv\
values.  The most updated list of structural parameters is given by
Trager \etal\ (1995) for 125 galactic globular clusters.

The studies constrained simultaneously by density and velocity
dispersion profiles provide the most reliable estimates of stellar
densities (the central mass density \rhoo, the mean mass
density \rhoh\ inside the half-mass radius, and the mean mass density
\rhot\ inside the tidal radius), with the total mass of the cluster
and its central and global \mlv.  Table~7.3 gives the list of the main
studies using multi-mass King-Michie type models simultaneously
constrained by density and velocity dispersion profiles.  In these
studies, all radial velocities have been acquired by single-object
spectrometers. 

{\sevenrm 
{{\hsize=12truecm
{
$$
\table
\tablewidth{10.0truecm}
\tablespec{\l\l\l}
\body{
\header{{\bf Table~7.3:} Dynamical studies using King-Michie type models }
\skip{10pt}
\hdoubleline
\skip{1pt}
\skip{1pt}
& cluster  & authors  & reference \end
\skip{1pt}
\skip{1pt}
\hline
\skip{1pt}
\skip{1pt}
& M3	  & Gunn \& Griffin	& 1979, AJ, 84, 752  		\end
\skip{1pt}
& \tuca\  & Mayor \etal\	& 1984,	A\&A, 134, 118  	\end
\skip{1pt}
& M92 	  & Lupton \etal\ 	& 1985,	IAU Symp. 113, p. 327   \end
\skip{1pt}
& M2 	  & Pryor \etal\  	& 1986, AJ, 91, 546	 	\end
\skip{1pt}
& M13 	  & Lupton \etal\ 	& 1987, AJ, 93, 1114	 	\end
\skip{1pt}
& \cent\  & Meylan 		& 1987,	A\&A, 184, 144	 	\end
\skip{1pt}
& \tuca\  & Meylan 		& 1988,	A\&A, 191, 215	 	\end
\skip{1pt}
& \tuca\  & Meylan 		& 1989,	A\&A, 214, 106	 	\end
\skip{1pt}
& M15 	  & Peterson \etal\	& 1989,	ApJ, 347, 251	 	\end
\skip{1pt}
& NGC~6397& Meylan \& Mayor	& 1991,	A\&A, 250, 113	 	\end
\skip{1pt}
& M15 	  & Grabhorn \etal\	& 1992, ApJ, 392, 86	 	\end
\skip{1pt}
& NGC~362 & Fischer \etal\ 	& 1993b, AJ, 106, 1508	 	\end
\skip{1pt}
& NGC~3201& Da Costa \etal\	& 1993, ASP Conf. Ser. 50, p. 81\end
\skip{1pt}
& \cent\  & Meylan \etal\ 	& 1995, A\&A, 303, 761	 	\end
\skip{1pt}
\skip{1pt}
\hline
}\endtable
$$
}
}}
}

The most updated list of dynamical parameters, like density, mass and
\mlv\ values, is given by Pryor \& Meylan (1993) for 56 galactic
globular clusters.  It is often said that such total masses and global
\mlv\ values are very model-dependent, partly, it is argued, because
the observations poorly constrain very low-mass populations.  In fact,
however, it was pointed out by Gunn \& Griffin (1979) that replacing
$N$ stars of low mass $m$ by, say, $2N$ stars of mass $m/2$ has little
effect on the model fits, since mass segregation implies that most of
these stars are at large radii, almost independent of their precise
mass.  In addition, it is argued that surface brightness and velocity
dispersion profiles do not uniquely determine the cluster mass.
Nevertheless the basis of the non-parametric methods advocated by
Merritt (see below) is that they {\sl do} constrain the mass
distribution uniquely, provided the distribution function is
isotropic.  Therefore it is the degree of anisotropy which may well be
the most important model-dependent assumption, and little is known
about its effect on the inferred total masses.

Using biweight estimators (Beers \etal\ 1990; these estimators are
insensitive to outliers) with the entire sample of 56 clusters yields a
mean \mlv\ of 2.3 and a dispersion about the mean of 1.1.  Similarly,
the mean central \mlv\ is 1.7 and the dispersion is 0.9.  The global
\mlv\ does not correlate significantly (absolute value of the
correlation coefficient $|r|$~$<$~0.22) with distance from the
galactic centre, distance from the galactic plane, metallicity,
concentration, or half-mass relaxation time.  Global \mlv\ is weakly
correlated with the total mass ($r$~=~0.31).  Independence of these
two quantities can be rejected at better than 95\% confidence, but
this conclusion and the correlation coefficient are compromised by the
correlation between the errors in mass and \mlv.  Mandushev \etal\
(1991) have found similar results with a sample of 32 clusters.
Whether \mlv\ really tends to increase with increasing mass is still
uncertain (Pryor \& Meylan 1993).

{\sevenrm 
{{\hsize=12truecm
{
$$
\table
\tablewidth{10.0truecm}
\tablespec{\l\l\l}
\body{
\header{{\bf Table~7.4:} Dynamical studies using non-parametric techniques }
\skip{10pt}
\hdoubleline
\skip{1pt}
\skip{1pt}
& cluster  & authors  & reference \end
\skip{1pt}
\skip{1pt}
\hline
\skip{1pt}
\skip{1pt}
&M15 		& Gebhardt \etal\ 	& 1994, AJ, 107, 2067 \end
\skip{1pt}
&~~~'' 		& Gebhardt \& Fischer 	& 1995, AJ, 109, 209  \end
\skip{1pt}
&\tuc\   	& Gebhardt \& Fischer 	& 1995, AJ, 109, 209  \end
\skip{1pt}
&NGC~362 	& Gebhardt \& Fischer 	& 1995, AJ, 109, 209  \end
\skip{1pt}
&NGC~3201 	& Gebhardt \& Fischer 	& 1995, AJ, 109, 209  \end
\skip{1pt}
&~~~''     	& C\^ot\'e \etal\ 	& 1995, ApJ, 454, 788 \end
\skip{1pt}
& \cent\ 	& Merritt \etal\  	& 1996, submitted     \end
\skip{1pt}
\skip{1pt}
\hline
}\endtable
$$
}
}}
}

The purpose of building models such as King's models and their
derivatives is to construct a simplified mathematical description of a
star cluster.  They provide a number of parameters (mass spectrum,
concentration, anisotropy radius, etc.) which can be adjusted to
optimise the fit with observations. Nevertheless they are based on
strict assumptions with regard to the form of the distribution
function. These assumptions take account of dynamical theory, though
in some respects they contradict or oversimplify it.  The results can
be strongly biased by the choice of the integrals of motion and the
form of the functional dependence.  Indeed it is true that, even when
the profiles of mass density and velocity dispersion are known, the
distribution function is still not uniquely determined (Dejonghe 1987,
Dejonghe \& Merritt 1992); for instance the anisotropy is still not
completely constrained.  For this and other reasons it is worth
considering methods which attempt to construct the distribution
function from the observations, with minimal assumptions.

There are quadratic programming techniques (Dejonghe 1989) which fall
into this class.  The distribution function is written in terms of
basis functions, but the possible risks of such an approach have been
discussed by Merritt \& Tremblay (1994) in a related context.  They
recommend a modification which serves to smooth the resulting estimate
of the distribution function.  The general aim is to infer the
gravitational potential $\Phi(r)$ and the phase-space distribution
function $f(\varepsilon)$, given the observations of the surface
density and velocity dispersion profiles of a ``tracer'' population.
Briefly, in the case of a globular cluster, (i) the projected density
$I(R)$ provides the space density $\nu(r)$, (ii) the projected
velocity dispersion $\sigma^2(R)$ provides the space velocity
dispersion $v^2(r)$, (iii) the Jeans equation provides the
gravitational potential $\Phi(r)$, and (iv) the Eddington equation
provides the phase-space distribution function $f(\varepsilon)$.
Nevertheless, a disadvantage of such techniques arises from the
delicate process of deprojection using Abel integrals (Merritt
1993a,b,c).

Table~7.4 gives a list of the non-parametric studies which have been
published for five globular clusters, using samples from a few hundred
up to a few thousand stars.  As a result, non-parametric mass density
and \mlv\ profiles can now be compared with more traditional
theoretical models for core-collapse clusters.  The two non-collapsed
globular clusters, viz., NGC~362 and NGC~3201, seem to exhibit
significant differences from the two possibly collapsed globular
clusters, \tuc\ and M15. The derived phase-space distribution
functions are not consistent with King models: NGC~362 and NGC~3201
have significantly more tightly-bound stars than King models, and
systematic differences appear between \tuc\ and M15 and either the
King models or the two less concentrated globular clusters. E.g.,
Fig.~7.7, which displays the non-parametric estimates of the \mlv\ for
the four clusters NGC~362, NGC~3201, M15, and \tuca, shows a
remarkable difference between the \mlv\ profiles of the two collapsed
and the two other clusters.  C\^ot\'e \etal\ (1995), using King,
King-Michie, and non-parametric models, present, for NGC~3201, an
interesting comparison between the different results, and a discussion
of how to disentangle the consequences of the assumptions and
disadvantages of each approach.

Incidentally, these methods do not attempt to construct a distribution
function for the entire cluster, but only one for a stellar species
for which both positional and kinematic data are available, and the
gravitational potential. A possible criticism of this technique is
that it goes too far in entirely ignoring dynamical theory, except
Jeans' theorem, and takes no account of the fact that the inner parts
of all galactic globular clusters should be nearly relaxed.  Also it
has not so far been extended to the construction of models with
anisotropic distribution functions, though axisymmetric systems can
now be treated (Merritt 1996).

\vskip 11truecm
\item\item{\sevenrm{{\bf Fig.~7.7.} \hskip 2mm Non-parametric estimates
of the \mlv\ (solid lines) and their 90\% confidence bands (dotted
lines), from Gebhardt \& Fischer (1995 Fig.~7).  The two possibly
collapsed globular clusters, \tuc\ and M15, differ significantly from
the two non-collapsed ones, NGC~362 and NGC~3201.  
}}
\vskip 0.5truecm

\vskip 30pt
\noindent
{\bf 8. Evolutionary models} 

\vskip 20pt
\noindent
The models described in the previous sections (\S\S7.5 and 7.7
especially) take account of a certain amount of dynamics, mainly the
assumption that the cluster is in dynamic equilibrium (Jeans'
theorem).  To some extent, but always approximately, some of these
models also take into account the effects of gravitational encounters.
The methods described in the present section are, however, needed if
the effects of these processes are to be modelled with any precision.
They can also incorporate a broad spectrum of important processes
which influence both the gross evolution of a cluster and the
evolution of its individual components (cf. Hut \etal\ 1992a for an
interesting general review).

\vskip 30pt
\noindent
{\sl 8.1 N-body integrations} 

\vskip 20pt
\noindent
Ideally, the dynamical evolution of a globular cluster would be
modelled with a direct $N$-body integration.  In fact this method is
inapplicable to globular clusters, because the required value of $N$
(of order $10^6$) is too large for a simulation to be completed within
a reasonable time (Fig.~8.1).  Larger values are used in cosmological
$N$-body simulations, but there one may exploit approximations in the
evaluation of forces which would lead to unacceptable errors in the
simulation of a star cluster, and the required number of time steps is
much smaller. Similar simplifications can be adopted in simulations of
star clusters for the investigation of certain kinds of phenomena,
namely, those that do {\sl not} involve two-body relaxation effects,
close encounters between binaries, etc., and here the fast methods
that have been developed for the study of galaxy dynamics may be
employed. This is too wide a subject for review here, and the
following paragraphs are devoted to $N$-body techniques which can
faithfully model all the gravitational processes that are relevant in
globular clusters.  Despite the limitation on $N$, direct $N$-body
simulations can teach us much about the dynamical evolution of
globular clusters, provided that the scaling with particle number is
understood.  Many developments have taken place since the review by
Aarseth \& Lecar (1975).

At present the best code is NBODY5 (Aarseth 1985a), and descendents
which are still under development.  In addition to a high-order
integrator (similar, in early versions, to that described in Wielen
1974b), it incorporates a number of subtle techniques which are
indispensable for adequate accuracy and efficiency, including the use
of individual time steps (so that the positions and velocities of
different particles are advanced with different frequencies),
computation of forces from near neighbors and distant stars with
different frequencies (the scheme of Ahmad \& Cohen 1973), and special
treatments (regularisation) of compact pairs (binaries) and other
few-body configurations (Mikkola 1985, Mikkola \& Aarseth 1990, 1993).
Modelling of star clusters with primordial binaries, for example,
would be impractical without these techniques. Even so, the simulation
of a cluster with only a few thousand stars and a few percent of
primordial binaries takes about 2,000 hours on a typical workstation
(Heggie \& Aarseth 1992).

Though the mix of techniques used in such codes as NBODY5 is well
tried and successful, it is always possible that improvements remain
to be discovered.  Makino (1991), for example, has examined such
aspects as the time step criterion and order of the integration
routine (cf. also Wielen 1967).  Other integration schemes have been
considered (Mann 1987, Press \& Spergel 1988); relatively recently it
has been found that so-called ``Hermite'' integration techniques
(which have since been incorporated into NBODY5) can offer significant
advantages (Makino \& Aarseth 1992), and new life has been breathed
into the humble leapfrog integrator by Hut \etal\ (1995); see also
Funato \etal\ (1996) for an application of the same ideas to
regularisation.  The leapfrog method is a so-called ``symplectic
integrator'', which refers to a class of methods which have some
attractions in studying $N$-body problems without dissipation, and
there has been considerable activity in this area (e.g., Ruth 1983,
Forest \& Ruth 1990).  In principle it is actually possible to express
the solutions of the $N$-body problem as infinite series (Wang 1991),
but so far this has not proved a useful guide to new numerical
methods.

\vskip 10truecm
\item\item{\sevenrm{{\bf Fig.~8.1.} \hskip 2mm The progress 
of $N$-body simulations.  Each plotted point gives the date of 
publication of the largest $N$-body simulation at that time which
extended well into core collapse (at least), except for the last two
points, which refer to preprints.  The largest value of $N$ has
increased by almost one decade per decade.  }}
\vskip 0.5truecm

Indirect methods of evaluating forces (tree or hierarchical schemes:
Appel 1983, Barnes \& Hut 1986, 1989, Ambrosiano \etal\ 1988,
Hernquist 1988, and especially McMillan \& Aarseth 1993) should also
become of increasing importance as the feasible values of $N$
increase.  Greengard (1990) provides an informal introduction to this.

Carrying out and analysing $N$-body simulations can be a laborious
task.  As in the data reduction phase of an observational project,
much time can be saved if the process is sufficiently automated (e.g.,
Carnevali \& Santangelo 1980).  It is especially convenient if this
can be done within a suitable, special-purpose software environment
(Hut \& Sussman 1986, Hut \etal\ 1993).

Hardware advances are also having a big impact.  The use of vector
machines is now routine, but the application of parallel computers in
this problem is still at a rather experimental level, at least for
star cluster problems (e.g., Makino \& Hut 1989a, Raine \etal\ 1989,
Warren \& Salmon 1995, Spurzem 1996).  An exception is the code
written for a transputer array by Sweatman (1990, 1991, 1993). Use of
the Connection Machine is described in Makino \& Hut (1989b), Brunet
\etal\ (1990), Theuns \& Rathsack (1993) and Hernquist \etal\ (1995),
while Katzenelson (1989) discusses parallelisation of a tree code.
The most exciting developments here, however, are in the field of
special-purpose hardware, designed and built by a group at Tokyo
University under the direction of D. Sugimoto (Sugimoto \etal\ 1990).
These devices fall into two classes.  One class (GRAPE-1 and GRAPE-3;
cf. Ito \etal\ 1990) compute forces with relatively low accuracy
(Makino \etal\ 1990, Okumura \etal\ 1992), but are still suitable for
problems which are not dominated by close two-body encounters,
binaries, or a high-density core (Hernquist \etal\ 1992).  The other
class (GRAPE-2 and GRAPE-4) are not yet quite so widespread, but are
ideal for all kinds of problems in star cluster dynamics (Ito \etal\
1991, 1993).

One of the great advantages of the $N$-body technique is that the
minimum of simplifying assumptions need be made.  By contrast with
other methods discussed in later subsections, no assumption is made of
spherical symmetry, or isotropy of the velocity distribution.  No
special steps need be taken to include gravitational interactions
involving pairs, triples (e.g., encounters between single stars and
binaries), quadruples (e.g.  binary-binary interactions), etc.  No
extra difficulties are created by the inclusion of a spectrum of
masses or a tidal field.  Indeed, some steps in the direction of
greater realism actually make $N$-body simulations easier.  It has
even been shown that binaries formed in three-body encounters, which
are usually regarded as a bottleneck in these studies, actually become
relatively unproblematic when $N$ becomes large enough (Makino \& Hut
1990).  However, {\sl primordial} binaries will remain time consuming
no matter how large $N$ is.

Aside from the scaling with respect to $N$, discussed below, the main
difficulty of the $N$-body technique is that the results are noisy,
because of the relatively small number of stars.  The implications of
this for the determination of core parameters has been studied in
detail by Casertano \& Hut (1985), whose work forms the basis for many
analyses of $N$-body results.  In fact the statistical noise can be
greatly alleviated by averaging results from many simulations (Giersz
\& Heggie 1993a).  More important from the astrophysical point of view,
it is very difficult to model rare species (e.g. stellar-mass black
holes) with a small $N$-body simulation.  A single massive object may
have a different qualitative effect on a small system than the same
proportion of objects of the same mass in a large system.

Though not thought to be of practical importance (otherwise it would
undermine the entire $N$-body modelling effort!), there is in
principle one further difficulty in the use of $N$-body techniques.
It stems from Miller's observation (Miller 1964, 1971) that the
solution of the $N$-body equations is highly unstable, on a time scale
much smaller than the typical length of a simulation (Kandrup \& Smith
1991, 1992; Goodman \etal\ 1993).  Though this means that the
positions and velocities of the stars in a simulation are almost
certainly quite wrong (Lecar 1968, Hayli 1970b), there is no reason to
believe that the statistical results are unreliable provided that the
total energy is well conserved (Smith 1977, 1979, Heggie 1991).
Numerical ``shadowing'' results (Quinlan \& Tremaine 1991, 1992)
provide some reassurance for this point of view.  Since energy
conservation is the main test available, it is probably wise to avoid
techniques which artificially {\sl force} a system to preserve its
total energy.

Now we turn to some examples of the requisite scaling of the results.
In $N$-body simulations, it is customary (H\'enon 1972, Heggie \&
Mathieu 1986) to use units in which $G$, the total mass and the virial
radius are unity.  Thus one has freedom to choose two of these units,
and then the third is determined by the value of $G$.  Equivalently,
one has freedom to scale the mass and the unit of time.

The scaling of time depends essentially on the mechanism to be
modelled.  Phenomena occurring on a crossing time scale (e.g., disk
shocking) could be modelled by scaling the crossing time of the
$N$-body model to that of a real cluster. Similarly, modelling of the
early evolution, especially the phase in which the cluster adjusts to
the rapid loss of mass from the evolution of its massive stars, could
be modelled by using the same scaling to determine the stellar
evolution time scale within the model.  In order to model relaxation
effects, including mass segregation, one would scale the mass and
half-mass relaxation time to those of an actual cluster.  Modelling of
the effects of stellar collisions within this context could be added
by suitable choice of the stellar radii (McMillan 1993).

Complications arise when the phenomenon to be modelled depends on two
or more processes whose time scales scale differently with $N$.  For
example, the time scale for formation of a single hard binary is of
order $Nt_r$, and so phenomena involving {\sl both} two-body
relaxation and binary formation in a globular cluster cannot easily be
modelled with a small $N$-body simulation. In fact, it was known
long ago (Hayli 1970a) that escape due to interactions involving
binaries does not scale in the same way as escape due to two-body
interactions.  Another consequence is that the density of the core at
the end of core collapse is $N$-dependent (Goodman 1987), and this is
one reason why gravothermal oscillations (cf. \S10.1) have only
recently come within reach of $N$-body models.  On the other hand, if
primordial binaries are present, then the {\sl formation} of binaries
can be neglected, and the evolution of the core can be modelled
successfully. (The $N$-dependence is logarithmic: Heggie \& Aarseth
1992).

Another important example is the escape rate, even in the absence of
interactions involving binaries.  In an isolated system the time scale
for escape scales with $t_r$, except for a logarithmic factor, but in
the presence of a tidal field, especially if it is time dependent (in
consequence of the orbital motion of a cluster through its parent
galaxy), phenomena which scale as the crossing time are also
important.  Therefore no straightforward $N$-body simulation will
correctly model these processes.

In view of these difficulties, a number of somewhat modified or ad hoc
$N$-body schemes have been devised.  Hybrid schemes (McMillan \&
Lightman 1984, Aarseth \& Bettwieser 1986) approach the problem
caused by the differing time scales of binary formation and two-body
relaxation by welding a simplified treatment of the latter with an
$N$-body treatment of the central parts (where almost all binaries are
formed); nevertheless, they offer only a rather modest advantage of
speed (Hut \etal\ 1988).  Similarly, tidal effects on escape have been
modelled (Oh \etal\ 1992) by a simplified treatment of relaxation and
careful modelling of the orbit in the tidal potential.  Disk shocking
has been modelled in a similar simplified way (assuming that
relaxation sets up a multi-mass King model between shocks) by
Capaccioli \etal\ (1993).  Stability can be studied in an especially
economical way by use of suitably modified $N$-body techniques
(Wachlin \etal\ 1992, Leeuwin \etal\ 1992), though these last two
topics belong to the domain of ``collisionless'' stellar dynamics,
where the scope for short-cuts is much richer.

What is noticeable about these issues is that the use of $N$-body
models in these contexts does not replace the use of theory.  Rather,
careful consideration of theoretical issues is required before
successful simulations can be devised.  One of the pitfalls, clearly,
may be that there is some slightly subtle and unsuspected interaction
between phenomena which scale differently with $N$.  These may remain
undiscovered until modelling efforts with the correct values of $N$
become feasible.

In the long run the $N$-body technique will become the method of
choice.  So far, however, $N$-body simulations have not yet been used
to model specific clusters, and even their use in the study of open
clusters has not made much progress since the work of Terlevich (1985,
1987), except for some remarkable recent developments by Aarseth
(1996a,b).  One issue that will have to be addressed is how one
compares the data from an $N$-body simulation with observations.  An
early attempt was von Hoerner's ``modulus of evolution'' (von Hoerner
1976); i.e., a single parameter whose time dependence is found from
$N$-body simulations and which can be determined observationally
(Kadla 1979).  This is a test of extreme simplicity, but nothing
better has been attempted since then.

\vskip 30pt
\noindent
{\sl 8.2 Fokker-Planck methods} 

\vskip 20pt
\noindent
One of the fundamental dynamical mechanisms in the evolution of
stellar systems is two-body relaxation (\S7.1). The theoretical
foundations were laid by Chandrasekhar, who introduced a description
in terms of a Fokker-Planck equation (Chandrasekhar 1943a,b).  His
formulation was improved by Rosenbluth \etal\ (1957) and the effect of
orbital motion was added by Kouzmine (1957).  The resulting
orbit-averaged Fokker-Planck equation was first put to practical use
by H\'enon (1961, 1965), whose two papers in this area are
long-standing classics.  In this formulation the equation resembles
the heat conduction equation, and it can be solved numerically by a
variety of methods which have all been of importance.  They divide
into several classes, which we discuss in turn.

First we discuss the Monte Carlo models, which again divide into two
types.  One was pioneered by Spitzer and his students (Spitzer \& Hart
1971a,b, Spitzer \& Shapiro 1972, Spitzer \& Thuan 1972, Spitzer \&
Chevalier 1973, Spitzer \& Shull 1975a,b, Spitzer \& Mathieu 1980),
and developed in important ways by Shapiro and his collaborators
(Shapiro \& Marchant 1978, Marchant \& Shapiro 1979,1980, Duncan \&
Shapiro 1982, Shapiro 1985).  The other method was devised by H\'enon
(1966, 1972, 1973, 1975) and later improved by Stod\'o\l kiewicz
(1982, 1986).  The essential difference in these models is that the
former followed the stars around their orbits, and was (in principle)
capable of modelling processes occurring on both relaxation and
crossing time scales, though in the phenomena actually studied with
these models processes of the latter kind were unimportant.  Models of
H\'enon's type, on the other hand, assumed dynamical equilibrium, and
that the distribution function depends only on integrals of motion.

Spitzer's method was used to explore a variety of important phenomena,
including mass segregation, anisotropy of the velocity distribution,
tidal shocking, the role of primordial binary stars, etc., and the
above sequence of papers is often a good starting point for
information on these areas.  At first, H\'enon's method was used to
explore somewhat more idealised problems -- for example it was the
first to break through the impasse of core collapse (H\'enon 1975),
but it was brought to an amazing level of realism by Stod\'o\l kiewicz
(1984, 1985).  Indeed from this point of view his papers remain
unsurpassed: they included such processes as the formation of binaries
by two- and three-body encounters, mass loss from stellar evolution,
tidal shocking, etc.

In view of the success of the Monte Carlo method it is surprising that
it has been ignored in the last few years.  One reason is that it
faced vigorous competition from a direct numerical (finite-difference)
solution of the Fokker-Planck equation, along lines pioneered by Cohn
(1979, 1980).  Similar methods had been developed for a fixed
potential by Ipser (1977) and by Cohn \& Kulsrud (1978), and since
then codes like Cohn's have been written independently by Inagaki \&
Wiyanto (1984) and by Chernoff \& Weinberg (1990).  Like the Monte
Carlo methods, Cohn's formulation assumes spherical symmetry, though
codes which can handle a rotating cluster have been devised by Goodman
(1983a) and by Einsel \& Spurzem (1996). More importantly, it is usual
to assume that the distribution of velocities is isotropic, which was
not customary in the Monte Carlo models.  One of the main reasons for
this simplification is that there exists in this case a numerically
very well behaved scheme due to Chang \& Cooper (1970).  Over the
years there have been several unsuccessful attempts to develop
something comparable for anisotropic models, whose numerical behavior
was therefore less satisfactory as judged by energy conservation (Cohn
1985). Recently, however, Takahashi (1995, 1996) has demonstrated a
new approach, which definitely appears to have cured finite difference
methods of this long-standing problem.

It is still true that the introduction of anisotropy greatly increases
the computational effort, as it resembles the transfer from
one-dimensional to two-dimensional diffusion.  Similarly, introduction
of unequal masses greatly increases the time taken to compute a model.
Equally demanding computationally is the introduction of further
physical processes, e.g., binaries, whether those formed by two body
encounters (Statler \etal\ 1987, Lee 1992, Lee \& Ostriker 1993) or
those present initially (Gao \etal\ 1991).  In the latter case, for
example, it was necessary to assume that the distribution of the
energies of the binaries was independent of their spatial
distribution, whereas $N$-body models of different kinds show that
more energetic binaries are found at larger radii (McMillan \etal\
1990, Hut \etal\ 1992b, Heggie \& Aarseth 1992).  Without the
inclusion of such processes it is unlikely that any satisfactory
models for the advanced evolution of globular clusters can be
constructed (cf. Drukier \etal\ 1992). For all these reasons the most
realistic Fokker-Planck models, like the largest $N$-body models,
require large-scale computing facilities.

A third technique for solving the Fokker-Planck equation has been
under development for a number of years now.  Based on the variational
formulation of Inagaki \& Lynden-Bell (1990), it now appears to be
roughly competitive with finite difference methods.  To begin with it
was developed and used successfully for isotropic models (Takahashi \&
Inagaki 1992; Takahashi 1992, 1993).  Recently, however, it has been
rapidly developed to the point where accurate anisotropic models can
be used to follow the evolution through core collapse into the
post-collapse regime (Takahashi 1995, 1996).

It is also worth mentioning a fourth formulation, in terms of
path-integrals (Horwitz \& Dagan 1988, Dagan \& Horwitz 1988), which
yet again offers different (and unexplored) possibilities for
numerical work.  Finally, Luciani \& Pellat (1987b) have presented a
form of the Fokker-Planck equation which makes minimal assumptions
about symmetry and the distribution function, though it has not yet
been put to practical use.

A final question mark over Fokker-Planck models is their validity.
Work by H\'enon (1975) revised the theory of the relaxation time which
had been used in the classic paper of Aarseth \etal\ (1974), and found
that the results of Fokker-Planck and $N$-body models were brought
into satisfactory agreement.  This refers to the case of isolated
models with equal or unequal masses, and recent work by Giersz \&
Heggie (1993a, 1994, 1996a) and Giersz \& Spurzem (1994; cf. Fig.~7.1
above) has strengthened and refined H\'enon's conclusion.  For more
realistic models the situation is less satisfactory.  The formula that
is often used for the rate of energy generation by binaries in
multi-component models (e.g., Lee \etal\ 1991) rests on a very slender
foundation.  The Fokker-Planck treatment of tidal effects is
necessarily spherically symmetric and often simplified to imposition
of a tidal cutoff (e.g., Chernoff \& Weinberg 1990), though in this
case a somewhat more realistic formulation has been devised (Lee \&
Ostriker 1987).  Fukushige \& Heggie (1995) find that the lifetime of
$N$-body models in the face of tidal disruption can greatly exceed
that found using a Fokker-Planck model.  In this case it is the
assumption of dynamical equilibrium which seems to be at fault.

\vskip 9.5truecm
\item\item{\sevenrm{{\bf Fig.~8.2.} \hskip 2mm Main sequence mass 
function index as a function of projected radius (in pc), at various
times during the evolution of a Fokker-Planck model (from Chernoff \&
Weinberg 1990 Fig.~35).  The index is defined by $dN\propto
m^{-\alpha} dm$. Times are given in years. Initial conditions are a
King model with $W_0$ = 7, $\alpha$ = 2.5 for 0.4 \msun $<$ m $<$ 15
\msun, total mass 2.82 \x\ 10$^5$ \msun, at galactocentric distance
$10$kpc.  }}
\vskip 0.5truecm

Despite these shortcomings, the direct Fokker-Planck method is at
present the most important and widely used source of information on a
wide range of essential phenomena, including mass segregation (Inagaki
\& Wiyanto 1984), and the additional effects of tidally-induced escape
and mass loss from stellar evolution (Weinberg \& Chernoff 1989,
Chernoff \& Weinberg 1990).  Their 1990 paper, with an update in
Chernoff (1993), is another excellent starting point for learning what
the Fokker-Planck model can teach us about the evolution of model star
clusters (cf. Figs.~7.3 and 8.2).

As was mentioned at the beginning of this section, the Fokker-Planck
equation treats relaxation as a diffusion process.  It can also be
treated in a manner more closely resembling the Boltzmann equation,
i.e., by a formulation in which the distribution function evolves by
discrete changes in the energies of the stars.  The equation to be
solved is then an integral-differential equation (Petrovskaya 1969a,b,
1971; Kaliberda \& Petrovskaya 1970).  Appropriate numerical
techniques have been devised (Goodman 1983b), but are much less well
developed than for the Fokker-Planck equation (see \S7.1 for
references to the basic theory).

\vskip 30pt
\noindent
{\sl 8.3 Conducting gas models}

\vskip 20pt
\noindent
The resemblance between a star consisting of huge numbers of atoms and
a star cluster or galaxy containing large numbers of stars becomes
clear at quite a simple level, e.g., the virial theorem.  It is more
surprising, but still true, that the resemblance extends to phenomena
such as heat transport, energy generation, and core-halo evolution.
For the study of the dynamical evolution of stellar systems this
resemblance was first exploited by Larson (1970a), whose work was the
first to exhibit the time-dependence of core collapse.

Models of this kind can be divided into two classes.  Larson's, which
was taken up by Angelleti \& Giannone in an important but unjustly
neglected series of papers (Angeletti \& Giannone 1976, 1977a,b, 1978,
1979, Angeletti \etal\ 1980), and has been developed further recently
by Louis \& Spurzem (1991), should really be regarded as an
approximate solution of the Fokker-Planck equation, obtained by
studying the moments of the velocity distribution.  Since these are
essentially the density, the mean velocity of the stars, the (kinetic)
energy density, etc., the resulting equations resemble those of
stellar evolution.  The other method, somewhat more phenomenological,
starts with the equations of stellar evolution and corrects the
physics: no radiative energy transport, and conduction altered to suit
the effects of two-body relaxation.  Except for the last point
(Lynden-Bell \& Eggleton 1980) the method was introduced by Hachisu
\etal\ (1978), and it was subsequently developed and exploited by
Bettwieser (1983, 1985a,b) and others.

The first remarkable point about these models is that they work, i.e.,
they give results which closely resemble those of Fokker-Planck and
$N$-body models in many respects (Aarseth \etal\ 1974, Bettwieser \&
Sugimoto 1985, Giersz \& Heggie 1993a, 1994, 1996a, cf. Fig.~7.1
above).  It is not obvious why this should be so, because the
treatment of conduction is so different: at each radius it is governed
by local conditions, whereas the orbital motion of stars implies that
encounters at one radius can and should affect the distribution
functions at widely different radii.  One phenomenon where this seems
to be important is in the growth of anisotropy (Giersz \& Spurzem
1994).  A speculative reason for the success of the gas model is that
so many aspects of the evolution of stellar systems have a
thermodynamic basis, and this is accurately described in these simple
models.

The foregoing remarks show that the gas model of stellar systems has
been developed to include a variety of phenomena, though not quite to
the same extent as the Fokker-Planck model (cf. Table~7.1 above).  A
spectrum of stellar masses can be included (Heggie \& Aarseth 1992,
Spurzem \& Takahashi 1995), despite unsatisfactory results of an
earlier attempt (Bettwieser \& Inagaki 1985).  Other developments
include simple treatments of the effects of stellar mass loss
(Angeletti \& Giannone 1977c, 1980), binary formation and evolution
(Heggie 1984, Heggie \& Ramamani 1989), even including stochastic
(random) effects (Giersz \& Heggie 1994) or primordial binaries
(Heggie \& Aarseth 1992). At one time the gas model led the field in
producing a dynamical evolutionary model of a specific cluster in
which the photometric properties of different kinds of star were
included for the production of multi-color surface density profiles
(Angeletti \etal\ 1980).

In general it may be said that the gas model has the same merits and
demerits as the Fokker-Planck model, except for two points: its
advantage is that it is quicker, but each new problem must be
approached with caution, as it is not really clear why it works as
well as it does, and its treatment of relaxation is inferior.  Its
main use is as a quick tool for exploring an area which can be
followed up later by better methods.

Most of the N-body, Fokker-Planck, and conducting gas studies have
been theoretically oriented, having in mind the numerical
investigation of questions linked to the dynamical evolution.  Their
results are presented in \S9.  Only a few Fokker-Planck models have
been directly fitted to observations.  Recent exceptions include the
work of Grabhorn \etal\ (1992) on M15 and NGC~6624, that of Phinney
(1993), also on M15, and especially the detailed study of NGC~6397 by
Drukier (1993, 1995, and cf. \S9.2 below).

\vskip 30pt
\noindent
{\bf 9. Towards catastrophic phases of evolution~?}  

\vskip 20pt
\noindent
In the present section we resume our discussion of dynamical
evolution.  We have already discussed the early evolution governed by
processes on the time scale of the crossing time and that of mass loss
from the evolution of massive stars (cf. \S5).  Assuming that the
cluster has then settled into a state of quasi-static dynamical
equilibrium, we discussed suitable models in \S7.5.  Now we turn to
the effects of two-body relaxation on time scales of several
relaxation times, i.e., what is sometimes called ``secular
evolution''.  Another way of expressing the position is that we now
turn to evolution on a ``thermal'' time scale (cf. the discussion of
the gas model in \S8.3), whereas parts of \S5 concern processes acting
on a dynamical time scale.  Many of the phenomena we discuss can only
be modelled adequately using the techniques of the previous section,
but now we concentrate on the results.

\vskip 30pt
\noindent
{\sl 9.1 The gravothermal instability and core collapse}

\vskip 20pt
\noindent
For many years (between about 1940 and 1960) secular evolution was
understood in terms of the ``evaporative model'' of Ambartsumian
(1938) and Spitzer (1940).  In this model it is assumed that two-body
relaxation attempts to set up a maxwellian distribution of velocities
on the time scale of a relaxation time, but that stars with velocities
above the escape velocity promptly escape.  If the escaping stars
carry off little energy, this model predicts that the number of stars
in the cluster varies approximately as $(t_0 - t)^{2/7}$, where $t_0$
is a constant representing the time at which the entire cluster will
have evaporated.  The evolution of its size, velocity dispersion,
etc., can be estimated equally simply.  For example the density varies
as:
$$
\rho\propto(t_0-t)^{-10/7}.\eqno(9.1)	
$$
The next major step in understanding came when it was discovered that
evolution arises also when stars escape from the inner parts of the
cluster to larger radii, without necessarily escaping altogether.
Antonov (1962) realised that these internal readjustments need not
lead to a structure in thermal equilibrium, because thermal
equilibrium may be unstable in self-gravitating systems.  The
explanation of Lynden-Bell (1968) is worth repeating.  Consider a
conducting, self-gravitating gas enclosed by a spherical wall. (In a
real system the inner parts are kept in by the pressure of the outer
layers, but this does not change anything qualitatively.) If the core
is warmer than the part inside the wall, thermal energy flows
outwards. The outer region, which is held in by the wall, heats up.
But the inner part also heats up because it is pressure-supported:
loss of thermal energy reduces the pressure, and in the subsequent
slight collapse gravitational potential energy is converted into
kinetic energy.  Whether the temperature difference between the inner
and outer parts is greater than it was before depends, among other
things, on the heat capacity, and therefore the mass, of the outer
layer.  If it is sufficiently great (i.e., the core is sufficiently
compact), the initial temperature excess of the core is enhanced, and
with it the conduction of heat and the collapse of the core.

In view of the central part played by Antonov's instability, it has
been investigated from various points of view. Lynden-Bell \& Wood
(1968) reworked Antonov's theory, related it to other stability
criteria (depending on the boundary conditions), and followed up some
consequences for the evolution of stellar systems.  They also carried
out a similar analysis for the King and Woolley sequences of models,
as well as for the isothermal case.  Though Taff \& van Horn (1975)
claimed that this analysis was faulty, similar results were obtained
(by different techniques) in a series of papers by Horwitz \& Katz
(1977, 1978), Katz \& Horwitz (1977) and Katz \etal\ (1978); see also
Ipser (1974), Larson (1970a), Katz (1978, 1980) and Padmanabhan
(1989a, 1990) for yet other approaches.  A particularly readable
account of the thermodynamic basis of the instability was provided by
Hachisu \& Sugimoto (1978) in the context of gaseous models.  Nakada
(1978), Inagaki (1980) and Luciani \& Pellat (1987a) provided
stability analyses on the basis of the conducting gas model, the
isotropic Fokker-Planck equation, and the anisotropic Fokker-Planck
equation, respectively.  The role of boundary conditions in the
stability of $N$-body systems was explored by Miller (1973), and the
stability of (singular) anisotropic models was investigated by Spurzem
(1991).

The ``gravothermal instability'', as it is often called, is
considerably complicated in the case of unequal masses. A classical
paper by Spitzer (1969) showed that the scope for thermal equilibria
(which requires equipartition of energies) is even more restricted
than in the case of equal masses (see also Lightman 1977, Yoshizawa
\etal\ 1978, Katz \& Taff 1983).  Addition of rotation adds further
features (Inagaki \& Hachisu 1978, Tajima 1981, Lagoute \& Longaretti
1996, Longaretti \& Lagoute 1996a,b).

What the above discussion does not make clear is the dynamical
consequence of the instability.  The well known process of ``core
collapse'' is usually interpreted as a manifestation of the
gravothermal instability.  It is best studied using the techniques
discussed in \S8, but various simplified and more or less instructive
models have been devised to illustrate the link between the two.
Several authors (Da Costa \& Pryor 1979, Da Costa \& Lightman 1979,
Danilov 1989) have constructed such models of the evolution of core
and halo in terms of the energy and mass exchanged between them.
Already Lightman \& Fall (1977) had provided a theory along similar
lines for two-component systems. It turns out that the way in which
the relaxation time depends on density and velocity dispersion is
crucial to the way the instability develops (Makino \& Hut 1991).

Now we summarise some results of more detailed models, mostly made
using the Fokker-Planck method (cf. also Spitzer 1984, 1985 for other
summaries).  Even in systems of stars with equal mass, and assuming an
isotropic distribution of velocities, the time-dependence is a little
complicated.  Expressed in terms of $t_{rc}$, the relaxation time in
the core, the $e$-folding time for the evolution of the central
density varies from about $5$ in the early stages (assumed here to be
a Plummer model) to about $330$ in late phases (Cohn 1980).  The time
scale for the evolution of the velocity dispersion is generally much
longer, as this quantity varies much less than the central
density. However, the increase of the projected central velocity
dispersion is sufficient to show up in even quite small $N$-body
simulations (Struble 1979).  The time scales in late core collapse are
somewhat longer if the distribution of velocities is allowed to be
anisotropic (Takahashi 1995, who obtained a result contrary to that of
Cohn 1979).

Late in the process of core collapse the evolution of the central
density, velocity dispersion, etc., becomes simple, and resembles that
in the evaporative model (Eq.~9.1), but with somewhat different
indices.  The reason for this is that the deep evolution is driven by
interactions within the inner parts of the system, and so the
influence of the outer parts of the cluster become negligible
(Lynden-Bell 1975).  The corresponding self-similar evolution was
revealed first by Lynden-Bell \& Eggleton (1980) using an isotropic
gas model, then by Heggie \& Stevenson (1988) with the Fokker-Planck
model, and finally by Louis (1990), who used an anisotropic gas code.

For an isolated cluster (without a tidal field) the time scale for the
entire evolution of the core (when the density has formally become
infinite) is about 15.7 $t_{rh}$(0), when expressed in terms of the
initial half-mass relaxation time (Cohn 1980).  This result is for an
isotropic code starting from a Plummer model with stars of equal mass,
and for an anisotropic code the time extends to 17.6 $t_{rh}(0)$
(Takahashi 1995).  Results for collapse from a King model, with or
without a steady tide, are given by Wiyanto \etal (1985), Inagaki
(1986a) and Quinlan (1996).

\vskip 10truecm
\item\item{\sevenrm{{\bf Fig.~9.1.} \hskip 2mm Evolution of Lagrangian
radii in an $N$-body model with a mass spectrum, a steady tide and
mass loss from stellar evolution (from Aarseth \& Heggie, in
preparation).  Initial conditions: $dN\propto m^{-2.5}dm$, King model
with $W_0$ = 7, mass 1.5 \x\ 10$^5$ \msun, galactocentric radius
4~kpc, orbital speed 220~\kms.  Time is in Myr, the unit of length is
6 pc.  Though the model has $N$ = 8192 stars initially, results are
scaled to the above initial conditions.  The radii plotted are (from
the top) tidal radius ($r_t$), and Lagrangian radii corresponding to
fractions 0.5, 0.1, 0.01 and 0.001 of the mass inside $r_t$. The initial
rise takes place in the early phase of intensive mass loss by stellar
evolution.  Core collapse is complete at about 10~Gyr, which compares
very well with the value 9.6~Gyr obtained by Chernoff \& Weinberg
(1990) using Fokker-Planck techniques.  The evolution around core
bounce and beyond would probably be significantly altered if an
appropriate population of primordial binaries had been included.
Tidal shocks were not included.
}}
\vskip 0.5truecm

The collapse time is generally shorter in the presence of unequal
masses (Inagaki \& Wiyanto 1984, Chernoff \& Weinberg 1990).  For
example, a tidally limited cluster starting from a King model with
scaled central potential $W_0 = 3$, with stellar masses having a
power-law distribution with index $x$ = $2.5$ over a range from 0.4 to
15 \msun, takes about 0.9 $t_{rh}(0)$ for complete collapse
(cf. Fig.~7.3).  Up to a point this reduction can simply be understood
because the equipartition time scale of the most massive stars varies
inversely with mass (Eq.~7.6).  At any rate, results such as these
contradict the conclusion of Sygnet \etal\ (1984), who asserted that
stellar systems cannot have suffered a gravothermal catastrophe
because the time scale is too great.

Murphy \& Cohn (1988) give surface brightness and velocity dispersion
profiles at various times during collapse, for a system with a
reasonably realistic present-day mass spectrum.  Addition of effects
of stellar evolution, modelled as instantaneous mass loss at the end
of main sequence evolution, delays the onset of core collapse
(Angeletti \& Giannone 1977c, 1980, Applegate 1986, Chernoff \&
Weinberg 1990, Kim \etal\ 1992).

As already mentioned, most of the foregoing results stem from
Fokker-Planck studies.  Goodman (1983b) has shown that late core
collapse proceeds in much the same way if a better model is used
(which does not make the same assumption of small-angle scattering).
Examples of $N$-body models which illustrate various aspects of core
collapse include Aarseth (1988), where $N = 1,000$, Giersz \& Heggie
(1993a) ($N\le2,000$), Spurzem \& Aarseth (1996) ($N = 10,000$), and
Makino (1996a,b; see Fig.~10.1 below) ($N\le32,000$) and Fig.~9.1.
 
The effect of a {\sl time-dependent} tidal field can be to accelerate
core collapse (Spitzer \& Chevalier 1973).  This may be a significant
point for the interpretation of observations (cf. \S\S9.2, 9.3, and
9.8), which show that galactic globular clusters with collapsed
cores are concentrated towards the galactic centre, where disk
shocking is indeed stronger and more frequent.  Note, however, that
the mean density of tidally limited clusters is higher towards the
galactic centre, and so the collapse time for clusters of a given mass
would be shorter near the galactic centre even in a steady tidal
field.

Although the mass of the core decreases during core collapse, the
inner parts of the cluster do flow inwards throughout most of the
collapse phase, and it can be understood from energy conservation that
the outer parts of the cluster move outwards, unless the cluster is
limited by the tidal field of the galaxy.  The half-mass radius is
relatively fairly static (Figs.~7.1 and 9.1).

\vskip 30pt
\noindent
{\sl 9.2 Observational evidence of core collapse through the density
profile and concentration} 

\vskip 20pt
\noindent
In the eighties, CCD observations allowed a systematic investigation
of the inner surface brightness profiles (within $\sim$ 3\arcm) of 127
galactic globular clusters (Djorgovski \& King 1986, Chernoff \&
Djorgovski 1989, Trager \etal\ 1995).  These authors sorted the
globular clusters into two different classes illustrated in Fig.~9.2:
(i) the ``King model clusters'', whose surface brightness profiles
resemble a single-component King model with a flat isothermal core and
a steep envelope, and (ii) the ``collapsed-core clusters'', whose
surface brightness profiles follow an almost pure power law with an
exponent of about --1.  In the Galaxy, about 20\% of the globular
clusters belong to the second type, exhibiting in their inner regions
apparent departures from King-model profiles.  Consequently, they are
considered to have collapsed cores.

Similar independent surveys in the Magellanic Clouds (Meylan \&
Djorgovski 1987, Mateo 1987) show possible signs of a collapsed core
in three high-concentration old clusters in the Large Magellanic
Cloud, viz., NGC~1916, NGC~2005, and NGC~2019, all three of which are
projected on the bar close to the centre of the LMC.  No such cluster
is observed in the Small Magellanic Cloud.  Bendinelli \etal\ (1993)
and Fusi Pecci \etal\ (1994) announce the first detection, thanks to
the high spatial resolution of HST, of a collapsed-core globular
cluster in M31, viz., G~105 $\equiv$ Bo~343.  See also Grillmair
\etal\ (1996) and Jablonka \etal\ (1996) for other clusters in M31..

For quite a few of the 125 globular clusters for which CCD
observations of their cores have been obtained (Trager \etal\ 1995),
within $\sim$ 3\arcm\ from the centre, there exists also aperture
photometry at intermediate radii, and star counts at large radii which
allow the construction of surface brightness profiles extending from
the core out to 30-50\arcm.  Nevertheless, discrimination between the
two different classes --- the ``King model clusters'' and the
``collapsed-core clusters'' --- may often be less clear-cut than it
might seem from Fig.~9.2.  There are two main reasons for this:

\vskip 7.5truecm
\item\item{\sevenrm{{\bf Fig.~9.2.} \hskip 2mm Examples of surface 
brightness profiles (Djorgovski \etal\ 1986 Fig.~1). Left: NGC~6388
resembles a King-model with a flat isothermal core and a steep
envelope. \hskip 2mm Right: Terzan 2 is an example of a collapsed-core
cluster whose light profile follows an almost pure power law with an
exponent of --1.  }}
\vskip 0.5truecm

{\sl (i) Statistical noise.}  Integrated surface brightnesses measured
for small areas in the cores of globular clusters are strongly
dominated by statistical fluctuations in the small numbers of bright
stars within the aperture.  Such fluctuations are conspicuous, e.g.,
in the case of NGC~6397, whose ground-based surface brightness
profile, in its inner 100\arcs, increases toward the centre in a very
bumpy way, especially when observed through $B$ or $U$ filters,
because of a high central concentration of bright blue stragglers
(Djorgovski \& King 1986, Auri\`ere \etal\ 1990, Meylan \& Mayor 1991,
Drukier 1995).  The power-law shape of the inner observable part of a
high-concentration profile may be difficult to observe because of
statistical fluctuations in the small numbers of bright stars (Sams
1995).  One way to alleviate this problem consists of using HST data,
allowing star counts (King \etal\ 1995).

{\sl (ii) Similarity between high-concentration King models and
power-law profiles.}  The inner parts of King models of very high
concentrations (cf. King 1966 Fig.~1) have profiles which resemble
that of a singular isothermal sphere.  See also \S7.2.  Consequently,
high-concentration King profiles can be successfully adjusted to the
surface brightness profiles of so-called ``collapsed-core clusters'',
as illustrated in Meylan (1994 Fig.~2).  For example, multi-mass
King-Michie models fit the surface brightness profile of NGC~6397
reasonably well and have very high concentrations, viz., \conc\
$\simeq$ 2.5 (Meylan \& Mayor 1991, Drukier 1995, King \etal\ 1995).
In a similar way, but see also Illingworth \& King (1977) and King
(1989), Grabhorn \etal\ (1992) are able to fit successfully a
multi-mass King model of even higher concentration, viz., \conc\
$\simeq$ 3.0, to the surface brightness profile of M15, the prototype
of the collapsed-core globular clusters.  There is no evidence that
fits of King models to post-collapse clusters are any less
satisfactory than those to uncollapsed clusters.

This indicates that the general dynamical status (is it collapsed or
not?) of a cluster may be straightforwardly deduced from the value of
its concentration, without careful study of the power-law shape of its
surface brightness profile.  Consequently, any globular cluster with a
concentration \conc\ \gsim\ 2.0-2.5 may be considered as collapsed, or
on the verge of collapsing, or just beyond.  It is worth mentioning
that the pre-, in-, and post-collapse terminology encountered in the
literature has only a theoretical meaning, since (so far) observations
are unable to differentiate these three different phases.  The
dynamical status evaluated from the concentration value also has the
advantage of alleviating the outstanding problem implied by the many
clusters --- like \tuca\ --- which show no trace of ``collapsed-core
cluster'' morphology, but have short enough dynamical times to have
collapsed in a small fraction of the Hubble time.

Although ground-based data are essential for most globular clusters,
any study aiming at resolving the core of the densest galactic
globular clusters is possible only with the refurbished HST.  For
example, detailed photometric studies have been published on the core
of NGC~6397 by King \etal\ (1995), of NGC~6624 by Sosin \& King
(1995), of NGC~6752 by Shara \etal\ (1995), and of M30 by Yanny \etal\
(1994b).  Although most clusters have a resolved core, a few clusters
--- e.g. M15 and NGC~6624 --- show inner star-count profiles which do
not suggest any sign of levelling off.

As already briefly mentioned in \S6.1, the globular cluster M15 has
long been considered as a prototype of the collapsed-core star
clusters.  Early electronographic determinations of its luminosity
profile by Newell \& O'Neil (1978), confirmed by further photographic
and CCD studies (e.g., Auri\`ere \& Cordoni 1981a,b), reveal a central
excess of light.  Newell, Da Costa, \& Norris (1976) found that these
observations were consistent with the existence of a central massive
object, possibly a black hole of about 800 \msun, while Illingworth \&
King (1977) were able to successfully fit dynamical models to the
entire surface-brightness profile, invoking a centrally-concentrated
population of neutron stars instead of a black hole.

\vskip 9truecm
\item\item{\sevenrm{{\bf Fig.~9.3.} \hskip 2mm A 5.6\arcs\ $\times$
3.5\arcs\ part of an FOC image of the centre of M15 (from King 1996
Fig.~1). The pixel size is 0.014\arcs.  The white areas are FOC
saturation in the brightest stars of this field.  The three bright
stars --- AC 214, 215, and 216 --- forming a near equilateral triangle
near the centre of the image are the main contributors to the former
bright luminosity cusp.  But see Fig.~9.4 for the faint stars radial
density profile. }}
\vskip 0.5truecm

Now, high-resolution imaging of the centre of M15 has resolved the
luminosity cusp into essentially three bright stars (\S6.1 and
Auri\`ere \etal\ 1984, Racine \& McClure 1989, Lauer \etal\ 1991,
Yanny \etal\ 1993, 1994a, and Sosin \& King 1996).  On the one hand,
Lauer \etal\ (1991) show that the surface-brightness profile of the
residual light, obtained after subtracting the bright resolved stars,
does not continue to rise at subarcsecond radii.  They determine a
core radius of 2.2\arcs\ = 0.13 pc.  On the other hand, also from
pre-refurbishment HST data, Yanny (1993) and Yanny \etal\ (1993,
1994a) find that a flat core is not apparent for $r$ \gsim\ 1.5\arcs.
They find the radial distribution consistent with a number of
scenarios, including: $i$) a central black hole of mass a few times
$10^3$ \msun; $ii$) a collapsed core with steep central profile of
slope $<$ -- 0.75, and $iii$) a small flat core of radius \lsim\
1.5\arcs\ = 0.09 pc.  Earlier reports of a weak color gradient in the
centre of M15 (Bailyn \etal\ 1988) are confirmed in the sense that
bright red giants are depleted in the centre relative to subgiants,
but the depletion of very blue horizontal-branch stars counteracts
this bluing (Stetson 1991, De Marchi \& Paresce 1994b).

In such a subtle matter, for which data processing methods are not
free of influence, star counts are far superior in quality to any
surface brightness measurement. Post-refurbishment HST star-count data
confirm that the 2.2\arcs\ core radius observed by Lauer \etal\
(1991), and questioned by Yanny \etal\ (1994a), is observed neither by
Guhathakurta \etal\ (1996b) with WFPC2 data nor by Sosin \& King
(1996) with FOC data.

A 5.6\arcs\ $\times$ 3.5\arcs\ area of an FOC image centered on the
core of M15 is displayed in Fig.~9.3.  The completeness-corrected
surface-density profile of stars with $V$ magnitude between 18.5 (just
above the main-sequence turn-off) and 20.0 is shown in Fig.~9.4 from
Sosin \& King (1996).  All the 839 stars have nearly the same mass.
This surface-density profile clearly continues to climb steadily
within 2\arcs.  A maximum-likelihood method rules out a 2\arcs\ core
at the 95\% confidence level. It is not possible to distinguish at
present between a pure power-law profile and a very small core.

\vskip 8truecm
\item\item{\sevenrm{{\bf Fig.~9.4.} \hskip 2mm The 
completeness-corrected surface-density profile of stars with $V$
magnitude between 18.5 and 20.0 (from Sosin \& King 1996 Fig.~1).  The
value of 2.2\arcs\ obtained by Lauer \etal\ (1991) for the core radius
is illustrated by the vertical dotted line.  }}
\vskip 0.5truecm

Density profiles of M15, very similar to those of Sosin \& King
(1996), deduced from star counts obtained with WFPC2 data are given by
Guhathakurta \etal\ (1996b) for two independent magnitude-limited
samples: $V$ $<$ 18.3 and 18.3 $<$ $V$ $<$ 20.0.  These two radial
profiles are the same, within Poisson errors.  This is to be expected,
since the difference between the average masses of the stars ($\sim$
0.75 \msun\ for $V$ = 18.3-20.0 and $\sim$ 0.78 \msun\ for $V$ $<$
18.3) is expected to be too small to give rise to significant effects
due to mass segregation.  Guhathakurta \etal\ (1996b) provide three
different approaches to measuring the surface density distribution:
binned star counts, parametric fits, and non-parametric estimates.
The density profile appears to rise smoothly towards the centre of the
cluster, with no suggestion of levelling off.  It can be approximated,
in the range from 0.3\arcs\ to 6\arcs, by a power law $r^\alpha$ with
$\alpha$ = --0.82 \pmm\ 0.12.  This value is similar to that expected
for the stellar distribution around a black hole with a mass of a few
times 10$^3$ \msun\ (Bahcall \etal\ 1975, Bahcall \& Wolf 1976, 1977)
although it is fully consistent with core-collapse models, which offer
a somewhat simpler, less exotic, explanation (see, e.g., Murphy \&
Cohn 1988, Murphy \etal\ 1990, Grabhorn \etal\ 1992; see also Goodman
1993 and references therein).

The most recent, best quality HST data show no evidence of any
levelling off of density profiles in the cores of the most
concentrated globular clusters like M15 and M30.  This can be
interpreted as a direct evidence of core collapse, from the density
profile.  The best current studies are limited by the uncertainties in
the cluster centroid position, in the correction factor for crowding
problems in star counts, and by Poisson error in the counts, which
restrict the analysis of the surface density profile to $r >$
0.3\arcs.

\vskip 30pt
\noindent
{\sl 9.3 Observational evidence of core collapse through the velocity
dispersion profile} 

\vskip 20pt
\noindent
Contrary to the spatial resolution of imaging techniques, which has
improved by about an order of magnitude (from $\sim$ 1.0\arcs\ to
$\sim$ 0.1\arcs) with the advent of HST, the spatial resolution of
spectroscopic capabilities applied to the measurement of the velocity
dispersion in globular clusters to about 1 \kms\ is still of the order
of $\sim$ 1.0\arcs.  Consequently, the search for a cusp in velocity
dispersion profiles is far less advanced than the search for cusps in
density profiles.  This is really a pity, since the definitive way to
distinguish between core-collapse and black-hole models consists of
measuring, as a function of the radius, the dispersion of the radial
velocities of as many stars as possible in the central regions.

Once again, M15 looks like the most interesting candidate, being the
only globular cluster which has been studied carefully for the
presence of a central cusp in velocity dispersion.  Cudworth's (1976)
proper motion study gave the first estimate of velocity dispersion in
M15, \sigp\ $\sim$ 10 \kms, based on stars between 1.5\arcm\ and
12\arcm\ from the centre.  Peterson \etal\ (1989) published the first
velocity dispersion profile in M15, derived from two different kinds
of data: (i) from individual radial velocities for 120 cluster members
scattered between 0.1\arcm\ and 4.6\arcm\ from the centre and (ii)
from integrated-light spectra of the central luminosity cusp. The
radial velocities of the 27 stars within 20\arcs\ of the centre give
\sigp\ = 14.2 \pmm\ 1.9 \kms, while the integrated-light spectra 
suggest a cusp in the velocity dispersion profile, with \sigpo\ of at
least 25 \kms, a unique case among globular clusters.  This central
value does not match the predicted velocity dispersion profile from
Fokker-Planck models (Grabhorn \etal\ 1992).

\vskip 10truecm
\item\item{\sevenrm{{\bf Fig.~9.5.} \hskip 2mm Upper panel:
superposition of two images of the central 10.5\arcs\ $\times$
10.5\arcs\ region of M15. The contour plot comes from a $V$-band image
of angular resolution of 0.35\arcs\ taken with HRCam at the CFHT
(Racine \& McClure 1989 Fig.~1).  The black dots are stars from an HST
FOC image obtained with the F342W filter and displayed with a high
low-cutoff so as to show only the sharp cores (0.08\arcs\ FWHM) of the
bright star.  The near equilateral triangle, mentioned in Fig.~9.3,
formed by three bright stars near the centre of the image is easily
recognisable.  The five dashed-line rectangles show the different
positions of the 1\arcs\ \x\ 8\arcs\ slit during five high-resolution
spectroscopic observations of the cluster core by Dubath \& Meylan
(1994).  For the purpose of illustration of the simulations by Dubath
\etal\ (1994), an integrated-light spectrum has been extracted from
each of the three particular areas indicated by three solid-line
rectangles and labelled I, II, and III.  \hskip 2mm Lower panel: the
three cross-correlation functions corresponding to the three
integrated-light spectra from areas I, II, and III, respectively.
From Dubath \& Meylan (1994 Fig.~2).  }}
\vskip 0.5truecm

As part of a long-term program to determine the central velocity
dispersion in the cores of high-concentration and collapsed-core
globular clusters, Dubath \etal\ (1993a,b, 1994) obtained an
integrated-light spectrum of the core of M15, over a central 6\arcs\
\x\ 6\arcs\ area, leading (see Eq.~6.1 above) to a projected velocity
dispersion \sigpo\ = 14.0 \kms.  It is worth mentioning that, because
of a larger sampling area, Dubath \etal\ (1994) would have probably
missed any central cusp in velocity dispersion.

Totally unexpectedly, and despite the high signal-to-noise ratio of
the observed spectrum, the cross-correlation function of the M15
spectrum is bumpy, as if it were the sum of two different gaussians.
This large departure from the usual gaussian function (see Fig.~6.1
and Fig.~6.2 above) is larger than the deviations produced by the
spectrum noise.  Such a behavior (also present in the
cross-correlation function displayed in Fig.~10 of Peterson \etal\
1989) is expected only if the integrated-light spectrum is completely
dominated by the contribution of the few brightest stars lying inside
the sampling area (slit) of the spectrograph.

A quantitative estimate of the small number statistics from a few
bright stars affecting the central velocity dispersion measurements of
M15 is absolutely necessary for further interpretations of any
results.  Detailed and exhaustive numerical simulations, with
different sampling apertures (1\arcs\ \x\ 1\arcs\ in the case of
Peterson \etal, and 6\arcs\ \x\ 6\arcs\ in the case of Dubath \etal),
of the cross-correlation functions of integrated-light spectra in the
core of M15 have been carried out by Dubath \etal\ (1993, 1994).  See
also Zaggia \etal\ (1992a,b, 1993) for similar simulations,
(originally related to their observation of another globular cluster,
viz. M30, but adapted to Peterson \etal's observations of M15).

The results of these simulations may be summarized by two points: (i)
when the light in the sampling area is dominated by one bright star,
the observed cross-correlation function is narrow, similar to that of
a standard star, and the derived velocity dispersion is too small (see
Fig.~9.5, area I, in lower-left panel); (ii) when the light in the
sampling area is dominated by two bright stars with unusually
different radial velocities, the observed cross-correlation function
is broadened because of its double dip, and the derived velocity
dispersion is too large (see Fig.~9.5, area III, in lower-right
panel).  The noisy shapes of Peterson \etal's and Dubath \etal's
observed cross-correlation functions of M15 are qualitatively easily
reproduced by the simulations.  They show that any velocity
dispersion obtained from integrated-light measurements over small
central areas suffers from large statistical errors due to the small
numbers of bright stars present in the integration area.

Two complementary observational studies have given a partial (maybe
not final) solution to the conundrum presented by the velocity
dispersion in the core of M15.

First, Gebhardt \etal\ (1994, 1995) have used the Rutgers Imaging
Fabry-Perot Spectrophotometer to measure radial velocities with
uncertainties of less than 5 \kms\ for 216 stars within 1.5\arcm\ of
the centre of M15, with a technique which can alleviate the problems
due to crowding and sampling.  Their velocity dispersion profile is
plotted in Fig.~9.6.  The small dots are the absolute values of each
star's deviation from the cluster velocity plotted versus the distance
from the centre.  The solid line and the open circles, with error
estimates, are the velocity dispersion estimated by two different
techniques: (i) the open circles are the maximum likelihood estimate
of the dispersion in bins of 22 stars; (ii) the solid line is a
locally weighted scatterplot smoothing fit to the velocity deviations
squared.  The velocity dispersion profile shows a sharp rise from 7 to
12 \kms\ at 0.6\arcm\ (1.8 pc) and then appears to flatten into the
innermost point at 0.15\arcm.  The dispersion at 30\arcs\ obtained by
Gebhardt \etal\ (1994) is 10 \kms, and it reaches its maximum, of only
12 \kms, at 20\arcs.  At smaller radii the dispersion remains
constant.  Their data confirm the rise in velocity dispersion seen by
Peterson \etal\ (1989) between 0.7\arcm\ and 0.4\arcm, but give a
velocity dispersion estimate about 1.7 standard deviations lower in
the region between 0.1\arcm\ and 0.4\arcm.  Because of their 1.8\arcs\
seeing Gebhardt \etal\ (1994) cannot determine the dispersion
accurately within the central few seconds of arc.

\vskip 8.5truecm
\item\item{\sevenrm{{\bf Fig.~9.6.} \hskip 2mm Velocity dispersion 
as a function of the radius, for stars in M15 (from Gebhardt \etal\
1994 Fig.~5).  The dots are the absolute deviations from the cluster
velocity of the individual radial velocity measurements.  The open
circles are the velocity dispersion estimates, with uncertainties, in
bins of 22 stars.  The solid line is a locally weighted scatterplot
smoothing fit to the velocity deviations squared and the dashed lines
are its 90\% confidence interval.  The central determination of the
velocity dispersion \sigp\ = 11.7 \pmm\ 2.6 \kms\ (Dubath \& Meylan
1994) is represented by the large filled circle. }}
\vskip 0.5truecm

Second, using the ESO New Technology Telescope, Dubath \& Meylan
(1994) have obtained five high-resolution integrated-light echelle
spectra over the core of the M15.  They used a 1\arcs\ \x\ 8\arcs\
slit, with a 1\arcs\ offset between each exposure in order to cover a
total central area of 5\arcs\ \x\ 8\arcs\ (Fig.~9.5, upper panel). By
taking advantage of the spatial resolution along the slit, they
extracted spectra at 120 different locations over apertures $\sim$
1\arcs\ square. The Doppler velocity broadening of the
cross-correlation functions of these integrated-light spectra is
always $\leq$ 17 \kms\ (6.5 $\leq$ \sigp\ $\leq$ 17.0 \kms), at all
locations in the 5\arcs\ \x\ 8\arcs\ area.  These observations confirm
that the cross-correlation functions of integrated-light spectra taken
over such small apertures are mostly dominated by the contribution of
one or two bright stars, leading to unreliable estimates of the
velocity dispersion.  This bias can be reduced by taking the average
of all 120 cross-correlation functions, normalized in intensity, over
the whole 5\arcs\ \x\ 8\arcs\ central area: this gives a velocity
dispersion \sigp\ = 11.7 \pmm\ 2.6 \kms.  This value is independently
confirmed by comparing these new observations with numerical
simulations.  The individual radial velocities of the 14 best-resolved
(spatially or spectroscopically) bright stars are also determined;
they give \sigp\ = 14.2 \pmm\ 2.7 \kms.  These measurements therefore
provide no evidence for the velocity dispersion cusp $\geq$ 25 \kms\
(8.4 $\leq$ \sigp\ $\leq$ 30.0 \kms) observed by Peterson \etal\
(1989).

The above three observed values of the velocity dispersion are
complementary to, and consistent with, the work by Gebhardt \etal\
(1994).  In addition, they are all consistent with the predictions of
theoretical dynamical models of M15: viz. \sigpo\ = 12-17 \kms\ from
Illingworth \& King (1977) using a King-Michie dynamical model,
\sigpo\ = 13-15 \kms\ from Phinney \& Sigurdsson (1991) and Phinney
(1992, 1993) using the observed decelerations of the two pulsars in
the core of M15, and \sigpo\ = 14 \kms\ from Grabhorn \etal\ (1992)
who fitted a Fokker-Planck model to surface brightness and velocity
dispersion profiles. Consequently, the presence of any massive black
hole or of some non-thermal stellar dynamics in the core of M15 is not
required in order to explain the present observations.  However, the
detection of a moderate velocity cusp --- M15 would be the place
to look for --- would require a better spatial resolution and a
sensitivity which are not available yet.

The conclusion of this section is that, contrary to density profiles
which provide clear indication of a central density cusp in a few very
concentrated globular clusters, no such evidence has been obtained so
far from velocity dispersion profiles.  So far, core collapse
diagnostics is based on density profiles only.

\vskip 30pt
\noindent
{\sl 9.4 Physical Interactions}

\vskip 20pt
\noindent
For a long time the study of the dynamics of globular star clusters
was one of the ``cleanest'' theoretical problems in astrophysics,
involving nothing more than the gravitational interaction of point
masses, albeit in very large numbers.  In the 1970s it was held that
this was a basic distinction between the evolution of globular
clusters and that of galactic nuclei (Saslaw 1973, Bisnovatyi-Kogan
1978).  At about the same time it was being realised (Fabian \etal\
1975, Finzi 1977) that stellar collisions and close encounters, though
they might not affect the overall evolution, could be of importance in
understanding the unusual stellar populations in globular clusters,
especially the X-ray sources.  In more recent years, however, the
realisation has grown (Statler \etal\ 1987) that inclusion of direct
physical interactions between individual pairs of stars may be
necessary if theoretical models are to remain reasonably realistic
approximations of the dynamical behavior of the entire system.  The
observational evidence for this is presented in \S9.8.  Here we
summarise recent work on the theory of collisions and other close
stellar interactions, though a great deal of other relevant work can
be found in literature devoted to galactic nuclei.

To begin with, much of the theory of these processes was
semi-analytical, following the technique introduced by Press \&
Teukolsky (1977); see, for example, Giersz (1986), Lee \& Ostriker
(1986) and McMillan \etal\ (1987).  The straightforward part of these
calculations is the computation of the energy lost in a single
encounter, which is what is relevant for the computation of the
capture probability.  One of the long-standing issues, however, is how
to account for the effect of this energy on the internal structure of
the participants (Kochanek 1992, Podsiadlowski 1996).  If it leads to
an expansion of either star, then the effect of subsequent encounters
may be collision and coalescence rather than capture.  If the tide
raised by one star on the other in the first encounter is not
dissipated quickly enough (see Kumar \& Goodman 1996), then the
subsequent evolution of the orbit may be chaotic rather than simply
dissipative (Mardling 1995a,b, 1996).  A two-body effect that is
certainly of importance for close binaries is gravitational radiation
(e.g., Buitrago \etal\ 1994); it might even be detectable in clusters
because of its effect on pulsar timings (Sazhin \& Saphanova 1993).

In the last few years considerable effort has been expended in the
detailed numerical modelling of encounters between stars of different
types, in order to measure the cross sections for merging and binary
formation, and to measure the amount of mass loss, etc.  (Shara \&
Regev 1986, Soker \etal\ 1987, Rozyczka et al 1989, Ruffert \& Muller
1990, Benz \& Hills 1992, Davies \etal\ 1991, 1992, 1993, Lai \etal\
1993, Rasio 1993, 1996a,b, Ruffert 1993, Chen \& Leonard 1993, Zwart \&
Meinen 1993, Lee \etal\ 1996).  In some cases the role of stellar
nuclear reactions can be substantial (Benz \etal\ 1989).

One way in which the importance of encounters may be enhanced is in
the context of binary stars.  As shown in the next section,
interactions involving binaries almost certainly have a major role to
play in the overall dynamical evolution, but these interactions
themselves will be significantly altered by the finite radii of the
stars (Hut \& Inagaki 1985).  This has been modelled in some detail by
Davies \etal\ (1994).  Another circumstance in which the effects of
encounters may be greatly enhanced is in cases where the stars are
accompanied by circumstellar disks (Murray \etal\ 1991, Murray \&
Clarke 1993).  At one time, when the evidence for a large population
of binaries in clusters was lacking, it was even suggested (Hills
1984) that any primordial binaries might not have survived to the
present time because of encounters between the components of a binary
at a time when their relative orbit had a high eccentricity.

An important issue in discussions of stellar interactions and mergers
is the nature of the objects which will be produced.  Such
astrophysical implications, and their possible relation with different
kinds of more-or-less exotic stellar populations, have been considered
by Bailyn (1988, 1989), Krolik (1983, 1984), Krolik \etal\ (1984), Ray
\etal\ (1987), and Chen \& Leonard (1993).  Particular attention has
been given to the possible formation by these processes of blue
stragglers (Leonard 1989, 1996, Leonard \& Fahlman 1991, Leonard \&
Linnell 1992, Hut 1993b, D'Antona \etal\ 1995, Lombardi \etal\ 1995,
Eggleton 1996, Rasio 1996a,b), X-ray binaries, recycled pulsars (Ray \&
Kembhavi 1988), and hot subdwarfs (Bailyn \& Iben 1989).  The
relationships between these different classes of objects are still
being argued about.  It is possible that the millisecond pulsars in
clusters are not descendents of X-ray binaries, for example (Chen
\etal\ 1993).  For blue stragglers, the astrophysical issues include
whether the encounters can adequately mix the material of the two
stars (Rasio 1996a,b), and the luminosity functions of the merger
products (Bailyn \& Pinsonneault 1995).

How these processes affect a given cluster depends in part, of course,
on its density and other parameters.  Hills \& Day (1976) give a
useful tabulation of the expected rates of collisions in a large
sample of galactic globular clusters, on the basis of the data
available at that time.  A more recent study along these lines,
emphasising the encounters which could lead to the formation by tidal
capture of low mass X-ray binaries, is reported by Verbunt \& Hut
(1987).  When, as in this case, the encounters involve stars of
different mass, the results are heavily dependent on dynamical
modelling of the clusters (van der Woerd \& van den Heuvel 1984,
Verbunt \& Meylan 1988, Hut \etal\ 1991).  A detailed theoretical
study of the rate of formation of cataclysmic variables by tidal
capture in two cluster environments is presented by Di Stefano
\& Rappaport (1994).  Sigurdsson \& Phinney (1995) have carried out a
remarkably thorough investigation of binary-single encounters which
takes account of the orbital evolution of the reactants.  The results
(in terms of the relative and absolute numbers of interesting
products, such as blue stragglers, cataclysmic variables and
millisecond pulsars) depend on the concentration of the cluster
model.

As already stated in the introduction to this section, the other main
consequence of these processes is their effect on the cluster itself.
Simplified models, based on the energetics of the interactions, were
devised by Milgrom \& Shapiro (1978), Alexander \& Budding (1979),
Dokuchaev \& Ozernoi (1981a,b), and Giersz (1985a,b).

\vskip 30pt
\noindent
{\sl 9.5 Dynamics and formation of binaries}

\vskip 20pt
\noindent
The fact that the central density is predicted to rise to infinity at
the end of core collapse (Eq.~9.1) is clear proof of a serious
deficiency in the theory.  The most likely missing ingredient is
binary stars.  Even if these are not present initially (primordial
binaries) they would form by one or other of the processes which we
shall discuss below.  Historically, it is these processes of formation
which have received most attention, because it was thought for a long
time that primordial binaries are essentially absent from globular
clusters.  This change of perspective is a fundamental revolution
which has not yet been fully absorbed, though the review by Hut \etal\
(1992a) makes the facts plain.

Binaries are important because of the energy which can be imparted to
single stars or other binaries in interactions.  This can be
understood from several points of view, including thermodynamic
arguments (e.g., Horwitz 1981, Padmanabhan 1989b); statistical
analyses (e.g., Mansbach 1970, Monaghan 1976a,b, Nash \& Monaghan
1978, 1980), since three-body interactions have many chaotic aspects
(Boyd 1993); and the study of close triple encounters by the
analytical techniques of celestial mechanics (e.g., Marchal 1980) and
atomic scattering theory (Gruji\'c \& Simonovi\'c 1988, Heggie \&
Sweatman 1991).  One of the main tools is the numerical study of the
three-body problem.  Much interesting data can be found in the papers
of the groups at St Petersburg (recent references including Anosova
1986, 1990, 1991, Anosova \& Orlov 1988, and Anosova \& Kirsanov
1991), at Austin (e.g., Szebehely 1972), and at Turku (e.g., Valtonen
1975, 1976, 1988a; Huang \& Valtonen 1987, Valtonen \& Mikkola 1991)
and in the references mentioned below.  Not all of the above data is
suitable for statistical analysis of the effects of encounters,
however, and sometimes the results are restricted in various ways,
e.g., to head-on encounters (zero impact parameter), or circular
binaries, though this is an important special case.  An ingenious and
freely available computational tool for few-body interactions is
described by McMillan (1996).

Our knowledge of the energetic effects of three-body interactions in
the case of equal masses is reasonably complete (Hills 1975a, Hut
1983, 1993a, Heggie \& Hut 1993), and there is extensive information
on the case of unequal masses, especially those relevant in
applications to globular clusters (Heggie 1975, Hills \& Fullerton
1980, Hills 1990, Sigurdsson 1992, Sigurdsson \& Phinney 1993, 1995).
Less well studied, but also important for investigations of
millisecond pulsars, are the effects of encounters on the eccentricity
of binary orbits (Hut \& Paczynski 1984, Rappaport \etal\ 1989,
D'Amico \etal\ 1993, Rasio \& Heggie 1995, Heggie \& Rasio 1996).  A
fairly comprehensive cross section for exchange involving hard
binaries with stars of unequal mass has been provided by Heggie \etal\
(1996).  Of course knowledge of the way in which binaries and stars of
different masses behave can only be applied satisfactorily in clusters
in which the spatial distribution of the different species is well
enough known (Hut \etal\ 1991).

Energetically these three-body processes are subsidiary to four-body
encounters (i.e., binary-binary collisions), where our knowledge of
the relevant reaction rates is more patchy, partly because of the
greater range of relevant parameters.  The most extensive published
data has been provided by Mikkola (1983a,b, 1984a,b), and other
significant studies have been carried out by Hoffer (1983, 1986) and
Hut (1992).

Though those binaries that are dynamically effective are almost
certainly outnumbered by single stars in globular clusters,
binary-binary encounters are still dominant energetically, for two
reasons: (i) the cross section for an energetic interaction with a
single star is considerably smaller than that for interaction with
another binary; and (ii) it is likely that the mean total mass of the
components of a binary exceeds the mean stellar mass, so that, by mass
segregation, they soon become preferentially concentrated in the core
(cf. Spitzer \& Mathieu 1980), where almost all the energetic
interactions take place.

The main effect of these interactions is to halt the collapse of the
core.  This was demonstrated long ago by Hills (1975b) by means of a
very simplified model, though doubt was cast on the effectiveness of
binary-binary collisions as a mechanism for halting core collapse by
the Fokker-Planck models of Spitzer \& Mathieu (1980); their results
implied that the halt was only temporary, and that the destruction of
binaries in binary-binary encounters (which had not been taken into
account by Hills) led quite quickly to continued collapse.  However
the general picture painted by Hills has subsequently been confirmed,
with much additional detail, using a different Fokker-Planck code from
that adopted by Spitzer \& Mathieu (Gao \etal\ 1991), though the
reasons for the disagreement have never been unearthed. We shall
return to the possibility of binary depletion and further core
collapse in our discussion of post-collapse evolution (cf. \S10.1
below).

These processes can be modelled satisfactorily with $N$-body
simulations (Aarseth 1980, Giannone \etal\ 1984, Giannone \& Molteni
1985, McMillan \etal\ 1990, 1991, McMillan 1993, McMillan \& Hut 1994,
Heggie \& Aarseth 1992, Aarseth \& Heggie 1993, Kroupa 1996), all of
which have confirmed that core collapse can indeed be brought to an
end, and have demonstrated how the point at which collapse is halted
is affected by the parameters of the binary distribution (mainly the
range of internal energies, and their numbers).  One of the
uncertainties here, however, is the distribution of masses of binary
components.

We have pointed out that binary-binary interactions are effective in
halting the collapse of the core.  The other main effect of
interactions involving binaries is the effect on the participating
stars and binaries, and even on putative planetary companions of
pulsars (Sigurdsson 1992)!  As already mentioned, collisions are
effective in destroying binaries (the outcome of most binary-binary
interactions being the destruction of one participant), in hardening
those that remain, and in ejecting them to the outer parts of the
cluster.  The effect of three-body interactions on the distribution of
(internal) binding energies was studied, in the context of a
homogeneous stellar system, by Retterer (1980b) and, using better
scattering cross sections, by Goodman \& Hut (1993).  In fact,
however, the hardening of the binaries also influences their spatial
distribution, as was clearly demonstrated in a simplified model by Hut
\etal\ (1992a).  For example, such interactions may be needed in order
to understand the spatial distribution of pulsars (Phinney \&
Sigurdsson 1991) and blue stragglers (Sigurdsson \etal\ 1994).  In
slightly more extreme cases triple interactions may lead to
high-velocity escapers (cf. \S7.3), and it is just possible that two
rapidly moving stars in M3 (Gunn \& Griffin 1979) as well as in \tuca\
(Meylan \etal\ 1991a) originated in this way.  A further effect of
interactions involving binaries is exchange reactions, which is
thought to be the main channel for the formation of X-ray sources.
Examples of specific systems whose dynamical evolution has been
studied with regard to triple interactions are provided by Rasio
\etal\ (1995) and Sigurdsson (1993).

\vfill\eject 

\vphantom{ }

~~~

\vskip 13truecm
\item\item{\sevenrm{{\bf Fig.~9.7.} \hskip 2mm  Hydrodynamic 
simulation of a binary-binary encounter (from Goodman \& Hernquist
1991 Fig.~2d).  The two binaries are shown emerging from an
encounter, which so perturbs the upper pair that they coalesce.  The
merger remnant is described as a rapidly rotating spheroidal star
surrounded by a thick disk. No collision occurs in a simulation
without hydrodynamics.  Initial conditions: binary orbits are circular
with random orientations, and semi-major axis $a$; relative orbit of
the binaries is parabolic; stars are equal-mass polytropes of index
$n$ = 3/2 and radius $a$/6.}}
\vskip 0.5truecm

Though one of the participants is likely to be destroyed in any close
encounter between two binaries, this is a significant route for the
formation of relatively long-lived multiple systems (Kiseleva \etal\
1996).  The stability of such systems is a long-standing issue in
dynamical astronomy (see, for this and many other aspects of the
three-body problem, the book by Marchal 1990), and clearly can have
profound implications for the internal evolution of the member stars.
This is one way in which the dynamical evolution of globular clusters
becomes bound up with the way in which the stars themselves evolve,
and it has stimulated renewed interest in the stability and stellar
evolution of triple systems (Kiseleva \etal\ 1994, Eggleton \&
Kiseleva 1995).

The main unexplored complication in all this is the finite radii of
the stars taking part in these interactions.  It has been pointed out
(Aarseth, pers.  comm.) that the $N$-body models predict their own
downfall by confirming theoretical expectations that stellar
collisions should be frequent.  As mentioned in the previous section,
the effects of collisions on the details of individual encounters are
dramatic (McMillan 1986a, Cleary \& Monaghan 1990, Goodman \&
Hernquist 1991; cf. Fig.~9.7).  As yet, however, little has been done
to follow through the consequences for core evolution, though this is
within the scope of suitable codes, using either ``sticky'' particles
or smooth particle hydrodynamics (SPH), except in the context of open
clusters (Aarseth 1992, 1996a,b).  Though the astrophysical complications
are great, so are the astrophysical pay-offs.  Possible consequences
of interactions involving binaries with stars of finite size are
enhanced mass transfer (Shull 1979), the formation of blue stragglers
(e.g., Bailyn 1992, Bacon \etal\ 1996), and the observed depletion of
red giants in cluster of high concentration (e.g., D'Antona \etal\
1995).  Davies (1995) has shown how a remarkable variety of
astrophysically realistic interactions may be modelled quite
economically, and this has been applied to \tuca\ and \cent\ by Davies
\& Benz (1995).

Now we briefly turn to some older studies involving the formation of
binaries in a system without primordial binaries.  Two mechanisms were
considered (see, for example, Giersz 1984 for unified treatments).
The less dominant in most conditions (Dokuchaev \& Ozernoi 1978,
Inagaki 1984) is the formation of binaries in three-body interactions,
which was essentially discovered in $N$-body simulations of van Albada
(1967) and Aarseth (1968, 1972, 1977, 1985b).  Quite simple estimates
(e.g., Heggie 1984) successfully predict that such binaries can arrest
core collapse, though usually some other process intervenes first: in
almost all real systems without primordial binaries it would be
dominated by the two-body (tidal) formation mechanism (Fabian \etal\
1975, Press \& Teukolsky 1977, Lee \& Ostriker 1986; this review,
\S9.4), except in the case of a cluster core dominated by degenerate
stars (Lee 1987a).  Thousands of binaries may form from this
mechanism, and part of its continuing importance is that a large
fraction would form before core collapse is well advanced.  Therefore
this is still an important mechanism even if core collapse is arrested
at relatively low densities by primordial binaries.  What complicates
the problem is that tidal binaries are extremely tight, and any
interaction is likely to lead to collision.  Furthermore the number of
pairs is likely to be matched by the number of stellar collisions.
Therefore the production of coalesced stars may be the main outcome of
this mechanism, but, despite its potential importance for giving rise
to exotic types of star (cf. \S9.7 below) its interaction with
dynamics has been little studied in recent years, except in the
context of post-collapse expansion (\S10.1).

\vskip 30pt
\noindent
{\sl 9.6 Observational evidence of binaries in globular clusters
(compared to the field)} 

\vskip 20pt
\noindent
Although there is now clear and plentiful observational evidence of
binaries in globular clusters, this has not always been the case. In
spite of intensive searches for some decades preceding the mid
eighties, i.e., before the use of CCDs in astronomy, there was no
known photometric or spectroscopic binary in globular clusters.  Gunn
\& Griffin (1979), who did not discover any spectroscopic binaries
among the 111 stars they observed in M3, with radial velocities to an
accuracy of $\sim$ 1 \kms, concluded that binarism involving stars
with separations in the range 0.3-10 AU is either very rare or absent
in globular clusters, in stark contrast to the situation in the solar
neighborhood and in open clusters.  Fifteen years later, it is now
clear that the usual sort of photometric and spectroscopic binaries,
with periods from one day to one year, do exist in globular clusters,
in addition to more exotic objects like binary milli-second pulsars.

Binary star formation scenarios may be categorised in a variety of
ways, but the most fundamental distinction is between those in which
stars form singly within a cluster and subsequently pair up (as
described in \S\S9.4 and 9.5 above), and those in which stars form as
binaries as a result of a splitting into two during the star formation
process (cf. \S5.4 above).  From a purely observational point of view,
it is impossible to disentangle the class of model (capture and
fragmentation) by which a given observed binary star has been formed,
although soft (long-period) binaries may be preferentially primordial
and hard (short-period) binaries may be of more recent formation.

This section describes the observational evidence for binaries
provided by photometric and radial-velocity surveys, without
mentioning the possible origin scenario.  The ultimate aims of such
studies are (i) to establish the frequency of binaries and (ii) to
determine their radial distribution within the cluster, two quantities
which are intimately linked to the internal dynamics.  See also the
reviews by Hut \etal\ (1992a) and Phinney (1996).

\vskip 20pt
\noindent
{\sl Observational evidence of binaries from photometry.}~~ For a
variety of practical reasons, most searches for photometric binaries
in globular clusters have been restricted to the study of short-period
(\lsim\ 5 days) eclipsing binaries and cataclysmic variables only.
For several decades, all eclipsing binaries observed in globular
clusters turned out not to be members, once the appropriate radial
velocity information was acquired.  

The first clear case of a genuine member was obtained by Niss \etal\
(1978), who identified in \cent\ one certain eclipsing binary (NJL~5,
with P = 1.38 day), whose membership was subsequently confirmed, on
the basis of radial velocity measurements, by Jensen \& J\o rgensen
(1985) and Margon \& Cannon (1989).  Since then, the list of such
stars has increased steadily, with potential of giving insight into
the frequency of binary stars. For example, Mateo \etal\ (1990)
discovered, from a long series of CCD exposures, three eclipsing
systems among the population of blue stragglers in NGC~5466, one of
them being of Algol type, the two others of W~UMa type.  Yan \& Mateo
(1994) found, in the centre of M71, five binaries, one of them being
of Algol type, the four others of W~UMa type.  See Yan \& Reid (1996)
for a search in M5 and Yan \& Cohen (1996) for NGC~5053.  Since
binaries are expected to be more numerous in the centre, due to mass
segregation, photometric searches from the ground are limited to the
loose globular clusters like \cent\ or to the outer parts of
concentrated clusters like \tuca.  E.g., Rubenstein \& Bailyn (1996)
have discovered a W~UMa binary in the globular cluster NGC~6397, at
about 2\arcm\ from the centre. A very interesting by-product of the
Optical Gravitational Lensing Experiment (OGLE) is the discovery,
apart from SX~Phe pulsating stars, of eclipsing binaries in \cent\ and
\tuca\ (Kaluzny \etal\ 1996a,b,c).

\vskip 10.5truecm
\item\item{\sevenrm{{\bf Fig.~9.8.} \hskip 2mm Cumulative radial 
distributions, of the binary stars, of the blue stragglers (BSs), and
of all main sequence, subgiant, and giant stars with the same
magnitude range (15.9 $<$ $U$ $<$ 20.4) as the binary stars detected
in \tuca\ (from Edmonds \etal\ 1996 Fig.~15).  }}
\vskip 0.5truecm

With HST, similar searches are possible even in the cores of the most
concentrated globulars.  Gilliland \etal\ (1995) and Edmonds \etal\
(1996) have monitored 20,000 stars in the core of \tuca\ using
differential time series $U$ photometry with the WFPC1.
Using aperture photometry, PSF fitting, and power spectrum techniques,
they discovered 2 W~UMa binaries in addition to 6 semi-detached or
detached binaries with periods between 0.41 and 1.5 days.

Fig.~9.8 displays, in the case of \tuca, the cumulative radial
distributions of the binaries discovered by Edmonds \etal\ (1996), the
cumulative radial distributions of the blue stragglers discovered by
Guhathakurta (1996a), and of all the other stars in the same range of
magnitude.  It appears that the binaries and the blue stragglers have
similar radial distributions and that both are more centrally
concentrated than the normal stars (see also Guhathakurta
\etal\ 1992).  This result implies that, in \tuca, the binaries,
compared to other stars, are more centrally concentrated than the
binaries in M71 (Yan \& Mateo 1995) and in NGC~4372 (Kaluzny
\& Krzeminski 1993).  This may be related to the different
concentrations and central relaxation times of these clusters,
although definitive statements will be made possible only when a
complete census of binary stars, down to a given limiting magnitude,
will be available.  

It is worth mentioning that the present samples obtained in different
clusters are very dissimilar, covering different period ranges, at
different distances from the centres of the clusters.

\vskip 20pt
\noindent
{\sl Observational evidence of binaries from radial velocities.}~~ So
far, virtually all radial velocity surveys of stars in globular
clusters have been related to the luminous giants.  Their large radii,
typically 0.1-0.4 AU, impose a bias on the periods detectable.  Binary
systems with periods shorter than about 40 days, i.e., with
separations less than about 0.25 AU, will not reach the luminosities
required to be included in the magnitude-limited samples because they
will have suffered mass transfer which either truncates the evolution
of the giant or leads to coalescence through a common-envelope stage
(Pryor \etal\ 1988).  This bias towards long periods means that
improved velocity precisions are needed to discover shorter period
binaries.

A typical giant primary of 0.8 \msun\ with a companion of 0.4 \msun,
separated by 0.25 AU, gives a binary with a period of 42 days; if the
orbit is circular, the giant star has an orbital velocity of 22 \kms.
Increasing the period by a factor of 10 increases the separation to
1.2 AU and decreases the velocity to 10 \kms.  Consequently, detecting
such binaries requires studies lasting years and velocity measurements
accurate to about 1 \kms.  Such velocity precisions have been
achieved, for about two decades now, by the two CORAVEL spectrometers,
the Dominion Astrophysical Observatory radial-velocity scanner, and by
the intensified Reticon system of the Smithsonian Astrophysical
Observatory.

The conclusion by Gunn \& Griffin (1979) that binarism is either very
rare or absent in globular clusters, was criticized by Harris \&
McClure (1983).  They pointed out that, giant stars in globular
clusters having lower masses and larger radii than field Population~I
giants, the results of Gunn \& Griffin (1979) were compatible with the
Abt \& Levy (1976) field binary frequency.  This prompted D. Latham
and T.  Pryor to undertake extensive new observations of the M3 giants
which resulted in the discovery of the first spectroscopic binary in a
globular cluster, vZ~164 (Latham \etal\ 1985).  The list of such stars
has increased steadily since then.

A major search for binaries in the globular cluster NGC~3201 has been
published by C\^ot\'e \etal\ (1994) who obtained multiple velocity
measurements for 276 stars, with a mean time span between observations
of 1.7 year, and with coverage up to about 6 years for the best
studied stars.  They find 21 stars with some signature of binarity,
although the radial velocity measurements of some of these binary
candidates, which are among the brightest cluster members, may suffer
from the so-called ``jitter'' due to stellar atmospheric motions,
first described by Gunn \& Griffin (1979) in M3; see also Mayor \etal\
(1984) in \tuca.

Another major study, from the point of view of both sample size and
time baseline, concerns the giant galactic globular cluster \cent\
(Mayor \etal\ 1996).  It is worth mentioning that \cent, being an old
globular cluster, is perfectly well suited to provide information on
primordial binaries since the characteristic time for dynamical
evolution of spectroscopic binaries (with periods $P$ $<$ 30~yr) is
much longer than the cluster age. The rather low central stellar
density of this loose globular cluster and its related large half-mass
relaxation time (26 $\leq$ \trh $\leq$ 46 \x\ \bily: Meylan \etal\
1995) ensure that dynamical influences on the primordial binary
population through close stellar encounters have not been great.
Actually, the present period range for primordial binaries among red
giants is limited on the short-period side by the onset of Roche-lobe
overflow, and on the long-period side by dynamical disruption.
Consequently, most of the primordial binaries among the giant of
\cent\ are expected to have periods from 200 to 4,000 days.

Between 1981-1993, the radial velocities of 310 giant stars which are
members of \cent\ were monitored by Mayor \etal\ (1996), with a mean
error of a radial velocity measurement of about 0.7 \kms.  All stars
have 3 or more measurements. The ``jitter'' observed in the radial
velocities of bright giant stars in globular clusters can be easily
disentangled from the variations due to spectroscopic binaries since
their effects on the observed cumulative distribution of the standard
deviations are quite distinct.  Two stars are definite binaries.

C\^ot\'e \& Fischer (1996) have undertaken a search for spectroscopic
binaries on the main sequence of the nearby globular cluster M4.  A
pair of radial velocities (median precision $\simeq$ 2 km/s) separated
by 11 months have been obtained for 33 turn-off dwarfs in the
magnitude range 16.9 $\le$ V $\le$ 17.4.

C\^ot\'e \etal\ (1996) report on search for long-period binaries in
M22.  They use observations accumulated between 1972 and 1994. This
22-year baseline is the longest available for any sample of globular
cluster stars. Using 333 repeat velocities for 109 cluster members,
they search for spectroscopic binaries with periods in the range 0.2
to 40 years and with mass ratios between 0.1 and 1.0.

\vfill\eject 

\vskip 20pt
\noindent
{\sl On the frequency of binaries.}~~ Contrary to what was thought in
the early 80's, we now know that binaries do exist in globular
clusters.  The observational evidence comes through two different
routes: (i) from photometric light curves, which are efficient in
discovering short-period (detached, semi-detached and contact)
binaries, with data acquired within a few nights, and (ii) from radial
velocity curves, which are efficient in discovering long-period
(primordial) binaries, with data acquired over more than a decade.

The ultimate aim of all searches for binary stars in globular clusters
is the knowledge of the frequency of binaries (is it higher, equal, or
smaller than in the solar neighborhood?), since, through their
formation and destruction, these stars play a fundamental role in the
dynamical evolution of these stellar systems, especially during core
collapse phases (cf. \S\S9.1 and 9.2 above).  Unfortunately, this is
not an easy task given the very large diversity of non exhaustive
surveys, sampling different period ranges, and the fact that the
binary frequency may vary from one cluster to the other, and from one
period range to the other.  The only recourse for estimating the
binary frequency consists of comparing data with large numbers of
simulations.

Considering the data from (short-period) eclipsing binary surveys and
adopting from Duquennoy \& Mayor (1991) a binary fraction in the solar
neighborhood of 65\%, Hut \etal\ (1992a) conclude that the overall
binary fraction in globular clusters is between 20\% and 35\%, i.e.,
significantly lower than the frequency in the solar neighborhood. In
the case of radial velocity surveys, for binaries with 0.2 $\leq$ $P$
$\leq$ 20~yr and $q \leq 0.22$, Hut \etal\ (1992a), quoting Pryor
\etal\ (1989b), argue for a fraction between 5\% and 12\%, i.e., at
most a small deficiency of binary stars in globular clusters when
compared to the solar neighborhood.

From their photometric survey, Yan \& Mateo (1994) determine a lower
limit of 1.3\% on the fraction of primordial binaries in M71 with
initial orbital periods in the range 2.5-5 days.  From their
simulations, they conclude that this implies an overall primordial
binary frequency $f$ = 22$+26 \atop -12$\% assuming d$f$/d$\log P$ =
Cst (the ``flat'' distribution) or $f$ = 57$+15 \atop -8$\% assuming
d$f$/d$\log P$ = 0.032$\log P$ + Cst (the ``sloped'' distribution) as
observed for G-dwarf binaries in the solar neighborhood (Duquennoy \&
Mayor 1991).  In the case of M5, Yan \& Reid (1996) estimate an
overall primordial binary frequency $f$ = 28$+11 \atop -5.8$\%
assuming d$f$/d$\log P$ = Cst (the ``flat'' distribution) for the
period range 2.5 days to 550 years.  Yan \& Cohen (1996) obtain a
binary frequency in NGC~5053 equal to 21-29\% with 3d $< P <$ 10yr,
0.125 $<$ q $<$ 1.75. This somewhat higher estimate is perhaps
related to the fact that NGC~5053 is relatively dynamically young when
compared to other clusters.  See also Yan (1996).

Edmonds \etal\ (1996), from considerations related to their observed
numbers of W~UMa and Algol systems and to orbital angular momentum
loss theory, conclude that the population of binaries detected
photometrically in the core \tuca\ appears fundamentally different
from populations discovered in other globular and open clusters.  They
argue that at least some of these binary systems have been formed in
the cluster core through stellar encounters.

In relation to their radial velocity survey, C\^ot\'e \etal\ (1994)
constrain the frequency of binaries in NGC~3201, from exhaustive Monte
Carlo simulations. Assuming a thermal distribution of eccentricities,
for periods 0.1 $\leq$ $P$ $\leq$ 5-10~yr and mass ratio 0.1 $\leq$
$q$ $\leq$ 1 they obtain a fraction of binaries
\lsim\ 15\% - 18\%.  For binaries with circular orbits, these limits
fall to 6\% - 10\%.  Consequently, the binary fraction in NGC~3201
appears equal to, or slightly higher than, that of the field which is
equal to 4\% - 8\% in a comparable range of period and mass ratio.

Assuming that \cent\ has a period distribution similar to the one
observed for the nearby G dwarfs (Duquennoy \& Mayor 1991), Mayor
\etal\ (1996) estimate the global binary frequency in \cent\ to be as
low as 3-4\%, significantly smaller than the 13\% of binaries with $P
<$ 3,000 days found among the nearby G dwarfs by Duquennoy \& Mayor
(1991).

From the turnoff main-sequence stars observed in M4, C\^ot\'e \&
Fischer (1996) find, using Monte-Carlo simulations, a binary fraction
of 15\% for systems with periods in the range from 2 days to 3 years
and mass ratios between 0.2 and 1.0.  From the giant stars observed in
M22, C\^ot\'e \etal\ (1996) find, using Monte-Carlo simulations, a
binary fractions between 1-3\%, results consistent with Mayor \etal\
(1996).  

The studies of Mayor \etal\ (1996) and C\^ot\'e \etal\ (1996) point
towards a fraction of primordial binaries in \cent\ which is
significantly smaller than the fraction of primordial binaries in the
solar vicinity and in open clusters.  This is at variance with, e.g.,
the results obtained by Pryor \etal\ (1989b), Hut \etal\ (1992a), and
C\^ot\'e \etal\ (1994) for other globular clusters.  This may be
either the results of intrinsic differences between the studied
clusters or the consequences of differences in the simulations and
their interpretation.  These simulations are in all cases
sophisticated, complicated by numerous assumptions and astrophysical
inputs, which make their analysis subtle and their comparison with the
observations not entirely straightforward.  C\^ot\'e \etal\ (1996)
speculate that both the relative abundances of short- and long-period
binaries in globular clusters and the large differences in measured
binary fractions for clusters with high binary ionization (i.e.,
disruption) rates (M22, Omega Cen) compared to those for clusters with
low ionization rates (M71, M4, NGC~3201) point to a frequency-period
distribution in which soft binaries have been disrupted by stellar
encounters.

\vskip 30pt
\noindent
{\sl 9.7 Influence of dynamical evolution on stellar populations} 

\vskip 20pt
\noindent
It is now commonly accepted that globular cluster stellar populations
exhibit numerous observable scars which betray the influence of
stellar dynamics on stellar populations.  The straightforward
observation concerns colors of stars and, consequently their positions
in the color-magnitude diagram.  The idea of linking macroscopic
(dynamical evolution of a cluster as a whole) and microscopic (stellar
evolution of a single star) phenomena is rather recent.  And it acts
both ways: the general dynamical evolution of the cluster can
influence the fate of a single star, but the presence, e.g., of a few
binaries in the core, can also influence the dynamical evolution of
the cluster as a whole.

\vskip 20pt
\noindent
{\sl Blue horizontal-branch stars and blue subdwarfs.}~~ Suspicions
that stellar dynamics may influence the stellar evolution in globular
clusters are more than a decade old (Renzini 1983).  E.g., Buonanno
\etal\ (1985a), in their study of the giant, asymptotic, and
horizontal branches in color-magnitude diagrams of globular clusters,
mentioned the very different dynamical status of M15 and NGC~5466 as
the possible reason for the presence or not of faint blue horizontal
branch stars.  From an age and metallicity point of view, these two
clusters are as similar as they could be, but they strongly differ
only in relation to their structural parameters and central densities,
as presented in Table~9.1.

%

{\sevenrm 
{{\hsize=12truecm
{
$$
\table
\tablewidth{11.0truecm}
\tablespec{\l\c\c}
\body{
\header{{\bf Table~9.1:} Comparison between M15 and NGC~5466 }
\skip{10pt}
\hdoubleline
\skip{1pt}
\skip{1pt}
& parameter  & NGC~7078 & NGC~5466 \end
&            & M15      &          \end
\skip{1pt}
\skip{1pt}
\hline
\skip{1pt}
\skip{1pt}
& age $\tau$  		& 16 \pmm\ 3 Gyr  	& 16 \pmm\ 3 Gyr   \end
\skip{1pt}
& helium abundance Y 	& 0.23 \pmm\ 0.02 	& 0.23 \pmm\ 0.02  \end
\skip{1pt}
& metal abundance \feh\ & --2.10 \pmm\ 0.20	& --2.05 \pmm\ 0.20\end
\skip{1pt}
&  [O/H]		& --1.3 \pmm 0.3	& --1.3 \pmm 0.3   \end 
\skip{1pt}
& concentration \conc\  &  2.8 			&   1.5            \end 
\skip{1pt}
& central density \roo\ & 1.6 $\times$ 10$^6$ \msun~pc$^{-3}$ & 6.3 $\times$ 10$^0$ \msun~pc$^{-3}$ \end 
\skip{1pt}
\skip{1pt}
\hline
}\endtable
$$
}
}}
}

M15 is a very concentrated globular cluster, considered as a prototype
of collapsed clusters (cf. \S\S9.1, 9.2, and 9.3 above), while NGC~5466
is a rather loose cluster.  The former should suffer large numbers of
stellar collisions in its core, contrary to the latter.  A high rate
of encounters and collisions should produce, through coalescence and
merging, numerous stars heavier and bluer than turn-off stars.  As
conspicuously visible in Fig.~9.9, the color-magnitude diagrams of M15
and NGC~5466 differ only by the presence of a large number of blue
stars at the left end of the horizontal branch.  They are called blue
subdwarfs and, in globular clusters, also referred to as extreme or
faint blue horizontal branch stars, since they form a vertical
continuation to the horizontal branch.

Given the fact that, in a color-magnitude diagram, the faintest blue
subdwarfs constitute a downwards extension of the horizontal branch,
they can be mixed and confused with the brightest blue stragglers,
which form an upwards extension of the main sequence, above the
turn-off.  Despite their proximity, blue subdwarfs and blue stragglers
are thought to be very different, the former being closely related to
horizontal branch stars.

From the examination of the horizontal branch structure of 53
clusters, Fusi Pecci \etal\ (1993a) find that the length of the blue
tail of the horizontal branch correlates with cluster density.  Recent
HST observations with WFPC2 of ten galatic globular clusters have
resulted in the first discovery of hot horizontal-branch stars in two
metal-rich clusters, NGC~6388 and NGC~6441 (Rich /etal/ 1996) and in
the discovery of an intriguing multimodal horizontal branch in
NGC~2808 (Sosin /etal/ 1996).

\vskip 14.0truecm
\item\item{\sevenrm{{\bf Fig.~9.9.} \hskip 2mm Upper panel:
Color-magnitude diagram for all the stars in M15 (NGC~7078) brighter
than $B$ = 18.6 in the annulus with radii 1.9\arcm\ $<$ $r$ $<$
5.0\arcm.  Variables and fields stars have been omitted.  \hskip 2mm
Lower panel: Color-magnitude diagram for all the stars in NGC~5466
brighter than $B$ = 19.0 in the annulus with radii 0\arcm\ $<$ $r$ $<$
5.5\arcm.  Variables and fields stars have been omitted.  All data are
from Buonanno \etal\ (1985a).  }}
\vskip 0.5truecm

In \cent, there is evidence of segregation, towards the cluster
centre, of blue subdwarfs with respect to other stars (Bailyn \etal\
1992), but Drukier \etal\ (1989) find in M71 that blue subdwarfs are
less concentrated than other giants.  A decrease in the frequency of
blue subdwarfs towards the core of M15 is also observed by Buonanno
\etal\ (1985a) and De Marchi \& Paresce (1994b).  Mass segregation
only cannot be used to account for such a diversity of behavior.  Ways
of producing blue subdwarfs from binaries have been investigated by
Iben \& Tutukov (1986) and Bailyn \& Iben (1989).

\vskip 20pt
\noindent
{\sl Color gradients.}~~ Early reports of color gradients in globular
clusters have illustrated the difficulty in providing conclusive
observations of such phenomena.  E.g., the results by Chun \& Freeman
(1979), who found cluster integrated light becoming redder towards the
centre, were subsequently shown to be the consequences of clumps of
red giants stars more or less centered on the aperture, whose position
can also slightly vary between two different bands (Buonanno \etal\
1981).  Since then, CCDs have allowed the clear observation of color
gradients in some of the galactic globular clusters, although these
gradients appear in the sense that cores are bluer than the outer
regions.  See Djorgovski \& Piotto (1993) for the most important
review on this subject.  

Two different and largely complementary methods have been used:

\noindent
$\bullet$ Piotto \etal\ (1988) employed a generalization of a standard
surface photometry technique (described by Djorgovski 1988) to
simultaneously measure multiple color frames of the collapsed cluster
M30.  They found that this cluster becomes bluer towards the centre.
This gradient is present in different sets of data, is significant at
a 10-$\sigma$ level, and amounts typically to $\Delta (B - V)/\Delta
{\rm log} r$ $\sim$ 0.15 mag and $\Delta (B - R)/\Delta {\rm log} r$
$\sim$ 0.25.  This direct surface photometry technique is mainly
sensitive to effects present among the bright stars, which contribute
most of the total light.  Djorgovski \etal\ (1989) found a color
gradient in NGC~6624, another collapsed cluster, but not in NGC~6093,
a King-model cluster.  Djorgovski \etal\ (1991) extended this study to
12 clusters, confirming the trend that color gradients are present in
collapsed clusters but not in King-model clusters. This points towards
a link between color gradients and the general dynamical evolution of
a cluster as a whole.

\noindent
$\bullet$ Bailyn \etal\ (1988, 1989), using a complementary technique
based on pixel histograms, discovered a gradient in the collapsed
cluster M15.  They claimed that this gradient is caused by some
intrinsically faint stellar population, since their technique is
mainly sensitive to the effects present among the faint and numerous
stars which cover most of the detector area.

This apparent contradiction between the results from the two different
techniques is believed to be due to some genuine difference in the
nature of gradients in different clusters.  Cederbloom \etal\ (1992)
and Stetson (1991, 1994), with excellent CFHT data, confirm the color
gradient in M15. The latter concludes that it is due to three
different effects: (i) a deficiency of the brightest red giants in the
cluster centre, (ii) the giant branch shifts towards the blue as the
centre of the cluster is approached, (iii) the very centre of the
cluster contains a large population of blue stragglers, many of them
with a significant ultraviolet excess (see also Auri\`ere \etal\ 1990,
Djorgovski \etal\ 1991, Djorgovski \& Piotto 1992).  The presence of
very blue stars in the core of M15 and NGC~6397 may also have some
causal links to stellar dynamics in the core of this cluster (De
Marchi \& Paresce 1994a,b, 1995b, 1996); this idea was already
mentioned by Dupree \etal\ (1979) and Djorgovski \& Piotto (1992).

Although exhibiting a large variety in their characteristics, color
gradients are observed in all collapsed/high-concentration globular
clusters in which they have been looked for, but not in King-model
clusters.  When present, a gradient is always in the sense of the
color becoming bluer towards the centre, and it starts in radius at
about 20\arcs\ to 100\arcs\ from the centre.  In some clusters, e.g.,
M30, color gradients seem to be only the consequences of differences
in the distributions of bright stars (Piotto \etal\ 1988, Djorgovski
\etal\ 1989, and Burgarella \& Buat 1996).  In other clusters, e.g.,
M15, the gradients seem to be due mainly to the fainter unresolved
stars (Bailyn \etal\ 1988, 1989 and Cederbloom \etal\ 1992).  And in a
third group of clusters, e.g., NGC~6397, there are color gradients in
the light from both bright and faint stars (Lauzeral \etal\ 1993 and
Djorgovski \& Piotto 1993).  In some clusters, e.g., M30, M15, and
NGC~6397, there is a clear depletion of bright red giants near the
cluster centre rather than an increase in the numbers of horizontal
branch stars.  See also Bailyn (1994) in the case of \tuca. The
morphology of the horizontal branch is correlated with the cluster
central density and/or concentration (Renzini 1983), in the sense that
denser and more concentrated clusters, like M15 (see Fig.~9.9 above)
tend to have a more extended horizontal branch with a faint blue tail
(Fusi Pecci \etal\ 1993a).  A possible explanation of this phenomenon
may be that the red horizontal branch stars near the centre could be a
progeny of blue stragglers (Fusi Pecci \etal\ 1992).

\vskip 30pt
\noindent
{\sl 9.8 Observational evidence of possible products of stellar
encounters (blue stragglers, high-velocity stars,  X-ray
sources, and pulsars.)}

\vskip 20pt
\noindent
{\sl Blue stragglers.}~~ Blue stragglers were first observed by
Sandage (1953) in the globular cluster M3, as a bunch of stars forming
an upwards extension of the main sequence, above the turn-off, in the
usual color-magnitude diagram.  During the following 3 decades, blue
stragglers were discovered, although at a rather slow pace, among the
halo field stars, in young and old open clusters, as well as in
globular clusters.  A resurgence of interest in the search for blue
stragglers in globular clusters was initiated by Nemec \& Harris
(1987), who, using CCDs and software for photometry in crowded fields
(cf. \S6.1), discovered blue stragglers in NGC~5466. Among others, two
important studies are by Auri\`ere \etal\ (1990), who observed blue
stragglers in the dense collapsed core of NGC~6397, and by Mateo
\etal\ (1990), who discovered three eclipsing binaries, with periods
between 0.298 and 0.511 day, among the nine variable blue stragglers
in NGC~5466.  But it is essentially since the launch of the HST, whose
spatial resolution allows the easy detection of blue stragglers even
in the crowded cores of globular clusters, that there has been a
flurry of discovery papers, starting with Paresce \etal\ (1991), who
observed blue stragglers in the core of \tuca.

Every appropriate search in any globular cluster has unveiled blue
stragglers: they are ubiquitous.  Any new search unveils more blue
stragglers (e.g., Burgarella \etal\ 1995 in M3).  Catalogs of blue
stragglers in globular clusters have been published by Fusi Pecci
\etal\ (1992, 1993b), and lists of globular clusters with blue
stragglers are given in Sarajedini (1993) and Ferraro \etal\ (1995a).
There are now more than 800 blue stragglers known in more than 30
globular clusters.  Unfortunately, statistics about blue stragglers
are difficult to extract from the data which vary strongly from
cluster to cluster: photometric filters (from $UV$ to $IR$), location
in the cluster with respect to its centre, area surveyed, limiting
magnitudes, are as different as they can be, given the intrinsic
differences between ground-based and HST data.

It is now convincingly demonstrated, from radial cumulative
distributions, that blue stragglers are more centrally concentrated
than the other stars of same magnitudes (see Fig.~9.8 above).  This
effect was first observed in NGC~5466 by Nemec \& Harris (1987), and
subsequently in numerous other clusters (e.g., Lauzeral \etal\ 1993;
see Bailyn 1995 for a review).  The central concentration of the blue
stragglers is considered as a consequence of mass segregation, since
the central relaxation time is always significantly smaller than the
lifetime of stars of masses in the range 1.0-1.5 \msun.  Consequently,
when present in a cluster, such stars should be more centrally
concentrated than the most luminous stars -- giants and subgiants --
whose masses are about 0.8 \msun.  Nemec \& Harris (1987) derived, by
comparison with multi-mass King models, a mean mass of 1.3 \pmm\ 0.3
\msun\ for the blue stragglers in NGC~5466, in agreement with what
would be expected from their position in the color-magnitude
diagram. There is, however, the noticeable exception of M3, where
there is an excess of blue stragglers in the inner and outer regions,
and a lack of blue stragglers at intermediate radii (Ferraro \etal\
1993, Bolte \etal\ 1993, Guhathakurta \etal\ 1994).  This could be due
to the existence of two different populations of blue stragglers
within the same cluster or to segregation effects in the production
and/or survival of blue stragglers (see Davies \etal\ 1994, Sigurdsson
\etal\ 1994).

It is clear, from the paper by Mateo \etal\ (1990), that blue
stragglers represent a very heterogeneous family: e.g., in NGC~5466, a
fraction of the blue stragglers are variable in luminosity, some being
pulsating stars and others being eclipsing binaries of W~UMa and Algol
types.  Consequently, more than one scenario may be at work in order
to provide this diversity (Livio 1993).  The numerous different models
can be divided into two groups: (i) models involving single stars and
(ii) those involving binaries (Livio 1993, Stryker 1993, Ouellette \&
Pritchet 1996).  Hereafter we mention only the most plausible
scenarios:

\noindent
{\bf ~~~(i-a)} Multiple bursts of star formation: this scenario may be
at work in young populations ($\tau$ \lsim\ 10$^8$~yr) for which
isochrones in a color-magnitude diagram reveal gaps which could
indicate that stars with masses $M$ $\geq$ 5 \msun\ were formed in a
more recent burst of star formation (Eggen \& Iben 1988).  In older
open and globular clusters, however, such a delayed formation scenario
would require implausibly large quantities of gas long after the first
generation of stars.

\noindent
{\bf ~~~(i-b)} Internal mixing: Wheeler (1979) suggested internal
mixing as a mechanism to extend the main sequence lifetime of stars.
The reason for mixing is not clear, although rotation and magnetic
fields have been mentioned (Maeder 1987).  From preliminary results
from stellar evolutionary codes including rotation, the lifetime of a
main sequence star could be significantly increased (Maeder \& Meynet
pers. comm.). Tidal interaction may be another way to induce mixing,
although it is related to scenarios involving binaries instead of
single stars. An interesting method to test for the mixing hypothesis
through lithium abundance has been suggested by Pritchet \& Glaspey
(1991).

\noindent
{\bf ~~~(ii-a)} Mass transfer in binaries without coalescence: the blue
stragglers would increase their masses via mass transfer in close
binaries (McCrea 1964, van den Heuvel 1994).  A clear prediction of
this scenario is that all blue stragglers should be in binaries, in
clear contradiction with observations (e.g., Milone \& Latham 1992).

\noindent
{\bf ~~~(ii-b)} Coalescence in binaries: the suggestion that some blue
stragglers are coalesced binaries is due to Zinn \& Searle
(1976). This would be the end result of contact binaries (van den
Heuvel 1994) or evolution through a common envelope phase (Meyer \&
Meyer-Hofmeister 1980).  In the case of NGC~5466 (Mateo \etal\ 1990),
comparison between the numbers of close binaries which are blue
stragglers and the expected numbers, based on the time scale required
for contact binaries to merge and the lifetime of blue stragglers,
leads to the conclusion that not all blue stragglers in NGC~5466
formed by coalescence.  See Livio (1993) for a summary of a few
observational facts leading to this conclusion.

\noindent
{\bf ~~~(ii-c)} Mergers during dynamical interactions (encounters and
collisions): the improvements during the last two decades in the
understanding of the dynamical evolution of globular clusters have
emphasized the essential role of tidal captures and collisions in the
core of these stellar systems (cf. \S\S9.4 and 9.5 above).  On the
basis of these mechanisms, Krolik (1983) predicted that globular
clusters should contain substantial numbers of close binaries, contact
binaries, and blue stragglers, with their origins in encounters or
collisions.  Mergers induced by collisions have been studies by Benz
\& Hills (1987), Leonard (1989), Leonard \& Fahlman (1991), Leonard \&
Linnell (1992), Lombardi \etal\ (1995), and Leonard \& Livio (1995),
among others (cf. Fig.~9.7).  There is still ambiguity about whether
blue stragglers are single or double stars simply because of the
possibility that some of them have merged, although it is clear that
the abundance of binaries among blue stragglers is unusually high
(Mateo 1996).  Ambiguity is also present in the prediction of the
rotational speed of a merger product: Leonard \& Livio (1995) find
that rapid rotation is not a signature of a collisionally merged blue
straggler.  Equally ambiguous is the degree of mixing, since blue
stragglers formed by direct stellar collisions are not necessarily
fully mixed and not expected to have anomalously high helium
abundances in their envelopes, or to have their cores replenished with
fresh hydrogen fuel (Lombardi \etal\ 1995, Procter \etal\ 1996).  When
applied to the Yale Rotating Evolution Code in order to explain the
six central bright blue stragglers in the core of NGC~6397, these
models predict that the collision products must be either more than
twice the turn-off mass or mixed by some process subsequent to the
initial collision and merger (Sills \etal\ 1996; see also Bailyn \&
Pinsonneault 1995).

There is a body of evidence which favor the idea that binaries, via
various mechanisms (viz., interaction, capture, coalescence, merging),
are related to the origin of some blue stragglers in galactic globular
clusters, although, from both observational and theoretical points of
view, the picture is still vague.  Fusi Pecci \etal\ (1992) and
Ferraro \etal\ (1995a) attempted to extract synthetic information from
the currently available surveys of blue stragglers in all observed
clusters.  E.g., in loose clusters the number of blue stragglers
detected so far seems to increase almost linearly with the amount of
sampled light, while the trend changes abruptly for clusters having
intermediate and high concentrations.  The fact that highly
concentrated globular clusters have far fewer blue stragglers per unit
of luminosity than loose globulars may be simply the consequence of
the greater difficulty in detecting them in dense cores (Ferraro
\etal\ 1995a, Kaluzny \etal\ 1996c).  If it is a genuine effect, it may
be that blue stragglers in loose clusters originate from primordial
binaries while those in high density clusters are produced by stellar
interactions (Bailyn 1992; Bailyn \& Pinsonneault 1995; Ferraro \etal\
1995a).

\vskip 20pt
\noindent
{\sl High-velocity stars.}~~ Two independent studies, based on radial
velocities of individual stars in globular clusters, have discovered
stars with unexpectedly high velocities.

Gunn \& Griffin (1979) were puzzled by two stars that they called
``interlopers''.  These are two high-velocity stars (viz. von Zeipel
764 and 911) located in the core of M3 $\equiv$ NGC~5272, both about
20\arcs\ from the centre.  They have radial velocities relative to the
cluster of +17.0 \kms\ and --22.9 \kms, corresponding to 3.5 and 4.5
times the velocity dispersion in the core, which is \sigp(core) = 4.9
\kms.  These radial velocities are still close enough to the mean
radial velocity of the cluster to carry a strong implication of
membership, since the cluster velocity is \Vr\ $\simeq$ --147 \kms,
high enough to make contamination by field stars very unlikely.  In a
similar way, Meylan \etal\ (1991a) discovered two high-velocity stars
in the core of the globular cluster \tuca. Located respectively at
about 3\arcs\ and 38\arcs\ from the centre, they have radial
velocities relative to the cluster of --36.7 \kms\ and +32.4 \kms,
corresponding to 4.0 and 3.6 times the core velocity dispersion of
\sigp(core) = 9.1 \kms.  The 1.5-yr time baseline during repeated
observations and the constancy of the radial velocity values indicate
that neither of these two stars is a binary or a pulsating star.
Unfortunately, the relatively low mean radial velocity of \tuca\ (\Vr\
$\simeq$ --19 \kms) does not allow an immediate discrimination between
field stars and members of the cluster.  But the positions of these
two stars in the color-magnitude diagram and the rather high galactic
latitude of \tuca\ (b = --44\deg) both argue for membership.  The
simplest way to eliminate the remaining tiny doubt about membership of
the two stars is to obtain high-resolution spectroscopic observations
and deduce the luminosity classes of these two objects.

A plausible mechanism to explain these interlopers is ejection from
the core by the recoil from an encounter between a single star and a
binary, or between two binary stars.  One problem with the ejection
mechanism is that most of the stars ejected will be moving across our
line of sight and so will not be noticed.  Thus even a few observed
high-velocity stars imply an uncomfortably large population of stars
on radial orbits (Sigurdsson 1991, Phinney \& Sigurdsson 1991).  A
possible solution is to have the encounter create a giant and Davies
\etal\ (1993) have studied this in detail by simulating encounters
involving a neutron star and a tidal-capture binary, the latter
consisting of a white dwarf and a main-sequence star.  Most exchange
encounters produced a single merged object with the white dwarf and
neutron star engulfed in a common envelope of gas donated by the
main-sequence primary of the original binary.  But a small fraction of
the exchanges caused a merger of the white dwarf and the main-sequence
star, with this object (presumably a giant) and the neutron star being
unbound and having large relative velocities at infinity.

Radial velocity observations of 548 stars within two core radii of the
centre of \tuca\ by Gebhardt \etal\ (1995) have increased the sample
of high-velocity stars from 2 to 8, although decreasing their fraction
(2/50 in Meylan \etal\ 1991, and 8/548 in Gebhardt \etal\ 1995).  With
velocities more than 32 km/s from the cluster mean (about three times
the core velocity dispersion of $\simeq$ 10 km/s), these stars are
moving at close to the cluster escape velocity.  Such velocities raise
the question of membership, but the Galaxy model by Bahcall \& Soneira
(1981) shows that the probability of these stars being foreground halo
objects is very low.  While this new larger sample shows a lower
frequency of high-velocity stars, theoretical studies (e.g., Phinney
\& Sigurdsson 1991) have demonstrated that the number of high-velocity
stars is a powerful tool for probing conditions in the core and
testing the dynamical models.  However, larger samples are needed to
exploit this tool.  Such high-velocity stars should also be detected
through proper motion studies.  A long-term HST program for precise
astrometry in the core of \tuca\ has started and will provide a
complete census of high-velocity stars from their proper motions
(Meylan \etal\ 1996).

\vskip 20pt
\noindent
{\sl X-ray sources.}~~ Some of the X-ray sources in globular clusters
are among the brightest ones in the sky and were easily discovered
with the earliest X-ray facilities.  More recently, the highly
sensitive X-ray satellites EINSTEIN and especially ROSAT allowed the
study of less luminous sources in globular clusters (see, for reviews,
Grindlay 1993, Verbunt 1993, 1996a,b, and Bailyn 1996).

There are twelve bright (\lx\ \gsim\ 10$^{35}$ \ergs) X-ray sources
observed in the galactic globular clusters.  The X-ray bursts seen in
most of these sources provide compelling evidence that such sources
are neutron stars, rather than black holes, accreting matter from a
low-mass ($M$ $<$ 1 \msun) companion filling its Roche lobe (Verbunt
1993).  They are called low-mass X-ray binaries (LMXBs), in contrast
to the high-mass X-ray binaries in which the donor is an O or B
star. Pointed EINSTEIN and ROSAT observations led to the discovery of
one such bright X-ray source in thirty globular clusters of M31
(Trinchieri \& Fabbiano 1991, Magnier 1994).  The most recent
additions were detected by ROSAT in NGC~6652 and Terzan~6 (Verbunt
\etal\ 1995).  ROSAT~High-Resolution-Imager (HRI) positions show that
all bright X-ray sources are in, or close to, the core of the host
globular cluster (Johnston \etal\ 1995a).  

Two factors point towards a dynamical origin for LMXBs in globular
cluster.  First, these sources are highly overabundant in globular
clusters with respect to the galactic disk, and, second, many of the
globular clusters which contain LMXBs have collapsed cores with high
stellar densities.  Consequently, tidal captures (Clark 1975, Fabian
\etal\ 1975) and encounters between a binary and a neutron star (Hills
1976), have been invoked to explain the origin of LMXBs, two dynamical
mechanisms which do not operate in the lower stellar density of the
galactic disk (see Verbunt 1993 for formation and evolution
scenarios).

However, the two known orbital periods of globular cluster LMXBs
point towards the fact that the companion of the neutron stars may
not be a main-sequence star.  The first LMXB is 4U\thinspace 1820--30,
located in NGC~6624,which has an orbital period of about 11 minutes
(Stella \etal\ 1987).  Although the variability is observed only in
X-ray, King \etal\ (1993) have found an UV and visible counterpart of
4U\thinspace 1820--30.  The Roche-lobe-filling companion can only be a
white dwarf, although direct captures of white dwarfs are unlikely
given their small cross section.  The collision between a neutron star
and a red giant, which would lose its envelope and leave its bare
core, may be a solution (Verbunt 1987).  The second LMXB is AC211 in
M15. It has an orbital period of about 17 hours both in X-ray and
optical wavelengths (Ilovaisky \etal\ 1993).  In this case the
secondary must be a subgiant rather than a main-sequence star, given
its high optical luminosity.  The above two cases show that the simple
model of LMXBs, made of a neutron star and a main-sequence star, may
need to be refined.

The same dynamical process which, in globular clusters, creates the
excess of accreting neutron-star binaries with respect to the field
should also produce large numbers of accreting white-dwarf binaries,
called cataclysmic variables (CVs) (see, e.g., Di Stefano \& Rappaport
1994, Livio 1996b).  Cataclysmic variables in the field are known to
be X-ray sources.

Observations with the ROSAT~HRI have resolved the core of
several galactic globular clusters.  They show multiple faint sources
in, e.g., NGC~6397, NGC~6752, and \tuca\ (Cool \etal\ 1993, Johnston
\etal\ 1994, Hasinger \etal\ 1994).  A total of about 30 dim sources,
either single or multiple, have been detected in or close to the cores
of 18 galactic globular clusters, with luminosities in the range
10$^{31}$ \lsim\ \lx\ \lsim\ 10$^{34}$ \ergs\ (Johnston \& Verbunt
1996).

Repeated observations of the core of \tuca\ show that the dim sources
are highly variable.  In Fig.~9.10, of the four sources detected
in April 1992, only two are again detected in April 1993, together
with one new source (Hasinger \etal\ 1994), and one of the sources
missing in April 1993 appears again in December 1994 (Verbunt 1996b).
The absolute positional accuracy of the ROSAT~HRI is about 5\arcs, a
value which precludes certain identification of any of the ROSAT
sources with either the single EINSTEIN X-ray source (Hertz \&
Grindlay 1983) or with any of the UV variable stars (Auri\`ere \etal\
1989 from the ground; Paresce \etal\ 1992, Paresce \& De Marchi 1994,
and Meylan \etal\ 1996 with HST).  

\vskip 6truecm
\item\item{\sevenrm{{\bf Fig.~9.10.} \hskip 2mm Three different 
ROSAT~HRI observations of the core of \tuca, separated by more than
one year.  The area is 100\arcs\ $\times$ 100\arcs.  All X-ray
detected photons are displayed; circles in the left and middle panels
encircle photons of the four sources detected in the first observation
(left panel) (from Verbunt 1996b Fig.~3).
}}
\vskip 0.5truecm

Contrary to the bright (\lx\ \gsim\ 10$^{35}$ \ergs) X-ray sources,
the real nature of the low-luminosity (\lx\ \lsim\ 10$^{35}$ \ergs)
X-ray sources, first observed with the EINSTEIN satellite by Hertz \&
Grindlay (1983), is still unknown.  They have been suggested to be
cataclysmic variables by Hertz \& Grindlay (1983) and Grindlay \etal\
(1984).  Unfortunately, unambiguous detections of CVs in globular
clusters have proven to be extremely difficult (see, e.g., Shara
\etal\ 1994, 1995).  It is only with the high spatial resolution of
HST that candidates have been found: the dwarf nova outburst in \tuca\
(Paresce \& De Marchi 1994) and the optical counterpart for the
historical nova in M80 (Shara \& Drissen 1995).  Grindlay \etal\
(1995) report, from observation with HST, the first spectra of three
stars, well below the main-sequence turn-off, located near the centre
of the dense collapsed globular cluster NGC~6397.  These spectra
confirm the suspicion from photometry with HST that these three stars
may be the long-sought cataclysmic variables in globular clusters.  If
so, they are likely to be the counterparts of some of the five dim
X-ray sources observed by ROSAT in this cluster (Cool \etal\ 1993,
1995, Grindlay \etal\ 1995).

However, in addition to the possibility of being CVs, and partly on
the basis of their luminosity distribution, it has also been suggested
that the low-luminosity X-ray sources are (i) soft X-ray transients,
i.e., neutron stars accreting mass from a companion, but at a low rate
(Verbunt \etal\ 1984), (ii) conglomerates of RS~CVn binaries or single
binaries (Bailyn \etal\ 1990, Verbunt \etal\ 1993), or (iii) radio
pulsars (Verbunt \& Johnston 1996).

Fig.~9.11 displays the X-ray luminosity distributions (\lx\ between
0.5 and 2.5 keV) for chromospherically active binaries (RS~CVn),
non-magnetic cataclysmic variables (CV), recycled millisecond radio
pulsars (ms PSR), soft X-ray transients in the galactic disk (SXT),
and dim X-ray sources in globular clusters (Glob. Cl.) (Verbunt \etal\
1994 and Johnston \etal\ 1995b).  It is conspicuous that the
distribution of the dim X-ray sources in globular clusters (Glob. Cl.)
overlaps with the distribution of soft X-ray transients in the
galactic disk (SXT) and with the bright end of the distribution of
cataclysmic variables in the galactic disk (CV).  The fact that some
of the dim X-ray sources in globular clusters, which were previously
detected as single sources, appear now to be multiple, has been taken
into account (Verbunt 1996b).  E.g., the sources found by Hasinger
\etal\ (1994) in the core of \tuca\ (see Fig.~9.10)  have
luminosities \lx\ \gsim\ 10$^{33}$ \ergs, i.e., higher than those
observed with ROSAT for cataclysmic variables but compatible with
those observed for soft X-ray transients in quiescence.  Given the
status of our present knowledge, it is reasonable to think that the
dim X-ray sources in globular clusters with \lx\ \gsim\ 10$^{33}$
\ergs\ may be soft X-ray transients in their low state, while the dim
X-ray sources with \lx\ \lsim\ 10$^{33}$ \ergs\ may be cataclysmic
variables created by dynamical processes in the dense cores of
collapsed clusters (see also Livio 1994 and van den Heuvel 1994 for
interesting reviews).

\vskip 6.5truecm
\item\item{\sevenrm{{\bf Fig.~9.11.} \hskip 2mm X-ray luminosity
distributions for chromospherically active binaries (RS CVn),
non-magnetic cataclysmic variables in the galactic disk (CV), recycled
millisecond radio pulsars (ms PSR), soft X-ray transients in the
galactic disk (SXT), and dim X-ray sources in globular clusters
(Glob. Cl.) (from Verbunt 1996b Fig.~4)
}}
\vskip 0.5truecm

\vskip 20pt
\noindent
{\sl Pulsars.}~~ Since the discovery by Hulse \& Taylor (1975) of the
first binary radio pulsars (see Taylor 1994 for a review), these
astronomical rotational clocks and their companions have been
intensively used in fields as different as relativistic gravity,
nuclear equations of state, neutron star magnetospheres and masses,
planet formation, and the dynamical evolution of globular clusters
(Phinney 1992).  See also Phinney (1993), Verbunt (1993), Phinney \&
Kulkarni (1994), and Phinney (1996) for recent reviews on pulsars in
globular clusters.

Most of the roughly 700 pulsars discovered in our galaxy are single
neutrons stars.  Observational evidence points towards their origin in
supernova explosions, by the collapse of the cores of massive stars,
with initial masses $M_i$ \gsim\ $M_{core}$ $\simeq$ 8 \msun.  Such
massive stars have not existed in galactic globular clusters for more
than 10 Gyr, although pulsars much younger than this age are found in
these stellar systems.  Observations have unveiled the presence of
more than 30 radio pulsars in galactic globular clusters (see Table~1
in Phinney 1996), out of which eight are members of M15 and eleven are
members of \tuca, two high-concentration clusters.  The closest
galactic analogues to the globular cluster pulsars are the 34 binary
pulsars and the 27 millisecond pulsars, which have pulse periods
shorter than 10 ms: these have a distribution of pulse period and
spin-down rate very different from that of the bulk of field pulsars,
and very similar to that for globular cluster pulsars (see Phinney
1996 Fig.~1).  The high rate of occurrence of pulsars in globular
clusters is conspicuous when it is considered that globular clusters
contain only about 0.05\% of the mass of the galaxy.  Pulsars, in a
way similar to LMXBs, originate mostly in high stellar-density
environments.

Soon after the discovery by Backer \etal\ (1982) of the first
millisecond binary pulsar in the field, and by McKenna \& Lyne (1988)
of the first millisecond binary pulsar in a globular cluster, it was
suggested that they resulted from the spin-up of an old neutron star
by accretion of matter from a companion star as it overflowed its
Roche lobe during its giant phase (Smarr \& Blandford 1976).  It is
the same process witnessed in low-mass X-ray binaries, and parallels
have been drawn between (i) high-mass X-ray binaries and high-mass
binary radio pulsars, and (ii) low-mass X-ray binaries and low-mass
binary radio pulsars.  See, e.g., Table~1 in Verbunt (1993) for
similar properties between members of these four families (Kulkarni
\etal\ 1990; see Phinney \& Kulkarni 1994, and Lyne 1995 for reviews).

Since potential donors in globular clusters have masses smaller than
about 0.8 \msun, globular cluster pulsars are only of the low-mass
type.  They are interpreted as old neutron stars or white dwarfs
recycled into pulsars by accretion from a companion in a binary
system, i.e., they are the descendants of low-mass X-ray binaries.
The name ``recycled pulsars'' designates the members of the class of
low magnetic field strength, short period, and frequently binary
pulsars.  It is worth mentioning that, given the typical distance of a
globular cluster, $\sim$ 5 kpc, only the brightest pulsars have been
detected.  Consequently, any calculations of the rate of formation and
total number of recycled pulsars in a given cluster require
considerable extrapolation.

The formation and properties of cluster pulsars are inextricably
linked to the dynamical histories of their host globular clusters. Our
understanding of cluster evolution has recently benefited from the
discovery of primordial and newly-formed binaries in globular clusters
(Hut \etal\ 1992a) and from improved computer simulations (Murphy
\etal\ 1990, Gao \etal\ 1991, Heggie \& Aarseth 1992, and Sigurdsson
\& Phinney 1995).

Globular clusters must be very efficient in recycling their old
pulsars, i.e., neutron stars.  There are three types of binaries
containing neutron stars: (i) primordial binaries, which survived the
supernova explosion (they are the only important source of recycled
pulsars in the galactic field; (ii) tidal-capture binaries (made
possible because of the high stellar density in globulars) in which a
neutron star captured or disrupted a non-degenerate star during a
close encounter occurring during a 2-body fly-by or an interaction
between a single or binary star with another binary; (iii) exchange
binaries (also made possible because of the high stellar density in
globulars) in which a neutron star has been substituted for one of the
original members of a binary which did not initially contain any
neutron stars (Phinney 1996).  A majority of globular cluster pulsars
are single, at variance with galactic field recycled pulsars, since
they may have lost their companions through a variety of scenarios:
exchange (Phinney \& Sigurdsson 1991, Sigurdsson \& Phinney 1993),
giant capture (Romani \etal\ 1987, Rappaport \etal\ 1989), main
sequence collisions (Krolik \etal\ 1984), and evaporation (Ruderman
\etal\ 1989).

Much of the above description of primordial binaries and tidal
captures applies if the neutron stars are replaced by white dwarfs.
If the donor transfers enough mass to the accreting white dwarf,
bringing it above the Chandrasekhar limit, the latter may transmute
into a neutron star by ``Accretion Induced Collapse'' (Canal \etal\
1990, Nomoto \& Kondo 1991).  One advantage of this scenario is its
capability of producing neutron stars with or without recoil
velocities and with or without weak magnetic fields.  The absence of
recoil means that very little mass is lost during the supernova event,
allowing all binary systems to survive this delicate phase.  This has
prompted Bailyn \& Grindlay (1990) to suggest ``Accretion Induced
Collapse'' as an efficient way to produce neutron stars and hence
pulsars in globular clusters.

The detailed physics of all the formation mechanisms briefly described
above is not well understood.  Even general points remain unknown.
E.g., in the case of ``Accretion Induced Collapse'', does the white
dwarf explode or implode to form a millisecond pulsar?  Most models
are too vague in their predictions and most observations are too scant
to allow meaningful comparison with theory.  Recently, the timing
measurements of three cluster pulsars show that they are young,
single, with a strong magnetic field, and clearly members of galactic
globular clusters; all three objects have properties typical of
pulsars in the galactic field, and the origin of such apparently young
objects in very old stellar systems is not understood (Lyne \etal\
1996).

It is worth mentioning that the first of the eight pulsars discovered
in M15, viz. PSR~2127+11A, has a period of 0.111 second and provided
quite a surprise. It was the first of the 500 known pulsars at that
time to have a negative period derivative (Wolszczan \etal\ 1989).  As
this pulsar is clearly not in a binary system, its change in period is
attributed to the acceleration of the pulsar towards the Earth as it
moves through the gravitational potential of the cluster.  Another
such pulsar, viz. PSR~2127+11D, has been discovered in M15.  Both have
\pdot/$P$ = --2 $\times$ 10$^{-16}$ s$^{-1}$, a value which has been
used to get otherwise unobtainable information on the density and the
masses of the stellar remnants in the core of this globular cluster
(Phinney 1992).

\vskip 30pt
\noindent
{\bf 10. Late phases of evolution and disruption} 

\vskip 30pt
\noindent
{\sl 10.1 Gravothermal oscillations; post-collapse evolution}

\vskip 20pt
\noindent
At one time, a review of the dynamical evolution of globular star
clusters might have ended after \S9.1.  It was not at all certain that
a cluster could survive beyond the end of core collapse, and indeed
empirical studies of the distribution of central relaxation times of
galactic globular clusters (Lightman \etal\ 1977) were consistent with
the idea that clusters somehow suddenly disappeared.  Thus, for a long
period in the history of cluster studies many experts doubted whether
the study of post-collapse clusters had any relevance to the
interpretation of observations.  Modelling of M15 (see \S\S9.2 and
9.3) forced a change of attitude, and now a significant proportion of
clusters are interpreted as exhibiting the structural characteristics
of post-core collapse evolution (cf. \S9.2).  Even before this was
realised, however, it was already becoming clear that the statistics
of core parameters could {\sl not} in fact be understood if it was
assumed that the entire present population of galactic globular
clusters was still undergoing core collapse (Cohn \& Hut 1984).

Because of the role of stellar collisions, and other factors, the
theoretical behavior of a star cluster after core collapse is subject
to some uncertainty, but by now several simplified models exist.  All
of them depend on providing a flow of energy from the central parts of
the cluster, and they differ essentially only in the main physical
mechanism which is assumed to be responsible for this.  Several
processes have been in favor at one time or another, including
different kinds of binaries (primordial, tidal and three-body), a
massive central black hole (Shapiro 1977) and mass loss from evolution
of merger products.  At present, probably the favored mechanism is
that provided by primordial binaries, a relatively old idea which was
revived in recent times by Goodman \& Hut (1989).  The main
uncertainty is the way in which the dynamical behavior of binaries is
affected by finite-size effects.  Despite such gross uncertainties,
post-collapse evolution is worth studying in some detail because, as
pointed out by H\'enon (1975), many aspects of the evolution appear to
be independent of details of the mechanism of energy generation, and
we concentrate on these.

In an {\sl isolated} system the outpouring of energy from one of these
mechanisms leads to an overall expansion of the cluster, first
modelled by H\'enon (1965).  Relatively little mass is lost on the
expansion time scale, and the size of the system varies nearly as
$r_h\propto (t-t_0)^{2/3}$, where $t_0$ is a constant.  This follows
from equating the expansion time scale to the half-mass relaxation
time (Eq.~7.2), if the mass is constant; Goodman (1983c) has
described models in which this is relaxed slightly.  When the cluster
is tidally limited, the outpouring of energy drives mass across the
tidal boundary, and the half-mass radius decreases to maintain
constant mean density (the usual condition for tidally limited stellar
systems).  Thus $r_h\propto (t_0 - t)^{1/3}$, where $t_0$ is a
different constant, and the system contracts.  Mass is lost nearly
linearly with time (H\'enon 1961).

All these results stem from simple theoretical ideas, and are most
easily developed for systems of stars of equal mass.  Detailed
numerical investigations have been used to explore more realistic
models, with suitable assumptions about the mechanism of energy
generation, where appropriate. $N$-body models consisting of point
masses of equal mass, in which the mechanism is binary formation by
three-body encounters, confirm the post-collapse expansion predicted
by simplified models (e.g., Giersz \& Heggie 1994).  The presence of a
spectrum of masses does not appear to complicate the evolution, even
though continued mass segregation might be expected to occur. This is
shown by recent $N$-body models (Giersz \& Heggie 1996a; see also
Inagaki 1986c).

Even when the variety of mechanisms considered by Stod\'o\l kiewicz
(1985) is included, one still observes the same linear dependence of
total mass on time, as predicted by H\'enon. Since this result
determines the lifetime of a cluster, it is worth recording the
numerical value.  In the case of equal masses (H\'enon) the lifetime
equals approximately $22.4$ current half-mass relaxation times.  For
the models of Stod\'o\l kiewicz (1982), which have unequal masses and
many other features, the corresponding value is about $8.7$.  There
are also considerable differences in the structure.  For H\'enon's
model the ratio of the tidal and half-mass radii is
$r_t/r_h\simeq6.9$, whereas for the models of Stod\'o\l kiewicz the
corresponding number is nearer $2.4$.  Similar values (around $2.5$)
are found in $N$-body models in post-collapse evolution by Giersz \&
Heggie (1996b), but such values are greatly at variance with typical
observational determinations for post-collapse clusters. For instance
Meylan \& Mayor (1991) found $r_t/r_h\simeq 12$ for most of their
models of NGC~6397.  It is possible that this is a manifestation of
the dependence of the tidal radius on the orbital phase, because all
of the theoretical values assume that the tidal environment is static
(except for disk shocking, in the case of the models of Stod\'o\l
kiewicz).

When it comes to the evolution of the core, the nature of the
mechanism for generating energy is all-important.  If we assume that
there are no primordial binaries and that the post-collapse evolution
is powered by binaries formed in three-body interactions, then
extremely high core densities are required, just as at the close of
core collapse (if this is arrested by three-body binaries).  These are
circumstances in which the central parts of the cluster may be
gravothermally unstable.  Now, however, the presence of binaries
prevents indefinite collapse of the core, but the emission of energy
from their evolution can cause a drop in the temperature of the core,
which drives the gravothermal instability in reverse, i.e., it drives
an expansion of the core.  What happens next, at least for systems
with more than a few thousand stars (Goodman 1987, Heggie \& Ramamani
1989, Breeden \etal\ 1994), is a complicated succession of collapses
and expansions, called ``gravothermal oscillations'' by their
discoverers (Sugimoto \& Bettwieser 1983, Bettwieser \& Sugimoto 1984;
see also Fall \& Malkan 1978 for a curious precursor of this
discovery, and Heggie 1994 for a recent review).  The oscillations are
superimposed on an overall expansion which approximately follows
simple theoretical relationships such as those summarised above
(Bettwieser \& Fritze 1984).  After deep core collapse, however, the
early expansion should follow a somewhat different scaling described
by Inagaki \& Lynden-Bell (1983).  Earlier studies of post-collapse
evolution (Heggie 1984, 1985) missed the oscillations for numerical
reasons.

\vskip 12truecm
\item\item{\sevenrm{{\bf Fig.~10.1.} \hskip 2mm  Core collapse in 
systems with equal masses (from Makino 1996b Fig.~1).  The logarithm
of the central density is plotted against time, scaled in proportion
to the initial half-mass relaxation time.  The successive curves,
which correspond to different values of $N$, have been displaced
vertically for clarity.  }}
\vskip 0.5truecm

Quite apart from their relevance in nature (see below), these
oscillations are interesting in their own right, as an example of
chaotic dynamics.  From this point of view they have been studied by
Allen \& Heggie (1992), Breeden \& Packard (1994), and Breeden \& Cohn
(1995).

Whether these investigations imply that such oscillations should occur
in nature is not clear, for a variety of reasons. For several years
after their discovery, the oscillations were studied almost entirely
with the aid of simplified models, i.e., gas models and Fokker-Planck
models (e.g., Hut \etal\ 1989, Cohn \etal\ 1989, Spurzem \& Louis
1993), and it has been argued that the subtle thermal effects which
are responsible are masked, in real systems, by fluctuations (Inagaki
1986b, 1988). Nevertheless, growing evidence from $N$-body simulations
was already pointing in the opposite direction (Bettwieser \& Sugimoto
1985, Makino \etal\ 1986, Makino \& Sugimoto 1987, Heggie 1989, Makino
1989, Heggie \etal\ 1994).  In 1995 the genuine occurrence of
gravothermal oscillations in $N$-body systems was spectacularly
demonstrated by Makino (Makino 1996a,b; see Fig. 10.1).  These results
confirm that the nature of post-collapse evolution in $N$-body systems
is far more stochastic than in the simplified continuum models on
which so much of our understanding rests at present.  It has been
known for a long time that the formation and evolution of individual
binaries in small $N$-body systems makes the evolution of the core
quite erratic after core collapse (e.g., Sugimoto 1985, McMillan
1986b), and one might have thought that the effects of individual
binaries would have been of less significance in much larger systems.
But now it is known that gravothermal oscillations make the evolution
of such large systems equally erratic.  Indeed the interaction between
these two processes had already been studied by Takahashi \& Inagaki
(1991), in a paper which develops an earlier model by Inagaki \& Hut
(1988); cf. also Spurzem \& Giersz (1996).

Even with simplified models it is known that the oscillations tend to
be suppressed by the presence of a mass spectrum (Murphy \etal\ 1990;
see also Bettwieser 1985a).  The main source of doubt about the
significance of these oscillations, however, is concerned with the
mechanism of energy generation.  Though oscillations also occur if
this is caused by tidal-capture binaries (Cohn \etal\ 1986, Cohn
1988), they may be suppressed by the presence of primordial
binaries. These have the effect of preventing the phases of extremely
high central density which are necessary in well developed
oscillations, just as collapse of the core is ended at much lower
densities if primordial binaries are present (McMillan \etal\ 1990,
1991, Heggie \& Aarseth 1992).  On the other hand, the steady
exhaustion of primordial binaries as an energy source, caused by their
destruction in mutual interactions, gradually erodes their
effectiveness. The Fokker-Planck models of Gao \etal\ (1991) suggest
that gravothermal oscillations do eventually occur even if the initial
abundance of primordial binaries is as high as 20\%.  These models are
approximate in some important ways, however, and do not include a
spectrum of masses, but the possibility that clusters with primordial
binaries may exhibit oscillations late in the post-collapse phase
cannot be ruled out.  On the other hand, if the post-collapse
evolution is assumed to be steady, the results of Vesperini \&
Chernoff (1994) can be used to estimate the likely size of the core.
The general theoretical issues involved in core size are considered by
Hut (1996b).

Whether or not post-collapse oscillations occur is not simply an
academic question, as it is directly related to the observable
structure of a post-collapse cluster, especially with regard to the
presence or absence of a resolved core.  If oscillations occur, then
the cluster is likely to be observed close to an expansion phase, when
the core may be large enough to be resolved, whereas a much smaller
core is expected if the post-collapse evolution is steady (Bettwieser
1985b, Grabhorn \etal\ 1992).  These questions are also involved in
the interpretation of the statistics of core parameters.

As already mentioned, another possible mechanism for powering
post-collapse expansion is runaway coalescence by two-body
interactions (see Lee 1987b).  As Goodman (1989) pointed out, the
simplest resulting scenario requires a core luminosity which is quite
inconsistent with observations, and some more elaborate scenario, such
as one involving gravothermal oscillations, is required.

\vskip 30pt
\noindent
{\sl 10.2 Disruption} 

\vskip 20pt
\noindent
Globular clusters are subject to several disruptive processes, both
internal and external.  In fact this distinction is not quite
clear-cut, as the rate of escape by evaporation depends on the tidal
field.  Still, among the main internal disruptive processes we include
evaporation, either by two-body interactions or those involving
binaries (cf. \S7.3).  A second internal process, of importance mainly
for young clusters, is mass loss from stellar evolution (cf. \S5.5).
External influences include time-dependent tidal fields, among which
are disk and bulge shocking (cf. \S7.4), and interactions with giant
molecular clouds (for which a main reference is still the classic
paper Spitzer 1957).  Another external destructive mechanism is
dynamical friction, which acts on the entire cluster as it ploughs
through the Galaxy (Tremaine \etal\ 1975, Tremaine 1976).  Not all
mechanisms affecting the population of galactic globular clusters are
destructive: successive capture from satellite galaxies may well be a
complicating factor (e.g., van den Bergh 1993a,b,c, Fusi Pecci \etal\
1995).

Many papers (in the sections which are referred to in the above
paragraph) tell us about the effects of these processes on a single
cluster, but much is to be learned by analysing the way in which their
effect alters the system of galactic globular clusters as a whole.
The aim of Fall \& Rees (1977; cf. also Fall \& Rees 1988) was to show
that the range of cluster masses could be accounted for, given a
suitable initial correlation of cluster mass and density, if the
population evolved by cluster-cluster interactions, disk shocking and
(internal) evaporation.  Dynamical friction was, they concluded,
relatively unimportant, though its efficiency depends on the nature of
the galactic potential (Pesce \etal\ 1992) and it will be important
for massive clusters at small radii (Surdin 1978, Capriotti \etal\
1996).  The effect of mass loss by internal stellar evolution was not
considered until relatively recently (cf. \S5.5), but was included (with
several other processes) in the work of Chernoff \& Shapiro (1987).
They studied the effect of these processes by assuming that the
evolution of individual clusters took place along the King sequence.
It has recently been found (Fukushige \& Heggie 1995) that previous
estimates of the lifetimes of globular clusters, which had been based
on Fokker-Planck modelling, may be underestimates.  The error may be
as much as a factor of ten in the case of systems which are destroyed
quickly.

\vskip 11.5truecm
\item\item{\sevenrm{{\bf Fig.~10.2.} \hskip 2mm The effect of several 
destructive mechanisms on the distribution of galactic globular
clusters (from Gnedin \& Ostriker 1996 Fig.~20)a.  Those clusters for
which, at the stated galactocentric radius, the combined theoretical
destruction time scale exceeds a Hubble time, are predicted to lie
within the corresponding curve.  Data for 119 galactic globular
clusters are plotted, with symbols determined by the mean
galactocentric distance of each cluster, which in turn was estimated
according to a simple kinematical model of the cluster system (``OC
isotropic'').  The other labels indicate the main destructive
mechanisms in each domain of the diagram.  }}
\vskip 0.5truecm

Aguilar \etal\ (1988) gave a more detailed assessment of several of
these processes, including also shocks due to the galactic bulge, but
concentrated more on determining their {\sl current} effect on the
population of galactic globular clusters.  They also showed how the
relative effectiveness of the processes they considered depended on
the orbital characteristics of the clusters.  Okazaki \& Tosa (1995)
have considered the influence of three of the main processes of
dissolution on the {\sl luminosity} function, and it would be
interesting to study the ``fundamental plane'' of cluster properties
(Djorgovski 1995, cf. also \S4.6 of this review, van den Bergh 1994,
Covino \& Pasinetti and Fracassini 1993) from this point of view.
Using the minimum of theory, Hut \& Djorgovski (1992) have estimated
that, in our Galaxy, globular clusters are dying at a rate of about 5
per Gyr.  This is not dissimilar to the theoretical prediction of
Gnedin \& Ostriker (1996) that more than half of the present
population may disappear within the next Hubble time (Fig.~10.2).

Provided that the effects of these known destruction mechanisms are
well understood, they can be used to make inferences about galactic
structure.  For example the likely effects of a galactic bar were
studied by Long \etal\ (1992), while Surdin (1993) has shown how
studies of the galactic globular cluster system from the point of view
of disk shocking might be used to constrain the structure of the disk.
More speculative destructive mechanisms which could be indirectly
studied in this way include hypothetical massive black holes (Wielen
1987, Moore 1993, Charlton \& Laguna 1995, Klessen \& Burkert 1996).
There are other reasons why the study of the dynamical evolution of
globular cluster systems may have far-reaching implications.  For
example, provided that the various mechanisms are well enough
understood, they can be applied to external cluster systems, and may
well help to explain the variation of the specific cluster frequency
with the mass of the parent galaxy (Murali \& Weinberg 1996).

\vskip 30pt
\noindent
{\bf 11. Future directions} 

\vskip 20pt
\noindent
In the realm of modelling, major advances may be expected in the next
few years. While Fokker-Planck models will continue to provide much
information on problems of interest, an increasing role will be played
by $N$-body methods.  At present these suffer from two major
deficiencies, as already mentioned in \S\S8.1 and 9.5, i.e. (i) the
fact that $N$ is still much too small, and (ii) the absence of an
adequate treatment of stellar collisions.

The first problem will {\sl eventually} be solved by advances in
computer speed.  Unfortunately, the computational effort (Hut \etal\
1988) grows with $N$ roughly as $N^\alpha$ with $2<\alpha<3$.  If we
suppose that computing speed roughly doubles each year, then it is
clear that the step from the largest simulation which is feasible at
present on a general-purpose computer ($N \sim 10^4$, Spurzem \&
Aarseth 1996 -- with great effort!) to a sizeable globular cluster ($N
\sim\ 10^6$) could not be taken within the next decade.  The
development of special-purpose hardware, however, is transforming the
picture.  The GRAPE/HARP project, successfully developed at the
University of Tokyo over the last few years (Makino \etal\ 1993,
Makino 1996a,b) now provides the ability to model systems of at least
3 \x\ 10$^4$ stars in a reasonable time.  Within five years it would,
in principle, be straightforward to build a hardware which would
increase this by another order of magnitude.

Striking as such advances are, little can be done with a single model;
some further time will elapse before the computation of such models
becomes routine, and this is necessary if it is desired to investigate
the effects of different parameters on the evolution.  Therefore there
can be no doubt that the other simpler methods mentioned in \S8 will
continue to provide much of our detailed information on the evolution
of star clusters for some years to come.  Here one of the most
promising developments is in Fokker-Planck models which incorporate
aspects of Monte Carlo methods.  Such a method was already used over
10 years ago by Stod\'o\l kiewicz (1985) to produce some of the most
realistic models of globular clusters that have yet been published,
and they included an astonishingly wide range of physical processes.
Their chief restriction was in the small number of ``stars'' that
could be handled at that time, but the subsequent ten years of
developments in general-purpose hardware should make possible a
dramatic improvement (Giersz, pers. comm.)  The cost of increasing the
number of stars by a given factor is considerably smaller than in
direct $N$-body models, and there are several reasons why it is
important to use larger $N$.  For example, it is difficult to study
the evolutionary effects of rare species (e.g., stellar-mass black
holes), because none may be present in a scaled-down model!

The other main obstacle to progress, within the context of $N$-body
models, is the handling of non-point mass effects.  Already much could
be done with existing codes, in which the outcome of a collision is
determined by a simple prescription, such as might be suggested by
results of simulations using smooth particle hydrodynamics (\S9.4).
The next step is to incorporate the SPH within an $N$-body code, so
that the simple prescription is replaced by a detailed modelling of
the particular collision that is occurring.  Indeed small-scale test
calculations of this kind have been carried out (McMillan,
pers. comm.).  Greater difficulties will occur in modelling
interactions between stars in binary systems, at the point where they
exchange mass over long periods of time, and in incorporating the
effects of stellar evolution on this and other processes (Livio 1996a,
Zwart 1996, Hut 1996a).  Aarseth (1996a,b) is making great progress
here.

There are essential points at which observations are needed to supply
more reliable parameters for the $N$-body models.  In relation to the
primordial binaries, what is the distribution of the masses of the
components, and that of the semi-major axes?  These are significant
factors in determining how effective the binaries can be at powering
the evolution of a cluster for its entire life, from birth to
dissolution.  What is the present fraction of binaries?  What is their
spatial distribution? Though much has been learned from radial
velocities of giants (\S9.6), thanks to the advent of multiple-fiber
devices, it is just now becoming possible to extend the
statistics to the upper main sequence (C\^ot\'e \& Fischer 1996),
where binaries of shorter periods will become observable.

The same multiple-fiber devices have provided, during these last few
years, increasingly large samples of stellar radial velocities, but
the next important improvement should come from the observation of
proper motions, whose large potential of dynamical information has not
been exploited yet.  E.g., the data obtained in \cent\ (Reijns \etal\
1993 and Seitzer, pers. comm.) --- proper motions for about 7,000
stars and stellar radial velocities for about 3,500 stars --- will
permit investigation of the 3-D space velocity distribution and
rotation.  The same spectra are also being used to determine
metallicity, to investigate the correlation between metallicity,
radius, and kinematics.  A quantum jump in the understanding of the
internal dynamics of this globular cluster will result from the
interpretation of these data.

Extensive and deep multicolor imaging should be obtained at different
radii from the centre of the clusters, in order to provide a clear and
more detailed view on the influence of dynamics on stellar evolution,
from color-magnitude diagrams and color gradients.

An essential parameter to be supplied for the $N$-body models concerns
the luminosity and mass functions, and especially their lower parts,
possibly lower ends.  E.g., in the case of NGC~6397, proper motions
from HST should soon provide the first clear observation of a globular
mass function close to the hydrogen-burning limit by allowing a clear
distinction between cluster members and field stars (King,
pers. comm.).

\vskip 10pt
The above few points illustrate that theorists are gradually turning
from the rather ``pure'' types of stellar dynamical calculations,
which have tended to dominate the subject in recent decades, to more
realistic simulations, e.g., from equal-mass systems to those with a
mass spectrum, and from isolated systems to ones which are tidally
truncated.  The same is true for observers who, as the instruments
improved, have gradually abandoned the idealised vision of a dormant
swarm of stars, whose members were thought to evolve individually.
Not only are the theoretical dynamical simulations gradually becoming
more and more realistic, but they are increasingly directed to the
questions posed by observations, e.g., the influence of dynamics on
the mass spectrum.

\vskip 30pt
\noindent
{\bf Acknowledgments}

\vskip 20pt
\noindent
DCH thanks E. Vesperini and GM thanks G. Djorgovski and F. Verbunt for
their advice on certain topics. GM gratefully acknowledges the
hospitality of the Aspen Center for Physics, from which this work has
benefited.  Many thanks also to E. Janssen for efficient help in the
preparation of the figures.

\vskip 30pt
\noindent
{\bf References} 

\noindent
\vskip 20pt
{\sevenrm
{\parskip=0pt
\ref{Aaronson M., Schommer R.A., Olszewski E.W., 1984, ApJ, 276, 221}
\ref{Aarseth S.J., 1968, Bull. Astron., 3, 105}
\ref{Aarseth S.J., 1972, in Gravitational $N$-Body Problem, 
	IAU Coll. 10, ed. Lecar M., (Dordrecht: Reidel), p. 88}
\ref{Aarseth S.J., 1977, Rev. Mexicana. Astron. Astrof., 3, 199}
\ref{Aarseth S.J., 1980, in Star Clusters, 
	IAU Symp. 85, ed. Hesser J.E., (Dordrecht: Reidel), p. 325}
\ref{Aarseth S.J., 1985a, in Multiple Time Scales, 
	eds. Brackbill J.U., Cohen B.I., (New York: Academic Press), p. 377}
\ref{Aarseth S.J., 1985b, in Dynamics of Star Clusters, 
	IAU Symp. 113, eds. Goodman J. \& Hut P., (Dordrecht: Reidel), p. 251}
\ref{Aarseth S.J., 1988, Bol. Acad. Nac. Cienc. Cordoba, 58, 189}
\ref{Aarseth S.J., 1992, in Binaries as Tracers of Stellar Formation,
	eds. Duquennoy A., \& Mayor M., 
	(Cambridge: Cambridge Univ. Press), p. 6}
\ref{Aarseth S.J., 1996a, in The Origins, Evolution, and Destinies of 
	Binary Stars in Clusters, eds. Milone G. \& Mermilliod J.-C., 
	ASP Conference Series Vol. 90, (San Francisco: ASP), p. 423}
\ref{Aarseth S.J., 1996b, in  Dynamical Evolution of Star Clusters:
	Confrontation of Theory and Observations, IAU Symp. 174, 
	eds. Hut P. \& Makino J. (Dordrecht: Kluwer), p. 161}
\ref{Aarseth S.J., Bettwieser E., 1986, in  The Use of
	Supercomputers in Stellar Dynamics, eds. Hut P. \& McMillan S.L.W., 
        (Berlin: Springer-Verlag), p. 201}
\ref{Aarseth S.J., Heggie D.C., 1993,  
 	in The Globular Cluster - Galaxy Connection, 
	ASP Conference Series Vol. 48, 
	eds. Smith G.H. \& Brodie J.P., (San Francisco: ASP), p. 701}
\ref{Aarseth S.J., H\'enon M., Wielen R., 1974, A\&A, 37, 183}
\ref{Aarseth S.J., Hills J.G., 1972, A\&A, 21, 255}
\ref{Aarseth S.J., Lecar M., 1975, ARA\&A, 13, 1}
\ref{Aarseth S.J., Lin D.N.C., Papaloizou J.C.B., 1988, ApJ, 324, 288}
\ref{Aarseth S.J., Woolf N.J., 1972, Astrophys. Lett., 12, 159}
\ref{Abraham R.G., van den Bergh S., 1995, ApJ, 438, 218}
\ref{Abt H.A., Levy S.G., 1976, ApJS, 30, 273}
\ref{Adams F.C., Fatuzzo M., 1996, ApJ, 464, 256}
\ref{Agekian T.A., 1958, Soviet. Astron., 2, 22 (Astr. Zh., 35, 26)}
\ref{Agekian T.A., 1959, Astr. Zh., 36, 41}
\ref{Aguilar L.A., Hut P, Ostriker, J.P., 1988, ApJ, 335, 720}
\ref{Aguilar L.A., White S.D.M., 1985, ApJ, 295, 374}
\ref{Ahmad A., Cohen L., 1973, J. Comp. Phys., 12, 389}
\ref{Akiyama K., 1991, Earth, Moon and Planets, 54, 203}
\ref{Akiyama K., Sugimoto D., 1989, PASJ, 41, 991}
\ref{Alexander M.E., Budding E., 1979, A\&A, 73, 227}
\ref{Alladin S.M., Sastry K.S., Potdar A., 1976, Bull. Astr. Soc. India, 4, 85}
\ref{Allen A.J., Papaloizou J., Palmer P.L., 1992, MNRAS, 256, 695}
\ref{Allen F.S., Heggie D.C., 1992, MNRAS, 257, 245}
\ref{Ambartsumian V.A., 1938,  Ann. Leningrad State Univ., 22, 19;
	English translation: Ambartsumian V.A., 1985, 
	in  Dynamics of Star Clusters, 
	IAU Symp. 113, eds. Goodman J. \& Hut P., (Dordrecht: Reidel), p. 521}
\ref{Ambrosiano J., Greengard L., Rokhlin V., 1988, Comput. Phys.
        Comm., 48, 117}
\ref{Anderson J., King I.R., 1996, 
	in  Formation of the Galactic Halo...Inside and Out, 
	ASP Conference Series Vol. 92, 
	eds. Morrison H. \& Sarajedini A., (San Francisco: ASP), p. 257}
\ref{Angeletti L., Capuzzo-Dolcetta R., Giannone P., 1983, A\&A, 121, 183}
\ref{Angeletti L., Dolcetta R., Giannone P., 1980, Ap\&SS, 69, 45}
\ref{Angeletti L., Giannone P., 1976, Mem. Soc. Astron. Ital., 47, 245}
\ref{Angeletti L., Giannone P., 1977a, Ap\&SS, 46, 205}
\ref{Angeletti L., Giannone P., 1977b, Ap\&SS, 50, 311}
\ref{Angeletti L., Giannone P., 1977c, A\&A, 58, 363}
\ref{Angeletti L., Giannone P., 1978, A\&A, 70, 523}
\ref{Angeletti L., Giannone P., 1979, A\&A, 74, 57}
\ref{Angeletti L., Giannone P., 1980, A\&A, 85, 113}
\ref{Angeletti L., Giannone P., 1983, A\&A, 121, 188}
\ref{Angeletti L., Giannone P., 1984, A\&A, 138, 396}
\ref{Anosova J.P., 1986, Ap\&SS, 124, 217}
\ref{Anosova J.P., 1990, Celes. Mech. Dyn. Astron., 48, 357}
\ref{Anosova J.P., 1991, Celes. Mech. Dyn. Astron., 51, 15}
\ref{Anosova J.P., Kirsanov N.O., 1991, Comments on Astrophysics, 15, 283}
\ref{Anosova J.P., Orlov V.V., 1988, in  The Few Body Problem, 
	IAU Coll. 96., ed. Valtonen M.J., (Dordrecht: Kluwer), p. 253}
\ref{Antonov V.A., 1962, Vest. leningr. gos. Univ., 7, 135; English
	translation: Antonov, V.A., 1985, in  Dynamics of Star Clusters, 
	IAU Symp. 113, eds. Goodman J. \& Hut P., (Dordrecht: Reidel), p. 525}
\ref{Appel A.W., 1983, SIAM J. Sci. \& Stat. Comput., 6, 85}
\ref{Applegate J.H., 1986, ApJ, 301, 132}
\ref{Armandroff T.E., 1988, AJ, 96, 588}
\ref{Armandroff T.E., 1989, AJ, 97, 375}
\ref{Armandroff T.E., Olszewski E.W., Pryor C., 1995, AJ, 110, 2131} 
\ref{Armandroff T.E., Zinn, R., 1988, AJ, 96, 92}
\ref{Ashman K.M., 1990, MNRAS, 247, 662}
\ref{Ashman K.M., Zepf S.E., 1992, ApJ, 384, 50}
\ref{Auri\`ere M., Cordoni J.P., 1981a, A\&A, 100, 307}
\ref{Auri\`ere M., Cordoni J.P., 1981b, A\&AS, 46, 347}
\ref{Auri\`ere M., Koch-Miramond L., Ortolani S., 1989, A\&A, 214, 113}
\ref{Auri\`ere M., Le F\`evre O., Terzan, A., 1984, A\&A, 138, 415}
\ref{Auri\`ere M., Ortolani S., Lauzeral C., 1990, Nature, 344, 638}
\ref{Baade W., 1958, in  Stellar Populations, ed. O'Connell D.J.K.,
	(Amsterdam: North Holland), p. 303}
\ref{Backer D.C., Kulkarni S.R., Heiles C., Davis M.M., Goss W.M.,
	1982, Nature, 300, 615}
\ref{Bacon D., Sigurdsson S.,  Davies M.B., 1996, MNRAS, 281, 830}
\ref{Bagin V.M., 1976a, SvA, 19, 613}
\ref{Bagin V.M., 1976b, SvA, 20, 54}
\ref{Bagin V.M., 1979, SvA, 23, 416}
\ref{Bahcall J.N., Bahcall N.A., Weistrop D., 1975, Astrophys. Lett, 16, 159} 
\ref{Bahcall J.N., Soneira R.M., 1981, ApJS, 47, 357}
\ref{Bahcall J.N., Wolf R.A., 1976, ApJ, 209, 214}
\ref{Bahcall J.N., Wolf R.A., 1977, ApJ, 216, 883}
\ref{Bailey S.I., 1893, Astronomy and Astro-Physics, XII, 689} 
\ref{Bailyn C.D., 1988, Nature, 332, 330}
\ref{Bailyn C.D., 1989, in  Dynamics of Dense Stellar Systems,
	ed. Merritt D., (Cambridge: Cambridge University Press), p. 167}
\ref{Bailyn C.D., 1992, ApJ, 392, 519}
\ref{Bailyn C.D., 1994, AJ, 107, 1073}
\ref{Bailyn C.D., 1995, ARA\&A, 33, 133}
\ref{Bailyn C.D., 1996, in  The Origins, Evolution, and Destinies of 
	Binary Stars in Clusters, eds. Milone G. \& Mermilliod J.-C., 
	ASP Conference Series Vol. 90, (San Francisco: ASP), p. 320}
\ref{Bailyn C.D., Grindlay J.E., 1990, ApJ, 353, 159}
\ref{Bailyn C.D., Grindlay J.E., Cohn H., Lugger P.M., 1988, ApJ, 331, 303}
\ref{Bailyn C.D., Grindlay J.E., Cohn H., Lugger P.M., 
	Stetson P.B., Hesser J.E., 1989, AJ, 98, 882}
\ref{Bailyn C.D., Grindlay J.E., Garcia M.R., 1990, ApJ, 357, L35}
\ref{Bailyn C.D., Iben I., 1989, ApJ, 347, L21}
\ref{Bailyn C.D., Pinsonneault M.H., 1995, ApJ, 439, 705}
\ref{Bailyn C.D., Sarajedini A., Cohn H., Lugger P.M., Grindlay J.E., 
	1992, AJ, 103, 1564}
\ref{Baranne A., Mayor M., Poncet J.L., 1979, Vistas in Astronomy, 23, 279}
\ref{Barnes J.E., Hut P., 1986, Nature, 324, 446
\ref{Barnes J.E., Hut P., 1989, ApJS, 70, 389}
\ref{Bassino L.P., Muzzio J.C., Rabolli M., 1994, ApJ, 431, 634}
\ref{Batt J., Faltenbacher W., Horst E., 1986, Arch. Rat. Mech. Anal., 93, 159}
\ref{Beers T.C., Flynn, K., Gebhardt, K., 1990, AJ, 100, 32}
\ref{Bellazzini M., Vesperini E., Fusi Pecci F., Ferraro F.R.,
	1996, MNRAS, 279, 337}
\ref{Bender R., Burstein D., Faber S.M., 1992, ApJ, 399, 462}
\ref{Bendinelli O., Cacciari C., Djorgovski S.G., \etal, 1993, ApJ, 409, L17}
\ref{Benz W., Hills J.G., 1987, ApJ, 323, 614}
\ref{Benz W., Hills J.G., 1992, ApJ, 389, 546}
\ref{Benz W., Thielemann F., Hills J.G., 1989, ApJ, 342, 986}
\ref{Bettwieser E., 1983, MNRAS, 203, 811}
\ref{Bettwieser E., 1985a, MNRAS, 215, 499}
\ref{Bettwieser E., 1985b, in 
         Dynamics of Star Clusters, 
        IAU Symp. 113, eds. J. Goodman \& P. Hut, 
        (Dordrecht: Reidel), p. 219}
\ref{Bettwieser E., Fricke K.J., Spurzem R., 1985, in 
         Dynamics of Star Clusters, 
        IAU Symp. 113, eds. J. Goodman \& P. Hut, 
        (Dordrecht: Reidel), p. 301}
\ref{Bettwieser E., Fritze U., 1984, PASJ, 36, 403}
\ref{Bettwieser E., Inagaki S., 1985, MNRAS, 213, 473}
\ref{Bettwieser E., Spurzem R., 1986, A\&A, 161, 102}
\ref{Bettwieser E., Sugimoto D., 1984, MNRAS, 208, 493}
\ref{Bettwieser E., Sugimoto D., 1985, MNRAS, 212, 189}
\ref{Bica E., Ortolani S., Barbuy B., 1994, A\&A, 283, 67}
\ref{Binney J., 1982, in  Morphology and Dynamics of Galaxies, 
	Saas-Fee Advanced Course 12, eds. Martinet L. \& Mayor M.,   
	(Geneva: Geneva Observatory), p. 1}
\ref{Binney J., Tremaine S., 1987,  Galactic Dynamics,
	(Princeton: Princeton University Press)}
\ref{Bisnovatyi-Kogan G.S., 1978, SvA, 4, L69}
\ref{Blanco V., Terndrup D., 1989, AJ, 98, 843}
\ref{Boccaletti D., Pucacco G., Ruffini R., 1991, A\&A, 244, 48}
\ref{Bolte M., 1989, AJ, 97, 1688}
\ref{Bolte M., 1993, in  The Globular Cluster - Galaxy Connection, 
	ASP Conference Series Vol. 48, 
	eds. Smith G.H. \& Brodie J.P., (San Francisco: ASP), p. 60}
\ref{Bolte M., Hesser J.E., Stetson P.B., 1993, ApJ, 408, L89}
\ref{Bolte M., Hogan C.J., 1995, Nature, 376, 399}
\ref{Boltzmann L., 1896,  Vorlesungen \"uber Gastheorie,
	(Leipzig: J. A. Barth)} 
\ref{Bouvier P., 1972, A\&A, 21, 441}
\ref{Boyd P.T., 1993, PhD Thesis, Drexel University}
\ref{Brandl B., Sams B.J., Bertoldi F., \etal, 1996, ApJ, 466, 254}
\ref{Breeden J.L., Cohn H.N., Hut P., 1994, ApJ, 421, 195}
\ref{Breeden J.L., Packard N.H., 1994, Int. J. Bifurcations and Chaos, 4, 311}
\ref{Breeden J.L., Cohn H.N., 1995, ApJ, 448, 672}
\ref{Brosche P., 1973, A\&A, 23, 259}
\ref{Brosche P., Lentes, F.-T., 1984, A\&A, 139, 474}
\ref{Brunet J.-P., Mesirov J.P., Edelman A., 1990, in 
	 Supercomputing 90, 
	IEEE, (New York: Computer Society Press), p. 748}
\ref{Buitrago J., Moreno-Garrido C., Mediavilla E., 1994, MNRAS, 268, 841}
\ref{Buonanno R., 1993, in
  	 The Globular Cluster - Galaxy Connection, 
	ASP Conference Series Vol. 48, 
	eds. Smith G.H. \& Brodie J.P., (San Francisco: ASP), p. 131}
\ref{Buonanno R., Castellani V., Corsi C.E., Fusi Pecci F., 1981, A\&A, 101, 1}
\ref{Buonanno R., Corsi C.E., Fusi Pecci F., 1985a, A\&A, 145, 97}
\ref{Buonanno R., Corsi C.E., Fusi Pecci F., Hardy E., Zinn R., 1985b,
	A\&A, 152, 65}
\ref{Buonanno R., Corsi C.E., Fusi Pecci F., Fahlman G.G., Richer H.B.,
	1994, ApJ, 430, L21}
\ref{Buonanno R., Corsi C.E., Fusi Pecci F., Richer H.B., Fahlman G.G., 
	1995a, AJ, 109, 650}
\ref{Buonanno R., Corsi C.E., Pulone L., \etal, 1995b, AJ, 109, 663}
\ref{Burgarella D., Buat V., 1996, A\&A, 313, 129}
\ref{Burgarella D., Paresce F., Quilichini V., 1995, A\&A, 301, 675}
\ref{Calzetti D., De Marchi G., Paresce F., Shara M., 1993, ApJ, 402, L1}
\ref{Camm G., 1952, MNRAS, 112, 155}
\ref{Canal R., Isern J., Labay J., 1990, ARA\&A, 28, 183}
\ref{Capaccioli M., Ortolani S., Piotto G., 1991, A\&A, 244, 298}
\ref{Capaccioli M., Piotto G., Stiavelli M., 1993, MNRAS, 261, 819}
\ref{Capriotti E.R.,  Hawley S., Hamlin M., 1996, 
	in  Formation of the Galactic Halo...Inside and Out, 
	ASP Conference Series Vol. 92, 
	eds. Morrison H. \& Sarajedini A., (San Francisco: ASP), p. 487}
\ref{Carnevali P., Santangelo P., 1980, Mem. Soc. Astron. Ital., 51, 529}
\ref{Casertano S., Hut P., 1985, ApJ, 298, 80}
\ref{Cederbloom S.E., Moss M.J., Cohn H.N., \etal, 1992, AJ, 103, 480}
\ref{Chaboyer B., 1995, ApJ, 444, L9}
\ref{Chaboyer B., Demarque P., Kernan P.J., Krauss L.M., 
	1996a, Sciences, 271, 957}
\ref{Chaboyer B., Demarque P., Kernan P.J., Krauss L.M., Sarajedini A.,
	1996b, MNRAS, in press}
\ref{Chaboyer B., Demarque P., Sarajedini A., 1996c, ApJ, 459, 558}
\ref{Chandrasekhar S., 1942,  Principles of Stellar Dynamics,
	(Chicago: Univ. of Chicago Press)}
\ref{Chandrasekhar S., 1943a, ApJ, 97, 255}
\ref{Chandrasekhar S., 1943b, ApJ, 97, 263}
\ref{Chandrasekhar S., 1943c, ApJ, 98, 54}
\ref{Chang J.S., Cooper G., 1970, J. Comp. Phys., 6, 1}
\ref{Charlton J.C., Laguna P., 1995, ApJ, 444, 193}
\ref{Chen K.Y., Leonard P.J.T., 1993, ApJ, 411, L75}
\ref{Chen K.Y., Middleditch J., Ruderman M., 1993, ApJ, 408, 17}
\ref{Chernoff D.F., 1993, in 
	Structure and Dynamics of Globular Clusters, 
	ASP Conference Series Vol. 50, 
	eds. Djorgovski S.G. \& Meylan G., (San Francisco: ASP), p. 245}
\ref{Chernoff D.F., Djorgovski S.G., 1989, ApJ, 339, L904}
\ref{Chernoff D.F., Kochanek C.S., Shapiro S.L., 1986, ApJ, 309, 183}
\ref{Chernoff D.F., Shapiro, S.L., 1987, ApJ, 322, 113}
\ref{Chernoff D.F., Weinberg M.D., 1990, ApJ, 351, 121}
\ref{Chi\`eze J.-P., 1987, A\&A, 171, 225}
\ref{Chi\`eze J.-P., Pineau des For\^ets G., 1987, A\&A, 183, 98}
\ref{Chi\`eze J.-P., Pineau des For\^ets G., 1989, A\&A, 221, 89}
\ref{Chu Y.-H., Kennicutt R.C., 1994, ApJ, 425, 720}
\ref{Chun M.S., Freeman K.C., 1979, ApJ, 227, 93}
\ref{Churchwell E., 1975, in HII Regions and Related Topics, 
	eds. Wilson T.L. \& Downs D., (Berlin: Springer), p. 245.}
\ref{Clark G., 1975, ApJ, 199, L143}
\ref{Cleary P.W., Monaghan J.J., 1990, ApJ, 349, 150}
\ref{Cohn H., 1979, ApJ, 234, 1036}
\ref{Cohn H., 1980, ApJ, 242, 765}
\ref{Cohn H., 1985, in Dynamics of Star Clusters, IAU Symp. 113, 
	eds. Goodman J. \& Hut P., (Dordrecht: Reidel), p. 161}
\ref{Cohn H., 1988, in  The Harlow-Shapley
	Symposium on Globular Cluster Systems in Galaxies, IAU Symp. 126,
	eds. Grindlay J.E. \& Philip A.G.D., 
	(Dordrecht: Kluwer), p. 379}
\ref{Cohn H., Hut P., 1984, ApJ, 277, L45}
\ref{Cohn H., Hut P.,  Wise M.W., 1989, ApJ, 342, 814}
\ref{Cohn H., Kulsrud R.M., 1978, ApJ, 226, 1087}
\ref{Cohn H., Wise M.W., Yoon T.S., \etal,
	1986, in The Use of Supercomputers in Stellar Dynamics, 
	eds. Hut P. \& McMillan S.L.W., (Berlin: Springer-Verlag),
	p. 206}
\ref{Cool A.M., Grindlay J.E., Cohn H.N., Lugger P.M., Slavin, S.D.,
	1995, ApJ, 439, 695}
\ref{Cool A.M., Grindlay J.E., Krockenberger M., Bailyn C.D., 
	1993, ApJ, 410, L103}
\ref{Cool A.M., Piotto G., King I.R., 1996, ApJ, in press}
\ref{C\^ot\'e P., Fischer P., 1996, AJ, 112, 565}
\ref{C\^ot\'e P., Pryor C., McClure R.D., Fletcher J.M., Hesser J.E., 1996, 
	AJ, 112, 574}
\ref{C\^ot\'e P., Welch D.L., Fischer P., \etal, 1994, ApJS, 90, 83}
\ref{C\^ot\'e P., Welch D.L., Fischer P., Gebhardt K., 1995, ApJ, 454, 788}
\ref{Couchman H.M.P, Rees M.J., 1986, MNRAS, 221, 53}
\ref{Covino S., Pasinetti Fracassini L.E., 1993, A\&A, 270, 83}
\ref{Cudworth K.M., 1976a, AJ, 81, 519}
\ref{Cudworth K.M., 1976b, AJ, 81, 975}
\ref{Cudworth K.M., 1979a, AJ, 84, 1312}
\ref{Cudworth K.M., 1979b, AJ, 84, 1866}
\ref{Cudworth K.M., 1985, AJ, 90, 65}
\ref{Cudworth K.M., 1986, AJ, 92, 348}
\ref{Cudworth K.M., 1988, AJ, 96, 105}
\ref{Cudworth K.M., Hanson R.B., 1993, AJ, 105, 168}
\ref{Cudworth K.M., Monet D.G., 1979, AJ, 84, 774}
\ref{Cudworth K.M., Rauscher B.J., 1987, AJ, 93, 856}
\ref{Cudworth K.M., Rees R.F., 1990, AJ, 99, 1491}
\ref{Cudworth K.M., Smetanka J.J., Majewski S.R., 1992, AJ, 103, 1252}
\ref{Da Costa G.S., 1979, AJ, 84, 505}
\ref{Da Costa G.S., 1995, PASP, 107, 937}
\ref{Da Costa G.S., Freeman K.C., 1976, ApJ, 206, 128}
\ref{Da Costa G.S., Freeman K.C., Kalnajs A.J., Rodgers A.W., 
	Stapinsky T.E., 1977, AJ, 82, 810}
\ref{Da Costa G.S., \etal, 1993, 
	in Structure and Dynamics of Globular Clusters, 
	ASP Conference Series Vol. 50, 
	eds. Djorgovski S.G. \& Meylan G., (San Francisco: ASP), p. 81}
\ref{Da Costa L.N., Lightman A.P., 1979, ApJ, 228, 543}
\ref{Da Costa L.N., Pryor C., 1979, ApJ, 233, 694}
\ref{Dagan E., Horwitz G., 1988, J. Phys. A, 21, 1001}
\ref{D'Amico N., Bailes M., Lyne A.G., Manchester R.N., Johnston
        S., Fruchter A.S., Goss W.M., 1993, MNRAS, 260, L7}
\ref{Danilov V.M., 1973, SvA, 17, 346}
\ref{Danilov V.M., 1978, Afz, 14, 37}
\ref{Danilov V.M., 1985, SvA, 29, 413}  
\ref{Danilov V.M., 1989, SvA, 32, 374}
\ref{D'Antona F., Mazzitelli I., 1996, ApJ, 456, 329}
\ref{D'Antona F., Vietri M., Pesce E., 1995, MNRAS, 272, 730}
\ref{Dauphole B., Geffert M., Colin J., \etal, 1996,  A\&A, 312, 119}
\ref{Davies M.B., 1995, MNRAS, 276, 887}
\ref{Davies M.B., Benz W., Hills J.G., 1991, ApJ, 381, 449}
\ref{Davies M.B., Benz W., Hills J.G., 1992, ApJ, 401, 246}
\ref{Davies M.B., Benz W., Hills J.G., 1993, ApJ, 411, 285}
\ref{Davies M.B., Benz W., Hills J.G., 1994, ApJ, 424, 870}
\ref{Davies M.B., Benz W., 1995, 276, 876}
\ref{Davoust E., 1977, A\&A, 61, 391}
\ref{Davoust E., 1986, A\&A, 166, 177}
\ref{de Boisanger C., Chi\`eze J.-P., 1991, A\&A, 241, 581}
\ref{de Carvalho R., Djorgovski S.G., 1992, ApJ, 389, L49}
\ref{Dehnen W., Gerhard O.E., 1993, MNRAS, 261, 311}
\ref{Dejonghe H., 1984, A\&A, 133, 225}
\ref{Dejonghe H., 1987, MNRAS, 224, 13}
\ref{Dejonghe H., 1989, ApJ, 343, 113}
\ref{Dejonghe H., de Zeeuw T., 1988, ApJ, 333, 90}
\ref{Dejonghe H., Merritt D., 1988, ApJ, 328, 93}
\ref{Dejonghe H., Merritt D., 1992, ApJ, 391, 531}
\ref{Deliyannis C.P., Demarque P., Kawaler S.D., 1990, ApJS, 73, 21}
\ref{De Marchi G., Paresce F., 1994a, A\&A, 281, L13} 
\ref{De Marchi G., Paresce F., 1994b, ApJ, 422, 597}  
\ref{De Marchi G., Paresce F., 1995a, A\&A, 304, 211} 
\ref{De Marchi G., Paresce F., 1995b, A\&A, 304, 202} 
\ref{De Marchi G., Paresce F., 1996, ApJ, 467, 658}   
\ref{de Zeeuw T., ed., 1987,  
	Structure and Dynamics of Elliptical Galaxies, 
	IAU Symp. 127, (Dordrecht: Reidel)}
\ref{de Zeeuw T., 1988, in  Integrability in Dynamical Systems, 
	Ann. NY Acad. Sci., p. 15}
\ref{Dickens R.J., Bell R.A., 1976, ApJ, 207, 506}
\ref{Dickens R.J., Croke B.F.W., Cannon R.D., Bell R.A., 1991,
	Nature, 351, 212}
\ref{Dickens R.J., Woolley R., 1967, Royal Obs. Bull. \# 128}
\ref{Dickman R.L., Horvath M.A., Margulis M., 1990, ApJ, 365, 586}
\ref{Di Fazio A., 1986, A\&A, 159, 49}
\ref{Di Stefano R., Rappaport S., 1994, ApJ, 423, 274}
\ref{Djorgovski, S.G., 1988, in  The Harlow-Shapley
	Symposium on Globular Cluster Systems in Galaxies, IAU Symp. 126,
	eds. Grindlay J.E. \& Philip A.G.D., 
	(Dordrecht: Kluwer), p. 333}
\ref{Djorgovski, S.G., 1991, in 
	 Formation and Evolution of Star Clusters, 
	ASP Conference Series Vol. 13, 
	ed. Janes K., (San Francisco: ASP), p. 112}
\ref{Djorgovski S.G., 1993a, in 
	 The Globular Cluster - Galaxy Connection, 
	ASP Conference Series Vol. 48, 
	eds. Smith G.H. \& Brodie J.P., (San Francisco: ASP), p. 496}
\ref{Djorgovski S.G., 1993b, in 
	 Structure and Dynamics of Globular Clusters, 
	ASP Conference Series Vol. 50, 
	eds. Djorgovski S.G. \& Meylan G., (San Francisco: ASP), p. 373}
\ref{Djorgovski S.G., 1995, ApJ, 438, L29}
\ref{Djorgovski S.G., Davis M., 1987, ApJ, 313, 59}
\ref{Djorgovski S.G., Gal R.R., McCarthy J.K., \etal, 1996,
	ApJ, submitted}
\ref{Djorgovski S.G., King I.R., 1986, ApJ, 305, L61}
\ref{Djorgovski S.G., King I.R., Vuosalo C., Oren A., Penner H., 1986,
	in Instrumentation and Research Programmes for Small Telescopes,
	IAU Symp. 118, 
	eds. Hearnshaw J.B. \& Cottrell P.L., (Dordrecht: Reidel), p. 281}
\ref{Djorgovski S.G., Meylan G., eds., 1993a,  Structure and Dynamics of 
	Globular Clusters, ASP Conference Series Vol. 50, 
	(San Francisco: ASP)}
\ref{Djorgovski S.G., Meylan G., 1993b, in  Structure and Dynamics of 
	Globular Clusters, ASP Conference Series Vol. 50, 
	eds. Djorgovski S.G. \& Meylan G., (San Francisco: ASP), p. 325}
\ref{Djorgovski S.G., Meylan G., 1994, AJ, 108, 1292}
\ref{Djorgovski S.G., Piotto G., 1992, AJ, 104, 2112}
\ref{Djorgovski S.G., Piotto G., 1993, in 
	 Structure and Dynamics of Globular Clusters, 
	ASP Conference Series Vol. 50, 
	eds. Djorgovski S.G. \& Meylan G., (San Francisco: ASP), p. 203}
\ref{Djorgovski S.G., Piotto G., Capaccioli M., 1993, AJ, 105, 2148}
\ref{Djorgovski S.G., Piotto G., King I.R., 1989, 
	in  Dynamics of Dense Stellar Systems,
	ed. Merritt D., (Cambridge: Cambridge University Press), p. 147}
\ref{Djorgovski S.G., Piotto G., Phinney E.S., Chernoff D.F., 1991,
	ApJ, 372, L41}
\ref{Djorgovski S.G., Santiago B.X., 1993, ApJ, 391, L85}
\ref{Dokuchaev V.I., Ozernoi L.M., 1978, SvA, 22, 15}
\ref{Dokuchaev V.I., Ozernoi L.M., 1981a, SvA Letters, 7, 155}
\ref{Dokuchaev V.I., Ozernoi L.M., 1981b, SvA Letters, 7, 158}
\ref{Drukier G.A., Fahlman G.G., Richer H.B., 1989, ApJ, 342, L27}
\ref{Drukier G.A., 1993, MNRAS, 265, 773}
\ref{Drukier G.A., 1995, ApJS, 100, 347}
\ref{Drukier G.A., 1996, MNRAS, 280, 498}
\ref{Drukier G.A., Fahlman G.G., Richer H.B., 1992, ApJ, 386, 106}
\ref{Drukier G.A., Fahlman G.G., Richer H.B., Searle L., Thompson I.B.,
	1993, AJ, 106, 2335}
\ref{Dubath P., Mayor M., Meylan G., 1993a, 
        in  Structure and Dynamics of Globular Clusters,
        ASP Conference Series Vol. 50,
        eds. Djorgovski S.G. \& Meylan G., (San Francisco: ASP), p. 69}
\ref{Dubath P., Mayor M., Meylan G., 1993b, 
	in The Globular Cluster - Galaxy Connection, 
	ASP Conference Series Vol. 48, 
	eds. Smith G.H. \& Brodie J.P., (San Francisco: ASP), p. 557}
\ref{Dubath P., Meylan G., 1994, A\&A, 290, 104}
\ref{Dubath P., Meylan G., Mayor M., 1990, A\&A, 239, 142}
\ref{Dubath P., Meylan G., Mayor M., 1992, ApJ, 400, 510}
\ref{Dubath P., Meylan G., Mayor M., 1994, ApJ, 426, 192} 
\ref{Dubath P., Meylan G., Mayor M., 1996, A\&A, in press}
\ref{Duncan M.J., Shapiro S.L., 1982, ApJ, 253, 921}
\ref{Dupree A., Hartmann L., Black J., \etal, 1979, ApJ, 230, L89}
\ref{Duquennoy A., Mayor M., 1991, A\&A, 248, 485}
\ref{Eddington A.S., 1913, MNRAS, 74, 5}
\ref{Eddington A.S., 1915a, MNRAS, 75, 366}
\ref{Eddington A.S., 1915b, MNRAS, 76, 37}
\ref{Eddington A.S., 1916, MNRAS, 76, 572}
\ref{Edmonds P.D., Gilliland R.L., Guhathakurta P., \etal, 
	1996, ApJ, 468, 241}
\ref{Einsel C., Spurzem R., 1996, preprint}
\ref{Eggen O.J., Iben I., 1988, AJ, 96, 635}
\ref{Eggen O.J., Lynden-Bell D., Sandage A.R., 1962, ApJ, 136,
	748}
\ref{Eggleton P.P.,  Kiseleva L.G., 1995, preprint, astro-ph/9510110}
\ref{Eggleton P.P., 1996, in  The Origins, Evolution, and Destinies of 
	Binary Stars in Clusters, eds. Milone G. \& Mermilliod J.-C., 
	ASP Conference Series Vol. 90, (San Francisco: ASP), p. 257}
\ref{\'Eigenson A.M., Yatsuk O.S., 1986, SvA, 30, 390}
\ref{\'Eigenson A.M., Yatsuk O.S., 1989, SvA, 33, 280}
\ref{Elmegreen B.G., 1985, in  Birth and Infancy of Stars, 
	Les Houches, eds. Lucas R. Omont A., \& Stora R.,  
	(Amsterdam: Elsevier), p. 257}
\ref{Elmegreen B.G., 1992, in  The Galactic Interstellar Medium, 
	Saas-Fee Advanced Course 21, eds. Pfenniger D. \& Bartholdi P., 
	(Berlin: Springer-Verlag), p. 157}
\ref{Elson R.A.W., 1991, ApJS, 76, 185}
\ref{Elson R.A.W., 1992, MNRAS, 256, 515}
\ref{Elson R.A.W., Gilmore G.F., Santiago B.X., Casertano S.,  
	1995, AJ, 110, 682}
\ref{Elson R.A.W., Fall S.M., 1988, AJ, 96, 1383}
\ref{Elson R.A.W., Fall S.M., Freeman K.C., 1987a, ApJ, 323, 54}
\ref{Elson R.A.W., Freeman K.C., Lauer T.R., 1989, ApJ, 347, L69}
\ref{Elson R.A.W., Hut P., Inagaki S., 1987b, ARA\&A, 25, 565}
\ref{Faber S.M., Dressler A., Davies R.L., \etal, 1987, 
	in  Nearly Normal Galaxies, 
	ed. S.M. Faber, (New York: Springer-Verlag), p. 175}
\ref{Faber S.M., Jackson R.E., 1976, ApJ, 204, 668}  
\ref{Fabian A.C., Nulsen P.E.J. Canizares C.R., 1984, Nature, 310, 733}
\ref{Fabian A.C., Pringle J.E., Rees M.J., 1975, MNRAS, 172, 15P}
\ref{Fahlman G.G., Richer H.B., Searle L., Thompson I.B., 1989, ApJ, 343, L49}
\ref{Falgarone E., Phillips T.G., Walker C.K., 1991, ApJ, 378, 186}
\ref{Fall S.M., Frenk C.S., 1983, AJ, 88, 1626}
\ref{Fall M., Frenk C.S., 1985, in 
	 Dynamics of Star Clusters, IAU Symp. 113, 
	eds. Goodman J. \& Hut P., (Dordrecht: Reidel), p. 285}
\ref{Fall S.M., Malkan M.A., 1978, MNRAS, 185, 899}
\ref{Fall S.M., Rees M.J., 1977, MNRAS, 181, 37P}
\ref{Fall S.M., Rees M.J., 1985, ApJ, 298, 18}
\ref{Fall S.M., Rees M.J., 1988, in  The Harlow-Shapley
	Symposium on Globular Cluster Systems in Galaxies, IAU Symp. 126,
	eds. Grindlay J.E. \& Philip A.G.D., 
	(Dordrecht: Kluwer), p. 323}
\ref{Farouki R.T., Salpeter E.E., 1982, ApJ, 253, 512}
\ref{Feast M.W., Thackeray A.D., Wesselink A.J., 1961, MNRAS, 122, 433}
\ref{Ferraro F.R., Fusi Pecci F., Cacciari C., 1993, AJ, 106, 2324}
\ref{Ferraro F.R., Fusi Pecci F., Bellazini M., 1995a, A\&A, 294, 80}
\ref{Ferraro F.R., Paresce F., 1993, AJ, 106, 154}
\ref{Ferraro I., Ferraro F.R., Fusi Pecci F., Corsi C.E., Buonanno R., 
	1995b, MNRAS, 275, 1057}
\ref{Finzi A., 1977, A\&A, 62, 149}
\ref{Fischer P., Welch D.L., C\^ot\'e P., Mateo M., Madore B.F.,
	1992a, AJ, 103, 857}  
\ref{Fischer P., Welch D.L., Mateo M., 1992b, AJ, 104, 1086} 
\ref{Fischer P., Welch D.L., Mateo M., 1993a, AJ, 105, 938} 
\ref{Fischer P., Welch D.L., Mateo M., C\^ ot\'e P., 1993b, AJ, 106, 1508}
\ref{Flechter A.B., Stahler S.W., 1994a, ApJ, 435, 313}
\ref{Flechter A.B., Stahler S.W., 1994b, ApJ, 435, 329}
\ref{Flechter J.M., Harris H.C., McClure R.D., Scarfe C.D., 1982,
	PASP, 94, 1017}
\ref{Forbes D.A., Franx M., Illingworth G.D., Carollo C.M., 
	1996, ApJ, 467, 126}
\ref{Forest E., Ruth R.D., 1990, Physica D, 43, 105}
\ref{Fran\c cois P., Spite M., Spite F., 1988, A\&A, 191, 267}
\ref{Freeman K.C., Norris J., 1981, ARA\&A, 19, 319}
\ref{Freeman K.C., 1990, in  Dynamics and Interactions of Galaxies, 
	ed. Wielen R., (Berlin: Springer-Verlag), p. 36}
\ref{Freeman K.C., 1993, in  The Globular Cluster - Galaxy Connection, 
	ASP Conference Series Vol. 48, 
	eds. Smith G.H. \& Brodie J.P., (San Francisco: ASP), p. 608}
\ref{Frenk C.S., Fall S.M., 1982, MNRAS, 199, 565}
\ref{Frenk C.S., White S.D.M., 1980, MNRAS, 193, 295}
\ref{Frenk C.S., White S.D.M., 1982, MNRAS, 198, 173}
\ref{Frogel J.A., Kuchinski L.E., Tiede G.P., 1995, AJ, 109, 1154}
\ref{Fujimoto M., Noguchi M., 1990, PASJ, 42, 505}
\ref{Fukushige T., Heggie D.C., 1995, MNRAS, 276, 206}
\ref{Funato Y., Hut P., McMillan S., Makino J., 1996, AJ, in press, 
	astro-ph/9604025}
\ref{Funato Y., Makino J., Ebisuzaki T., 1992, PASJ, 44, 613}
\ref{Fusi Pecci F., Battistini P., Bendinelli O., \etal, 1994, A\&A, 284, 349}
\ref{Fusi Pecci F., Bellazzini M., Cacciari C.,  Ferraro F.R., 1995,
	AJ, 110, 1664}
\ref{Fusi Pecci F., Bellazzini M., Ferraro F., Buonanno R., Corsi C.E., 1996, 
	in  Formation of the Galactic Halo...Inside and Out, 
	ASP Conference Series Vol. 92, 
	eds. Morrison H. \& Sarajedini A., (San Francisco: ASP), p. 221}
\ref{Fusi Pecci F., Ferraro F., Bellazzini M., \etal, 1993a, AJ, 105, 1145}
\ref{Fusi Pecci F., Ferraro F., Cacciari C., 1993b, in  Blue Stragglers, 
	ASP Conference Series Vol. 53,  ed. Saffer, R.A., 
	(San Francisco: ASP), p. 97}
\ref{Fusi Pecci F., Ferraro F., Corsi, C.E., Cacciari C., Buonanno R.,
	1992, AJ, 104, 1831}	
\ref{Gao B., Goodman J., Cohn H., Murphy B., 1991, ApJ, 370, 567}
\ref{Gascoigne S.C.B, Burr E.J., 1956, MNRAS, 116, 570}
\ref{Gebhardt K., Fischer P., 1995, AJ, 109, 209}
\ref{Gebhardt K., Pryor C., Williams T.B., Hesser J.E., 1994, AJ, 107, 2067}
\ref{Gebhardt K., Pryor C., Williams T.B., Hesser J.E., 1995, AJ, 110, 1699}
\ref{Geyer E.H., Hopp U., Nelles B., 1983, A\&A, 125, 359}
\ref{Giannone G., Lo Cascio L., Molteni D., 1984, Nuovo Cim., 7C, 856}
\ref{Giannone G., Molteni D., 1985, A\&A, 143, 321}
\ref{Giersz M., 1984, Post\c epy Astron., 32, 249}
\ref{Giersz M., 1985a, Acta Astron., 35, 119}
\ref{Giersz M., 1985b, Acta Astron., 35, 401}
\ref{Giersz M., 1986, Acta Astron., 36, 181}
\ref{Giersz M., Heggie D.C., 1993a, MNRAS, 268, 257}
\ref{Giersz M., Heggie D.C., 1993b, 
	in  The Globular Cluster - Galaxy Connection, 
	ASP Conference Series Vol. 48, 
	eds. Smith G.H. \& Brodie J.P., (San Francisco: ASP), p. 713}
\ref{Giersz M., Heggie D.C., 1994, MNRAS, 270, 298}
\ref{Giersz M., Heggie D.C., 1996a, MNRAS, 279, 1037}
\ref{Giersz M., Heggie D.C., 1996b, MNRAS, in preparation}
\ref{Giersz M., Spurzem R., 1994, MNRAS, 269, 241}
\ref{Gilliland R.L., Edmonds P.D., Petro  L., Saha A., Shara M.M., 
	1995, ApJ, 447, 191}
\ref{Gilmore G., 1989, in  The Milky Way as a Galaxy, 
	Saas-Fee Advanced Course 19, eds. Buser R. \& King I.R.,  
	(Geneva: Geneva Observatory)}
\ref{Gizis J.E., Mould J.R., Djorgovski S.G., 1993, PASP, 105, 871}
\ref{Gnedin O.Y., Ostriker J.P., 1996, preprint, astro-ph/9603042}
\ref{Goodman J., 1983a, PhD thesis, Princeton University}
\ref{Goodman J., 1983b, ApJ, 270, 700}
\ref{Goodman J., 1983c, ApJ, 280, 298}
\ref{Goodman J., 1987, ApJ, 313, 576}
\ref{Goodman J., 1989, in  Dynamics of Dense Stellar Systems,
	ed. Merritt D., (Cambridge: Cambridge University Press), p. 183}
\ref{Goodman J., 1993,
	in  Structure and Dynamics of Globular Clusters, 	
	ASP Conference Series Vol. 50, 
	eds. Djorgovski S.G. \& Meylan G., (San Francisco: ASP), p. 87}
\ref{Goodman J., Heggie D.C., Hut P., 1993, ApJ, 415, 715}
\ref{Goodman J., Hernquist L., 1991, ApJ, 378, 637}
\ref{Goodman J., Hut P., eds., 1985,  Dynamics of Star Clusters, 
	IAU Symp. 113, (Dordrecht: Reidel)}
\ref{Goodman J., Hut P., 1989, Nature, 339, 40}
\ref{Goodman J., Hut P., 1993, ApJ, 403, 271}
\ref{Goodwin S.P., 1996, 
	in  Formation of the Galactic Halo...Inside and Out, 
	ASP Conference Series Vol. 92, 
	eds. Morrison H. \& Sarajedini A., (San Francisco: ASP), p. 495}
\ref{Gorti U., Bhatt H.C., 1996, MNRAS, 278, 611}
\ref{Grabhorn R.P., Cohn H.N., Lugger P.M., Murphy B.W., 1992,
	ApJ, 392, 86}
\ref{Green E.M., Norris J.E., 1990, ApJ, 353, L17}
\ref{Greengard L., 1990, Computers in Phys., 3, 142-52}
\ref{Griffin R.F., 1967, ApJ, 148, 465}
\ref{Griffin R.F., 1974, ApJ, 191, 545}
\ref{Grillmair C.J., Ajahr E.A., Faber S.M., \etal, 1996, AJ, in press}
\ref{Grillmair C.J., Freeman K.C., Irwin M., Quinn P.J., 1995a, AJ, 109, 2553}
\ref{Grillmair C.J., Quinn P.J., Freeman K.C., Salmon J., 1995b, 
	in preparation}
\ref{Grindlay J.E., 1993,
	in  Structure and Dynamics of Globular Clusters, 	
	ASP Conference Series Vol. 50, 
	eds. Djorgovski S.G. \& Meylan G., (San Francisco: ASP), p. 285}
\ref{Grindlay J.E., Cool A.M., Callanan P.J., Bailyn C.D., 
	Cohn H.N., Lugger P.M., 1995, ApJ, 455, L47}
\ref{Grindlay J.E., Hertz P., Steiner J.E., Murray S.S., Lightman A.P.,
	1984, ApJ, 282, L13}
\ref{Grindlay J.E., Philip A.G.D., eds., 1988,  The Harlow-Shapley
	Symposium on Globular Cluster Systems in Galaxies, IAU Symp. 126,
	(Dordrecht: Kluwer)}
\ref{Gruji\'c P., Simonovi\'c N., 1988, in  The Few Body Problem, 
	IAU Coll. 96., ed. Valtonen M.J., (Dordrecht: Kluwer), p. 249}
\ref{Guhathakurta P., Edmonds P.D., Gilliland R.L., 1996a, in preparation}
\ref{Guhathakurta P., Yanny B., Bahcall J.N., Schneider D.P.,
	1994, AJ, 108, 1786}
\ref{Guhathakurta P., Yanny B., Schneider D.P., Bahcall J.N., 1992, 
	AJ, 104, 1790}
\ref{Guhathakurta P., Yanny B., Schneider D.P., Bahcall J.N., 1996b, 
	AJ, 111, 267}
\ref{Gunn J.E., 1980, in  Globular Clusters, 
	eds. Hanes D. \& Madore B.F., 
	(Cambridge: Cambridge University Press), p. 301}
\ref{Gunn J.E., Griffin R.F., 1979, AJ, 84, 752}
\ref{Gurzadyan V.G., 1993,  
	in  Structure and Dynamics of Globular Clusters, 	
	ASP Conference Series Vol. 50, 
	eds. Djorgovski S.G. \& Meylan G., (San Francisco: ASP), p. 127}
\ref{Gurzadyan V.G., Kocharyan A.A., 1987, Ap\&SS, 135, 307}
\ref{Gurzadyan V.G., Savvidi G.K., 1984, Sov. Phys. Dokl., 29, 520}
\ref{Gurzadyan V.G., Savvidy G.K., 1986, A\&A, 160, 203}
\ref{Hachisu I., Nakada Y., Nomoto K., Sugimoto D., 1978, 
	Prog. Theor. Phys., 60, 393}
\ref{Hachisu I., Sugimoto D., 1978, Prog. Theor. Phys., 60, 123}
\ref{Han C., Ryden B.S., 1994, ApJ, 433, 80}
\ref{Harding G.A., 1965, Royal Obs. Bull. \# 99}
\ref{Harris H.C., McClure R.D., 1983, ApJ, 265, L77}
\ref{Harris W.E., 1976, AJ, 81, 1095}
\ref{Harris W.E., 1981, ApJ, 251, 497}
\ref{Harris W.E., 1991, ARA\&A, 29, 543}
\ref{Harris W.E., 1996, 
	in  Formation of the Galactic Halo...Inside and Out, 
	ASP Conference Series Vol. 92, 
	eds. Morrison H. \& Sarajedini A., (San Francisco: ASP), p. 231}
\ref{Harris W.E., Pudritz R.E., 1994, ApJ, 429, 177}
\ref{Harris W.E., Racine, R., 1979, ARA\&A, 17, 241}
\ref{Harris W.E., van den Bergh S., 1981, AJ, 86, 1627}
\ref{Hartwick F.D.A., Sargent W.L.W., 1978, ApJ, 221, 512}
\ref{Hasinger G.R., Johnston H.M., Verbunt F., 1994, A\&A, 288, 466}
\ref{Hayli A., 1967, Bull. Astron. (3), 2, 67}
\ref{Hayli A., 1970a, A\&A, 7, 17}
\ref{Hayli A., 1970b, A\&A, 7, 249}
\ref{Hayli A., ed., 1975,  Dynamics of Stellar Systems, 
	IAU Symp. 69, (Dordrecht: Reidel)}
\ref{Heggie D.C., 1975, MNRAS, 173, 729}
\ref{Heggie D.C., 1984, MNRAS, 206, 179}
\ref{Heggie D.C., 1985, in  Dynamics of Star Clusters, IAU Symp. 113, 
	eds. Goodman J. \& Hut P., (Dordrecht: Reidel), p. 139}
\ref{Heggie D.C., 1989, in  Dynamics of Dense Stellar Systems,
	ed. Merritt D., (Cambridge: Cambridge University Press), p. 195}
\ref{Heggie D.C., 1991, in  Predictability, Stability, and
        Chaos in N-Body Dynamical Systems, ed. Roy A.E., 
	(New York: Plenum), p. 47}
\ref{Heggie D.C., 1994, in 
	Proc. Workshop,  Astrophysics with Special Purpose Computers,
	(Tokyo: Tokyo University), p. 115; astro-ph/9312018}
\ref{Heggie D.C., Aarseth S.J., 1992, MNRAS, 257, 513}
\ref{Heggie D.C., Griest K., Hut P., 1993, 
	in  Structure and Dynamics of Globular Clusters, 	
	ASP Conference Series Vol. 50, 
	eds. Djorgovski S.G. \& Meylan G., (San Francisco: ASP), p. 137}
\ref{Heggie D.C., Hut P., 1993, ApJS, 85, 347}
\ref{Heggie D.C., Hut P., 1996, in  Dynamical Evolution of Star Clusters:
	Confrontation of Theory and Observations, IAU Symp. 174, 
	eds. Hut P. \& Makino J. (Dordrecht: Kluwer), p. 303}
\ref{Heggie D.C., Hut P., McMillan S.L.W., 1996, ApJ, 467, 359}
\ref{Heggie D.C., Inagaki S., McMillan S.L.W., 1994, MNRAS, 271, 706}
\ref{Heggie D.C., Mathieu R.D., 1986, 
	in  The Use of Supercomputers in Stellar Dynamics, 
	eds. Hut P. \& McMillan S.L.W., (Berlin: Springer-Verlag), p. 233}
\ref{Heggie D.C., Ramamani N., 1989, 237, 757}
\ref{Heggie D.C., Ramamani N., 1995, MNRAS, 272, 317}
\ref{Heggie D.C., Rasio F.A., 1996, MNRAS, in press}
\ref{Heggie D.C., Stevenson D., 1988, MNRAS, 230, 223}
\ref{Heggie D.C., Sweatman W.L., 1991, MNRAS, 250, 555}
\ref{H\'enon M., 1959, Ann. d'Astrophys., 22, 126}
\ref{H\'enon M., 1960a, Ann. d'Astrophys., 23, 668}
\ref{H\'enon M., 1960b, Ann. d'Astrophys., 23, 467}
\ref{H\'enon M., 1961, Ann. d'Astrophys., 24, 369}
\ref{H\'enon M., 1964, Ann. d'Astrophys., 27, 83}
\ref{H\'enon M., 1965, Ann. d'Astrophys., 28, 62}
\ref{H\'enon M., 1966, Comptes Rendus Acad. Sci., 262, 666}
\ref{H\'enon M., 1969, A\&A,  2, 151}
\ref{H\'enon M., 1972, in  Gravitational $N$-Body Problem, 
	IAU Coll. 10, ed. Lecar M., (Dordrecht: Reidel), p. 406}
\ref{H\'enon M., 1973, 
	in  Dynamical Structure and Evolution of Stellar Systems,
	Saas-Fee Advanced Course 3, eds. Martinet L. \& Mayor M.,  
	(Geneva: Geneva Observatory), p. 224}
\ref{H\'enon M., 1975, in  Dynamics of Stellar Systems, 
	IAU Symp. 69, ed. Hayli A., (Dordrecht: Reidel), p. 133}
\ref{Henry T.J., McCarthy, D.W., 1993, AJ, 106, 773}
\ref{Henriksen R.N., 1986, ApJ, 310, 189}
\ref{Hernquist L., 1988, Comp. Phys. Comm., 48, 107}
\ref{Hernquist L., Hut P., Makino J., 1992, ApJ, L85}
\ref{Hernquist L., Sigurdsson S., Bryan G.L., 1995, ApJ, 446, 717}
\ref{Herschel J., 1847, in  Results of Astronomical Observations
	made at the Cape of Good Hope, (London: Smith, Elder, \& Co.)}
\ref{Herschel W., 1814, Phil. Trans., CIV, 248}
\ref{Hertz P., Grindlay J.E., 1983, ApJ, 275, 105}
\ref{Hesser J.E., ed., 1980,  Star Clusters, 
	IAU Symp. 85, (Dordrecht: Reidel)}
\ref{Hesser J.E., 1993, 
	in  Structure and Dynamics of Globular Clusters, 
	ASP Conference Series Vol. 50, 
	eds. Djorgovski S.G. \& Meylan G., (San Francisco: ASP), p. 15}
\ref{Hesser J.E., Shawl S.J., Meyer J.E., 1986, PASP, 98, 403}
\ref{Hills J.G., 1975a, AJ, 80, 809}
\ref{Hills J.G., 1975b, AJ, 80, 1075}
\ref{Hills J.G., 1976, MNRAS, 175, 1P}
\ref{Hills J.G., 1980, ApJ, 235, 986}
\ref{Hills J.G., 1984, AJ, 89, 1811}
\ref{Hills J.G., 1990, AJ, 99, 979}
\ref{Hills J.G., Day C.A., 1976, Astrophys. Lett., 17, 87}
\ref{Hills J.G., Fullerton L.W., 1980, AJ, 85, 1281}
\ref{Hoffer J.B., 1983, AJ, 88, 1420}
\ref{Hoffer J.B., 1986, Astrophys. Lett., 25, 127}
\ref{Holtzman J.A., Faber S.M., Shaya E.J., \etal, 1992, AJ, 103, 6911}
\ref{Holtzman J.A., Watson A.M., Mould J.R., \etal, 1996, AJ, 112, 416}
\ref{Horwitz G., 1981, in Proc. Fifth G\"ottingen-Jerusalem-Symp.,
	eds. Fricke K.J. \& Shaham J., 
	(G\"ottingen: Vandenhoeck \& Ruprecht), p. 149}
\ref{Horwitz G., Dagan E., 1988, J. Phys. A., 21, 1017}
\ref{Horwitz G., Katz J., 1977, ApJ, 211, 226}
\ref{Horwitz G., Katz J., 1978, ApJ, 222, 941}
\ref{Huang S., Dubinski J., Carlberg R.G., 1992, ApJ, 404, 73}
\ref{Huang T.-Y., Valtonen M.J., 1987, MNRAS, 229, 333}
\ref{Hulse R.A., Taylor, J.H. Jr., 1975, ApJ, 195, L51}
\ref{Hunter D.A., Shaya E.J., Scowen P., \etal, 1995, ApJ, 444, 758}
\ref{Hut P., 1983, ApJ, 272, L29}
\ref{Hut P., 1985, in  Dynamics of Star Clusters, IAU Symp. 113, 
	eds. Goodman J. \& Hut P., (Dordrecht: Reidel), p. 231}
\ref{Hut, P., 1989, in  Dynamics of Dense Stellar Systems,
	ed. Merritt D., (Cambridge: Cambridge University Press), p. 229}
\ref{Hut P., 1992, in  X-Ray Binaries and Recycled Pulsars, 
	eds. van den Heuvel E.P.J. \& Rappaport S.A., 
	(Dordrecht: Kluwer), p. 317}
\ref{Hut P., 1993a, ApJ, 403, 256}
\ref{Hut P., 1993b, in  Blue Stragglers, 
	ASP Conference Series Vol. 53,  ed. Saffer, R.A., 
	(San Francisco: ASP), p. 44}
\ref{Hut P., 1996a, in  The Origins, Evolution, and Destinies of 
	Binary Stars in Clusters, eds. Milone G. \& Mermilliod J.-C., 
	ASP Conference Series Vol. 90, (San Francisco: ASP), p. 391}
\ref{Hut P., 1996b, in  Dynamical Evolution of Star Clusters:
	Confrontation of Theory and Observations, IAU Symp. 174, 
	eds. Hut P. \& Makino J. (Dordrecht: Kluwer), p. 121}
\ref{Hut P., Breeden J.L., Cohn H., Makino J., McMillan S.L.W., 1989,
	in  Dynamics of Dense Stellar Systems,
	ed. Merritt D., (Cambridge: Cambridge University Press), p. 237}
\ref{Hut P., Djorgovski S.G., 1992, Nature, 359, 806}
\ref{Hut P., Inagaki S., 1985, ApJ, 298, 502}
\ref{Hut P., Makino J., McMillan S.L.W., 1988, Nature, 336, 31}
\ref{Hut P., Makino J., McMillan S.L.W., 1995, Nature, 443, L93}
\ref{Hut P., Makino J., eds., 1996, Dynamical Evolution of Star Clusters: 
	Confrontation of Theory and Observation, 
	IAU Symp. 174, (Dordrecht: Kluwer)}
\ref{Hut P., McMillan S.L.W., Makino J., 1993, Starlab Primer,
	privately distributed}
\ref{Hut, P., McMillan, S.L.W., Goodman, J., \etal, 1992, PASP,
        104, 981}
\ref{Hut P., McMillan S.L.W., Romani R., 1992, ApJ, 389, 527}
\ref{Hut P., Murphy B.W., Verbunt F., 1991, A\&A, 241, 137}
\ref{Hut P., Paczynski B., 1984, ApJ, 284, 675}
\ref{Hut P., Sussman G.J., 1986, 
	in  The Use of Supercomputers in Stellar Dynamics, 
	eds. Hut P. \& McMillan S.L.W., (Berlin: Springer-Verlag), p.
	193}
\ref{Iben I., Renzini A., 1983, ARA\&A, 21, 271}
\ref{Iben I., Tutukov A., 1986, ApJ, 311, 742}
\ref{Ichikawa S., 1985, Ann. Tokyo Astron. Obs., 2$^{nd}$ Ser., 21, 1}
\ref{Illingworth G., Illingworth W., 1976, ApJS, 30, 227}
\ref{Illingworth G., King I.R., 1977, ApJ, 218, L109} 
\ref{Ilovaisky S.A., Auri\`ere M., Koch-Miramond L., \etal, 
	1993, A\&A, 270, 139}
\ref{Inagaki S., 1980, PASJ, 32, 213}
\ref{Inagaki S., 1983, PASJ, 36, 391}
\ref{Inagaki S., 1984, MNRAS, 206, 149}
\ref{Inagaki S., 1985, in  Dynamics of Star Clusters, IAU Symp. 113, 
	eds. Goodman J. \& Hut P., (Dordrecht: Reidel), p. 189}
\ref{Inagaki S., 1986a, in  The Harlow-Shapley
	Symposium on Globular Cluster Systems in Galaxies, IAU Symp. 126,
	eds. Grindlay J.E. \& Philip A.G.D., 
	(Dordrecht: Kluwer), p. 367}
\ref{Inagaki S., 1986b, PASJ, 38, 853}
\ref{Inagaki S., 1986c, Ap\&SS, 118, 367}
\ref{Inagaki S., 1988, Vistas in Astron., 31, 193}
\ref{Inagaki S., Hachisu I., 1978, PASJ, 30, 39}
\ref{Inagaki S., Hut P., 1988, in  The Few Body Problem, 
	IAU Coll. 96., ed. Valtonen M.J., (Dordrecht: Kluwer), p. 319}
\ref{Inagaki S., Lynden-Bell D., 1983, MNRAS, 205, 913}
\ref{Inagaki S., Lynden-Bell D., 1990, MNRAS, 244, 254}
\ref{Inagaki S., Saslaw W.C., 1985, ApJ, 292, 339}
\ref{Inagaki S., Wiyanto P., 1984, PASJ, 36, 391}
\ref{Inman R.T., Carney B.W., 1987, AJ, 93, 1166}
\ref{Innanen K.A., Harris W.E., Webbink R.F., 1983, AJ, 88, 338}
\ref{Ipser J.R., 1974, ApJ, 193, 463}
\ref{Ipser J.R., 1977, ApJ, 218, 846}
\ref{Ipser J.R., Semenzato R., 1983, ApJ, 271, 294}
\ref{Irwin M.J., Demers S., Kunkel W.E., 1995, ApJ, 453, L21}
\ref{Ito T., Makino J., Ebisuzaki T., Sugimoto D., 1990, Comp.
        Phys. Comm., 60, 187}
\ref{Ito T., Ebisuzaki T., Makino J., Sugimoto D., 1991, PASJ, 43, 547}
\ref{Ito T., Makino J., Fukushige T., \etal, 1993, PASJ, 45, 339}
\ref{Jablonka P., Sarajedini A., Bridges T., Meylan G., 1996,  in preparation}
\ref{Jacoby G.H., Branch D., Ciardullo R., \etal, 1992, PASP, 104, 599}
\ref{Janes K., ed., 1991,  The Formation and Evolution of Star Clusters, 
	ASP Conference Series Vol. 13, (San Francisco: ASP)}
\ref{Jeans J.H., 1913, MNRAS, 74, 109}
\ref{Jeans J.H., 1915, MNRAS, 76, 70}
\ref{Jeans J.H., 1916a, MNRAS, 76, 552}
\ref{Jeans J.H., 1916b, MNRAS, 76, 567} 
\ref{Jeans J.H., 1929,  Astronomy and Cosmogony, (Cambridge:
	Cambridge University Press; also New York: Dover Publications
	Inc., 1961)}
\ref{Jefferys, W.H.. 1976, AJ, 81, 983}
\ref{Jensen K.S., J\o rgensen H.E., 1985, A\&AS, 60 229}
\ref{Johnston H.M., Verbunt F., 1996, A\&A, 312, 80}
\ref{Johnston H.M., Verbunt F., Hasinger G.R., 1994, A\&A, 289, 763}
\ref{Johnston H.M., Verbunt F., Hasinger G.R., 1995a, A\&A, 298, L21}
\ref{Johnston H.M., Verbunt F., Hasinger G.R., Bunk W., 1995b,
	in  Compact Stars in Binaries, IAU Symp. 165,
	eds. van Paradijs J., van den Heuvel E.P.J., \& Kuulkers E, 
	(Dordrecht: Kluwer), p. 389}
\ref{Johnstone D., 1993,  AJ, 105, 155}
\ref{Kadla Z.I., 1979, SvA, 23, 555}
\ref{Kaliberda V.S., 1969, Afz, 5, 433}
\ref{Kaliberda V.S., Petrovskaya I.V., 1970, Afz, 6, 135}
\ref{Kaluzny J., Krzeminski W., 1993, MNRAS, 264, 785}
\ref{Kaluzny J., Kubiak M., Szymanski M., \etal, 
	1996a, in  The Origins, Evolution, and Destinies of 
	Binary Stars in Clusters, eds. Milone G. \& Mermilliod J.-C., 
	ASP Conference Series Vol. 90, (San Francisco: ASP), p. 38}
\ref{Kaluzny J., Kubiak M., Szymanski M., \etal, 1996b, A\&A, submitted}
\ref{Kaluzny J., Kubiak M., Szymanski M., \etal, 1996c, A\&A, submitted}
\ref{Kandrup H.E., 1980, Phys. Rep., 63, 1}
\ref{Kandrup H.E., 1983, Ap\&SS, 97, 435}
\ref{Kandrup H.E., 1987, MNRAS, 225, 995}
\ref{Kandrup H.E., 1988, MNRAS, 235, 1157}
\ref{Kandrup H.E., Smith H. Jr., 1991, ApJ, 374, 255}
\ref{Kandrup H.E., Smith H. Jr., 1992, ApJ, 386, 635}
\ref{Kandrup H.E., Willmes D.E., 1994, A\&A, 283, 59}
\ref{Kang H., Shapiro, P.R., Fall, S.M., Rees, M.J., 1990, ApJ, 363, 488}
\ref{Katz J., 1978, MNRAS, 183, 765}
\ref{Katz J., 1980, MNRAS, 190, 497}
\ref{Katz J., Horwitz G., 1977, ApJS, 33, 251}
\ref{Katz J., Horwitz G., Dekel A., 1978, ApJ, 223, 299}
\ref{Katz J., Taff L.G., 1983, ApJ, 264, 476}
\ref{Katzenelson J., 1989, SIAM J. Sci. Stat. Comput., 10, 787}
\ref{Keenan D.W., Innanen K.A., 1975, AJ, 80, 290}
\ref{Kennicutt R.C., Chu Y.-H., 1988, AJ, 95, 720}
\ref{Kim C.-H., Chun M.-S., Min K.W., 1992, As\&SS, 196, 191}
\ref{King I.R., 1958a, AJ, 63, 109}
\ref{King I.R., 1958b, AJ, 63, 114}
\ref{King I.R., 1958c, AJ, 63, 465}
\ref{King I.R., 1960, AJ, 65, 122}
\ref{King I.R., 1962, AJ, 67, 471}
\ref{King I.R., 1965, AJ, 70, 376}
\ref{King I.R., 1966, AJ, 71, 64}
\ref{King I.R., 1985, in  Dynamics of Star Clusters, IAU Symp. 113, 
	eds. Goodman J. \& Hut P., (Dordrecht: Reidel), p. 19}
\ref{King I.R., 1989, in  Dynamics of Dense Stellar Systems,
	ed. Merritt D., (Cambridge: Cambridge University Press), p. 157}
\ref{King I.R., 1996, in  Dynamical Evolution of Star Clusters:
	Confrontation of Theory and Observations, IAU Symp. 174, 
	eds. Hut P. \& Makino J. (Dordrecht: Kluwer), p. 29}
\ref{King I.R., Cool A.M., Piotto G., 1996a,
	in  Formation of the Galactic Halo...Inside and Out, 
	ASP Conference Series Vol. 92, 
	eds. Morrison H. \& Sarajedini A., (San Francisco: ASP), p. 277}
\ref{King I.R., Hedemann, E., Hodge, S.M., White, R.E., 1968, AJ, 73, 456}
\ref{King I.R., Piotto G., Cool A.M., Anderson J., Sosin C., 1996b,
	in the ESO/STScI workshop on
	Science with the Hubble Space Telescope - II, 
	eds. Benvenuti P., Macchetto F.D., \& Schreier E.J.,
	(Baltimore: STScI), p. 297}
\ref{King I.R., Sosin C., Cool A.M., 1995, ApJ, 452, L33}
\ref{King I.R., Stanford, S.A., \etal, 1993, ApJ, 413, L117}
\ref{Kinman T.D., 1959, MNRAS, 119, 559}
\ref{Kiseleva L.G., Eggleton P.P., Orlov V.V., 1994, MNRAS, 267, 161}
\ref{Kiseleva L.G., Aarseth S.J., Eggleton P.P., de la Fuente Marcos R.,
	1996, in  The Origins, Evolution, and Destinies of 
	Binary Stars in Clusters, eds. Milone G. \& Mermilliod J.-C.,
	ASP Conference Series Vol. 90, (San Francisco: ASP), p. 433}
\ref{Klessen R., Burkert A., 1996, MNRAS, 280, 735}
\ref{Knobloch E., 1976, ApJ, 209, 411}
\ref{Knobloch E., 1977, ApJ, 218, 406}
\ref{Kochanek C.S., 1992, ApJ, 385, 604}
\ref{Kondrat'ev B.P., Ozernoi L.M., 1982, Ap\&SS, 84, 431}
\ref{Kormendy J., 1985, ApJ, 295, 73}
\ref{Kouzmine G.G., 1957, Pub. Obs. Tartu, 33, No.2}
\ref{Kraft R.P., 1989, PASP, 101, 1113}
\ref{Krolik J.H., 1983, Nature, 305, 506}
\ref{Krolik J.H., 1984, ApJ, 282, 452}
\ref{Krolik J.H., Meiksin A., Joss P.C., 1984, ApJ, 282, 466}
\ref{Kron G.E., Gordon K.C., 1986, PASP, 98, 544 ~~(A ``Catalogue of
	Concentric Aperture  UBVRI Photoelectric Photometry of 
	Globular Clusters'' is available from the Astronomical Data 
	Center, Goddard Space Flight Center, Code 633.8, Greenbelt, 
	MD 20771, USA)}
\ref{Kron G.E., Hewitt A.V. Wasserman L.H., 1984, PASP, 96, 198}
\ref{Kron G.E., Mayall N.U., 1960, AJ, 65, 581}
\ref{Kroupa P., 1996, MNRAS, 277, 1522}
\ref{Kulkarni S. R., Hut P., McMillan S., 1993, Nature, 364, 421}
\ref{Kulkarni S. R., Narayan R., Romani R.W., 1990, ApJ, 356, 174}
\ref{Kumar P., Goodman J., 1996, ApJ, 466, 946}
\ref{Kundic T., Ostriker J.P., 1995, ApJ, 438, 702}
\ref{Lada C.J., Lada E.A., 1991, in 
	 Formation and Evolution of Star Clusters, 
	ASP Conference Series Vol. 13, 
	ed. Janes K., (San Francisco: ASP), p. 3}
\ref{Lagoute C., Longaretti P.-Y., 1996, A\&A, 308, 441}
\ref{Lai D., Rasio F.A., Shapiro S.L., 1993, ApJ, 412, 593}
\ref{Laird J.B., Carney B.W., Rupen M.P., Latham D.W., 1988, AJ, 96, 1908}
\ref{Larson R.B., 1970a, MNRAS, 147, 323}
\ref{Larson R.B., 1970b, MNRAS, 150, 93}
\ref{Larson R.B., 1981, MNRAS, 194, 809}
\ref{Larson R.B., 1982, MNRAS, 200, 159}
\ref{Larson R.B., 1984, MNRAS, 210, 763}
\ref{Larson R.B., 1986, MNRAS, 218, 409}
\ref{Larson R.B., 1990, PASP, 102, 709}
\ref{Larson R.B., 1992, MNRAS, 256, 641}
\ref{Larson R.B., 1993, 
	in  The Globular Cluster - Galaxy Connection, 
	ASP Conference Series Vol. 48, 
	eds. Smith G.H. \& Brodie J.P., (San Francisco: ASP), p. 675}
\ref{Larson R.B., 1996, 
	in  Formation of the Galactic Halo...Inside and Out, 
	ASP Conference Series Vol. 92, 
	eds. Morrison H. \& Sarajedini A., (San Francisco: ASP), p. 241}
\ref{Latham D.W., 1985, in  Stellar Radial Velocities, 
	IAU Colloquium 88, eds. Philip A.G.D. \& Latham D.W., 
	(Schenectady: L. Davis Press), p. 21}
\ref{Latham D.W., Hazen M.L., Pryor C., 1985, 
	in  Stellar Radial Velocities, 
	IAU Colloquium 88, eds. Philip A.G.D. \& Latham D.W., 
	(Schenectady: L. Davis Press), p. 269}
\ref{Lauer T.R., Holtzman J.A., Faber S.M., \etal, 1991, ApJ, 369, L45}
\ref{Lauer T., Kormendy J., 1986, ApJ, 303, L1}
\ref{Lauzeral C., Ortolani S., Auri\`ere M., Melnick J., 1993, A\&A, 262, 63}
\ref{Layzer D., 1977, Gen. Rel. \& Grav., 8, 3}
\ref{Lecar M., 1968, Bull. Astron., 3, 91}
\ref{Lecar M., Cruz-Gonzalez C., 1971,  Ap\&SS, 13, 360}
\ref{Lee H.M., 1987a, ApJ, 319, 772}
\ref{Lee H.M., 1987b, ApJ, 319, 801}
\ref{Lee H.M., 1990, J. Korean Astron. Soc., 23, 97}
\ref{Lee H.M., 1992, J. Korean Astron. Soc., 25, 47}
\ref{Lee H.M., Fahlman G.G., Richer H.B., 1991, ApJ, 366, 455}
\ref{Lee H.M., Goodman J., 1995, ApJ, 443, 109}
\ref{Lee H.M., Kim S.S., Kang H., 1996, J. Korean Astron. Soc., in press}
\ref{Lee H.M., Ostriker J.P., 1986, ApJ, 310, 176}
\ref{Lee H.M., Ostriker J.P., 1987, ApJ, 322, 123}
\ref{Lee H.M., Ostriker J.P., 1993, ApJ, 409, 617}
\ref{Lee M.G., 1991, 
	in  The Magellanic Clouds, IAU Symp. 148,
	eds. Haynes R. \& Milne D., (Dordrecht: Kluwer), p. 207} 
\ref{Lee Y.W., Demarque P., Zinn R.J., 1990, ApJ, 350, 155}
\ref{Lee Y.W., Demarque P., Zinn R.J., 1994, ApJ, 423, 248}
\ref{Leeuwin F., Combes F., Binney J., 1993, MNRAS, 262, 1013}
\ref{Leonard P.J.T., 1989, AJ, 98, 217}
\ref{Leonard P.J.T., 1996, in  The Origins, Evolution, and Destinies of 
	Binary Stars in Clusters, eds. Milone G. \& Mermilliod J.-C., 
	ASP Conference Series Vol. 90, (San Francisco: ASP), p. 337}
\ref{Leonard P.J.T., Duncan M.J., 1988, AJ, 96, 222}
\ref{Leonard P.J.T., Duncan M.J., 1990, AJ, 99, 608}
\ref{Leonard P.J.T., Fahlman G.G., 1991, AJ, 102, 994}
\ref{Leonard P.J.T., Linnell, A.P., 1992, AJ, 103, 1928}
\ref{Leonard P.J.T., Livio M., 1995, ApJ, 447, L121}
\ref{Leonard P.J.T., Richer H.B., Fahlman G.G., 1992, AJ, 104, 2104}
\ref{Lightman A.P., 1977, ApJ, 215, 914}
\ref{Lightman A.P., Fall S.M., 1977, ApJ, 221, 567}
\ref{Lightman A.P., Press W.H., Odenwald S.F., 1977, ApJ, 219, 629}
\ref{Lightman A.P., Shapiro S.L., 1978, Rev. Mod. Phys., 50, 437}
\ref{Livio M., 1993, in  Blue Stragglers, 
	ASP Conference Series Vol. 53,  ed. Saffer, R.A., 
	(San Francisco: ASP), p. 3}
\ref{Livio M., 1994, in  Interacting Binaries, 
	Saas-Fee Advanced Course 22, eds. Nussbaumer H. \& Orr A.,   
	(Berlin: Springer-Verlag), p. 135}
\ref{Livio M., 1996a, in  The Origins, Evolution, and Destinies of 
	Binary Stars in Clusters, eds. Milone G. \& Mermilliod J.-C., 
	ASP Conference Series Vol. 90, (San Francisco: ASP), p. 291}
\ref{Livio M., 1996b, in  The Origins, Evolution, and Destinies of 
	Binary Stars in Clusters, eds. Milone G. \& Mermilliod J.-C., 
	ASP Conference Series Vol. 90, (San Francisco: ASP), p. 312}
\ref{Lombardi J.C., Rasio F.A., Shapiro S.L., 1995, ApJ, 445, L117}
\ref{Long K., Ostriker J.P., Aguilar L., 1992, ApJ, 388, 362}
\ref{Longaretti P.-Y., Lagoute C., 1996a, A\&A, 308, 453}
\ref{Longaretti P.-Y., Lagoute C., 1996b, A\&A, 308, 441}
\ref{Louis P.D., 1990, MNRAS, 244, 478}
\ref{Louis P.D., 1993, MNRAS, 261, 283}
\ref{Louis P.D., Spurzem R., 1991, MNRAS, 251, 408}
\ref{Luciani J.F., Pellat R., 1987a, ApJ, 317, 241}
\ref{Luciani J.F., Pellat R., 1987b, J. Physique, 48, 591}
\ref{Lupton R.H., 1985, in 
	 Dynamics of Star Clusters, IAU Symp. 113, 
	eds. Goodman J. \& Hut P., (Dordrecht: Reidel), p. 327}
\ref{Lupton R.H., Fall S.M., Freeman K.C., Elson R.A.W., 1989, ApJ, 347, 201}
\ref{Lupton R.H., Gunn J., 1987, AJ, 93, 1106}
\ref{Lupton R.H., Gunn J.E., Griffin R.F., 1985, in 
	 Dynamics of Star Clusters, IAU Symp. 113, 
	eds. Goodman J. \& Hut P., (Dordrecht: Reidel), p. 19}
\ref{Lupton R.H., Gunn J.E., Griffin R.F., 1987, AJ, 93, 1114}
\ref{Lyne A.G., 1995, in  Millisecond Pulsars - A Decade of Surprises,
	eds. Fruchter A.S., Tavani M., Backer D.C., 
	ASP Conference Series Vol. 72, (San Francisco: ASP), p. 35}
\ref{Lyne A.G., Manchester R.N., D'Amico, N., 1996, ApJ, 460, L41}
\ref{Lynden-Bell D., 1962, MNRAS, 124, 279}
\ref{Lynden-Bell D., 1967, MNRAS, 136, 101}
\ref{Lynden-Bell D., 1968, Bull. Astron., 3, 305}
\ref{Lynden-Bell D., 1975, in  Dynamics of Stellar Systems, 
        IAU Symp. 69, ed. Hayli A., (Dordrecht: Reidel), p. 27}
\ref{Lynden-Bell D., Eggleton P.P., 1980, MNRAS, 191, 483}
\ref{Lynden-Bell D., Wood R., 1968, MNRAS, 138, 495}
\ref{Madore B.F., Freedman W.L., 1989, ApJ, 340, 812}
\ref{Madsen J., 1996, MNRAS, in press} 
\ref{Maeder A., 1987, A\&A, 178, 159}
\ref{Magnier E., 1994, in  The Evolution of X-ray Binaries, 
	eds. Holt S.S. \& Day C., (New York: AIP), p. 640}
\ref{Makino J., 1989, in  Dynamics of Dense Stellar Systems,
	ed. Merritt D., (Cambridge: Cambridge University Press), p. 201}
\ref{Makino J., 1991, ApJ, 369, 200}
\ref{Makino J., 1996a, 	in  Dynamical Evolution of Star Clusters:
	Confrontation of Theory and Observations, IAU Symp. 174, 
	eds. Hut P. \& Makino J. (Dordrecht: Kluwer), p. 141 and p. 151}
\ref{Makino J., 1996b, ApJ, 470, in press}
\ref{Makino J., Aarseth S.J., 1992, PASJ, 44, 141}
\ref{Makino J., Hut P., 1989a, Comp. Phys. Rep. 9, 199}
\ref{Makino J., Hut P., 1989b, in  Applications of Computer 
	Technology to Dynamical Astronomy, IAU Coll. 109, ed. Davies
	M.S. (Dordrecht: Kluwer), p. 141}
\ref{Makino J., Hut P., 1990, ApJ, 365, 208}
\ref{Makino J., Hut P., 1991, ApJ, 383, 181}
\ref{Makino J., Ito T., Ebisuzaki T., 1990, PASJ, 42, 717}
\ref{Makino J., Kokubo E., Taiji M., 1993, PASJ, 45, 349}
\ref{Makino J., Sugimoto D., 1987, PASJ, 39, 589}
\ref{Makino J., Tanekusa J., Sugimoto D., 1986, PASJ, 38, 865}
\ref{Malumuth E.M., Heap S.R., 1994, AJ, 107, 1054}
\ref{Mandushev G., Spassova N., Staneva, A., 1991, A\&A, 252, 94}
\ref{Mann P.J., 1987, Comp. Phys. Comm., 47, 213}
\ref{Mansbach P., 1970, ApJ, 160, 135}
\ref{Marchal C., 1980, Acta Astron., 7, 123}
\ref{Marchal C., 1990,  The Three-Body Problem, (Amsterdam: Elsevier)}
\ref{Marchant A.B., Shapiro S.L., 1979, ApJ, 234, 317}
\ref{Marchant A.B., Shapiro S.L., 1980, ApJ, 239, 685}
\ref{Marcy G.W., Butler R.P., 1992, PASP, 104, 270}
\ref{Mardling R.A., 1995a, ApJ, 450, 722}
\ref{Mardling R.A., 1995b, ApJ, 450, 732}
\ref{Mardling R.A., 1996, in  The Origins, Evolution, and Destinies of 
	Binary Stars in Clusters, eds. Milone G. \& Mermilliod J.-C., 
	ASP Conference Series Vol. 90, (San Francisco: ASP), p. 399}
\ref{Margon B., Cannon R., 1989, Observatory, 109, 82}
\ref{Mateo M., 1987, ApJ, 323, L41}
\ref{Mateo M., 1994, in  Dwarf Galaxies, 
	ESO/OHP Workshop, ESO Conf. Vol. 49, 
	eds. Meylan G. \& Prugniel P., (Garching: ESO), p. 309}
\ref{Mateo M., 1996, in  The Origins, Evolution, and Destinies of 
	Binary Stars in Clusters, eds. Milone G. \& Mermilliod J.-C., 
	ASP Conference Series Vol. 90, (San Francisco: ASP), p. 346}
\ref{Mateo M., Harris H.C., Nemec J., Olszewski E.W., 1990, AJ, 100, 469}
\ref{Mateo M., Welch D.L., Fischer P., 1991, 
	in  The Magellanic Clouds, IAU Symp. 148,
	eds. Haynes R. \& Milne D., (Dordrecht: Kluwer), p. 191} 
\ref{Mathieu R.D., 1986, in  Highlights of Astronomy, 
	ed. Swings J.P., (Dordrecht: Kluwer), p. 481}
\ref{Maxwell J.C., 1860, Phil. Mag., January and July}
\ref{Mayor M., 1985, in  Stellar Radial Velocities, IAU Colloquium 88, 
	eds. Philip A.G.D. \& Latham D.W., 
	(Schenectady: L. Davis Press), p. 299}
\ref{Mayor M., Duquennoy A., Alimenti A., Andersen J., Nordstr\"om, B., 
	1996, in  The Origins, Evolution, and Destinies of 
	Binary Stars in Clusters, eds. Milone G. \& Mermilliod J.-C., 
	ASP Conference Series Vol. 90, (San Francisco: ASP), p. 190}
\ref{Mayor M., Imbert M., Andersen J., \etal, 1984, A\&A, 134, 118}
\ref{Mayor M., Queloz D., 1995, Nature, 378, 355}
\ref{Mazzitelli I., D'Antona F., Caloi V., 1995, A\&A, 302, 382}
\ref{McClure R.D., Fletcher J.M., Grundmann W.A., Richardson E.H., 1985, 
	in  Stellar Radial Velocities, IAU Colloquium 88, 
	eds. Philip A.G.D. \& Latham D.W.,
	(Schenectady: L. Davis Press), p. 49}
\ref{McClure R.D., VandenBerg D.A., Smith G.H., \etal, 1986, ApJ, 307, L49}
\ref{McCrea W.H., 1964, MNRAS, 128, 147}
\ref{McCrea W.H., 1982, in  Progress in Cosmology,
	ed. Wolfendale A.W., (Dordrecht: Reidel), p. 239}
\ref{McKenna J., Lyne A.G., 1988, Nature, 336, 226}
\ref{McLaughlin D.E., Pudritz R.E., 1996, ApJ, 457, 578}
\ref{McMillan S.L.W., 1986a, ApJ, 306, 552}
\ref{McMillan S.L.W., 1986b, ApJ, 307, 126}
\ref{McMillan S.L.W., 1993, 
	in  Structure and Dynamics of Globular Clusters, 
	ASP Conference Series Vol. 50, 
	eds. Djorgovski S.G. \& Meylan G., (San Francisco: ASP), p. 171}
\ref{McMillan S.L.W., 1996, in  The Origins, Evolution, and Destinies of 
	Binary Stars in Clusters, eds. Milone G. \& Mermilliod J.-C., 
	ASP Conference Series Vol. 90, (San Francisco: ASP), p. 413}
\ref{McMillan S.L.W., Aarseth S.J., 1993, ApJ, 414, 200}
\ref{McMillan S.L.W., Casertano S., Hut P., 1987, 
	in  The Few Body Problem,
	IAU Colloquium 96., ed. Valtonen M.J., (Dordrecht: Kluwer), p. 313}
\ref{McMillan S., Hut P., 1994, ApJ, 427, 793}
\ref{McMillan S.L.W., Hut P., Makino J., 1990, ApJ, 362, 522}
\ref{McMillan S.L.W., Hut P., Makino J., 1991, ApJ, 372, 111}
\ref{McMillan S.L.W., Lightman A.P., 1984, ApJ, 283, 801}
\ref{McMillan S.L.W., McDermott P.N., Taam R.E., 1987, ApJ, 318, 261}
\ref{Merritt D., 1981, AJ, 86, 318}
\ref{Merritt D., 1985a, AJ, 90, 1027}
\ref{Merritt D., 1985b, MNRAS, 214, 25p}
\ref{Merritt D., 1987a, 
	in  Structure and Dynamics of Elliptical Galaxies, 
	IAU Symp. 127, ed. de Zeeuw, T., (Dordrecht: Reidel), p. 315.}
\ref{Merritt D., 1987b, ApJ, 313, 121}
\ref{Merritt D., ed., 1989,  Dynamics of Dense Stellar Systems,
	(Cambridge: Cambridge University Press)}
\ref{Merritt D., 1990, in  Galactic Models,  
	Proc. 4$^{th}$ Florida Workshop on Nonlinear Dynamics, 
	NY Acad. Sci. Annals, 596, 150}
\ref{Merritt D., 1993, in  Structure and Dynamics of Globular Clusters, 
	ASP Conference Series Vol. 50, 
	eds. Djorgovski S.G. \& Meylan G., (San Francisco: ASP), p. 117}
\ref{Merritt D., 1993a, ApJ, 413, 79}
\ref{Merritt D., 1993b, ApJ, 409, 75}
\ref{Merritt D., 1996, AJ, submitted, astro-ph/9603160}
\ref{Merritt D., Hernquist L., 1991, ApJ, 376, 439}
\ref{Merritt D., Meylan G., Mayor M., 1996, submitted}
\ref{Merritt D., Saha P., 1993, ApJ, 409, 75}
\ref{Merritt D., Tremblay B., 1994, AJ, 108, 514}
\ref{Messier C. 1784,  Catalogue des N\'ebuleuses et des Amas 
	d'\'Etoiles Observ\'es \`a Paris, Connaissance des Temps, 
	(Paris: Acad\'emie Royale des Sciences), p. 227}
\ref{Meyer F., Meyer-Hofmeister E., 1980, 
	in  Close binary stars: Observations and Interpretation, 
	IAU Symp. 88, ed. Plavec M.J., Popper D.M., \& Ulrich R.K., 
	(Dordrecht: Reidel), p. 145.}
\ref{Meylan G., 1987, A\&A, 184, 144}
\ref{Meylan G., 1988a, A\&A, 191, 215}
\ref{Meylan G., 1988b, ApJ, 331, 718}
\ref{Meylan G., 1989, A\&A, 214, 106}
\ref{Meylan G., 1993, in  The Globular Cluster - Galaxy Connection, 
	ASP Conference Series Vol. 48, 
	eds. Smith G.H. \& Brodie J.P., (San Francisco: ASP), p. 588}
\ref{Meylan G., 1994, in  Ergodic Concepts in Stellar Dynamics, 
	Lecture Notes In Physics, Vol. 430, 
	eds. Gurzadyan V.G., \& Pfenniger D., (Berlin: Springer-Verlag), p. 22}
\ref{Meylan G., 1996, 
	in  Dynamical Evolution of Star Clusters:
	Confrontation of Theory and Observations, IAU Symp. 174, 
	eds. Hut P. \& Makino J. (Dordrecht: Kluwer), p. 61}
\ref{Meylan G., Djorgovski S.G., 1987, ApJ, 322, L91}
\ref{Meylan G., Dubath P., Mayor M., 1989, BAAS, 21, 711}
\ref{Meylan G., Dubath P., Mayor M., 1991a, ApJ, 383, 587}
\ref{Meylan G., Dubath P., Mayor M., 1991b, BAAS, 23, 833}
\ref{Meylan G., Dubath P., Mayor M., 1991c, 
	in  The Magellanic Clouds, IAU Symp. 148,
	eds. Haynes R. \& Milne D., (Dordrecht: Kluwer), p. 211} 
\ref{Meylan G., Mayor M., 1986, A\&A, 166, 122}
\ref{Meylan G., Mayor M., 1991, A\&A, 250, 113}
\ref{Meylan G., Mayor M., Duquennoy A., Dubath P., 1994, BAAS, 26, 956}
\ref{Meylan G., Mayor M., Duquennoy A., Dubath P., 1995, A\&A, 303, 761}
\ref{Meylan G., Minniti D., Pryor C., Tinney C., Phinney E.S., Sams B., 
	1996, in the ESO/STScI workshop on
	Science with the Hubble Space Telescope - II, 
	eds. Benvenuti P., Macchetto F.D., \& Schreier E.J.,
	(Baltimore: STScI), p. 316}
\ref{Meylan G., Pryor C., 1993, 
	in  Structure and Dynamics of Globular Clusters, 
	ASP Conference Series Vol. 50, 
	eds. Djorgovski S.G. \& Meylan G., (San Francisco: ASP), p. 31}
\ref{Meziane K., Colin J., 1996, A\&A, 306, 747}
\ref{Michie R.W., 1961, ApJ, 133, 781}
\ref{Michie R.W., 1963a, MNRAS, 125, 127}
\ref{Michie R.W., 1963b, MNRAS, 126, 269}
\ref{Michie R.W., 1963c, MNRAS, 126, 331}
\ref{Michie R.W., 1963d, MNRAS, 126, 499}
\ref{Michie R.W., 1964, ARA\&A, 2, 49} 
\ref{Mighell K.J., Rich R.M., Shara M., Fall S.M., 1996, AJ, in press}
\ref{Mikkola S., 1983a, MNRAS, 203, 1107}
\ref{Mikkola S., 1983b, MNRAS, 205, 733}
\ref{Mikkola S., 1984a, MNRAS, 207, 115}
\ref{Mikkola S., 1984b, MNRAS, 208, 75}
\ref{Mikkola S., 1985, MNRAS, 215, 171}
\ref{Mikkola S., Aarseth S.J., 1990, Celes. Mech. Dyn. Astron., 47,375}
\ref{Mikkola S., Aarseth S.J., 1993, Celes. Mech. Dyn. Astron., 57, 439} 
\ref{Milgrom M., Shapiro S.L., 1978, ApJ, 223, 991}
\ref{Miller G.E., Scalo J.M., 1979, ApJS, 41, 513}
\ref{Miller R.H., 1964, ApJ, 140, 250}
\ref{Miller R.H., 1971, J. Comp. Physics, 8, 449}
\ref{Miller R.H., 1973, ApJ 180, 759}
\ref{Milone A.A.E., Latham D.W., 1992,
	in  Evolutionary Processes in Interacting Binary Stars, 
	IAU Symp. 151, eds. Kondo Y., Sistero R.F., \& Polidan R.S., 
	(Dordrecht: Kluwer), p. 475} 
\ref{Milone E.F., Mermilliod J.-C., eds., 1996, 
	 The Origins, Evolution, and Destinies of 
	Binary Stars in Clusters, 
	ASP Conference Series Vol. 90, (San Francisco: ASP)}
\ref{Minniti D., 1995, AJ, 109, 1663}
\ref{Minniti D., Meylan G., Kissler-Patig M., 1996, A\&A, 312, 49}
\ref{Mitalas R., 1995, A\&A, 302, 924}
\ref{Monaghan J.J., 1976a, MNRAS, 176, 63}
\ref{Monaghan J.J., 1976b, MNRAS, 177, 583}
\ref{Moore B., 1993, ApJ, 413, L93}
\ref{Moore B., 1996, ApJ, 461, L13}
\ref{Morgan S., Lake G., 1989, ApJ, 339, 171}
\ref{Morgan W.W., 1959, AJ, 64, 432}
\ref{Mukherjee K., Anthony-Twarog B., Twarog B.A., 1992, PASP, 104, 561}
\ref{Murali C., Weinberg M.D., 1996, MNRAS, submitted}
\ref{Murphy B.W., Cohn H.N., 1988, MNRAS, 232, 835}
\ref{Murphy B.W., Cohn H.N., Hut P., 1990, MNRAS, 245, 335}
\ref{Murray S.D., Clarke C.J., Pringle J.E., 1991, ApJ, 383, 192}
\ref{Murray S.D., Clarke C.J., 1993, MNRAS, 265, 169}
\ref{Murray S.D., Lin, D.N.C., 1989a, ApJ, 339, 933}
\ref{Murray S.D., Lin, D.N.C., 1989b, ApJ, 346, 155}
\ref{Murray S.D., Lin, D.N.C., 1990a, ApJ, 357, 105}
\ref{Murray S.D., Lin, D.N.C., 1990b, ApJ, 363, 50}
\ref{Murray S.D., Lin, D.N.C., 1991, ApJ, 367, 149}
\ref{Murray S.D., Lin, D.N.C., 1992, ApJ, 400, 265}
\ref{Murray S.D., Lin, D.N.C., 1993, in
	 The Globular Cluster - Galaxy Connection, 
	ASP Conference Series Vol. 48, 
	eds. Smith G.H. \& Brodie J.P., (San Francisco: ASP), p. 738}
\ref{Murtagh F., Heck, A., 1987,  Multivariate Data Analysis,
	(Dordrecht: Reidel)}
\ref{Muzzio J.C., 1987, PASP, 99, 245}
\ref{Myers P.C., Fuller G.A., 1993, ApJ, 402, 635}
\ref{Nakada Y., 1978, PASJ, 30, 57}
\ref{Nash P.E., Monaghan J.J., 1978, MNRAS, 184, 119}
\ref{Nash P.E., Monaghan J.J., 1980, MNRAS, 192, 809}
\ref{Nemec J.M., Harris H.C., 1987, ApJ, 316, 172}
\ref{Newell B., Da Costa G.S., Norris J.E., 1976, ApJ, 208, L55}
\ref{Newell B., O'Neil E.J., 1978, ApJS, 37, 27} 
\ref{Ninkovi\'c S., 1985, Astron. Nachr., 306, 237}
\ref{Niss B., J\o rgensen H.E., Lautsen S., 1978, A\&AS, 32, 387}
\ref{Nomoto K., Kondo Y., 1991, ApJ, 367, L19}
\ref{Norris, J.E., Da Costa G.S., 1995, ApJ, 441, L81}
\ref{O'Connell R.W., Gallagher J.S., Hunter D.A., 1994, ApJ, 433, 65}
\ref{O'Connell R.W., Gallagher J.S., Hunter D.A., Colley W.N.,
	1995, ApJ, 446, L1}
\ref{Oh K.S., Lin D.N.C., 1992, ApJ, 386, 519}
\ref{Oh K.S., Lin D.N.C., Aarseth S.J., 1992, ApJ, 386, 506}
\ref{Okazaki T., Tosa M., 1995, MNRAS, 274, 48}
\ref{Okumura S.K., Makino J., Ebisuzaki T., Ito T.,
        Fukushige T., Sugimoto D., Hashimoto E., Tomida K.,
        Miyakawa N., 1992, in Proc. 25$^{th}$ Hawaii Intl. Conf. 
	IEEE Comput. Soc. Press, vol.1, 151}
\ref{Olszewski E.W., Pryor C., Armandroff T.E., 1996, AJ, 111, 750}
\ref{Oort J.H., 1977, ApJ, 218, L97}
\ref{Oort J.H., van Herk, G., 1959, Bull. Astron. Inst. Neth., 14, 299}
\ref{Ortolani S., Bica E., Barbuy B., 1993, A\&A, 273, 415}
\ref{Ortolani S., Bica E., Barbuy B., 1995, A\&A, 300, 726}
\ref{Ostriker J.P., Spitzer L. Jr., Chevalier, R.A., 1972, ApJ, 176, L51}
\ref{Osipkov L.P., 1979, Pis'ma. Astr. Zh., 5, 77}
\ref{Ouellette J., Pritchet C., 1996, in  
	The Origins, Evolution, and Destinies of 
	Binary Stars in Clusters, eds. Milone G. \& Mermilliod J.-C., 
	ASP Conference Series Vol. 90, (San Francisco: ASP), p. 356}
\ref{Padmanabhan T., 1989a, ApJS, 71, 651}
\ref{Padmanabhan T., 1989b, ApJ, 344, 848}
\ref{Padmanabhan T., 1990, Phys. Rep., 188, 285}
\ref{Palmer P.L., 1994,  Stability of Collisionless Stellar Systems: 
	Mechanisms for the Dynamical Structure of Galaxies, 
	(Kluwer: Dordrecht)}
\ref{Palmer P.L., Papaloizou J., 1987, MNRAS, 224, 1043}
\ref{Palmer P.L., Papaloizou J., Allen A.J., 1990, MNRAS, 243, 282}
\ref{Paresce F., De Marchi G., 1994, ApJ, 427, L33}
\ref{Paresce F., De Marchi G., Ferraro F.R., 1992, Nature, 360, 46}
\ref{Paresce F., De Marchi G., Romaniello M., 1995, ApJ, 440, 216}
\ref{Paresce F., Shara M.M., Meylan G., \etal, 1991, Nature, 352, 297}
\ref{Parisot J.-P., Severne G., 1979, Ap\&SS, 61, 121}
\ref{Patel K., Pudritz R.E., 1994, ApJ, 424, 688}
\ref{Paturel G., Garnier R., 1992, A\&A, 254, 93}
\ref{Peebles P.J.E., 1984, ApJ, 277, 470}
\ref{Peebles P.J.E., Dicke R.H., 1968, ApJ, 154, 891}
\ref{Peng W., Weisheit J.C., 1992, MNRAS, 258, 476}
\ref{Pesce E., Capuzzo-Dolcetta R., Vietri M., 1992, MNRAS, 254, 466}
\ref{Peterson C.J., King I.R., 1975, AJ, 80, 427}
\ref{Peterson C.J., 1993a, 
	in  Structure and Dynamics of Globular Clusters, 
	ASP Conference Series Vol. 50, 
	eds. Djorgovski S.G. \& Meylan G., (San Francisco: ASP), p. 337}
\ref{Peterson R.C., 1993b, 
	in  Structure and Dynamics of Globular Clusters, 
	ASP Conference Series Vol. 50, 
	eds. Djorgovski S.G. \& Meylan G., (San Francisco: ASP), p. 65}
\ref{Peterson R.C., Latham D.W., 1986, ApJ, 305, 645}
\ref{Peterson R.C., Seitzer P., Cudworth K.M., 1989, ApJ, 347, 251}
\ref{Peterson R.C., Cudworth K.M., 1994, ApJ, 420, 612}
\ref{Peterson R.C., Rees R.F., Cudworth K.M., 1995, ApJ, 443, 124}
\ref{Petrou M., 1983a, MNRAS, 202, 1195} 
\ref{Petrou M., 1983b, MNRAS, 202, 1209}
\ref{Petrovskaya I.V., 1969a, AZh, 46, 824}
\ref{Petrovskaya I.V., 1969b, AZh, 46, 1220}
\ref{Petrovskaya I.V., 1971, AZh, 48, 309}
\ref{Pfenniger D., 1986, A\&A, 165, 74}
\ref{Phinney E.S., 1992, Phil. Trans. R. Soc. London, A, 341, 39}
\ref{Phinney E.S., 1993, 
        in  Structure and Dynamics of Globular Clusters,
        ASP Conference Series Vol. 50,
        eds. Djorgovski S.G. \& Meylan G., (San Francisco: ASP), p. 141}
\ref{Phinney E.S., 1996, in  The Origins, Evolution, and Destinies of 
	Binary Stars in Clusters, eds. Milone G. \& Mermilliod J.-C., 
	ASP Conference Series Vol. 90, (San Francisco: ASP), p. 163}
\ref{Phinney E.S., Kulkarni S.R., 1994, ARA\&A, 32, 591}
\ref{Phinney E.S., Sigurdsson S., 1991, Nature, 349, 220}
\ref{Picard A., Johnston H.M., 1994, A\&A, 283, 76}
\ref{Picard A., Johnston H.M., 1996, A\&A, in press}
\ref{Pickering E.C., 1897, Ann. Harvard Coll. Obs., 26, 213} 
\ref{Piotto G., Cool A.M., King I.R., 1996a, in preparation}
\ref{Piotto G., Cool A.M., King I.R., 1996b, 
	in  Dynamical Evolution of Star Clusters:
	Confrontation of Theory and Observations, IAU Symp. 174, 
	eds. Hut P. \& Makino J. (Dordrecht: Kluwer), p. 71}
\ref{Piotto G., King I.R., Djorgovski S.G., 1988, AJ, 96, 1918}
\ref{Plummer H.C., 1911, MNRAS, 71, 460}
\ref{Plummer H.C., 1915, MNRAS, 76, 107}
\ref{Plummer W.E., 1905, MNRAS, 65, 801}
\ref{Podsiadlowski P., 1996, MNRAS, 279, 1104}
\ref{Podsiadlowski P., Price N.M., 1992, Nature, 359, 305}
\ref{Prata S.W., 1971a, AJ, 76, 1017}
\ref{Prata S.W., 1971b, AJ, 76, 1029}
\ref{Prendergast K.H., Tomer E., 1970, AJ, 75, 674}
\ref{Press W.H., Spergel D.N., 1988, ApJ, 325, 715}
\ref{Press W.H., Teukolsky S.A., 1977, ApJ, 213, 183}
\ref{Pritchet C.J., Glaspey J.W., 1991, ApJ, 373, 105}
\ref{Procter A., Bailyn C., Demarque P., 1996, in 
	 The Origins, Evolution, and Destinies of 
	Binary Stars in Clusters, eds. Milone G. \& Mermilliod J.-C., 
	ASP Conference Series Vol. 90, (San Francisco: ASP), p. 380}
\ref{Pryor C., 1994, in  Dwarf Galaxies, 
	ESO/OHP Workshop, ESO Conf. Vol. 49, 
	eds. Meylan G. \& Prugniel P., (Garching: ESO), p. 323}
\ref{Pryor C., Latham D.W., Hazen M.L., 1988, AJ, 96, 123}
\ref{Pryor C., McClure R.D., Fletcher J.M., Hartwick F.D.A., 
	Kormendy J., 1986, AJ, 91, 546}
\ref{Pryor C., McClure R.D., Fletcher J.M., Hesser J.E., 1989a,
	AJ, 98, 596}
\ref{Pryor C., McClure R.D., Fletcher J.M., Hesser J.E., 1989b,
	in  Dynamics of Dense Stellar Systems,
	ed. Merritt D., (Cambridge: Cambridge University Press), p. 175}
\ref{Pryor C., McClure R.D., Fletcher J.M., Hesser J.E., 1991,
	AJ, 102, 1026}
\ref{Pryor C., Meylan G., 1993, 
	in  Structure and Dynamics of Globular Clusters, 
	ASP Conference Series Vol. 50, 
	eds. Djorgovski S.G. \& Meylan G., (San Francisco: ASP), p. 357}
\ref{Quinlan G.D., 1996, New Astron., submitted; astro-ph/9606182}
\ref{Quinlan G.D., Tremaine S., 1991, in  Dynamics of Disk Galaxies,
	ed. Sundelius B., (G\"oteberg: G\"oteberg University Press), p. 143}
\ref{Quinlan G.D., Tremaine S., 1992, MNRAS, 259, 505}
\ref{Racine R., Harris W., 1989, AJ, 98, 1609}
\ref{Racine R., McClure R.D., 1989, PASP, 101, 731}
\ref{Raine A.R.C., Fincham D., Smith W., 1989, Comp. Phys. Comm., 55, 13}
\ref{Rappaport S., Putney A., Verbunt F., 1989, ApJ, 345, 210}
\ref{Rasio F.A., 1993, PASP, 105, 973}
\ref{Rasio F.A., 1996a, in  The Origins, Evolution, and Destinies of 
	Binary Stars in Clusters, eds. Milone G. \& Mermilliod J.-C., 
	ASP Conference Series Vol. 90, (San Francisco: ASP), p. 368}
\ref{Rasio F.A., 1996b, 
	in  Dynamical Evolution of Star Clusters:
	Confrontation of Theory and Observations, IAU Symp. 174, 
	eds. Hut P. \& Makino J. (Dordrecht: Kluwer), p. 253}
\ref{Rasio F.A., Heggie D.C., 1995, ApJ, 445, L133}
\ref{Rasio F.A., McMillan S.,  Hut P., 1995, ApJ, 438, L33}
\ref{Rauch K.P., Tremaine S., 1996, New Astron., submitted}
\ref{Ray A., Kembhavi A.K., Antia H.M., 1987, A\&A, 184, 164}
\ref{Ray A., Kembhavi A.K, 1988, Mod. Phys. Letters, 3, 229}
\ref{Rees M.J., Ostriker J.P., 1977, MNRAS, 179, 541}
\ref{Rees R.F., 1992, AJ, 103, 1573}
\ref{Rees R.F., 1993, AJ, 106, 1524}    
\ref{Rees R.F., Cudworth K.M., 1991, AJ, 102, 152}
\ref{Reijns R., Le Poole R., de Zeeuw T., Seitzer P., Freeman K.C., 1993, 
	in  Structure and Dynamics of Globular Clusters, 
	ASP Conference Series Vol. 50, 
	eds. Djorgovski S.G. \& Meylan G., (San Francisco: ASP), p. 79}
\ref{Renard M., Chi\`eze J.-P., 1993, A\&A, 267, 549}
\ref{Renzini A., 1983, Mem. Soc. Astron. Ital., 54, 335}
\ref{Renzini A., 1991, in  The Magellanic Clouds, IAU Symp. 148,
	eds. Haynes R. \& Milne D., (Dordrecht: Kluwer), p. 165} 
\ref{Retterer J.M., 1980a, PhD thesis, University of California, 
        Berkeley}
\ref{Retterer J.M., 1980b, AJ, 85, 249}
\ref{Retterer J.M., 1979, AJ, 84, 370}
\ref{Rich R.M., \etal, ApJL, 1996, in preparation}
\ref{Richer H.B., Crabtree D.R., Fabian A.C., Lin D.N.C., 1993, 
	AJ, 105, 877}
\ref{Richer H.B., Fahlman G.G., 1989, ApJ, 339, 178}
\ref{Richer H.B., Fahlman G.G., 1992, Nature, 358, 383}
\ref{Richer H.B., Fahlman G.G., Buonanno R., Fusi Pecci F., 1990,
	ApJ, 359, L11}
\ref{Richer H.B., Fahlman G.G., Buonanno R., \etal, 1991, ApJ, 381, 147}
\ref{Richer H.B., Fahlman G.G., Ibata R.A., \etal, 
	1995, ApJ, 451, L17}  
\ref{Richer H.B., Harris W.E., Fahlman G.G., \etal, 1996, ApJ, 463, 602}
\ref{Robinson C., Lyne A.G., Manchester R.N., \etal, 1995, MNRAS, 274, 547}
\ref{Rodgers A.W., Roberts W.H., 1994, AJ, 107, 1737}
\ref{Romani R.W., Kulkarni S.R., Blandford R.D., 1987, Nature, 329, 309}
\ref{Rosenblatt E.I., Faber S.M., Blumenthal G.R., 1988, ApJ, 330, 191}
\ref{Rosenbluth M.N., MacDonald W.M., Judd D.L., 1957, Phys. Rev., 107, 1}
\ref{Ross D.J., Mennim A., Heggie D.C., 1996, MNRAS, in press}
\ref{Rozyczka M., Yorke H.W., Bodenheimer P., Mueller E., Hashimoto M., 
	1989, A\&A, 208, 69}
\ref{Rubenstein E.P., Bailyn C.D., 1996, AJ, 111, 260}
\ref{Ruderman M., Shaham J., Tavani M., Eichler D., 1989, ApJ, 343, 292}
\ref{Ruffert M., 1993, A\&A, 280, 141}
\ref{Ruffert M., Mueller E., 1990, A\&A, 238, 116}
\ref{Ruth R.D., 1983, IEEE Trans. Nucl. Sci. NS, 30, 2669}
\ref{Ryden B.S., 1996, ApJ, 461, 146}
\ref{Saffer, R.A., ed., 1993,  Blue Stragglers, 
	ASP Conference Series Vol. 53, (San Francisco: ASP)}
\ref{Sagar R., Richtler T., 1991, A\&A, 250, 324}
\ref{Saha A., 1985, ApJ, 289, 310} 
\ref{Saha P., 1992, MNRAS, 254, 132}
\ref{Saito M., 1976, A\&A, 46, 171}
\ref{Saito M., Yoshizawa M., 1976, Ap\&SS, 41, 63}
\ref{Sakagami M., Gouda N., 1991, MNRAS, 249, 241}
\ref{Salpeter E.E., 1955, ApJ, 121, 161}
\ref{Sams B.J., 1995, ApJ, 445, 221}
\ref{Sandage A., 1953, AJ, 58, 61}
\ref{Sandage A., 1954, AJ, 59, 162}
\ref{Sandage A., 1990, JRASC, 84, 70}
\ref{Santiago B.X., Djorgovski S.G., 1993, MNRAS, 261, 753}
\ref{Santiago B.X., Elson R.A.W., Gilmore G.F., 1996, MNRAS, 281, 1363}
\ref{Sarajedini A., 1993, in  Blue Stragglers, 
	ASP Conference Series Vol. 53,  ed. Saffer, R.A., 
	(San Francisco: ASP), p. 14}
\ref{Sarajedini A., Demarque P., 1990, ApJ, 365, 219}
\ref{Sarajedini A., Lee Y.-W., Lee D.-H., 1995, ApJ, 450, 712}
\ref{Saslaw W.C., 1973, PASP, 85, 5}
\ref{Saslaw W.C., 1987,  Gravitational Physics of Stellar and 
        Galactic Systems, (Cambridge: Cambridge University Press)}
\ref{Sazhin M. V., Saphonova M.V., 1993, Ap\&SS, 208, 93}
\ref{Scalo J., 1986, Fund. Cosm. Phys., 11, 1}
\ref{Schaeffer R., Maurogordato S., Cappi A., Bernardeau F., 1993,
	MNRAS, 263, L21}
\ref{Schechter P.L., Mateo M., Saha A., 1993, PASP, 105, 1342}
\ref{Schweizer F., 1987, in  Nearly Normal Galaxies, 
	ed. Faber S.M., (New York: Springer-Verlag), p. 18} 
\ref{Schweizer F., Seitzer P., 1993, ApJ, 417, L29}
\ref{Searle L., Zinn R., 1978, ApJ, 225, 357}
\ref{Seitzer P., 1983, PhD Thesis, University of Virginia}
\ref{Seitzer P., 1991, in  The Magellanic Clouds, IAU Symp. 148,
	eds. Haynes R. \& Milne D., (Dordrecht: Kluwer), p. 213} 
\ref{Severne G., Luwel M., 1984, Phys. Lett., 104A, 127}
\ref{Shapiro S.L., 1977, ApJ, 217, 281}
\ref{Shapiro S.L., 1985, in  Dynamics of Star Clusters, 
        IAU Symp. 113, eds. Goodman J. \& Hut P.,
        (Dordrecht: Reidel), p. 373}
\ref{Shapiro S.L., Marchant A.B., 1976, ApJ, 210, 757}
\ref{Shapiro S.L., Marchant A.B., 1978, ApJ, 225, 603}
\ref{Shapley H., 1930, Star Clusters, (New York: McGraw-Hill)}
\ref{Shara M.M., Bergeron L.E., Moffat A.F.J., 1994, ApJ, 429, 767}
\ref{Shara M.M., Drissen L., 1995, ApJ, 448, 203}
\ref{Shara M.M., Drissen L., Bergeron L.E., Paresce F., 1995, ApJ, 441, 617}
\ref{Shara M.M., Regev O., 1986, ApJ, 306, 543}
\ref{Shi X., 1995, ApJ, 446, 637}
\ref{Shi X., Schramm D.N., Dearborn D.S.P., Truran J.W., 1995,
        Comm. Ap. 17, 343}
\ref{Shiveshwarkar S.W., 1936, MNRAS, 36, 749} 
\ref{Shull J.M., 1979, ApJ, 231, 534}
\ref{Shu F.H., 1978, ApJ, 225, 83}
\ref{Shu F.H., 1987, ApJ, 316, 502}
\ref{Shu F.H., Adams F.C., Lizano S., 1987, ARA\&A, 25, 23}
\ref{Sigurdsson S., 1991, PhD Thesis, California Institute of Technology}
\ref{Sigurdsson S., 1992, ApJ, 399, L95}
\ref{Sigurdsson S., 1993, ApJ, 415, L43}
\ref{Sigurdsson S., Davies M., Bolte M., 1994, ApJ, 431, L115}
\ref{Sigurdsson S., Hernquist L., 1993, Nature, 364, 423}
\ref{Sigurdsson S., Phinney E.S., 1993, ApJ, 415, 631}
\ref{Sigurdsson S., Phinney E.S., 1995, ApJS, 99, 609}
\ref{Silk J., 1977, ApJ, 210, 638}
\ref{Silk J., Wyse R.F.G., 1993, Physics Reports, 231, 293}
\ref{Sills A.P., Bailyn C.D., Demarque P., 1996, ApJ, 455, L163}
\ref{Smarr L.L., Blandford R., 1976, ApJ, 207, 574}
\ref{Smith G.H., 1986, ApJ, 306, 565}
\ref{Smith G.H., 1987, PASP, 99, 67}
\ref{Smith G.H., 1996, PASP, 108, 176}
\ref{Smith G.H., Brodie J.P., eds., 1993,  The Globular Cluster - Galaxy
	Connection, ASP Conference Series Vol. 48, 
	(San Francisco: ASP)}
\ref{Smith H. Jr., 1977, A\&A, 61, 305}
\ref{Smith H. Jr., 1979, A\&A, 76, 192}
\ref{Smith H. Jr., 1992, ApJ, 398, 519} 
\ref{Soker N., Regev O., Livio M., Shara M.M., 1987, ApJ, 318, 760}
\ref{Sosin C., King I.R., 1995, AJ, 109, 639}
\ref{Sosin C., King I.R., 1996, 
	in  Dynamical Evolution of Star Clusters:
	Confrontation of Theory and Observations, IAU Symp. 174, 
	eds. Hut P. \& Makino J. (Dordrecht: Kluwer), p. 343}
\ref{Sosin C., \etal, ApJL, 1996, in preparation}
\ref{Spergel D.N., Hernquist L., 1992, ApJ, 397, L75}
\ref{Spitzer L. Jr., 1940, MNRAS, 100, 396}
\ref{Spitzer L. Jr., 1957, ApJ, 127, 17}
\ref{Spitzer L. Jr., 1969, ApJ, 158, L139}
\ref{Spitzer L. Jr., 1984, Science, 225, 465}
\ref{Spitzer L. Jr., 1985, in 
	 Dynamics of Star Clusters, IAU Symp. 113, 
	eds. Goodman J. \& Hut P., (Dordrecht: Reidel), p. 109}
\ref{Spitzer L. Jr., 1987,  Dynamical Evolution of Globular Clusters,
	(Princeton: Princeton University Press)}
\ref{Spitzer L. Jr., Chevalier R.A., 1973, ApJ, 183, 565}
\ref{Spitzer L. Jr., H\"arm R., 1958, ApJ, 127, 544}
\ref{Spitzer L. Jr., Hart M.H., 1971a, ApJ, 164, 399}
\ref{Spitzer L. Jr., Hart M.H., 1971b, ApJ, 166, 483}
\ref{Spitzer L. Jr., Mathieu R.D., 1980, ApJ, 241, 618}
\ref{Spitzer L. Jr.,  Shapiro S.L., 1972, ApJ, 173, 529}
\ref{Spitzer L. Jr., Shull J.M., 1975a, ApJ, 200, 339}
\ref{Spitzer L. Jr., Shull J.M., 1975b, ApJ, 201, 773}
\ref{Spitzer L. Jr., Thuan T.X., 1972, ApJ, 175, 31}
\ref{Spurzem R., 1991, MNRAS, 252, 177}
\ref{Spurzem R., 1996, J. Comp. Phys., submitted}
\ref{Spurzem R., Aarseth S.J., 1996, MNRAS, 282, 19}
\ref{Spurzem R., Giersz M., 1996, MNRAS, in press}
\ref{Spurzem R., Louis P.D., 1993, 
	in  Structure and Dynamics of Globular Clusters, 
	ASP Conference Series Vol. 50, 
	eds. Djorgovski S.G. \& Meylan G., (San Francisco: ASP), p. 135}
\ref{Spurzem R., Takahashi K., 1995,  MNRAS, 272, 772}
\ref{Standish E.M., Aksnes K., 1969, ApJ, 158, 519}
\ref{Statler T.S., Ostriker J.P., Cohn H.N., 1987, ApJ, 316, 626}
\ref{Stebbins J., Whitford A.E., 1943, ApJ, 98, 20}
\ref{Stella L., Priedhorsky W., White N.E., 1987, ApJ, 312, L17}
\ref{Stetson P.B., 1987, PASP, 99, 191}
\ref{Stetson P.B., 1991, in  Precision Photometry: Astrophysics of
	the Galaxy, eds. Philip A.G.D., Upgren A.R., \& Janes K., 
	(Schenectady: L. Davis Press), p. 69}
\ref{Stetson P.B., 1994, PASP, 106, 250}
\ref{Stetson P.B., VandenBerg D.A., Bolte M., 1996, PASP, 108, 560}
\ref{Stetson P.B., West M.J., 1994, PASP, 106, 726}
\ref{Stiavelli M., 1990, in  Galactic Models, N.Y. Acad. Sci. 
        Annals, 596, 145}
\ref{Stiavelli M., Piotto, G., Capaccioli, M., Ortolani, S., 1991, 
	in  Formation and Evolution of Star Clusters, 
	ASP Conference Series Vol. 13, 
	ed. Janes K., (San Francisco: ASP), p. 449}
\ref{Stiavelli M., Piotto G., Capaccioli M., 1992, 
	in  Morphological and Physical Classification of Galaxies, 
	eds. Longo G., Capaccioli M., \& Busarello G., 
	(Dordrecht: Kluwer), p. 455}
\ref{Stod\'o\l kiewicz J.S., 1982, Acta Astron., 32, 63}
\ref{Stod\'o\l kiewicz J.S., 1984, in Ceskoslovenska Akad. Ved 
        Fifth Conf. on  Star Clusters and Associations and Their
        Relation to the Evolution of the Galaxy, p. 103}
\ref{Stod\'o\l kiewicz J.S., 1985, in 
	 Dynamics of Star Clusters, IAU Symp. 113, 
	eds. Goodman J. \& Hut P., (Dordrecht: Reidel), p. 361}
\ref{Stod\'o\l kiewicz J.S., 1986, Acta Astron., 36, 19}
\ref{Struble M.F., 1979, Ap\&SS, 64, 319}
\ref{Stryker L.L., 1993, PASP, 105, 1081}
\ref{Sugimoto D., 1985, in
	 Dynamics of Star Clusters, IAU Symp. 113, 
	eds. Goodman J. \& Hut P., (Dordrecht: Reidel), p. 207}
\ref{Sugimoto D., Bettwieser E., 1983, MNRAS, 204, 19P}
\ref{Sugimoto D., Chikada Y., Makino J., Ito T., Ebisuzaki T.,
        Umemura M., 1990, Nature, 345, 33}
\ref{Suntzeff N.B., 1993, in
	 The Globular Cluster - Galaxy Connection, 
	ASP Conference Series Vol. 48, 
	eds. Smith G.H. \& Brodie J.P., (San Francisco: ASP), p. 167}
\ref{Surdin V.G., 1978, Sov. Astron., 22, 401}
\ref{Surdin V.G., 1979, Sov. Astron., 23, 648}
\ref{Surdin V.G., 1983, in  Star Clusters and Associations and 
	Their Relation to the Evolution of the Galaxy, 
	eds. Ruprecht J. \& Palous J., (Praha: Publ. Astron. Inst.
        Czech. Acad. Sci.), No. 56, p. 117}
\ref{Surdin V.G., 1993,  in
	 The Globular Cluster - Galaxy Connection, 
	ASP Conference Series Vol. 48, 
	eds. Smith G.H. \& Brodie J.P., (San Francisco: ASP), p. 342}
\ref{Surdin V.G., 1995, Astron. Astrophys. Transact., 7, 147}
\ref{Sweatman W., 1990, in  Dynamics and Interactions of Galaxies, 
	ed. Wielen R., (Berlin: Springer-Verlag), p. 219}
\ref{Sweatman W.L., 1991, J. Comp. Phys., 111, 110}
\ref{Sweatman W.L., 1993, MNRAS, 261, 497}
\ref{Sygnet J.F., Des Forets G., Lachieze-Rey M., Pellat R., 
	1984, ApJ, 276, 737}
\ref{Szebehely V., 1972, Celes. Mech., 6, 84}
\ref{Szebehely V., 1973, A\&A, 22, 171}
\ref{Taff L.G., Van Horn H.M., 1975, ApJ, 197, L23}
\ref{Taillet R., Longaretti P.-Y., Salati P., 1995, Astroparticle Phys., 4, 87}
\ref{Taillet R., Salati P., Longaretti P.-Y., 1996, ApJ, 461, 104}
\ref{Tajima N., 1981, Prog. Th. Phys., 65, 1264}
\ref{Takahashi K., 1992, PASJ, 45, 233}
\ref{Takahashi K., 1993, PASJ, 45, 789}
\ref{Takahashi K., 1995, PASJ, 47, 561}
\ref{Takahashi K., 1996, PASJ, submitted}
\ref{Takahashi K., Inagaki S., 1991, PASJ, 43, 589}
\ref{Takahashi K., Inagaki S., 1992, PASJ, 44, 623}
\ref{Tanekusa J., 1987, PASJ, 39, 425}
\ref{Tavarez M., Friel E.D., 1995, AJ, 110, 223}
\ref{Taylor, J.H. Jr., 1994, Rev. Mod. Phys., 66, 711}
\ref{Terlevich E., 1985, in  Dynamics of Star Clusters, IAU Symp. 113, 
	eds. Goodman J. \& Hut P., (Dordrecht: Reidel), p. 471}
\ref{Terlevich E., 1987, MNRAS,  224, 193}
\ref{Theuns T., 1996, MNRAS, 279, 827}
\ref{Theuns T., Rathsack M.E., 1993, Computer Physics Comm., 76, 141}
\ref{Thomas P., 1989, MNRAS, 238, 1319}
\ref{Trager S.C., Djorgovski S.G., King I.R., 1993, 
	in  Structure and Dynamics of Globular Clusters, 
	ASP Conference Series Vol. 50, 
	eds. Djorgovski S.G. \& Meylan G., (San Francisco: ASP), p. 347}
\ref{Trager S.C., King I.R., Djorgovski S.G., 1995, AJ, 109, 218}
\ref{Tremaine S.D., 1976, ApJ, 203, 72}
\ref{Tremaine S.D., H\'enon M., Lynden-Bell D., 1986, MNRAS, 219, 285}
\ref{Tremaine S.D., Ostriker J.P., Spitzer L., Jr., 1975, ApJ, 196, 407}
\ref{Tremaine S., Richstone D.O., Byun Y.-I., \etal, 
	1994, AJ, 107, 634}
\ref{Tremaine S., Weinberg M.D., 1984, MNRAS, 209, 729}
\ref{Trinchieri G., Fabbiano G., 1991, ApJ, 382, 82}
\ref{Trumpler R.J., 1930, Lick Obs. Bull., 420}
\ref{Tutukov A.V., 1978, A\&A, 70, 57}
\ref{Valtonen M.J., 1975, Mem. RAS, 80, 77}
\ref{Valtonen M.J., 1976, As\&SS,  42, 331}
\ref{Valtonen M.J., 1988a, Vistas in Astronomy, 32, 23}
\ref{Valtonen M.J., ed., 1988b, The Few Body Problem, 
	IAU Colloquium 96, (Dordrecht: Kluwer)}
\ref{Valtonen M., Mikkola S., 1991, ARA\&A, 29, 9}
\ref{van Albada T.S., 1967, Bull. Astron., 2, 67}
\ref{VandenBerg D.A., Bolte M., Stetson P.B., 1990, AJ, 100, 445}
\ref{van den Bergh S., 1979, AJ, 84, 317}
\ref{van den Bergh S., 1981, A\&AS, 46, 79}
\ref{van den Bergh S., 1982, PASP, 94, 459}
\ref{van den Bergh S., 1984, PASP, 96, 329}
\ref{van den Bergh S., 1993a, MNRAS, 262, 588}
\ref{van den Bergh S., 1993b, AJ, 105, 971}
\ref{van den Bergh S., 1993c, ApJ, 411, 178}
\ref{van den Bergh S., 1993d, 
	in  Structure and Dynamics of Globular Clusters, 
	ASP Conference Series Vol. 50, 
	eds. Djorgovski S.G. \& Meylan G., (San Francisco: ASP), p. 1}
\ref{van den Bergh S., 1994, ApJ, 435, 203}
\ref{van den Bergh S., 1995a, Nature, 374, 215}
\ref{van den Bergh S., 1995b, ApJ, 450, 27}
\ref{van den Bergh S., 1995c, Science, 270, 1942}
\ref{van den Bergh S., 1996, AJ, in press}
\ref{van den Heuvel E.P.J., 1994, in  Interacting Binaries, 
	Saas-Fee Advanced Course 22, eds. Nussbaumer H. \& Orr A.,   
	(Berlin: Springer-Verlag), p. 263}
\ref{Vandervoort P.O., Welty D.E., 1981, ApJ, 248, 504}
\ref{van der Woerd H., van den Heuvel E.P.J., 1984, A\&A, 132, 361}
\ref{Veltmann U.-I.K., 1983, SvA, 27, 127}
\ref{Verbunt F., 1987, ApJ, 312, L23}
\ref{Verbunt F., 1993, ARA\&A, 31, 93}
\ref{Verbunt F., 1996a, in  R\"ontgenstrahlung from the Universe,
	MPE Report 263, eds. Zimmermann H.U., Tr\"umper J., \& Yorke H.,
	(Garching: MPE), p. 93}
\ref{Verbunt F., 1996b,  
	in  Dynamical Evolution of Star Clusters:
	Confrontation of Theory and Observations, IAU Symp. 174, 
	eds. Hut P. \& Makino J. (Dordrecht: Kluwer), p. 183}
\ref{Verbunt F., Bunk W., Hasinger G.R., Johnston H.M., 1995, A\&A, 300, 732}
\ref{Verbunt F., Elson R.A.W., van Paradijs J., 1984, MNRAS, 210, 899}
\ref{Verbunt F., Hasinger G.R., Johnston H.M., Bunk W., 
	1993, Adv. Space. Res., 13(12), 151}
\ref{Verbunt F., Hut P., 1987, in 
	 The Origin and Evolution of Neutron Stars, IAU Symp. 125,
	eds. Helfand D.J. \& Huang J.-H., (Dordrecht: Reidel), p. 187}
\ref{Verbunt F., Johnston H.M., 1996, 
	in  The Origins, Evolution, and Destinies of 
	Binary Stars in Clusters, eds. Milone G. \& Mermilliod J.-C., 
	ASP Conference Series Vol. 90, (San Francisco: ASP), p. 300}
\ref{Verbunt F., Johnston H.M., Hasinger G.R., Belloni T., Bunk W.,
	1994, in  Interacting Binary Stars, ed. Shafter A.W., 
	ASP Conference Series Vol. 56, (San Francisco: ASP), p. 244}
\ref{Verbunt F., Meylan G., 1988, A\&A, 203, 297}
\ref{Vesperini E., 1992, Europhys.  Lett.,  17, 661}
\ref{Vesperini E., Chernoff D.F., 1994, ApJ, 431, 231}
\ref{Vesperini E., Chernoff D.F., 1996, ApJ, 458, 178}
\ref{Vietri M., Pesce E., 1995, ApJ, 442, 618}
\ref{von Hoerner S., 1958, Z. f. A., 44, 221}
\ref{von Hoerner S., 1960, Z. f. A., 50, 184}
\ref{von Hoerner S., 1976, A\&A, 46, 293}
\ref{von Weizs\"acker C.F., 1955, Z. f. A., 35, 252}
\ref{von Zeipel M.H., 1908, Ann. Obs. Paris, 25, F1}
\ref{V'yuga A.A., Kaliberda V.S., Petrovskaya I.V., 1976, Trud. 
        Astron. Obs. Leningrad, 32, 105}
\ref{Wachlin F.C., Rybicki G.B., Muzzio J.C., 1992, MNRAS, 262, 1007}
\ref{Walker A.R., 1992, ApJ, 390, L81}
\ref{Wang Q.-D., 1991, Celes. Mech. Dyn. Astr., 50, 73}
\ref{Warren M.S., Salmon J.K., 1995, Comp. Phys. Comm., 87, 266}
\ref{Webbink R.F., 1981, ApJS, 45, 259}
\ref{Weinberg M.D., 1991, ApJ, 368, 66}
\ref{Weinberg M.D., 1993a, ApJ, 410, 543}
\ref{Weinberg M.D., 1993b, ApJ, 421, 481}
\ref{Weinberg M.D., 1993c,   in
	 The Globular Cluster - Galaxy Connection, 
	ASP Conference Series Vol. 48, 
	eds. Smith G.H. \& Brodie J.P., (San Francisco: ASP), p. 689}
\ref{Weinberg M.D., 1994a, AJ, 108, 1398}
\ref{Weinberg M.D., 1994b, AJ, 108, 1403}
\ref{Weinberg M.D., 1994c, AJ, 108, 1414}
\ref{Weinberg M.D., Chernoff D.F., 1989, 
	in  Dynamics of Dense Stellar Systems,
	ed. Merritt D., (Cambridge: Cambridge University Press), p. 221}
\ref{Weinberger R., 1995, PASP, 107, 58}
\ref{West M.J., 1993, MNRAS, 265, 755}
\ref{West M.J., C\^ot\'e P., Jones C., Forman W., Marzke R.O.,  1995, ApJ,
	453, L77}
\ref{Wheeler J.C., 1979, ApJ, 234,569}
\ref{White R.E., Shawl S.J., 1987, ApJ, 317, 246}
\ref{White S.D.M., 1976, MNRAS, 174, 19}
\ref{Whitmore B.C., Schweizer F., 1995, AJ, 109, 960}
\ref{Whitmore B.C., Schweizer F., Leitherer C., Borne K., Robert C. 
	1993, AJ, 106, 1354}
\ref{Wielen R., 1967, Ver\"off. des Astron. Rechen-Inst. Heidelberg, 19, 5}
\ref{Wielen, R. 1968, Bull. Astron., 3, III, 127}
\ref{Wielen R., 1974a, in Proc. 1st Europ. Astron. Meeting, 
        Athens, Sep 4-9, 1972, vol.2., (Berlin: Springer-Verlag), p. 326}
\ref{Wielen R., 1974b, 
	in  Numerical Solution of Ordinary Differential Equations,
	Lecture Notes in Math. Vol. 362., ed. Bettis D.G., 
	(Berlin: Springer-Verlag), p. 276}
\ref{Wielen R., 1975, in  Dynamics of Stellar Systems, 
        IAU Symp. 69, ed. Hayli A., (Dordrecht: Reidel), p. 119}
\ref{Wielen R., 1987, in  The Harlow-Shapley
	Symposium on Globular Cluster Systems in Galaxies, IAU Symp. 126,
	eds. Grindlay J.E. \& Philip A.G.D., 
	(Dordrecht: Kluwer), p. 393}
\ref{Wilson C.P., 1975,  AJ, 80, 175}
\ref{Wiyanto P., Kato S., Inagaki S., 1985, PASJ, 37, 715}
\ref{Wolszczan A., Kulkarni S.R., Middleditch J., Backer D.C.,
	Fruchter A.S., Dewey R.J., 1989, Nature, 337, 531}
\ref{Woltjer L., 1975, A\&A, 42, 109}
\ref{Woolley R.v.d.R., 1954, MNRAS, 114, 191}
\ref{Woolley R.v.d.R., Dickens R.J., 1962, ROE Bull., 54}
\ref{Woolley R.v.d.R., Robertson, D.A. 1956, MNRAS, 116, 288}
\ref{Wybo M., Dejonghe H., 1995, A\&A, 295, 347}
\ref{Wybo M., Dejonghe H., 1996, A\&A, 312, 649}
\ref{Yan L., 1996,  PhD Thesis, California Institute of Technology}
\ref{Yan L., Mateo M., 1994, AJ, 108, 1810}
\ref{Yan L., Reid I.N., 1996, MNRAS, 279, 751}
\ref{Yan L., Cohen J.G., 1996, AJ, in press, astro-ph/9607116}
\ref{Yanny B., 1993, PASP, 105, 969}
\ref{Yanny B., Guhathakurta P., Schneider D.P., Bahcall J.N., 1993, 
	in  Structure and Dynamics of Globular Clusters, 
	ASP Conference Series Vol. 50, 
	eds. Djorgovski S.G. \& Meylan G., (San Francisco: ASP), p. 275}
\ref{Yanny B., Guhathakurta P., Bahcall J.N., Schneider D.P., 
	1994a, AJ, 107, 1745}  
\ref{Yanny B., Guhathakurta P., Schneider D.P., Bahcall J.N.,
	1994b, ApJ, 435, L59}  
\ref{Yoshizawa M., Inagaki S., Nishida M.T., \etal, 1978, PASJ, 30, 279}
\ref{Zaggia S., Capaccioli M., Piotto G., 1992a, in  Star
	Clusters and Stellar Evolution, eds. Brocato E., Ferraro F.,
	\& Piotto G., Mem. Soc. Astron. Ital., 63, 211}
\ref{Zaggia S., Capaccioli M., Piotto G. 1993, A\&A, 278 415}
\ref{Zaggia S., Capaccioli M., Piotto G., Stiavelli, M., 1992b, 
	A\&A, 258, 302}
\ref{Zaggia S., Piotto G., Capaccioli M., 1991, 
	in  Formation and Evolution of Star Clusters, 
	ASP Conference Series Vol. 13, 
	ed. Janes K., (San Francisco: ASP), p. 458}
\ref{Zepf S.E., Ashman K.M., 1993, MNRAS, 264, 611}
\ref{Zepf S.E., Carter D., Sharples R.M., Ashman K.M., 1995, ApJ, 445, L19}
\ref{Zinn R., 1980, ApJ, 241, 602}
\ref{Zinn R., 1985, ApJ, 293, 424}
\ref{Zinn R., 1990, J. R. Astr. Soc. Can., 84, 89}
\ref{Zinn R., 1996, 
	in  Formation of the Galactic Halo...Inside and Out, 
	ASP Conference Series Vol. 92, 
	eds. Morrison H. \& Sarajedini A., (San Francisco: ASP), p. 211}
\ref{Zinn R., Searle L., 1976, ApJ, 209, 734}
\ref{Zinn R., West M.J., 1984, ApJS, 55, 45}
\ref{Zinnecker H., 1996, in  The Interplay between Massive-Star
	Formation, the ISM, and Galaxy Evolution, 11$^{th}$ IAP Conference,
	eds. Kunth D. \etal, (Paris: Editions Fronti\`eres), in press}
\ref{Zinnecker H., Keable C.J., Dunlop J.S., Cannon R.D.,
	Griffiths W.K., 1988, in  The Harlow-Shapley
	Symposium on Globular Cluster Systems in Galaxies, IAU Symp. 126,
	eds. Grindlay J.E. \& Philip A.G.D., 
	(Dordrecht: Kluwer), p. 603}
\ref{Zwart S.F.P., 1996, in  The Origins, Evolution, and Destinies of 
	Binary Stars in Clusters, eds. Milone G. \& Mermilliod J.-C., 
	ASP Conference Series Vol. 90, (San Francisco: ASP), p. 378}
\ref{Zwart S.F.P., Meinen A.T., 1993, A\&A, 280, 174}
}
}

\vfill\eject
\end
\bye